\documentclass[useAMS,usenatbib]{mn2e}
\usepackage{times}
\usepackage{natbib}
\usepackage{graphics}
\usepackage{graphicx}
\usepackage{color}
\usepackage{verbatim}
\usepackage{amsmath}
\usepackage{amssymb}
\usepackage{mathtools}
\usepackage{subfigure}
\usepackage{float}
\usepackage{tabularx}
\usepackage{supertabular}
\usepackage{rotating}
\usepackage{longtable}
\usepackage{dpfloat, booktabs}
\usepackage{color}
\usepackage{multicol}
\usepackage{graphicx}
\usepackage{multirow}               
\usepackage{amsfonts}

\usepackage{pifont}

\usepackage{wrapfig}
\usepackage{subfig}   

\definecolor{Mygrey}{gray}{0.75}

\title[Molecular Gas in NGC~0628] {Molecular Line Ratio Diagnostics Along the Radial Cut and Dusty UV-bright Clumps in a Spiral Galaxy NGC~0628}
\author[S.\ Topal]  {Sel\c{c}uk Topal$^{1,2}$\thanks{E-mail:
    selcuktopal@yyu.edu.tr}\\
    $^{1}$Van 100. Yil University, Department of Physics, Van, 65080,
  Turkey\\
  $^{2}$Sub-department of Astrophysics, University of Oxford, Denys
  Wilkinson Building, Keble Road, Oxford OX1~3RH, U.K.\\}

\begin{document}
\date{Accepted . Received ; in original form }
\pagerange{\pageref{firstpage}-\pageref{lastpage}} \pubyear{2019}
\maketitle
\label{firstpage}
%
%
\begin{abstract}
Molecular emission lines are essential tools to shed lights on many questions regarding star formation in galaxies. Multiple molecular lines are particularly useful to probe different phases of star-forming molecular clouds. In this study, we investigate the physical properties of giant molecular clouds (GMCs) using multiple lines of CO, i.e. CO(1--0, 2--1, 3--2) and $^{13}$CO(1--0), obtained at selected $20$ positions in the disc of NGC~0628. Eleven positions were selected over the radial cut, including the centre, and remaining nine positions were selected across the southern and northern arms of the galaxy. $13$ out of $20$ positions are brighter at $24\micron$ and ultraviolet (UV) emission and hosting significantly more H\,{\small II} regions compared to the rest of the positions indicating opposite characteristics. Our line ratio analysis shows that the gas gets warmer and thinner as a function of radius from the galaxy centre up to $1.7$~kpc, and then the ratios start to fluctuate. Our empirical and model results suggest that the UV-bright positions have colder and thinner CO gas with higher hydrogen and CO column densities. However, the UV-dim positions have relatively warmer CO gas with lower densities bathed in GMCs surrounded by less number of H\,{\small II} regions. Analysis of multi-wavelength infrared and UV data indicates that the UV-bright positions have higher star formation efficiency than that of the UV-dim positions.

\end{abstract}
%
%
\section{Introduction}
\label{sec:intro}
The interstellar medium (ISM) of galaxies are the sites for stars to be born and die, providing fingerprints of past and 
current star formation activities. One of the ways to track these fingerprints down is to study 
molecular emission lines as they are results of different physical conditions in molecular clouds, 
e.g. different temperature, density, and opacity. Molecules are necessary coolants for the gas clouds, 
helping the clouds to collapse and form stars. Molecules, therefore, have a vital role in better 
understanding the physics of the gas leading stars to be born and shaping the ISM of a galaxy. 

The Universe consists of many types of galaxies from giant ellipticals to spirals; 
each has a different amount of molecular gas and shows different levels of star formation activity. 
Spiral galaxies, like the subject of this study NGC~0628, have a substantial amount of molecular gas allowing us to 
probe the nature of star-forming gas in greater details. Carbon monoxide (CO) is 
the second most abundant molecule in the ISM after hydrogen molecule (H$_{2}$). 
At high densities, where stars are born, hydrogen tends to be molecular, but it is tough 
to observe directly due to its quantum structure. We, therefore, use proxies instead,  
such as CO, a well-known proxy to probe the physical properties of molecular gas indirectly. 
Since molecular gas is strongly correlated with star formation (\citealt{dob04} and references therein), 
studying the properties of the gas has an unignorable potential to answer many outstanding questions regarding 
star formation processes in galaxies and evolution of galaxies at large.

Effects of the ultraviolet (UV) radiation to giant molecular clouds (GMCs) are either 
destructive or advantageous in terms of star formation \citep[e.g.][]{elm77,deh10,bis11,med14,kim18}, 
but the UV radiation feedback overall has negative effects \citep[e.g.][]{mc07}. 
UV radiation field also plays an essential role in regulating the CO abundances \citep{van88}.
NGC~0628 has been reported to contain as many as $376$ H\,{\small II} complexes, 
one of the largest numbers in the nearby Universe \citep{fat07}. 
H\,{\small II} regions are regions of photoionized gas created by 
 ionising photons coming from O- and B-type stars, the source of UV radiation in the ISM. 
Photoionization is thought to be the primary cause of destruction for 
GMCs \citep[e.g.][]{whit79,matz02,krum06} and 
feedbacks from supernovae, and protostellar outflows are thought to be negligible 
compared to H\,{\small II} regions \citep[e.g.][]{krum06,mc07}.
While less massive GMCs ($~10^{4}-10^{5}~M_{\sun}$) with lower densities are weaker against the feedback from H\,{\small II} regions, denser and more massive clouds ($>10^{6}~M_{\sun}$) with larger escape velocities mostly remain dynamically unaffected by ionizing feedback \citep[e.g.][]{dale12,dale13}. H\,{\small II} regions, as an essential energy injection source for GMCs \citep[e.g.][]{matz02}, therefore, have a vital role in the fate of molecular gas and ultimately in star formation. 
 
NGC~0628 (M74) is a face-on nearby
spiral galaxy ($7.3$~Mpc; \citealt{kar04}) and is well studied in both
H${\alpha}$ and far-ultraviolet (FUV) emission \citep{adl99,her10},
both indicating intense star formation activity in the arms. 
In total, we observed $20$ selected positions throughout the disc of NGC~0628
 in $^{12}$CO(1--0) and $^{13}$CO(1--0) lines. 
 The positions consist of $9$ positions in the arms brighter in the UV and 
$11$ positions over the SE-NW cut, including the centre (see Figure~\ref{fig:n0628}). 
The UV-bright positions are also brighter at $24\micron$ and 
filled with many H\,{\small II} regions compared to the positions located 
over the SE-NW cut (see Figure~\ref{fig:n0628}). 
Using our new observations and additional data of $^{12}$CO(2--1) and $^{12}$CO(3--2) 
from the literature, the goals of our study are: i) to increase our understanding
of the star formation processes in external spiral galaxies, in
particular in spiral arms; ii) to determine for the first time the
properties of the molecular gas in the arms of this particular galaxy,
this by using multiple lines of CO and radiative transfer modelling; 
iii) to probe differences in star formation activity between the spiral arms and the centre of the
galaxy (and other galaxies); iv) to better understand the effects of young massive stars 
(hinted by intense UV radiation and H\,{\small II} regions) and dust
to surrounding molecular gas and finally v) to study radial variations in physical conditions of the gas 
over the disc.

The paper is organised as follows. Section~\ref{sec:obsdata} describes
the observations and data reduction, while \S~\ref{sec:iman} presents imaging 
and analysis. The results and discussions are presented in \S~\ref{sec:redis}, 
and we finally conclude briefly in \S~\ref{sec:conc}.
%
%
\begin{table}
  \begin{center}
    \caption{General properties of NGC~0628.}	
    \begin{tabular}{@{}l l r r r r r@{}} \hline
      Galaxy & Property & Value & Reference \\ \hline
      NGC~0628 & Type & SA(s)c & a\\
      & RA~(J2000) & $1^{\rm h}36^{\rm m}41.7^{\rm s}$ & a\\
      & Dec~(J2000) & $15^{\rm d}47^{\rm m}01^{\rm s}$ & a\\
      & V$_{\sun}$~(km~s$^{-1}$) & $657$ & a\\
      & Major diameter & $10\farcm5$ & a\\
      & Minor diameter & $9\farcm5$ & a\\
      & SFR$_{\rm H_\alpha}$~($M_\odot$~yr$^{-1}$) & $4.0$ & b\\
      & $\log(M_{{\rm H}_2}/M_\odot)$ & $9.49$ & b\\
      & Distance (Mpc) & $7.3$ & c\\
      & Position angle & $25\degr$ & d\\
      & Inclination & $19.8\degr$ & d\\		
      \hline
    \end{tabular}
    \label{tab:gprop}
  \end{center}
  References: 
  $^{\rm a}$ Nasa/Ipac Extragalactic Database (NED);
  $^{\rm b}$ \citealt{ken03};
  $^{\rm c}$ \citealt{kar04};
  $^{\rm d}$ \citealt{mak04}.
\end{table}

%
%
%
\section{Observations and Data Reduction}
\label{sec:obsdata}

\subsection{Observation}
\label{sec:obs}
We have acquired new single-dish observations of CO lines 
in the disc of NGC~0628 (see Table~\ref{tab:obs}).
We defined $9$ positions along the northern
and southern arms of the galaxy where the FUV emission shows a peak 
in Galaxy Evolution Explorer Survey (GALEX) map \citep{gil07}. 
Additionally, together with a central pointing, $11$ positions over the SE-NW 
cut were observed (see Figure~\ref{fig:n0628}).

The CO(1--0) and $^{13}$CO(1--0) observations were carried out using
the IRAM~30m telescope. The telescope is equipped with the Eight MIxer Receiver (EMIR) 
and a backend FTS200 providing $200$~kHz frequency resolution 
corresponding to a velocity resolution of about $0.5$~kms$^{-1}$ for the observing
frequency of $110.201$~GHz. We estimated the sensitivity level for the observation 
for average conditions ($7.0$~mm of pwv, T$_{\rm sys}$ = $168.8$~K (TA$^{\star}$) mean per pixel). 
Before applying further analysis, we binned the spectra to a velocity resolution of $5$~kms$^{-1}$. 
CO(1--0) was detected (i.e $S/N>3\sigma$, where $\sigma$ = $13$~mK) 
in all $20$ positions, while $^{13}$CO(1--0) was detected in $15$ positions 
($\sigma$ = $5.2$~mK). The beam size of CO(1--0) and $^{13}$CO(1--0) at IRAM~30m telescope is about $22^{\prime\prime}$ 
corresponding to a linear size of $778$~pc at the distance of $7.3$~Mpc for NGC~0628 \citep{kar04}, 
allowing us to study the physical conditions of the gas at sub-kpc resolution.
The spatial coverage of the observations is shown in Figure~\ref{fig:n0628}, overlaid on 
an optical image of the galaxy (Sloan Digitized Sky Survey, SDSS). 
\subsection{Literature CO Data}
\label{sec:lite}
The CO(2--1) data were taken from the
HERA CO Line Extragalactic Survey (HERACLES; \citealt{ler09}), 
while the data for CO(3--2) were taken from the James Clerk Maxwell Telescope (JCMT) Nearby
Galaxies Legacy Survey (NGLS; \citealt{war10}). Noise in the CO(2--1) and CO(3--2) data cubes 
are $22$~mK and $15.8$~mK, respectively (see Table~\ref{tab:obs}). 
For more details on the observational parameters, please see the related papers.

All in all, together with the literature data, there are four lines
detected at $13$ out of the $20$ positions and three lines at 
$4$ positions (i.e.\ either $^{13}$CO(1--0) or CO(3--2) were not detected 
at those positions) and finally two lines detected at the remaining $3$ positions 
(i.e.\ only CO(1--0) and CO(2--1) emission were detected).
\begin{table*}
	\begin{center}
	\caption{Main observational parameters for NGC~0628}
		\begin{tabular}{lcccccl} \hline
  			Transition &Rest Freq.& Obs. Date & Telescope & Beam & Noise$^{\star}$& Reference \\ 
			&(GHz)&&&(arcsec)&(mK)&\\ \hline \\
  			$^{12}$CO(1-0)& $115.271$ & \multirow{2}{*}{JULY~$2012$} &\multirow{2}{*}{IRAM 30m}&\multirow{2}{*}{22}&$13$&\multirow{2}{*}{This study} \\
			$^{13}$CO(1-0)& $110.201$ &&&&$5.2$&\\ \\
			$^{12}$CO(2-1) & $230.538$ &JAN~$2007$ &IRAM 30m&$13.4$&$22$&\citealt{ler09}\\
			$^{12}$CO(3-2) & $345.795$ &NOV~$2007$ - MAR $2008$&JCMT 15m&$14.5$&$15.8$&\citealt{war10}\\ \\		
 			 \hline
		\end{tabular}
		\label{tab:obs}
		\end{center}
		\parbox[t]{0.75\textwidth}{$^{\star}$For our single-dish data, noise was estimated using channels free of emission, while for the literature data cubes, 
		namely for CO(2--1) and CO(3--2), noise was taken from the related papers (see Section~\ref{sec:lite}).}
		\end{table*}

\subsection{Data Reduction}
\label{sec:redu}
Our IRAM 30m data were reduced using the Continuum
and Line Analysis Single-Dish Software (CLASS) software package in the
Grenoble Image and Line Analysis System (GILDAS). A baseline fit
(polynomial of order $0$ or $1$) was removed from
each scan before averaging all integrations. The antenna temperature 
scale ($T_{\rm A}^*$) was then transformed to the
main beam brightness temperature scale ($T_{\rm mb}$) by dividing by
the main beam efficiency, $\eta_{\rm mb}$, where $\eta_{\rm mb} = 0.83$ and $0.84$ 
for CO(1--0) and $^{13}$CO(1--0) respectively. The beam efficiencies were taken from 
the IRAM\footnote{http://www.iram.es/IRAMES/mainWiki/Iram30mEfficiencies}. 
For $\eta_{\rm mb}$ values of $^{13}$CO(1--0) not specified
there, we linearly interpolated between the two nearest values.
Each spectrum was then converted to the Flexible Image Transport System (FITS) format for 
further analysis in the Interactive Data Language (IDL) environment.

\section{Imaging and analysis}
\label{sec:iman}

\subsection{Emission regions and moment maps}
\begin{figure*}
  \includegraphics[width=8.5cm,clip=]{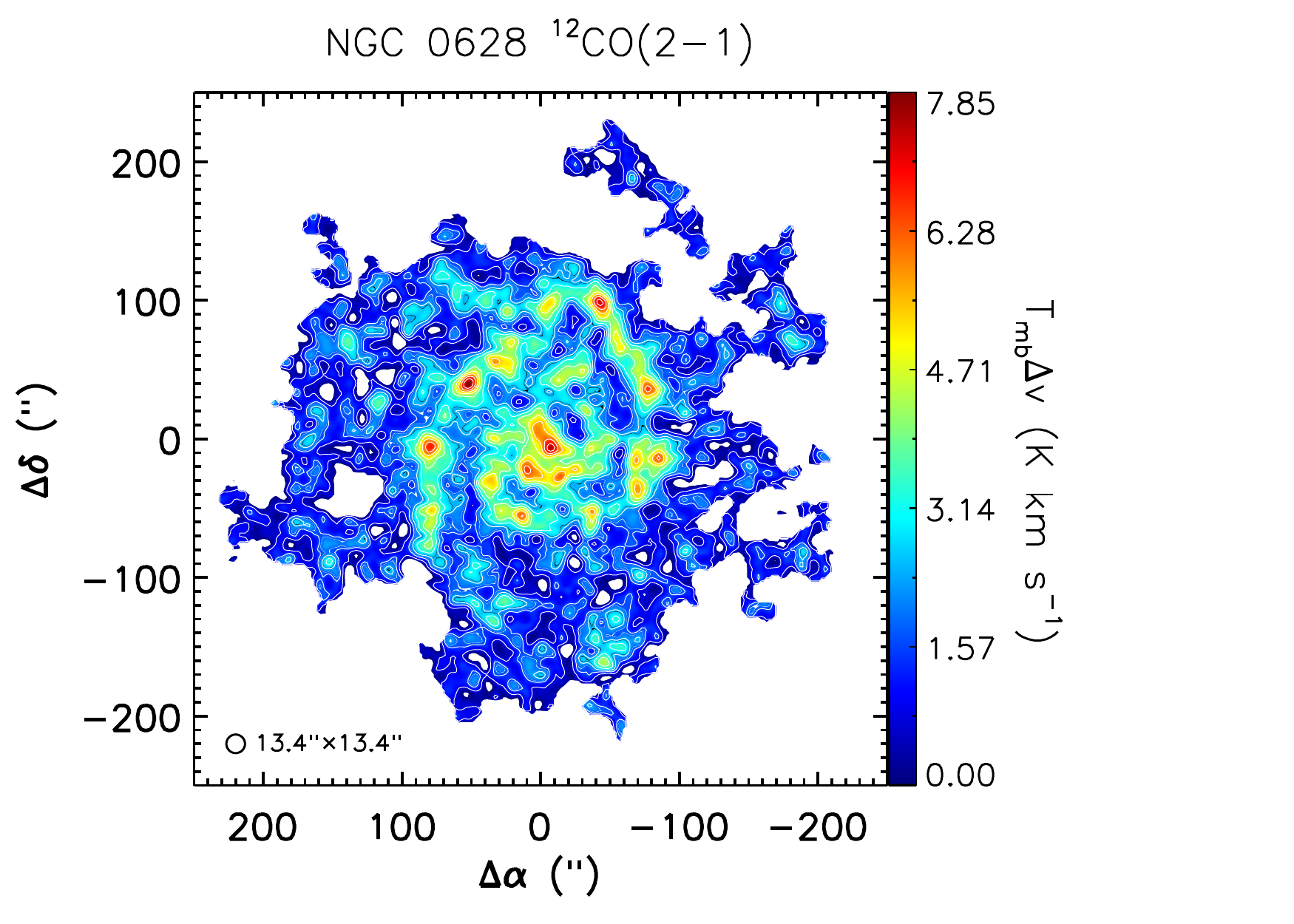}
   \includegraphics[width=8.5cm,clip=]{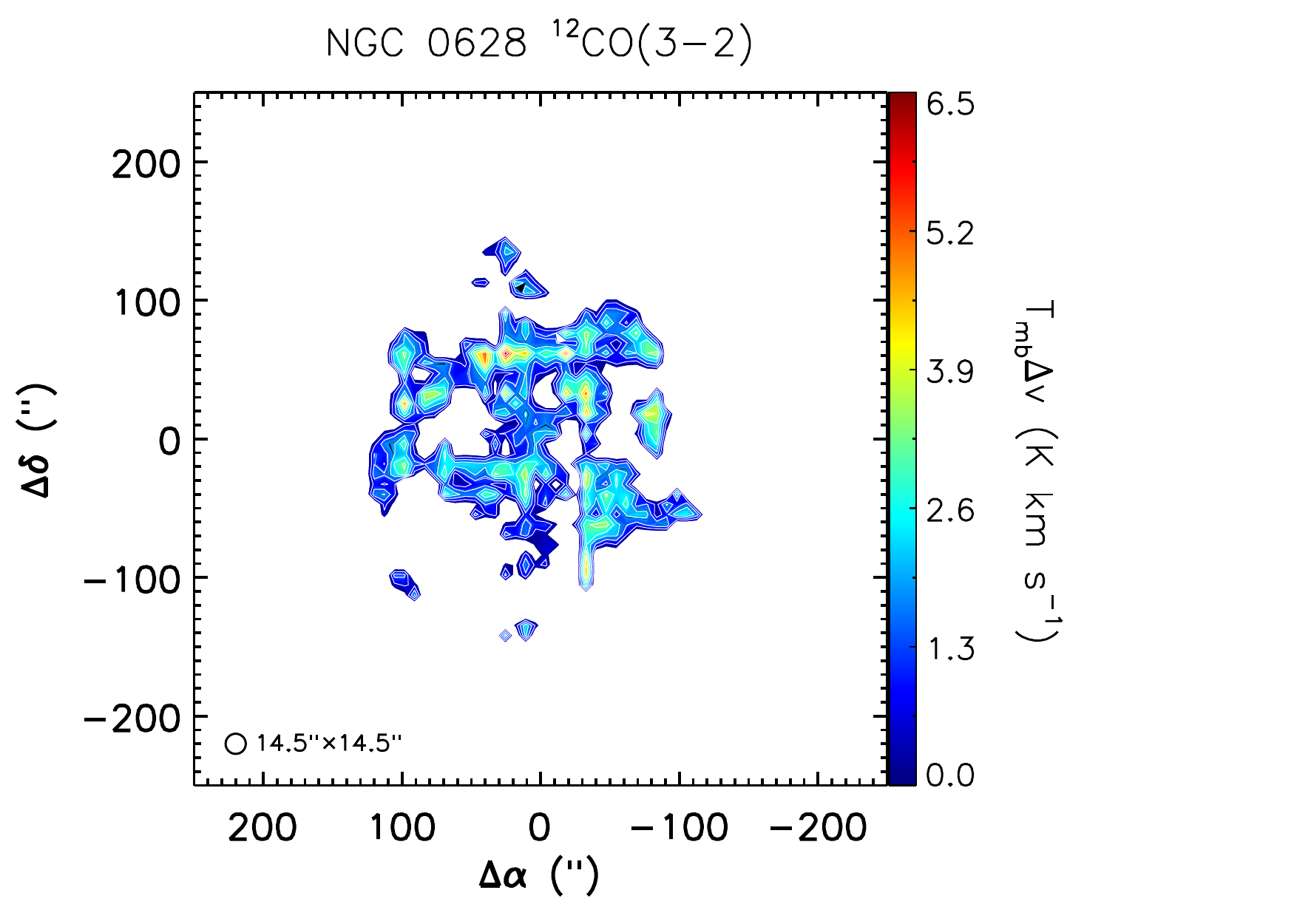}\\
  \includegraphics[width=8.5cm,clip=]{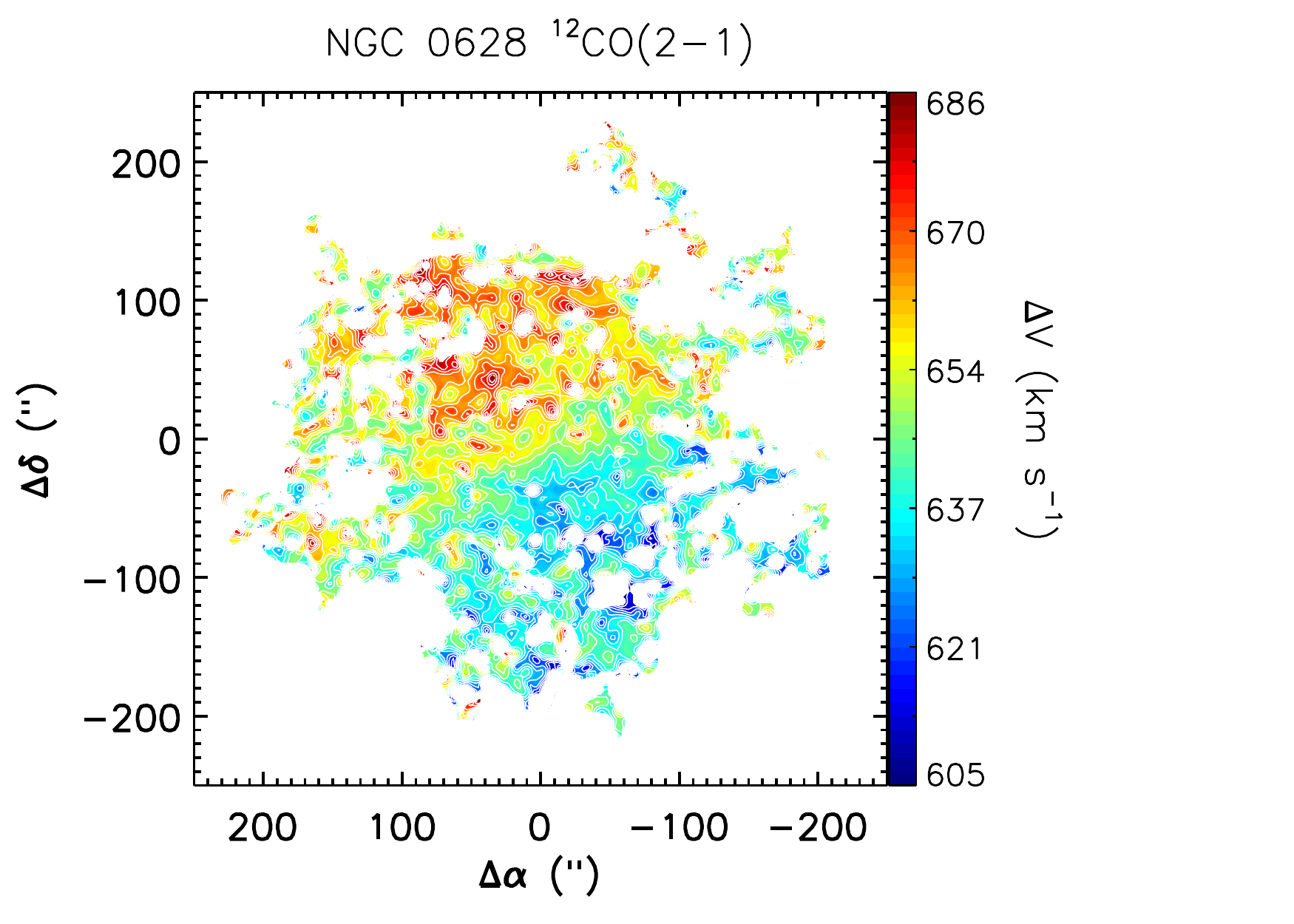}
   \includegraphics[width=8.5cm,clip=]{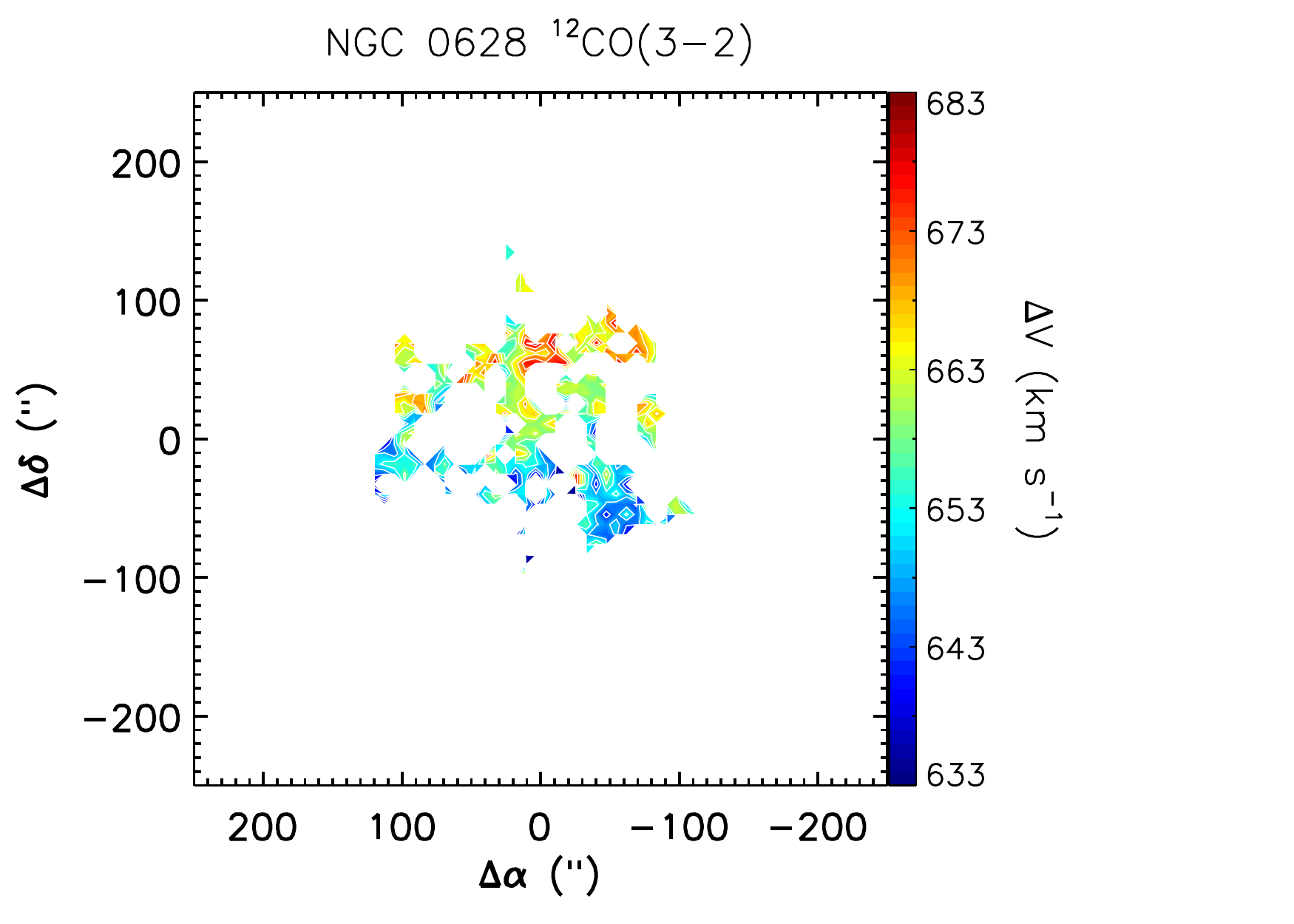}\\
    \includegraphics[width=8.5cm,clip=]{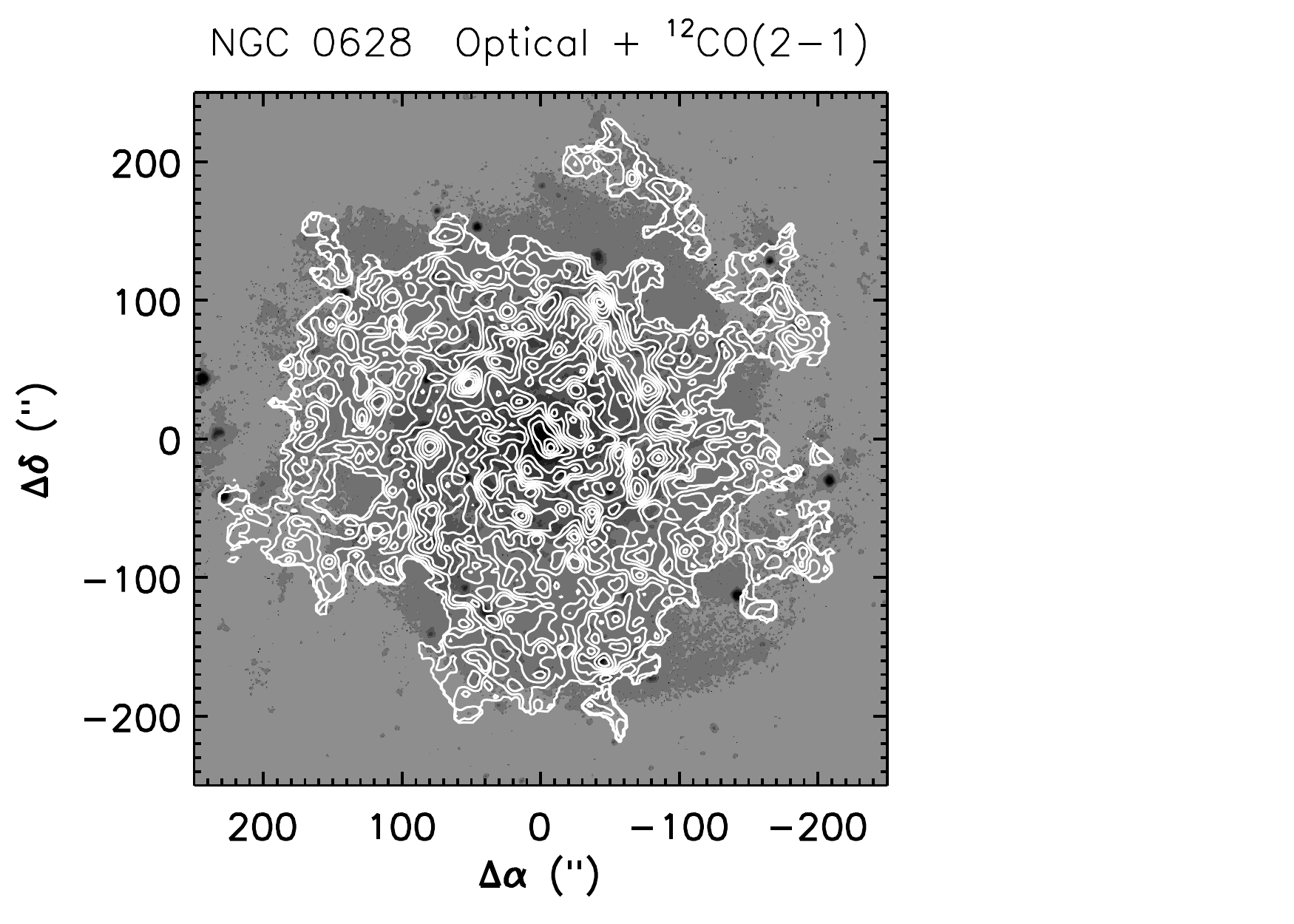}
   \includegraphics[width=8.5cm,clip=]{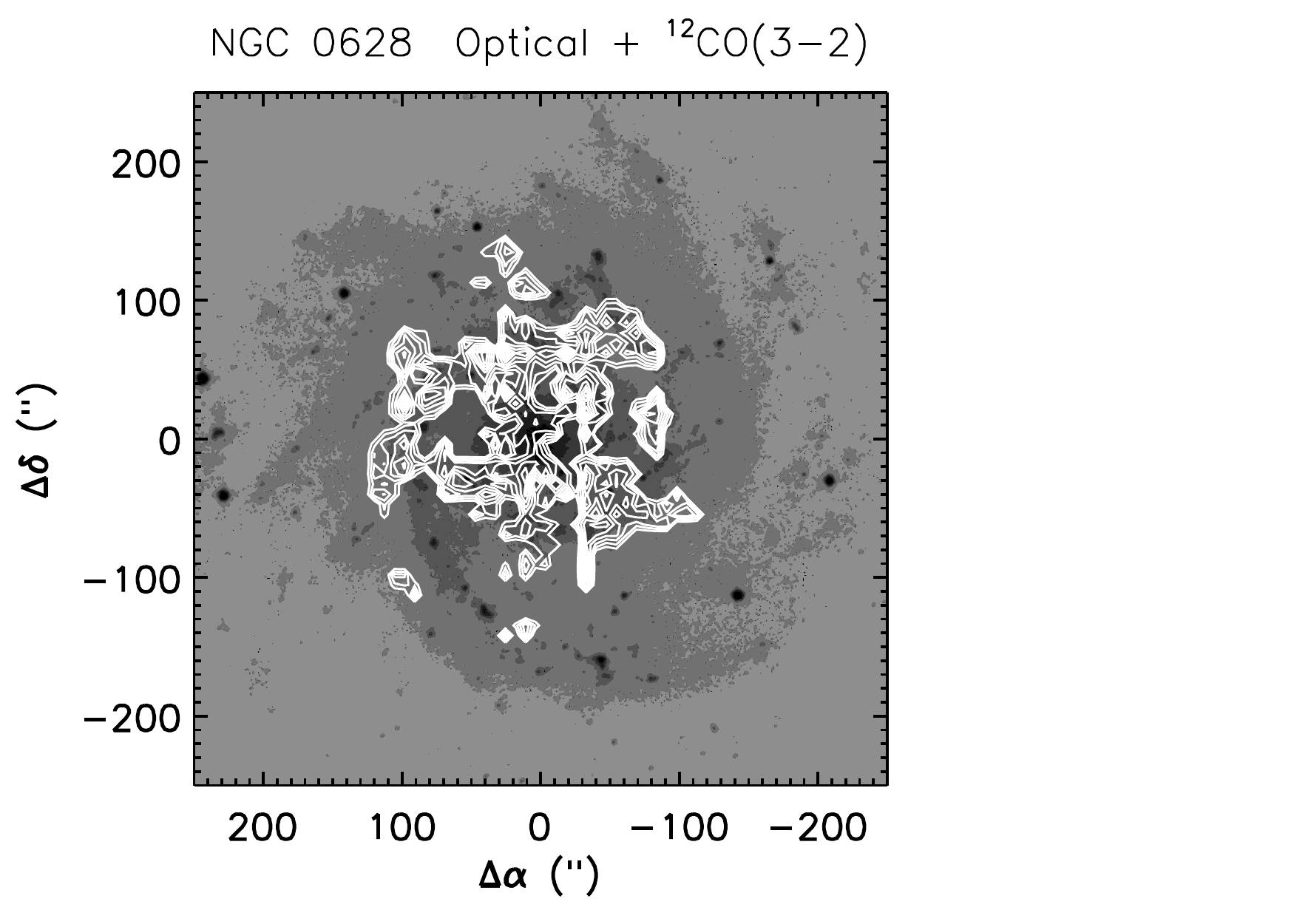}\\
  \caption{Moment maps from literature data. {\bf Top}: Moment 0 contour maps for CO(2--1) and CO(3--2) respectively. The beam sizes are also shown in the bottom left of each panel. Contour levels on the moment 0 maps are from $10$ to $100$ per cent of the peak integrated line intensity in steps of $10$ per cent. The moment 0 peaks are $7.85$~K~km~s$^{-1}$ and $6.50$~K~km~s$^{-1}$ for CO(2--1) and CO(3--2) respectively. {\bf Middle}: Moment 1 contour maps for CO(2--1) and CO(3--2) respectively. Contour levels on the moment 1 map are spaced by $5$~km~s$^{-1}$. {\bf Bottom}: Same moment 0 contour maps (white) of the CO(2--1) and CO(3--2) lines are overlaid on an optical image (greyscale) of the galaxy (SDSS). North is up and east to the left in all images.}
  \label{fig:moments}
\end{figure*}

We created moment maps, i.e. integrated intensity map (hereafter moment 0) 
and velocity maps (hereafter moment 1) over the disc of NGC~0628 
(see Figure~\ref{fig:moments}) by following the method described in \citet{topal16}.
To define the spatial extent of the emission in CO(2--1) and CO(3--2) transitions, 
we first defined a region of contiguous source emission in the fully-calibrated 
and cleaned data cubes \citep{ler09, war10}. The data cubes were first 
smoothed spectrally and then smoothed spatially with a full-width at half-maximum (FHWM) equals to that 
of the beams, i.e. $13\farcs4$ and $14\farcs5$ for CO(2--1) and CO(3--2) respectively. We then 
clipped each (smoothed) cube at $3\sigma$ threshold ($3\times$\,rms in the
smoothed cube) and (smoothed) moment maps were created. The region of contiguous 
emission for each line was then defined using {\small IDL} region-growing algorithm 
$label\_region$. The 3D masks were created from these 2D moment maps. The 3D masks were then applied 
to the original fully-calibrated CO(2--1) and CO(3--2) data cubes using the Multichannel 
Image Reconstruction, Image Analysis and Display (MIRIAD) package
\citep{st95} to obtain the moment maps. 
 Please see Section~3.1 in \citet{topal16} 
 for more details on defining a region of contiguous emission over a galaxy. 
The moment maps are shown in Figure~\ref{fig:moments}.

%
%
\begin{figure*}
  \includegraphics[width=9cm,clip=]{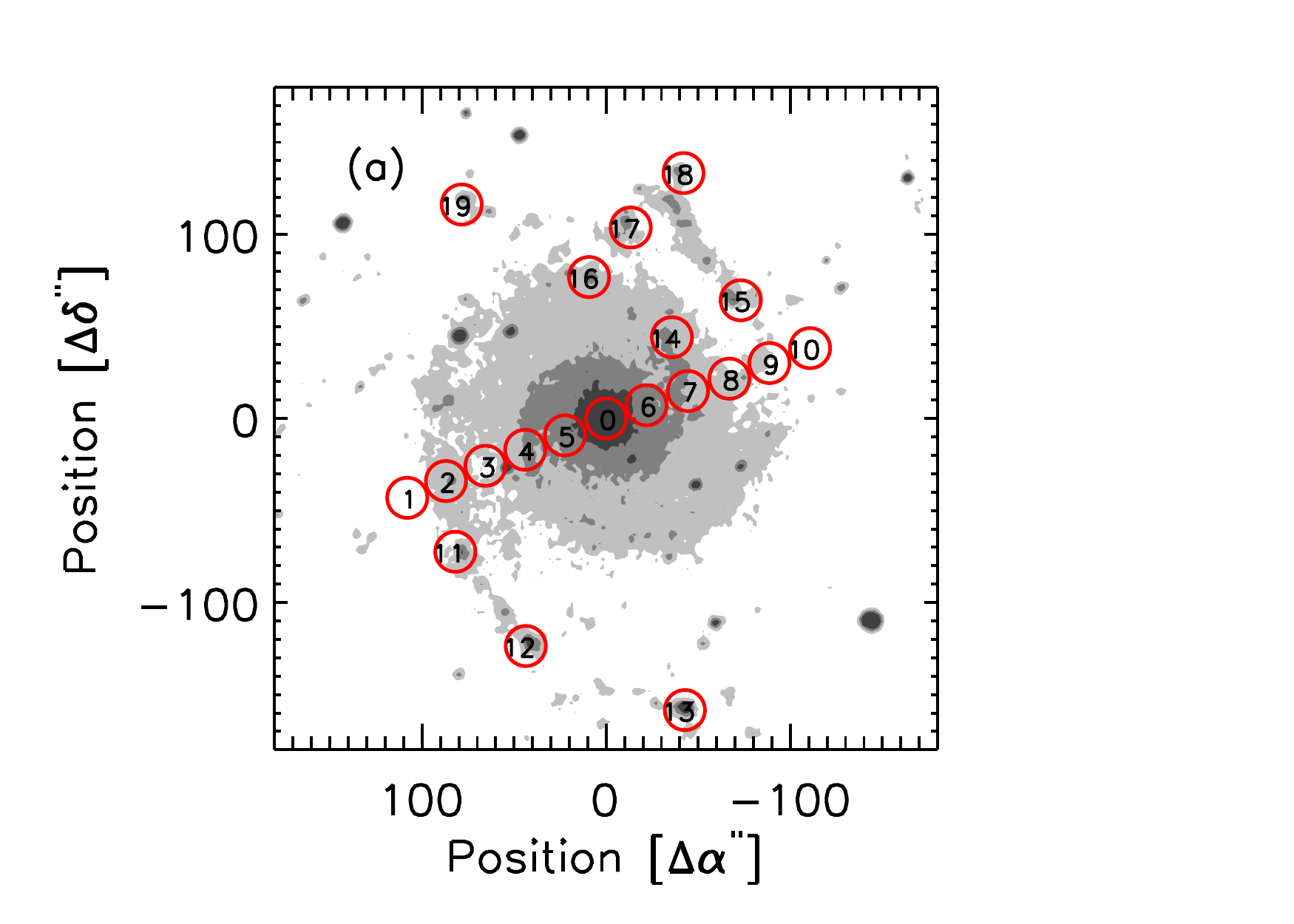}
  \hspace{-80pt}
  \includegraphics[width=9cm,clip=]{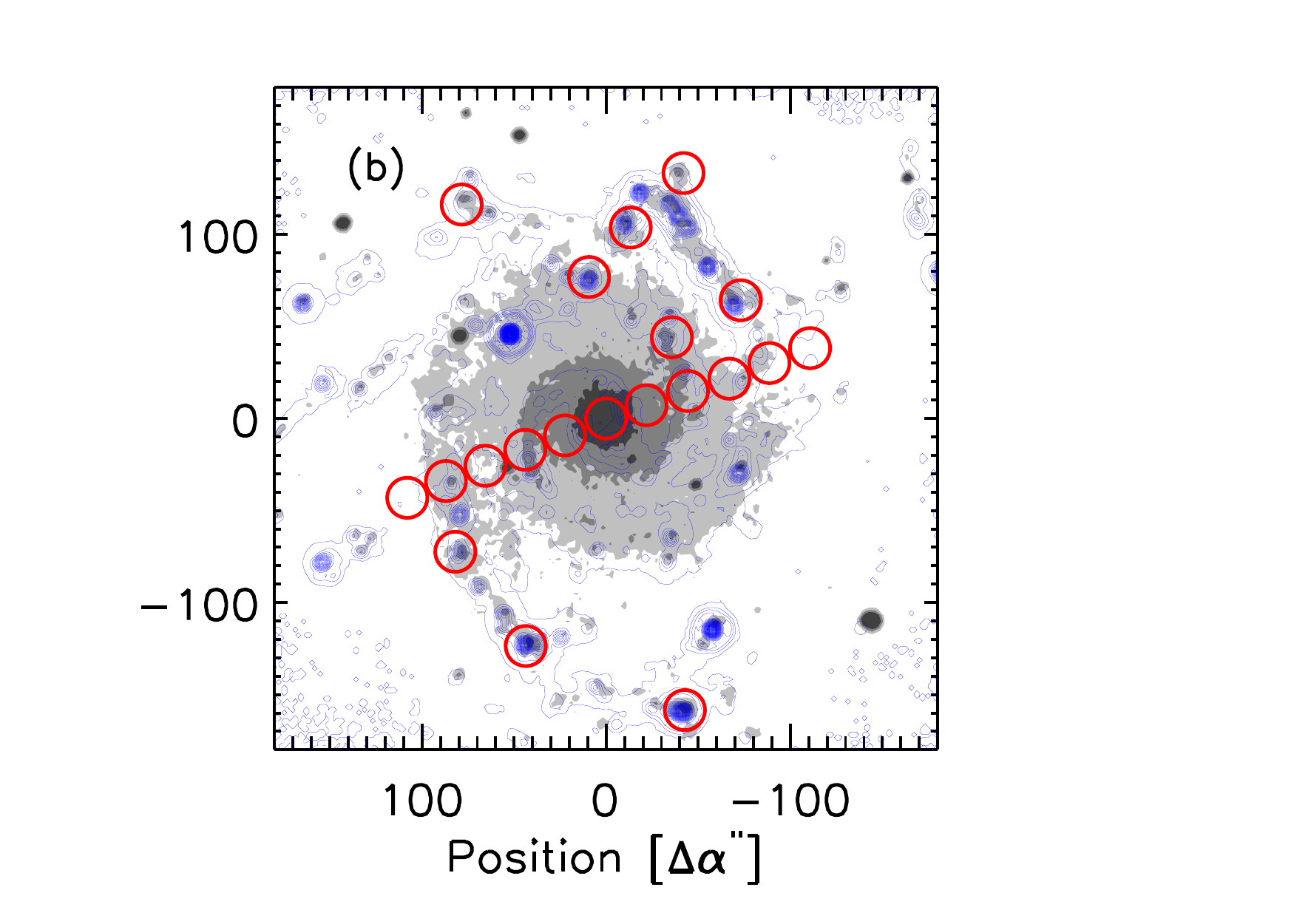} \\
  \includegraphics[width=9cm,clip=]{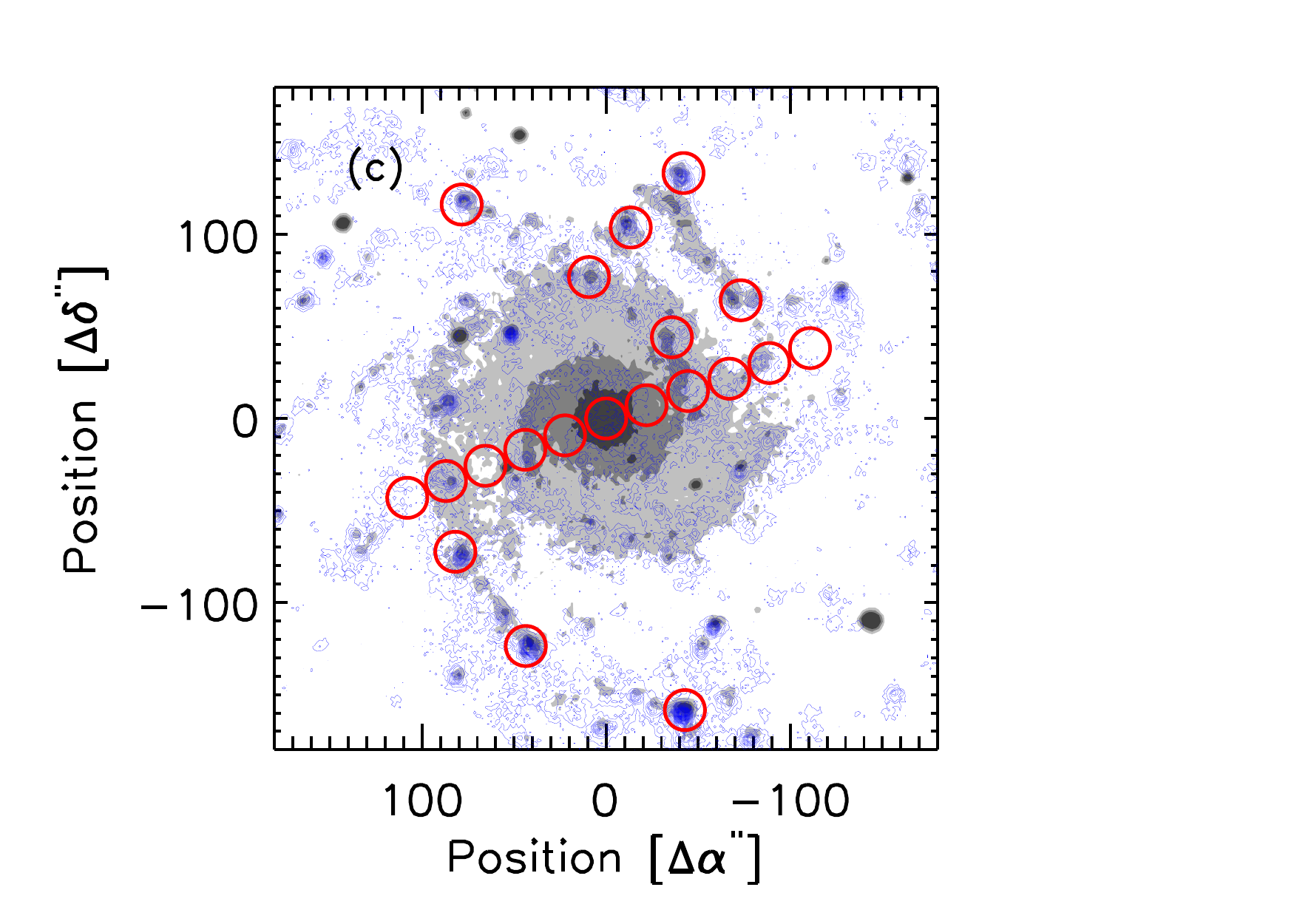}
  \hspace{-80pt}
   \includegraphics[width=9cm,clip=]{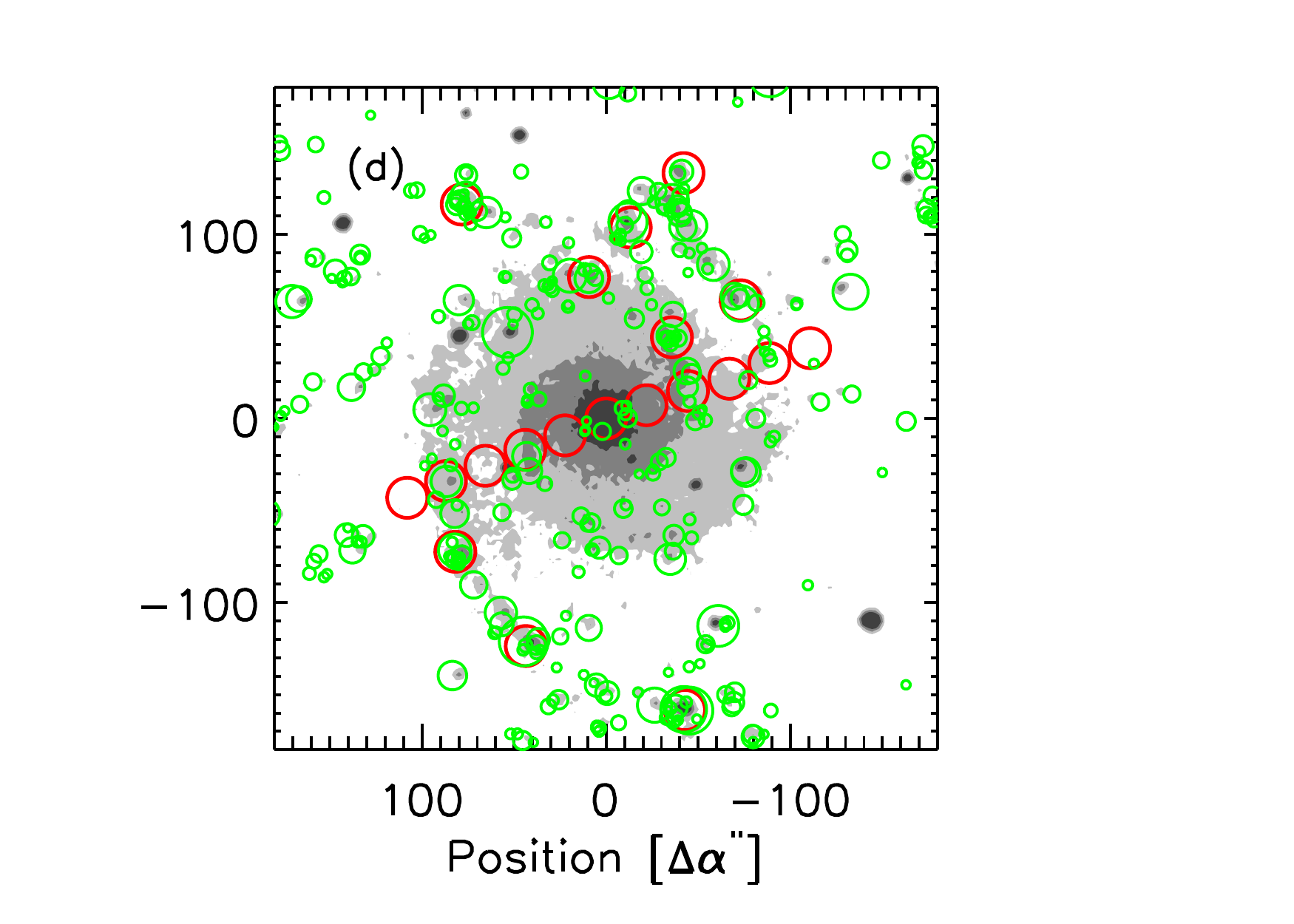}\\
  \caption{{\bf Panel a}: The targeted positions in the spiral galaxy 
  NGC~0628 are overlaid on an optical image of the galaxy (SDSS). 
  The red circles in each panel indicate the targeted positions with the IRAM~30m beam 
  of $22^{\prime\prime}$ at $115$~GHz. The index numbers 
  for the positions are also shown on the panel as listed in Table~\ref{tab:intensities}. 
   {\bf Panel b}: The positions are overlaid on both optical and 
   $24\micron$ images of the galaxy \citep{ken03}, shown by the greyscale and blue contour 
   respectively.  Contour levels on the $24\micron$ map are from $1$ to $100$ per cent of the peak emission 
  in steps of $1$ per cent. {\bf Panel c}: The targeted positions and 
  the FUV emission (GALEX; \citealt{gil07}) are overlaid on the same optical image. 
  Blue contours represent the FUV emission. Contour levels on the FUV map are from $1$ to $100$ per cent of the peak emission in steps of $2$ per cent. {\bf Panel d}: The targeted positions and H\,{\small II} regions \citep{fat07} 
  are overlaid on the same optical image. The green circles indicate the 
  H\,{\small II} regions. The radius of H\,{\small II} regions ranges from $80$~pc to $480$~pc at the distance of 
  $7.3$~Mpc for the galaxy \citep{kar04}. North is up and east to the left in both panels. The offsets are measured with respect to 
  the galaxy centre of $\alpha = 1^{\rm h}36^{\rm m}41.7^{\rm s}$ and $\delta = 15^{\rm d}47^{\rm m}1.0^{\rm s}$ (NED).}
  \label{fig:n0628}
\end{figure*}

\subsection{Beam-averaged quantities}
\subsubsection{Intensities, line ratios and fluxes}
\label{sec:inten}
Our CO(1--0) and $^{13}$CO(1--0) observations have a beam size of $22^{\prime\prime}$, 
while the literature data of CO(2--1) and CO(3--2) have a beam size of $13\farcs4$ and $14\farcs5$ respectively. 
We, therefore, convolved CO(2--1) and CO(3--2) data cubes to a common beam size of $22^{\prime\prime}$ 
using MIRIAD task \emph{convol}. We then extracted the spectra from the targeted positions in the cubes 
using MIRIAD task \emph{imspect}. Gaussian was then fit to the integrated spectra to derive
the velocity-integrated CO line intensities in the unit of K~km~s$^{-1}$. 
The single Gaussian function is given by

\begin{equation}
  \label{eq:SGfunction}
  f(v) = A\,{\rm e}^{\frac{-(v-v_{0})^{2}}{2\sigma^{2}}}\,\,\,
\end{equation} 

where $A>0$ is the flux of the peak at the central (and mean) velocity
$v_{0}$, and $\sigma>0$ is the width of the profile (root mean square
velocity). The fits were carried out with the package {\sc MPFIT} \citep{ma09}, that employs a
Levenberg-Marquardt minimisation algorithm. To avoid local minima, in
each case we ran {\sc MPFIT} several times with different initial
guesses. The fitting parameters with the smallest
$|1-\chi_{\textrm{red}}^2|$ value were taken as the best fit. 

CO(1--0) and CO(2--1) were detected (i.e. $S/N>3\sigma$, where $\sigma$ 
is the uncertainty in the integrated line intensities) 
at all targeted positions. CO(3-2) and $^{13}$CO(1-0) were not detected 
in $5$ out of $20$ positions, three positions are in common, i.e. positions 
$12$, $13$ and $18$ where an upper limit for the integrated 
intensities was estimated (see Table~\ref{tab:intensities}). 
We estimated the upper limit as $3\times\sigma\times\Delta V$, where $\Delta V$ is the line width of CO(1--0) detected 
at that position and $\sigma$ is the rms in the spectra calculated using the channels free of emission. 
The integrated intensities for all the lines at all positions are listed in Table~\ref{tab:intensities} and the spectra 
are shown in Figures~\ref{fig:spec1} -~\ref{fig:spec4} in Appendix~\ref{sec:Ap1}.

We estimated the beam-averaged total intensities per unit area $I$ (W~m$^{-2}$~sr$^{-2}$) 
and total fluxes $F$ (W~m$^{-2}$) by using the standard expressions below (see also
\citealt{ba04,topal16}):
\begin{equation}
  \frac{I}{{\rm W}\,{\rm m}^{-2}\,{\rm
      sr}^{-1}}=1.02\times10^{-18}\,\bigg(\frac{\nu}{{\rm
        GHz}}\bigg)^3\,\bigg(\frac{S}{{\rm K}\,{\rm km}\,{\rm s}^{-1}}\bigg)\, 
\end{equation}
\begin{equation}
  \frac{F}{{\rm W}\,{\rm m}^{-2}}=\bigg(\frac{I}{{\rm W}\,{\rm
        m}^{-2}\,{\rm sr}^{-1}}\bigg)\,\bigg(\frac{\Omega_{\rm B}}{\rm sr}\bigg)\,
\end{equation}
\begin{equation}	
  \frac{\Omega_{\rm B}}{\rm sr}=1.133\,\bigg(\frac{\theta_{\rm B}^2}{{\rm arcsec}^2}\bigg)\,\bigg(\frac{1}{206265^2}\bigg)\,
\end{equation}
\begin{equation}
  \frac{\theta_{\rm B}}{\rm rad}=1.22\,\,\frac{\lambda}{D}\, 
\end{equation}
where $\nu$ and $\lambda$ are the observed line frequency
and wavelength respectively, $\Omega_{\rm B}$ is the solid angle sustained by the
beam, $\theta_{\rm B}$ is the beam size (FWHM) and $D$ is the diameter
of the telescope. The beam-averaged total intensities and total fluxes are listed 
in Table~\ref{tab:intensities}. 

After obtaining the beam-averaged integrated intensities 
we calculated the ratios of the integrated line intensities at each position studied 
throughout the disc of NGC~0628, 
i.e. $^{12}$CO(1--0) / $^{12}$CO(2--1) (hereafter $R_{12}$), 
$^{12}$CO(1--0) / $^{12}$CO(3--2) (hereafter $R_{13}$) and 
$^{12}$CO(1--0) / $^{13}$CO(1--0) (hereafter $R_{11}$). The line ratios 
are listed in Table~\ref{tab:ratios}.
%
%
%
\begin{table*}
  \setlength{\tabcolsep}{4pt}
  \caption{Beam-corrected line quantities obtained for each position over the galaxy.}
  \begin{tabular}{ccccccc}
    \hline
    Position & Offset & Radius$^{\rm a}$ & Line & $\int T_{\rm mb}$ d$v$ & Total intensity & Total flux \\
    & ($\Delta\alpha^{\prime\prime}$, $\Delta\delta^{\prime\prime}$) & (arcsec $|$ kpc) & &(K~km~s$^{-1}$)&(W~m$^{-2}$~sr$^{-2}$)&(W~m$^{-2}$) \\
    \hline \\
    $0$ & $(0, 0)$ & $\phantom{00}0~|~0$ & $^{12}$CO(1-0) & $8.37\pm0.40$&$1.31\,\pm\,0.06\times10^{-11}$& $1.69\,\pm\,0.08\times10^{-19}$  \\
      &              &        & $^{12}$CO(2-1) & $4.52\pm0.21$&$5.65\,\pm\,0.27\times10^{-11}$& $7.28\,\pm\,0.35\times10^{-19}$ \\
    &   	&	 & $^{12}$CO(3-2) & $1.59\pm0.29$&$6.72\,\pm\,1.23\times10^{-11}$& $8.66\,\pm\,1.58\times10^{-19}$  \\
    &   	&	 & $^{13}$CO(1-0) & $1.12\pm0.14$&$1.53\,\pm\,0.19\times10^{-12}$& $1.97\,\pm\,0.25\times10^{-20}$ \\\\
    $1$ & $(108, 43.1)$ & $116.3~|~4.1$ & $^{12}$CO(1-0) & $1.70\pm0.16$&$2.66\,\pm\,0.24\times10^{-12}$& $3.42\,\pm\,0.32\times10^{-20}$  \\
      &              &        & $^{12}$CO(2-1) & $1.43\pm0.14$&$1.79\,\pm\,0.18\times10^{-11}$& $2.31\,\pm\,0.23\times10^{-19}$  \\
    &   	&	 & $^{12}$CO(3-2) & $0.53\pm0.20$&$2.25\,\pm\,0.84\times10^{-11}$& $2.90\,\pm\,1.08\times10^{-19}$\\
    &   	&	 & $^{13}$CO(1-0) & $0.22\pm0.08$&$3.04\,\pm\,1.18\times10^{-13}$& $3.92\,\pm\,1.52\times10^{-21}$\\\\
    $2$ & $(87, -34)$ & $\phantom{0}93.4~|~3.3$ & $^{12}$CO(1-0) &$4.74\pm0.14$& $7.41\,\pm\,0.22\times10^{-12}$& $9.55\,\pm\,0.28\times10^{-20}$  \\
      &              &        & $^{12}$CO(2-1) & $2.96\pm0.14$&$3.69\,\pm\,0.17\times10^{-11}$&$4.76\,\pm\,0.22\times10^{-19}$   \\
    &   	&	 & $^{12}$CO(3-2) & $0.61\pm0.24$&$2.59\,\pm\,1.01\times10^{-11}$& $3.34\,\pm\,1.30\times10^{-19}$\\
    &   	&	 & $^{13}$CO(1-0) & $0.43\pm0.06$&$5.90\,\pm\,0.89\times10^{-13}$& $7.61\,\pm\,1.15\times10^{-21}$  \\\\
    $3$ & $(65.6, -25.7)$ & $\phantom{0}70.4~|~2.5$ & $^{12}$CO(1-0) & $2.56\pm0.27$&$4.00\,\pm\,0.42\times10^{-12}$& $5.16\,\pm\,0.54\times10^{-20}$  \\
      &              &        & $^{12}$CO(2-1) & $2.46\pm0.15$&$3.08\,\pm\,0.19\times10^{-11}$& $3.96\,\pm\,0.25\times10^{-19}$\\
    &   	&	 & $^{12}$CO(3-2) & $0.85\pm0.24$&$3.60\,\pm\,0.99\times10^{-11}$& $4.64\,\pm\,1.28\times10^{-19}$\\
    &   	&	 & $^{13}$CO(1-0) & $\phantom{000}\le0.25$&$\phantom{0000}\le3.41\times10^{-13}$& $\phantom{0000}\le4.39\times10^{-21}$\\\\
       $4$ & $(44, -17)$ & $\phantom{0}47.1~|~1.7$ & $^{12}$CO(1-0) & $5.03\pm0.22$&$7.86\,\pm\,0.34\times10^{-12}$ & $1.01\,\pm\,0.04\times10^{-19}$  \\
      &              &        & $^{12}$CO(2-1) & $2.98\pm0.17$&$3.73\,\pm\,0.22\times10^{-11}$ & $4.80\,\pm\,0.27\times10^{-19}$ \\
    &   	&	 & $^{12}$CO(3-2) & $1.31\pm0.24$&$5.52\,\pm\,1.01\times10^{-11}$ & $7.11\,\pm\,1.30\times10^{-19}$ \\
    &   	&	 & $^{13}$CO(1-0) & $0.44\pm0.08$&$6.02\,\pm\,1.15\times10^{-13}$ & $7.76\,\pm\,1.49\times10^{-21}$\\\\
    $5$ & $(22.4, -9.2)$ & $\phantom{0}24.2~|~0.9$ & $^{12}$CO(1-0) &$5.22\pm0.20$&$8.16\,\pm\,0.31\times10^{-12}$ & $1.05\,\pm\,0.04\times10^{-19}$ \\
      &              &        & $^{12}$CO(2-1) & $3.73\pm0.17$&$4.67\,\pm\,0.21\times10^{-11}$ & $6.01\,\pm\,0.27\times10^{-19}$ \\
    &   	&	 & $^{12}$CO(3-2) & $1.63\pm0.36$&$6.86\,\pm\,1.50\times10^{-11}$ & $8.84\,\pm\,1.93\times10^{-19}$\\
    &   	&	 & $^{13}$CO(1-0) & $0.52\pm0.08$&$7.08\,\pm\,1.11\times10^{-13}$ & $9.12\,\pm\,1.44\times10^{-21}$\\\\
    $6$ & $(-21.9, 7.2)$ & $\phantom{0}23.1~|~0.8$ & $^{12}$CO(1-0) &$5.28\pm0.19$&$8.25\,\pm\,0.30\times10^{-12}$ & $1.06\,\pm\,0.04\times10^{-19}$ \\
      &              &        & $^{12}$CO(2-1) & $3.58\pm0.18$&$4.48\,\pm\,0.23\times10^{-11}$ & $5.77\,\pm\,0.29\times10^{-19}$\\
    &   	&	 & $^{12}$CO(3-2) & $0.97\pm0.19$&$4.09\,\pm\,0.79\times10^{-11}$ & $5.27\,\pm\,1.02\times10^{-19}$\\
    &   	&	 & $^{13}$CO(1-0) & $0.50\pm0.09$&$6.79\,\pm\,1.21\times10^{-13}$ & $8.75\,\pm\,1.56\times10^{-21}$\\\\
    $7$ & $(-44.4, 14.9)$ & $\phantom{0}46.8~|~1.7$ & $^{12}$CO(1-0) &$4.94\pm0.33$&$7.71\,\pm\,0.51\times10^{-12}$ & $9.94\,\pm\,0.66\times10^{-20}$\\
      &              &        & $^{12}$CO(2-1) & $2.78\pm0.15$&$3.47\,\pm\,0.19\times10^{-11}$ & $4.47\,\pm\,0.24\times10^{-20}$\\
    &   	&	 & $^{12}$CO(3-2) & $1.03\pm0.25$&$4.34\,\pm\,1.04\times10^{-11}$ & $5.60\,\pm\,1.34\times10^{-19}$\\
    &   	&	 & $^{13}$CO(1-0) & $0.41\pm0.10$&$5.64\,\pm\,1.36\times10^{-13}$ & $7.27\,\pm\,1.75\times10^{-21}$\\\\
    $8$ & $(-66.9, 21.6)$ & $\phantom{0}70.3~|~2.5$ & $^{12}$CO(1-0) &$2.69\pm0.25$&$4.20\,\pm\,0.38\times10^{-12}$ & $5.42\,\pm\,0.50\times10^{-20}$\\
      &              &        & $^{12}$CO(2-1) & $2.19\pm0.15$&$2.73\,\pm\,0.19\times10^{-11}$ & $3.52\,\pm\,0.25\times10^{-19}$\\
    &   	&	 & $^{12}$CO(3-2) & $1.07\pm0.37$&$4.53\,\pm\,1.55\times10^{-11}$ & $5.84\,\pm\,1.99\times10^{-19}$\\
    &   	&	 & $^{13}$CO(1-0) & $0.28\pm0.09$&$3.85\,\pm\,1.20\times10^{-13}$ & $4.96\,\pm\,1.55\times10^{-21}$\\\\
    $9$ & $(-88.6, 30.1)$ & $\phantom{0}93.6~|~3.3$ & $^{12}$CO(1-0) &$3.82\pm0.37$&$5.97\,\pm\,0.58\times10^{-13}$ & $7.70\,\pm\,0.75\times10^{-20}$\\
      &              &        & $^{12}$CO(2-1) & $2.35\pm0.17$& $2.93\,\pm\,0.22\times10^{-11}$ & $3.78\,\pm\,0.28\times10^{-19}$\\
    &   	&	 & $^{12}$CO(3-2) & $1.10\pm0.23$&$4.62\,\pm\,0.98\times10^{-11}$ & $5.96\,\pm\,1.26\times10^{-19}$\\
    &   	&	 & $^{13}$CO(1-0) & $0.36\pm0.13$&$4.91\,\pm\,1.78\times10^{-13}$ & $6.32\,\pm\,2.30\times10^{-21}$\\\\
    
    \hline   
  \end{tabular}
  \label{tab:intensities}
 \parbox[t]{0.80\textwidth}{$^{\rm a}$Radii are calculated with respect to the galaxy centre and given in the units of arcseconds and kpc (i.e. corresponding linear size for the offset with respect to the galaxy centre).}
\end{table*}

%
\addtocounter{table}{-1}
\begin{table*}
  \setlength{\tabcolsep}{4pt}
  \caption{{\bf Continued.} Beam-corrected line quantities obtained at each position over the galaxy.}
  \begin{tabular}{ccccccc}
    \hline
    Position & Offset & Radius$^{\rm a}$ & Line & $\int T_{\rm mb}$ d$v$ & Total intensity & Total flux \\
    & ($\Delta\alpha^{\prime\prime}$, $\Delta\delta^{\prime\prime}$) & (arcsec $|$ kpc) & &(K~km~s$^{-1}$)&(W~m$^{-2}$~sr$^{-2}$)&(W~m$^{-2}$) \\
    \hline \\
    $10$&$(-110.7, 38.2)$ & $117.1~|~4.1$ & $^{12}$CO(1-0) & $1.74\pm0.18$&$2.73\,\pm\,0.28\times10^{-12}$ & $3.51\,\pm\,0.36\times10^{-20}$ \\
     &              &        & $^{12}$CO(2-1) & $1.57\pm0.16$&$1.97\,\pm\,0.20\times10^{-11}$ & $2.54\,\pm\,0.26\times10^{-19}$ \\
    &   	&	 & $^{12}$CO(3-2) & $0.42\pm0.16$ &$1.79\,\pm\,0.70\times10^{-11}$ & $2.31\,\pm\,0.90\times10^{-19}$\\
    &   	&	 & $^{13}$CO(1-0) & $\phantom{000}\le0.20$&$\phantom{0000}\le2.71\times10^{-13}$ & $\phantom{0000}\le3.50\times10^{-21}$ \\\\
    $11$&$(81.9, -72.4)$ & $109.3~|~3.9$ & $^{12}$CO(1-0) & $3.52\pm0.19$ &$5.49\,\pm\,0.30\times10^{-12}$ & $7.08\,\pm\,0.39\times10^{-20}$\\
     &              &        & $^{12}$CO(2-1) & $2.00\pm0.15$&$2.50\,\pm\,0.18\times10^{-11}$ & $3.23\,\pm\,0.24\times10^{-19}$\\
    &   	&	 & $^{12}$CO(3-2) & $\phantom{000}\le0.63$&$\phantom{0000}\le2.68\times10^{-11}$ & $\phantom{0000}\le3.45\times10^{-19}$ \\
    &   	&	 & $^{13}$CO(1-0) & $0.25\pm0.08$&$3.41\,\pm\,1.14\times10^{-13}$ & $4.40\,\pm\,1.47\times10^{-21}$\\\\
 $12$&$(43.6, -123.6)$ & $131.1~|~4.6$ & $^{12}$CO(1-0) & $2.99\pm0.22$&$4.67\,\pm\,0.34\times10^{-12}$ & $6.02\,\pm\,0.44\times10^{-20}$\\
    &              &        & $^{12}$CO(2-1) & $1.67\pm0.16$&$2.09\,\pm\,0.20\times10^{-11}$ & $2.69\,\pm\,0.26\times10^{-19}$\\
 &   	&	 & $^{12}$CO(3-2) & $\phantom{000}\le0.40$ &$\phantom{0000}\le1.70\times10^{-11}$ & $\phantom{0000}\le2.20\times10^{-19}$\\
  &   	&	 & $^{13}$CO(1-0) & $\phantom{000}\le0.22$ &$\phantom{0000}\le2.69\times10^{-13}$ & $\phantom{0000}\le3.82\times10^{-21}$\\\\
    $13$&$(-42.7, -158.4)$ & $164.1~|~5.8$ & $^{12}$CO(1-0) & $1.79\pm0.27$&$2.80\,\pm\,0.42\times10^{-12}$ & $3.61\,\pm\,0.54\times10^{-20}$\\
      &              &        & $^{12}$CO(2-1) & $1.45\pm0.16$&$1.81\,\pm\,0.21\times10^{-11}$ & $2.33\,\pm\,0.26\times10^{-19}$\\
   &   	&	 & $^{12}$CO(3-2) & $\phantom{000}\le0.53$&$\phantom{0000}\le2.23\times10^{-11}$ & $\phantom{0000}\le2.87\times10^{-19}$\\
   &   	&	 & $^{13}$CO(1-0) & $\phantom{000}\le0.25$&$\phantom{0000}\le3.43\times10^{-13}$ & $\phantom{0000}\le4.42\times10^{-21}$\\\\
   $14$&$(-35.7, 44.1)$ & $\phantom{0}56.7~|~2.0$ & $^{12}$CO(1-0) & $4.38\pm0.17$&$6.85\,\pm\,0.27\times10^{-12}$ & $8.82\,\pm\,0.35\times10^{-20}$\\
&              &        & $^{12}$CO(2-1) & $2.45\pm0.16$&$3.06\,\pm\,0.20\times10^{-11}$ & $3.94\,\pm\,0.26\times10^{-19}$\\
 &   	&	 & $^{12}$CO(3-2) & $1.69\pm0.28$ &$7.13\,\pm\,1.18\times10^{-11}$ & $9.18\,\pm\,1.52\times10^{-19}$\\
&   	&	 & $^{13}$CO(1-0) & $0.38\pm0.08$&$5.21\,\pm\,1.16\times10^{-13}$ & $6.72\,\pm\,1.50\times10^{-21}$\\\\
   $15$&$(-73.1, 64.2)$ & $\phantom{0}97.3~|~3.4$ & $^{12}$CO(1-0) & $5.24\pm0.18$&$8.18\,\pm\,0.28\times10^{-12}$ & $1.05\,\pm\,0.04\times10^{-19}$\\
  &              &        & $^{12}$CO(2-1) & $3.01\pm0.15$&$3.76\,\pm\,0.19\times10^{-11}$ & $4.84\,\pm\,0.24\times10^{-19}$\\
 &   	&	 & $^{12}$CO(3-2) & $0.90\pm0.21$&$3.79\,\pm\,0.89\times10^{-11}$ & $4.88\,\pm\,1.15\times10^{-19}$\\
 &   	&	 & $^{13}$CO(1-0) & $0.46\pm0.09$ &$6.33\,\pm\,1.17\times10^{-13}$ & $8.16\,\pm\,1.51\times10^{-21}$\\\\
  $16$&$(9.3, 76.8)$ & $\phantom{0}77.3~|~2.7$ & $^{12}$CO(1-0) & $3.16\pm0.31$ &$4.94\,\pm\,0.49\times10^{-12}$ & $6.37\,\pm\,0.63\times10^{-20}$\\
 &              &        & $^{12}$CO(2-1) & $2.52\pm0.14$&$3.15\,\pm\,0.17\times10^{-11}$ & $4.05\,\pm\,0.23\times10^{-19}$\\
&   	&	 & $^{12}$CO(3-2) & $1.69\pm0.28$&$7.12\,\pm\,1.19\times10^{-11}$ & $9.18\,\pm\,1.53\times10^{-19}$\\
  &   	&	 & $^{13}$CO(1-0) & $0.68\pm0.18$ &$9.34\,\pm\,2.42\times10^{-13}$ & $1.20\,\pm\,0.31\times10^{-20}$\\\\
$17$&$(-13.4, 103.7)$ & $104.6~|~3.7$ & $^{12}$CO(1-0) & $4.52\pm0.23$&$7.06\,\pm\,0.36\times10^{-12}$ & $9.10\,\pm\,0.46\times10^{-20}$\\
&              &        & $^{12}$CO(2-1) & $2.68\pm0.15$&$3.35\,\pm\,0.18\times10^{-11}$ & $4.32\,\pm\,0.24\times10^{-19}$\\
&   	&	 & $^{12}$CO(3-2) & $0.62\pm0.18$&$2.60\,\pm\,0.74\times10^{-11}$ & $3.36\,\pm\,0.96\times10^{-19}$\\
&   	&	 & $^{13}$CO(1-0) & $0.34\pm0.09$&$4.69\,\pm\,1.27\times10^{-13}$ & $6.04\,\pm\,1.64\times10^{-21}$ \\\\
$18$&$(-42, 133.3)$ & $139.7~|~4.9$ & $^{12}$CO(1-0) & $1.00\pm0.20$&$1.56\,\pm\,0.31\times10^{-12}$ & $2.01\,\pm\,0.40\times10^{-20}$ \\
   &              &        & $^{12}$CO(2-1) & $0.88\pm0.11$&$1.10\,\pm\,0.14\times10^{-11}$ & $1.41\,\pm\,0.18\times10^{-19}$\\
  &   	&	 & $^{12}$CO(3-2) & $\phantom{000}\le0.49$&$\phantom{0000}\le2.05\times10^{-11}$ & $\phantom{0000}\le2.65\times10^{-19}$\\
&   	&	 & $^{13}$CO(1-0) & $\phantom{000}\le0.19$&$\phantom{0000}\le2.66\times10^{-13}$ & $\phantom{0000}\le3.42\times10^{-21}$\\\\
$19$&$(-78.4, 116.2)$ & $140.2~|~4.9$ & $^{12}$CO(1-0) & $3.73\pm0.17$&$5.83\,\pm\,0.27\times10^{-12}$ & $7.51\,\pm\,0.34\times10^{-20}$ \\
 &              &        & $^{12}$CO(2-1) & $2.03\pm0.13$&$2.54\,\pm\,0.16\times10^{-11}$ & $3.27\,\pm\,0.21\times10^{-19}$\\
&   	&	 & $^{12}$CO(3-2) & $\phantom{000}\le0.50$&$\phantom{0000}\le2.13\times10^{-11}$ & $\phantom{0000}\le2.74\times10^{-19}$\\
&   	&	 & $^{13}$CO(1-0) & $0.31\pm0.07$&$4.23\,\pm\,1.00\times10^{-13}$ & $5.46\,\pm\,1.28\times10^{-21}$\\\\
    
    \hline   
  \end{tabular}
  \label{tab:intensities}
  \parbox[t]{0.80\textwidth}{$^{\rm a}$Radii are calculated with respect to the galaxy centre and given in the units of arcseconds and kpc (i.e. corresponding linear size for the offset with respect to the galaxy centre).}
\end{table*}

\begin{table*}
 \caption{Beam-averaged line ratios over the disc of NGC~0628.}
 \begin{tabular}{cllccll}
Position & Ratio & Value && Position & Ratio & Value  \\ 
   \hline\\
  $0$&$R_{\rm 12}$& $\phantom{0}1.85\,\pm\,0.13$ && $10$&$R_{\rm 12}$& $\phantom{0}1.11\,\pm\,0.16$\\
  &$R_{\rm 13}$& $\phantom{0}5.25\,\pm\,0.99$ & &&$R_{\rm 13}$& $\phantom{0}4.11\,\pm\,1.65$\\
  &$R_{\rm 11}$& $\phantom{0}7.48\,\pm\,1.02$ & &&$R_{\rm 11}$& $\phantom{00000}\ge8.78$\\\\
  $1$&$R_{\rm 12}$& $\phantom{0}1.19\,\pm\,0.16$ && $11$&$R_{\rm 12}$& $\phantom{0}1.76\,\pm\,0.16$\\
  &$R_{\rm 13}$& $\phantom{0}3.19\,\pm\,1.22$ & &&$R_{\rm 13}$& $\phantom{00000}\ge5.54$\\
  &$R_{\rm 11}$& $\phantom{0}7.63\,\pm\,3.04$ & &&$R_{\rm 11}$& $14.07\,\pm\,4.77$\\\\
  $2$&$R_{\rm 12}$& $\phantom{0}1.60\,\pm\,0.09$ && $12$&$R_{\rm 12}$& $\phantom{0}1.79\,\pm\,0.22$\\
  &$R_{\rm 13}$& $\phantom{0}7.72\,\pm\,3.02$ & &&$R_{\rm 13}$& $\phantom{00000}\ge7.40$\\
  &$R_{\rm 11}$& $10.97\,\pm\,1.69$ & &&$R_{\rm 11}$& $\phantom{00000}\ge13.78$\\\\
  $3$&$R_{\rm 12}$& $\phantom{0}1.04\,\pm\,0.13$ && $13$&$R_{\rm 12}$& $\phantom{0}1.24\,\pm\,0.23$\\
  &$R_{\rm 13}$& $\phantom{0}3.00\,\pm\,0.89$ & &&$R_{\rm 13}$& $\phantom{00000}\ge3.39$\\
  &$R_{\rm 11}$& $\phantom{00000}\ge10.26$& &&$R_{\rm 11}$& $\phantom{00000}\ge7.13$\\\\
  $4$&$R_{\rm 12}$& $\phantom{0}1.69\,\pm\,0.12$ && $14$&$R_{\rm 12}$& $\phantom{0}1.79\,\pm\,0.14$\\
  &$R_{\rm 13}$& $\phantom{0}3.85\,\pm\,0.72$ & &&$R_{\rm 13}$& $\phantom{0}2.59\,\pm\,0.44$\\
  &$R_{\rm 11}$& $11.41\,\pm\,2.24$ & &&$R_{\rm 11}$& $11.48\,\pm\,2.60$\\\\
  $5$&$R_{\rm 12}$& $\phantom{0}1.40\,\pm\,0.08$ && $15$&$R_{\rm 12}$& $\phantom{0}1.74\,\pm\,0.11$\\
  &$R_{\rm 13}$& $\phantom{0}3.21\,\pm\,0.71$ & &&$R_{\rm 13}$& $\phantom{0}5.83\,\pm\,1.39$\\
  &$R_{\rm 11}$& $10.08\,\pm\,1.63$ & &&$R_{\rm 11}$& $11.30\,\pm\,2.12$\\\\
  $6$&$R_{\rm 12}$& $\phantom{0}1.47\,\pm\,0.09$ && $16$&$R_{\rm 12}$& $\phantom{0}1.26\,\pm\,0.14$\\
  &$R_{\rm 13}$& $\phantom{0}5.45\,\pm\,1.07$ & &&$R_{\rm 13}$& $\phantom{0}1.87\,\pm\,0.36$\\
  &$R_{\rm 11}$& $10.62\,\pm\,1.94$ & &&$R_{\rm 11}$& $\phantom{0}4.62\,\pm\,1.28$\\\\
  $7$&$R_{\rm 12}$& $\phantom{0}1.78\,\pm\,0.15$ && $17$&$R_{\rm 12}$& $\phantom{0}1.69\,\pm\,0.13$\\
  &$R_{\rm 13}$& $\phantom{0}4.79\,\pm\,1.19$ & &&$R_{\rm 13}$& $\phantom{0}7.32\,\pm\,2.12$\\
  &$R_{\rm 11}$& $11.94\,\pm\,2.98$ & &&$R_{\rm 11}$& $13.15\,\pm\,3.63$\\\\
  $8$&$R_{\rm 12}$& $\phantom{0}1.23\,\pm\,0.14$& & $18$&$R_{\rm 12}$& $\phantom{0}1.14\,\pm\,0.27$\\
  &$R_{\rm 13}$& $\phantom{0}2.50\,\pm\,0.88$ & &&$R_{\rm 13}$& $\phantom{00000}\ge2.05$\\
  &$R_{\rm 11}$& $\phantom{0}9.55\,\pm\,3.11$ & &&$R_{\rm 11}$& $\phantom{00000}\ge5.13$\\\\
  $9$&$R_{\rm 12}$& $\phantom{0}1.63\,\pm\,0.20$ && $19$&$R_{\rm 12}$& $\phantom{0}1.84\,\pm\,0.15$\\
  &$R_{\rm 13}$& $\phantom{0}3.49\,\pm\,0.81$ & &&$R_{\rm 13}$& $\phantom{00000}\ge7.40$\\
  &$R_{\rm 11}$& $10.64\,\pm\,4.00$ && &$R_{\rm 11}$& $12.03\,\pm\,2.88$\\\\ 	
   \hline
 \end{tabular}
 \parbox[t]{0.55\textwidth}{\textit{Notes: }The exact locations of the positions over the disc of NGC~0628 are shown in Figure~\ref{fig:n0628}.}
 \label{tab:ratios}
\end{table*}

\subsubsection{Gas mass and surface density}
As shown below, the beam-averaged total molecular gas mass at each position  
was estimated in the unit of solar mass ($M_\odot$) using the adopted beam, 
CO-to-H$_{2}$ conversion factor ($X_{\rm CO}$) and the distance to the galaxy,

\begin{equation}
  \frac{M_{{\rm H}_2}}{M_\odot}=5.5\times10^5\,\bigg(\frac{S_{\rm 1-0}}{{\rm K}\,{\rm km}\,{\rm s}^{-1}}\bigg)\,
  \label{eq:mass}
\end{equation}

where $S_{\rm 1-0}$ is the integrated intensity of $^{12}$CO(1--0) in the unit of K~km~s$^{-1}$. 
We also estimated a surface density, $\Sigma_{H_{2}}$, using $S_{\rm 1-0}$ 
and an adopted $X_{\rm CO}$ value. The $X_{\rm CO}$ value of $2\times10^{20}$ cm$^{-2}$~(K~km~s$^{-1}$)$^{-1}$ 
($\alpha_{\rm CO} = 4.35~M_{\sun}$(K~km~s$^{-1}$~pc$^{2}$)$^{-1}$, where $\alpha_{\rm CO}\,\equiv\,\Sigma_{H_{2}}$ / $S_{\rm 1-0}$) was adopted in this study \citep{dame01,bol13,sun18}. The $X_{\rm CO}$ is related to the beam-averaged H$_{2}$ column density ($N_{{\rm H}_2}$) as $X_{\rm CO}$ = $N_{{\rm H}_2}$ / $S_{\rm 1-0}$. The values of $M_{{\rm H}_2}$ and $N_{{\rm H}_2}$ enclosed within the adopted beam and also gas surface density in the unit of $M_{\sun}$~pc$^{-2}$ at the observed positions are listed in Table~\ref{tab:mass}.

\begin{table*}
 \caption{Total H$_{2}$ mass, column density and gas surface density obtained at each position studied.}
 \begin{tabular}{cccccccc}\hline
Position & $M_{\rm H_{2}}$ & $N_{\rm H_{2}}$ & $\Sigma_{H_{2}}$&Position & $M_{\rm H_{2}}$ & $N_{\rm H_{2}}$ & $\Sigma_{H_{2}}$ \\ 
&($M_{\sun}\times10^{5}$)&(cm$^{-2}\times10^{20}$)&($M_{\sun}$~pc$^{-2}$)&&($M_{\sun}\times10^{5}$)&(cm$^{-2}\times10^{20}$)&($M_{\sun}$~pc$^{-2}$)  \\ 
   \hline
  $0$&$45.9\,\pm\,1.3$&$16.7\,\pm\,0.8$&$36.4\,\pm\,1.8$&$\phantom{0}10^{\star}$&$\phantom{0}9.6\,\pm\,0.6$&$\phantom{0}3.5\,\pm\,0.4$&$\phantom{0}7.6\,\pm\,0.8$\\
 $\phantom{0}1^{\star}$&$\phantom{0}9.3\,\pm\,0.5$&$\phantom{0}3.4\,\pm\,0.3$&$\phantom{0}7.4\,\pm\,0.7$&$11$&$19.3\,\pm\,0.6$&$\phantom{0}7.0\,\pm\,0.4$&$15.3\,\pm\,0.8$\\
 $2$&$26.0\,\pm\,0.4$&$\phantom{0}9.5\,\pm\,0.3$&$20.6\,\pm\,0.6$&$12$&$16.4\,\pm\,0.7$&$\phantom{0}6.0\,\pm\,0.4$&$13.0\,\pm\,0.9$\\
 $\phantom{0}3^{\star}$&$14.1\,\pm\,0.8$&$\phantom{0}5.1\,\pm\,0.5$&$11.1\,\pm\,1.2$&$13$&$\phantom{0}9.8\,\pm\,0.8$&$\phantom{0}3.6\,\pm\,0.5$&$\phantom{0}7.8\,\pm\,1.2$\\
 $\phantom{0}4^{\star}$&$27.6\,\pm\,0.7$&$10.1\,\pm\,0.4$&$21.9\,\pm\,0.9$&$14$&$24.1\,\pm\,0.5$&$\phantom{0}8.8\,\pm\,0.3$&$19.1\,\pm\,0.8$\\
 $\phantom{0}5^{\star}$&$28.7\,\pm\,0.6$&$10.4\,\pm\,0.4$&$22.7\,\pm\,0.9$&$15$&$28.7\,\pm\,0.6$&$10.5\,\pm\,0.4$&$22.8\,\pm\,0.8$\\
 $\phantom{0}6^{\star}$&$29.0\,\pm\,0.6$&$10.6\,\pm\,0.4$&$22.9\,\pm\,0.8$&$16$&$17.4\,\pm\,1.0$&$\phantom{0}6.3\,\pm\,0.6$&$13.8\,\pm\,1.4$\\
 $7$&$27.1\,\pm\,1.0$&$\phantom{0}9.9\,\pm\,0.7$&$21.4\,\pm\,1.4$&$17$&$24.8\,\pm\,0.7$&$\phantom{0}9.0\,\pm\,0.5$&$19.7\,\pm\,1.0$\\
 $8$&$14.8\,\pm\,0.8$&$\phantom{0}5.4\,\pm\,0.5$&$11.7\,\pm\,1.1$&$18$&$\phantom{0}5.4\,\pm\,0.6$&$\phantom{0}2.0\,\pm\,0.4$&$\phantom{0}4.3\,\pm\,0.9$\\
 $\phantom{0}9^{\star}$&$21.0\,\pm\,1.2$&$\phantom{0}7.6\,\pm\,0.7$&$16.6\,\pm\,1.6$&$19$&$20.5\,\pm\,0.5$&$\phantom{0}7.4\,\pm\,0.3$&$16.2\,\pm\,0.7$\\
   \hline\\
 \end{tabular}
 \parbox[t]{\textwidth}{\textit{Notes: \rm{The positions from $0$ to $10$ are the positions along the SE-NW cut
  while the other positions are located in the northern and southern arms of the galaxy (see Fig.~\ref{fig:n0628}). The positions 
  indicated with a star symbol are the UV-dim positions, while the rest are the UV-bright positions (see Section~\ref{sec:redis}).}}}
 \label{tab:mass}
\end{table*}

\subsection{Modelling and the best model identification}

In addition to inferring the physical properties of the molecular gas from the line ratios only, 
we followed a second approach to quantitatively study the 
physical properties of the gas, such as density and temperature. 
We run a radiative transfer code, RADEX \citep{van07}. 
RADEX is a non-LTE radiative transfer code using the large-velocity gradient (LVG)
approximation \citep{sob60, cas70, gol74, jon75}. RADEX yields line intensities as a
function of a set of user-specified parameters: gas kinetic
temperature $T_{\rm K}$, molecular hydrogen number volume density
$n$(H$_2$) and CO number column density per unit line width
$N$(CO)/$\Delta v$. To create the model grids for $^{12}$CO and $^{13}$CO lines, 
we take the $\Delta v$ as an average FWHM; $18.5$ and $17$~kms$^{-1}$ for the CO(1--0) and $^{13}$CO(1-0) lines, respectively. 
Please note that the line widths only minimally affect the model results \citep{van07}.
We kept the $T_{\rm K}$,  $n$(H$_2$) and $N$(CO) as free parameters and created the model grids as follows. 
$T_{\rm K}$ ranges from $5$ to $20$~K in steps of $1$~K to sample the lower temperature 
regime better, and it ranges from $20$ to $250$~K in steps of $5$~K. $n$(H$_2$) ranges from $10^{2}$ to $10^{7}$~cm$^{-3}$ 
in steps of $0.25$~dex, while $N$(CO) ranges from $10^{13}$ to $10^{21}$~cm$^{-2}$ in steps 
of $0.25$~dex. This produces $42,966$ model grids in total for each adopted abundance ratio value (see below).

Stellar evolution and chemical process within the ISM 
can affect the [$^{12}$CO]/[$^{13}$CO] abundance ratio. In our analysis, we considered no difference 
between isotopic ($^{12}$C/$^{13}$C) abundance ratio and isotopologue ($^{12}$CO/$^{13}$CO) abundance ratios.
[$^{12}$C]/[$^{13}$C] abundance ratio ranges widely from the solar neighbourhood to starburst galaxies, 
from $20$ to $90$ \citep{wr94, ag89, hm93a, hm93b}. 
In the Large Magellanic Cloud it is $50$ \citep{w09}, while it is 
[$^{12}$C]/[$^{13}$C]$ > 40$ in starbursts \citep{mar10}. 
Since the abundance ratio shows a radial gradient in galaxies and ranges 
widely among galaxy types, we consider a range for the abundance ratio instead of taking one single value 
for all the positions studied. In our model calculations, the abundance ratio ranges from 
$20$ to $90$ in steps of $10$.

The best model was chosen by applying both $\chi^{2}$ and likelihood approaches. 
For the positions where four lines detected, i.e. 3 line ratios are available, 
reduced $\chi^{2}$ was defined for each set of model parameters as 

\begin{equation}
  \chi_{\rm r}^{2}\equiv\sum\limits_{i}\bigg(\frac{R_{i,{\rm
      mod}}-R_{i,{\rm obs}}}{\Delta R_{i,{\rm obs}}}\bigg)^2\,, 
  \label{eq:chi2}
\end{equation}

where $R_{\rm mod}$ is the modelled line ratio, $R_{\rm obs}$ is the
observed line ratio with uncertainty $\Delta R_{\rm obs}$, and the summation is over 
all independent line ratios $i$ (one fewer than the number of line ratios
available for that position). The model with the smallest $\chi_{\rm r}^2$ (i.e. $\chi_{\rm r, min}^2$)
was taken as the best model representing the observed line ratios best. 
For the positions with at least one line has an upper limit integrated 
intensity, the best model parameters were not obtained, 
as it would only introduce more uncertainty to actual physical conditions 
that we are after.

%
%
%

%
%
\section{Results \& Discussion}
\label{sec:redis}
Although our $22^{\prime\prime}$ beam ($778$~pc) typically 
contains many GMCs, we assume that 
the GMCs within our resolution element have the same average 
physical properties and the CO transitions are emitted 
from the same regions. Studying the GMCs using additional emission lines characteristic of star formation, 
such as UV and infrared (IR) emissions, allows us to probe the physical 
nature of the gas better. The UV radiation and stellar winds, coming
from young O and B stars, can heat the gas and dust, create 
photon-dominated regions (PDRs) between H\,{\small II} regions and molecular clouds 
\citep{hol99}. Also note that FUV radiation from 
young massive stars can heat the outer layers of a GMC 
and increase the gas temperature there, up to $\approx1000$~K. 
However, deeper in the cloud CO could have a lower temperature 
(e.g. due to dust shielding and H$_2$ self-shielding), 
and the hydrogen density should also relatively be higher \citep{hol99}. 
It is, therefore, worth to note that GMC complexes relatively brighter in FUV emission does not necessarily indicate a higher 
average temperature for the CO gas. Regions with active star formation are usually 
associated with strong infrared emission coming from the dust heated by massive stars \citep[e.g.][]{misi04,flo04}, 
and it is known that the $24\micron$ emission is sensitive to the star formation rate \citep[e.g.][]{gor04,cal05}.
However, shock waves from supernova explosions and strong stellar winds 
could make the regions of active star formation inhospitable for dust, 
causing only the foreground dust to survive \citep{cal96}.

The positions from $11$ to $19$, 
i.e. the locations in the northern and southern arms of the galaxy (see Fig.~\ref{fig:n0628}), 
were chosen based on being relatively brighter in the GALEX UV image (hereafter initially selected UV-bright positions)
compared to the positions over the cut (i.e. the positions from $0$ to $10$). 
We calculated the Gaussian weighted total FUV, 
$24\micron$, $70\micron$, and $160\micron$ fluxes at each position as follows. 
After applying unit conversions necessary to Spitzer and GALEX data, we obtained all the data in the unit of Jy. 
We then multiplied the flux in each pixel by a 2D Gaussian weighting function of 
FWHM equal to that of the adopted beam (and centred at each position). 
Finally, the total $24\micron$, $70\micron$, $160\micron$, and FUV fluxes for each position were calculated 
by summing Gaussian weighted fluxes in all pixels in each image. 

Based on the estimated total FUV fluxes, we defined the positions either UV-bright or UV-dim as follows. 
Any positions over the radial cut (i.e. the positions from $0$ to $10$) with an FUV flux lower than the lowest FUV flux found among the initially selected UV-bright positions (i.e. the positions from $11$ to $19$) are defined as the UV-dim positions. As a result, all positions outside the radial cut and the positions $0$, $2$, $7$, and $8$ are considered as UV-bright positions while the rest, namely the positions $1$, $3$, $4$, $5$, $6$, $9$, and $10$ are considered as UV-dim positions. The discussions will be made based on this grouping in the upcoming sections.

As seen from Figure~\ref{fig:n0628}, the UV-bright positions include more H\,{\small II} complexes 
compared to the UV-dim positions, and most positions brighter in the $24\micron$ image are also brighter in the UV. 
These observational characteristics, therefore, indicate that the star formation 
activity could be stronger at the UV-bright positions than that of the UV-dim positions. 
This allows us to probe the effects of UV radiation on molecular clouds. However, the effects of extinction to the observable 
UV should first be considered to understand better the reason(s) behind the difference in the 
UV-brightness seen among the positions studied (see Figure~\ref{fig:firfuv}). 
Additionally, probing the line ratios at the positions chosen side by side over the SE-NW cut enables us to study 
any radial variations in physical conditions of the gas from the centre to the outskirt of the galaxy without selection bias.

\subsection{Intensities and molecular line ratios}
\label{sec:ratios}

\subsubsection{Intensities}
\label{sec:intenco}

The change in the velocity-integrated CO line intensities and related 
line ratios across the positions studied
are shown in Figure~\ref{fig:ratios}. As seen from panel $a$ of Figure~\ref{fig:ratios}, 
the central region is the brightest in CO emission. 
The CO integrated line intensities show a decrease, with some fluctuations, 
 as a function of radius from the centre of the galaxy up to $165^{\prime\prime}$ or a linear distance of about $5.8$~kpc 
 (see panel $a$ in Figure~\ref{fig:ratios} and Table~\ref{tab:intensities}). 
 When we consider the positions over 
 the SE-NW cut only, the decrease is more clearly seen on both sides of the cut (see panel $b$ in Figure~\ref{fig:ratios})
 The CO(1-0) integrated intensities at some UV-bright positions 
 show an increase in respect to the general trend, causing more fluctuations in the ratios after about $70^{\prime\prime}$ 
from the centre (see panel $a$ of Fig.~\ref{fig:ratios}).

\subsubsection{$^{12}$CO / $^{12}$CO ratio}
\label{sec:ratioco}

The CO(1--0) has an upper-level energy temperature of $\approx5.5$~K, while 
it is $\approx16.5$~K and $\approx33$~K for CO(2--1) and CO(3--2) 
respectively. Since CO(3--2) requires a denser environment compared to 
CO(1--0) and CO(2--1), it traces not only warmer but also slightly more compact regions of the gas cloud. 
As seen from panel $c$ of Figure~\ref{fig:ratios} and Table~\ref{tab:ratios}, 
the $R_{13}$ ratios across the entire disc are always higher than the $R_{12}$ ratios indicating 
that CO(2--1) transition is brighter than CO(3--2) at all positions studied. 
Additionally, the $R_{13}$ ratios show larger fluctuations than that of 
the $R_{12}$ ratios. 

The $R_{12}$ ratios range from $1$ to $2$ at all positions studied except the positions $3$, $10$, and $18$, 
where the ratio could slightly get less than $1$. The average $R_{12}$ ratio over the disc of NGC~0628 
is $1.51\,\pm\,0.02$ (excluding the lower limits). If we consider the UV-bright positions only, the average $R_{12}$ ratio is 
$1.59\,\pm\,0.02$, while it is $1.36\,\pm\,0.04$ at the UV-dim positions (excluding the lower limits). 
These results indicate that the UV-dim positions have a lower $R_{12}$ ratio (relatively warmer gas) 
compared to that of the UV-bright positions (relatively colder gas).

While we detected CO(1--0) at all $20$ positions studied, the CO(3--2) was detected at $15$ positions over the galaxy, 
i.e. all UV-dim positions and $8$ UV-bright positions (see Table~\ref{tab:intensities} 
and panel $a$ in Figure~\ref{fig:n0628}). As a result, 
the $R_{13}$ ratio was obtained at $75\%$ of the positions studied, while the $R_{13}$ ratios in 
the remaining $5$ positions are just a lower limit. 
The range for the $R_{13}$ ratios is wider ($2 \le R_{13} \le 10.7$) than that for the $R_{12}$ ratios ($1 \le R_{12} \le 2$). The $R_{13}$ ratios at all positions studied are higher than $2$, except positions $1$, $8$ and $16$ where the ratios could get lower, given the error bars (see Table~\ref{tab:ratios}). The average $R_{13}$ ratio over the disc of NGC~0628 is $4.3\,\pm\,0.1$ (excluding the lower limits). 
The average ratio at the UV-bright positions is $R_{13} = 4.74\,\pm\,0.28$ while it is $3.76\,\pm\,0.12$ at the UV-dim positions. 
The lower average ratio at the UV-dim positions indicates the existence of relatively warmer 
gas compared to that of the UV-bright positions.

 The $R_{12}$ line ratio in the Milky Way and nearby galaxies 
 typically spans between $1 - 1.5$ \citep[e.g.][]{hase97, ler09}, similar to 
 the range we found in the disc of NGC~0628. \citet{mao10} obtained $R_{13}$ ratios 
 in the central region of $61$ galaxies (i.e., spirals, starbursts, Seyferts, 
 and luminous infrared galaxies) and found no correlation for $R_{13}$ ratio to Hubble types. 
 The range of $R_{13}$ ratios that we found in the disc of NGC~0628 is wider than that of $61$ galaxies 
 ($0.5 \le R_{13} \le 5$; \citealt{mao10}). In their sample, there are ten normal spirals with $R_{13} > 1$, while starbursts 
 can have a ratio as low as $0.5$ \citep{mao10}. The $R_{13}$ ratios 
 across the disc of NGC~0628 are, therefore, similar and even higher 
 compared to those found in the centre of normal spirals. 
 Since the lower the $R_{13}$ ratios, the more enhanced the molecular excitations, the positions we studied 
 in the disc of NGC~0628, have similar or even colder ISM 
 concerning the centre of normal spirals, and particularly that of starbursts. 

As discussed above, the UV-bright positions (hosting more H\,{\small II} regions) 
in NGC~0628 have higher $R_{12}$ and $R_{13}$ ratios 
(lower temperature) compared with 
the UV-dim positions over the cut, where the feedback from massive stars seems to affect the 
physics of the GMCs relatively weakly (see panel $c$ of Figure~\ref{fig:n0628}). 
This indicates the existence of colder gas 
in GMCs surrounded by more H\,{\small II} regions at the UV-bright positions. 
However, at the UV-dim positions, there is slightly warmer gas bathed in 
GMCs with less number of H\,{\small II} regions.

\subsubsection{$^{12}$CO / $^{13}$CO ratio}
\label{sec:ratio13co}

In molecular clouds, $^{13}$CO is generally less abundant and optically thin compared to 
its parent molecule $^{12}$CO. The $R_{11}$ ratios, therefore, trace diffuse gas, 
i.e. the larger the $R_{11}$ ratio, the thinner the CO gas (and vice-versa). 
Feedbacks from star formation activities, such as supernova explosions, strong UV radiation from massive stars and 
H\,{\small II} regions could make the gas more diffused. As a first thought, we might, therefore, expect to 
find higher $R_{11}$ ratios, so more diffused gas at the UV-bright positions where star formation activity seems 
to be stronger compared to the UV-dim positions.

\begin{figure*}
  \includegraphics[width=8.0cm,clip=]{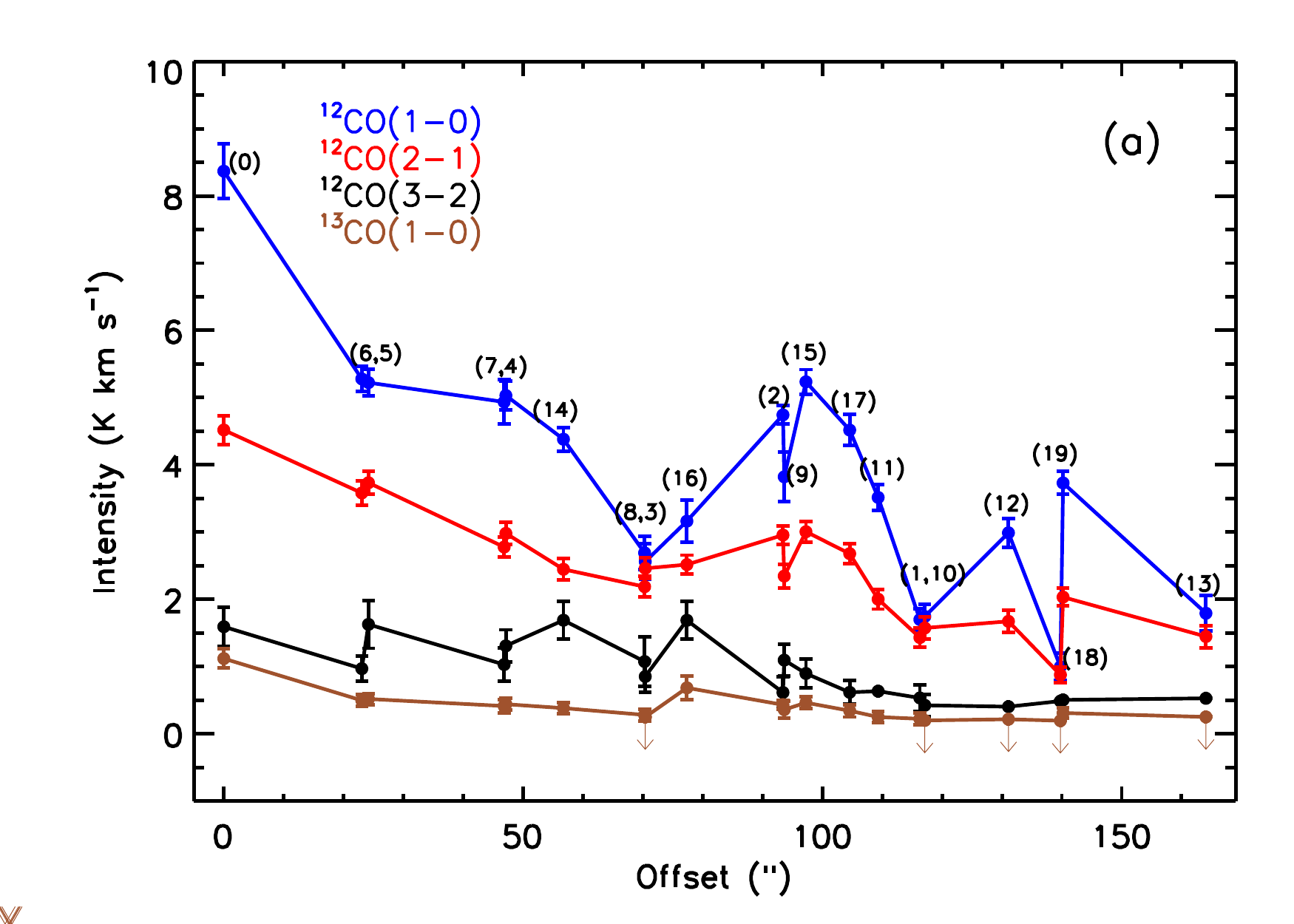}
  \includegraphics[width=8.0cm,clip=]{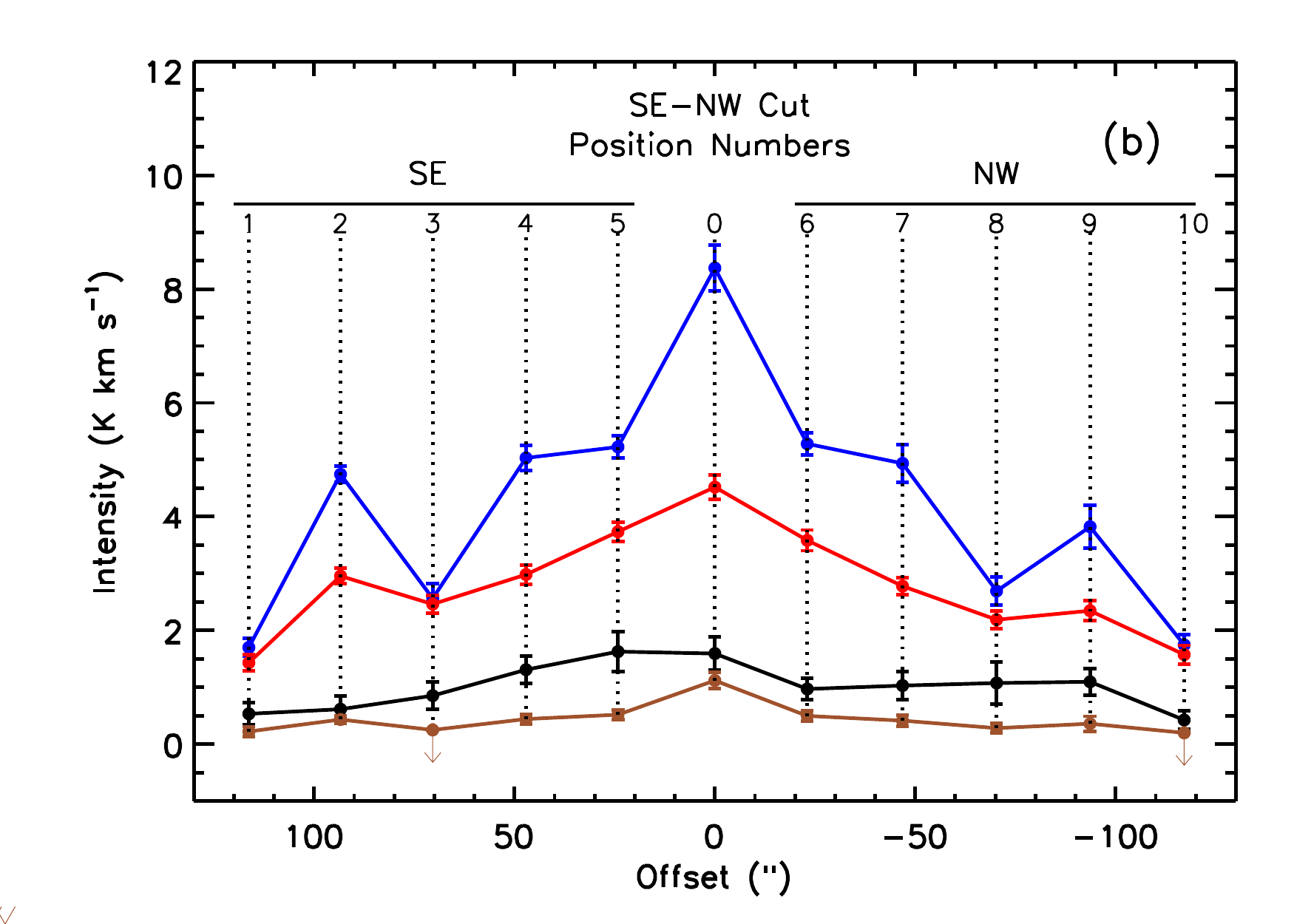}\\
   \includegraphics[width=8.0cm,clip=]{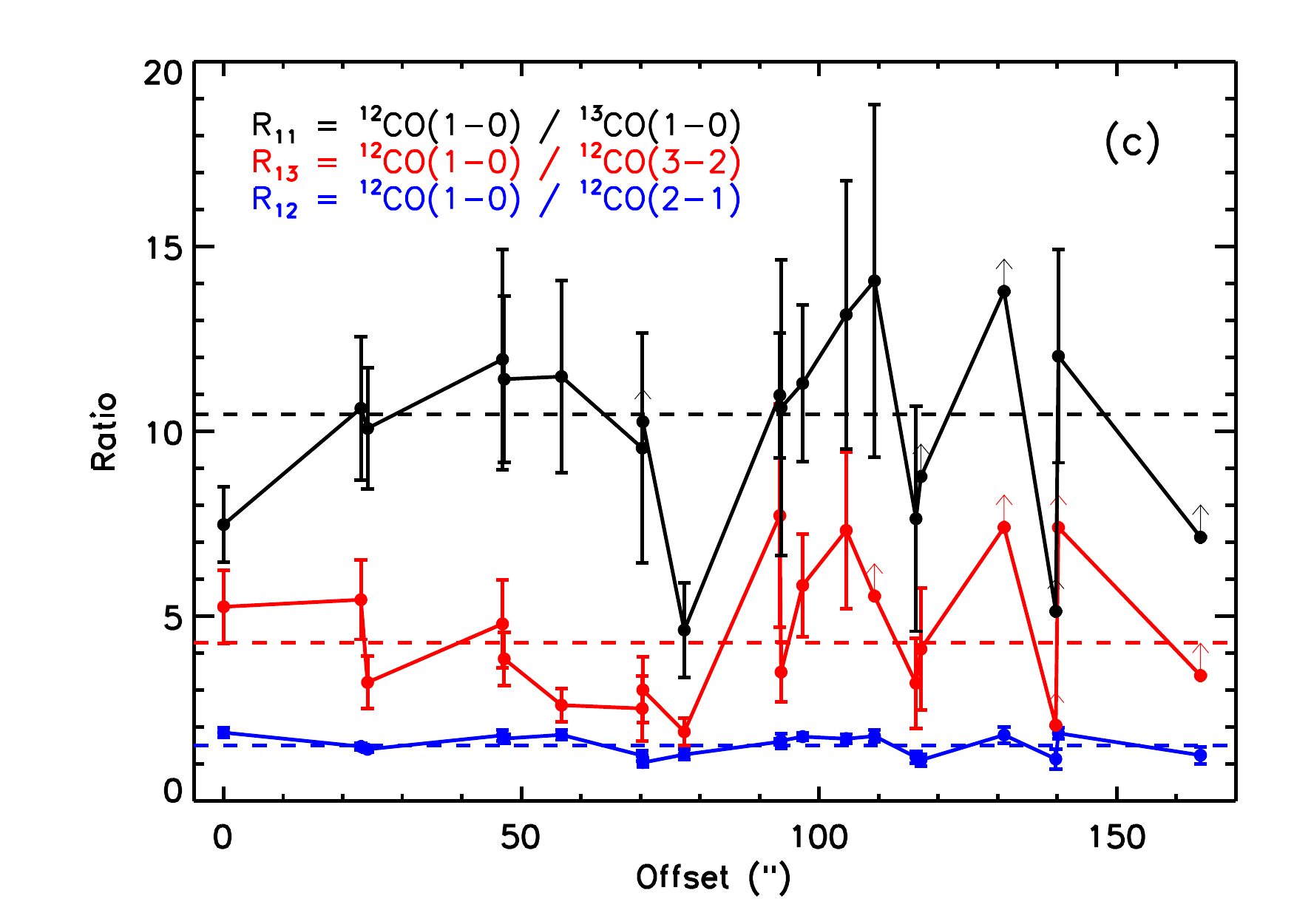}
   \includegraphics[width=8.0cm,clip=]{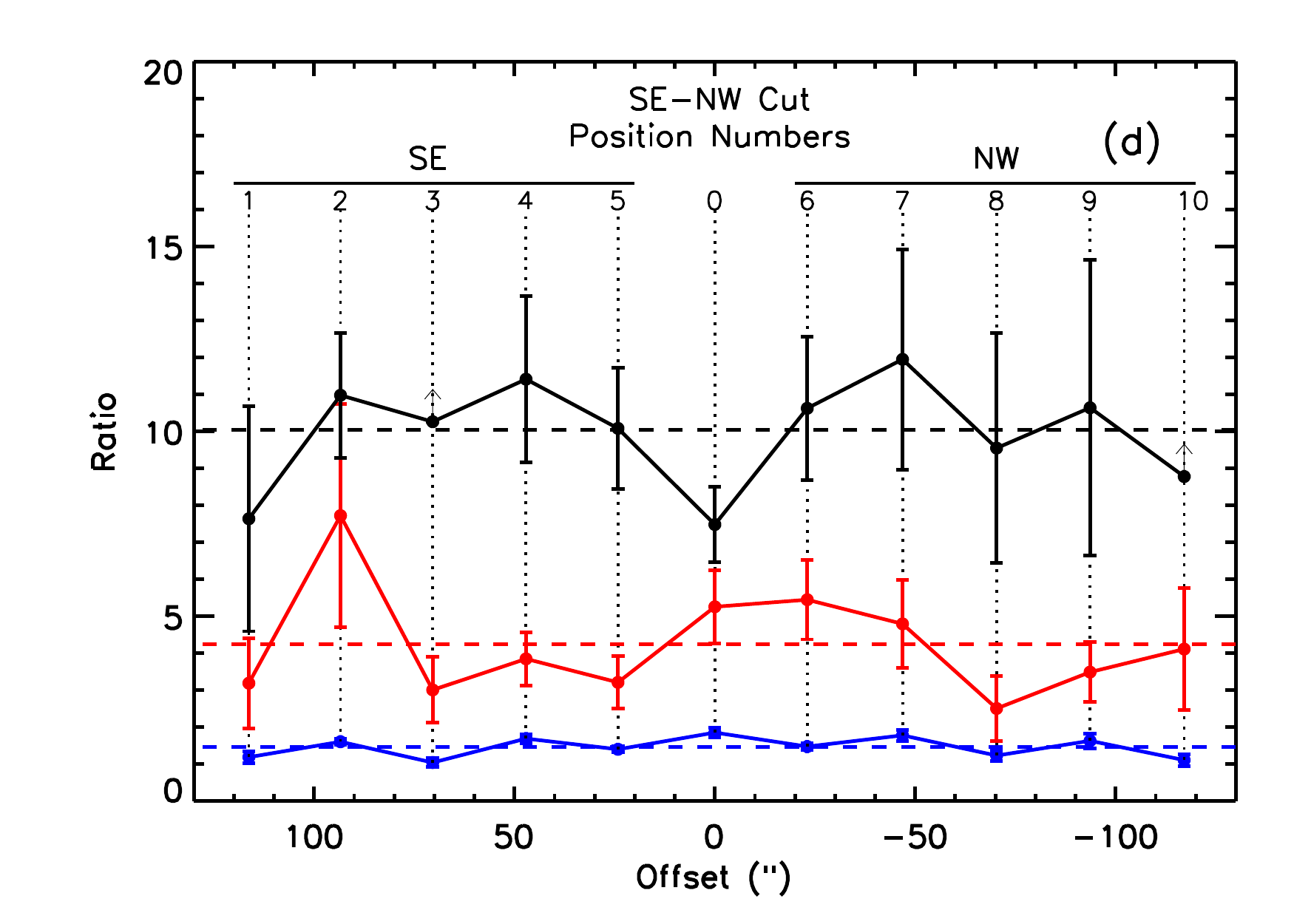}
  \caption{{\bf Panel a:} The integrated intensities at all positions are shown as a function of distance 
  from the galaxy centre. The position numbers (also listed in Table~\ref{tab:intensities})
  are also indicated in parentheses. {\bf Panel b:} The same as panel $a$ but for the positions 
  along the SE-NW cut only. The integrated intensities 
  across the SE-NW cut are shown (please see panel $a$ in Figure~\ref{fig:n0628} and Table~\ref{tab:intensities}). 
  {\bf Panel c:} The line ratios obtained at all positions along the disc of NGC~0628 are shown as a function 
  of distance from the galaxy centre. From the centre to the most distant position observed, the same 
  position numbers applied, as shown in the top left panel. The average value for each group of line ratios 
  (excluding the lower limits, see Table~\ref{tab:ratios}), namely $R_{12}$, $R_{13}$ and $R_{11}$, is also shown with dashed lines. {\bf Panel d:} The same as the panel $c$ but for the positions along the SE-NW cut only. The position numbers are also indicated for both SE and NW parts of the cut.}
  \label{fig:ratios}
\end{figure*}

As seen from Figure~\ref{fig:ratios}, the $R_{11}$ ratios show more fluctuations 
as a function of radius compared to that of $R_{12}$ and $R_{13}$ ratios. 
As listed in Table~\ref{tab:ratios}, the average $R_{11}$ ratio over the disc of 
NGC~0628 is $10.5\pm0.2$ (excluding the lower limits). The average $R_{11}$ ratio is $10.66\pm0.28$ at the UV-bright positions, while it is $10.08\pm0.29$ at UV-dim positions (excluding the lower limits), indicating a slightly higher ratio, so more diffused gas, at the UV-bright positions. Please note that the UV-bright positions include significantly more H\,{\small II} regions and brighter at $24\micron$ emission compared to the UV-dim positions (see Figure~\ref{fig:n0628}). It is, therefore, natural to suggest the existence of a high level of star formation activity and in turn, dominant effects of massive stars in the ISM of the UV-bright positions. Thereupon the intense star formation activity could cause the high $R_{11}$ ratio, so the more diffused gas, seen at the UV-bright positions.

The $R_{11}$ ratio increases up to about $50^{\prime\prime}$ (or equally $\approx1.7$~kpc) 
and then starts to fluctuate (see panel $c$ Fig.~\ref{fig:ratios}). 
However, the $R_{13}$ ratio steadily decreases up to about $50^{\prime\prime}$ as opposed to the $R_{11}$ ratio; 
it then starts to fluctuate. A closer look at the ratios on both sides of the SE-NW cut reveals the same trend; 
while the $R_{11}$ ratio increases up to about $50^{\prime\prime}$, the $R_{13}$ 
ratio decreases (please see panel $d$ in Figure~\ref{fig:ratios}).  
This indicates that the gas is getting thinner and warmer
from the centre through the outskirt up to $1.7$~kpc. 
However, the decrease in $R_{13}$ ratios through the NW part of the cut is happening more 
slowly compared to the SE part of the disc (see panel $d$ of Figure~\ref{fig:ratios}).

The range of $R_{11}$ ratios in NGC~0628 is $5 \le R_{11} \le 19$, given the error bars (see Table~\ref{tab:ratios}). 
This range is similar to the range found for the centre of spirals  
($5 \le R_{11} \le 20$, \citealt{pag01}) and Seyferts ($8 \le R_{11} \le 17$, \citealt{hen94,mei08,is09a,is09b}).
However, the range of $R_{11}$ ratios in NGC~0628 is narrower than what is found for the 
centre of lenticular galaxies ($3 \le R_{11} \le 30$; \citealt{kri10}, \citealt{cr12} and \citealt{topal16}) 
and starbursts ($10 \le R_{11} \le 33$; \citealt{aa95} and \citealt{ba08}). This supports the idea that 
when the star formation is high, the gas gets thinner, so the $R_{11}$ ratio gets higher, 
as we usually see in starbursts and some lenticulars with an unexpectedly high amount of 
molecular gas. However, although we found a narrower range for $R_{11}$ ratios compared to that of 
starbursts and some lenticulars, there are essentially some starbursts and lenticulars with similar line ratios 
at their centres compared to what we found in the disc of NGC~0628. 
These results lead us to the following conclusion. 
The level of star formation activity and feedbacks should be the main driver 
defining the line ratios, regardless of galaxy types, i.e. positions in the arms or inter-arms 
of any types of galaxies may reveal similar molecular line ratios to that found at different 
positions over different types of galaxies. However, the central region of different types of galaxies 
could have different line ratios as the physics of the ISM could be driven by a different phenomenon 
in their centres, e.g. the existence of an active galactic nucleus (AGN). 
The $R_{11}$ ratio at the centre of NGC~0628 is $7.48\pm1.02$ (see Table~\ref{tab:ratios}), which is indeed different and lower than the ratios seen in starbursts \citep[e.g.][]{aa95,ba08}.
%
%
%

\subsection{Total H$_{\rm 2}$ mass and extinction}
\label{sec:massuv}
\begin{figure*}
 \includegraphics[width=7.0cm,clip=]{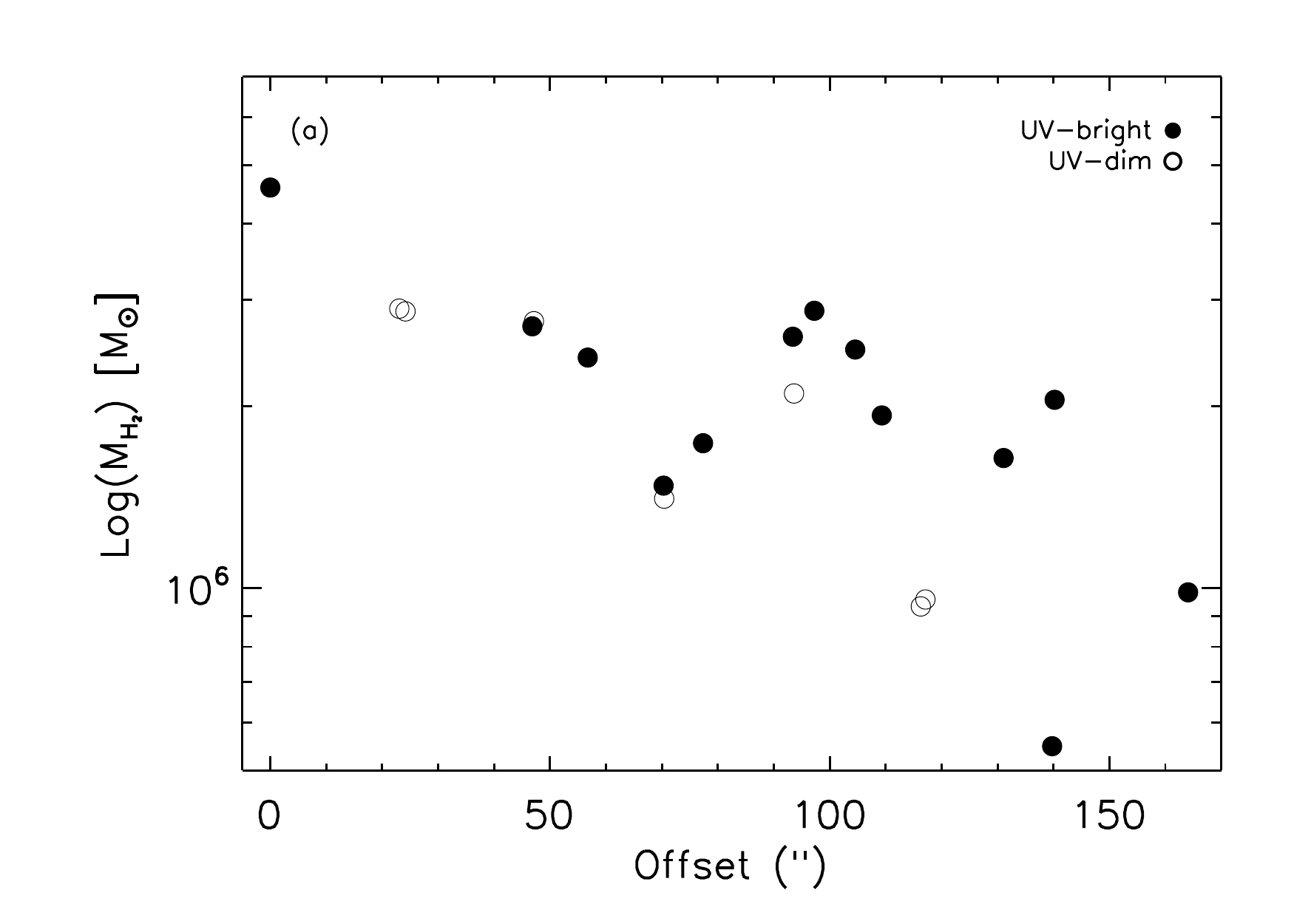}
 \hspace{-10pt}
  \includegraphics[width=7.0cm,clip=]{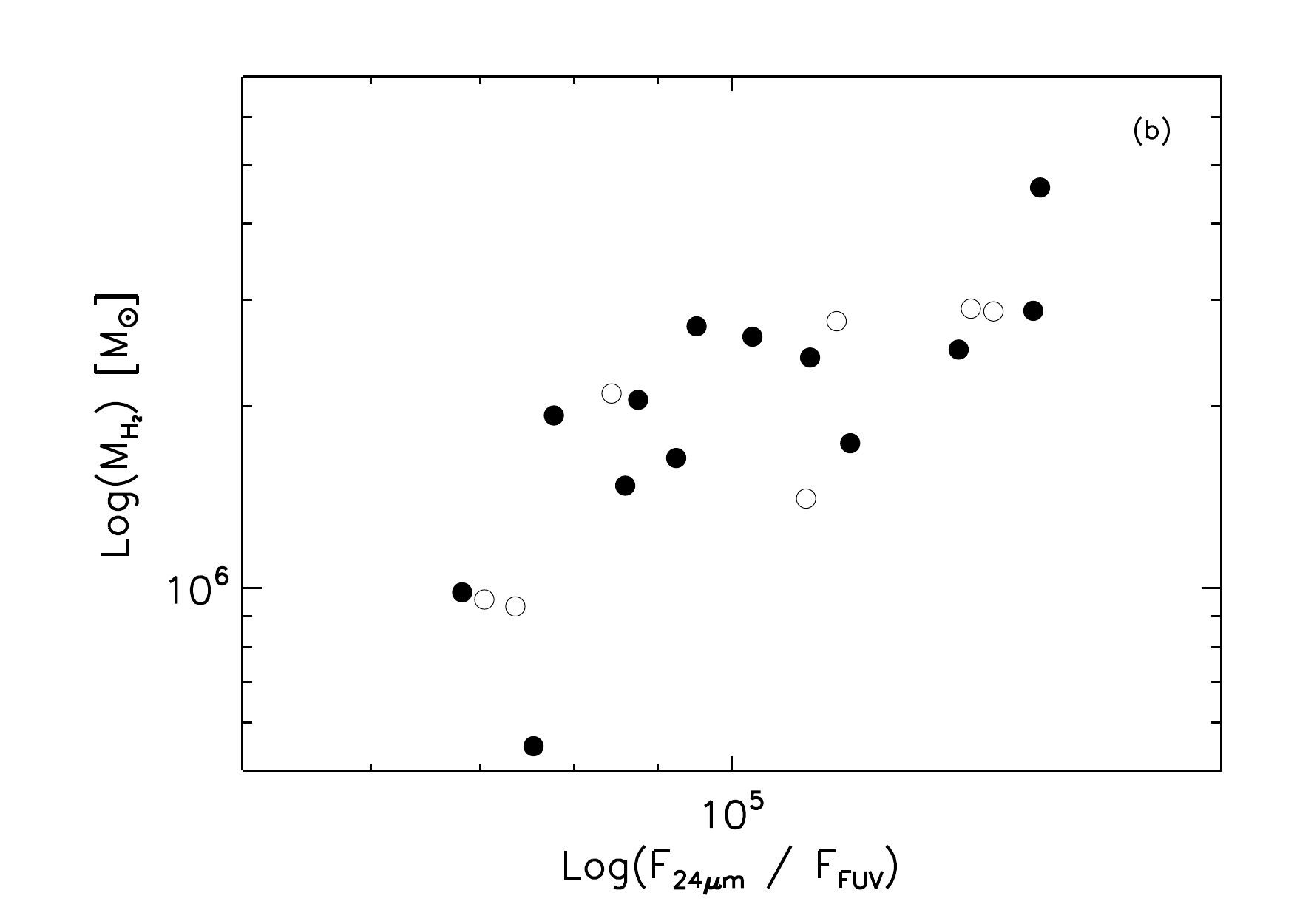}\\
  \includegraphics[width=7.0cm,clip=]{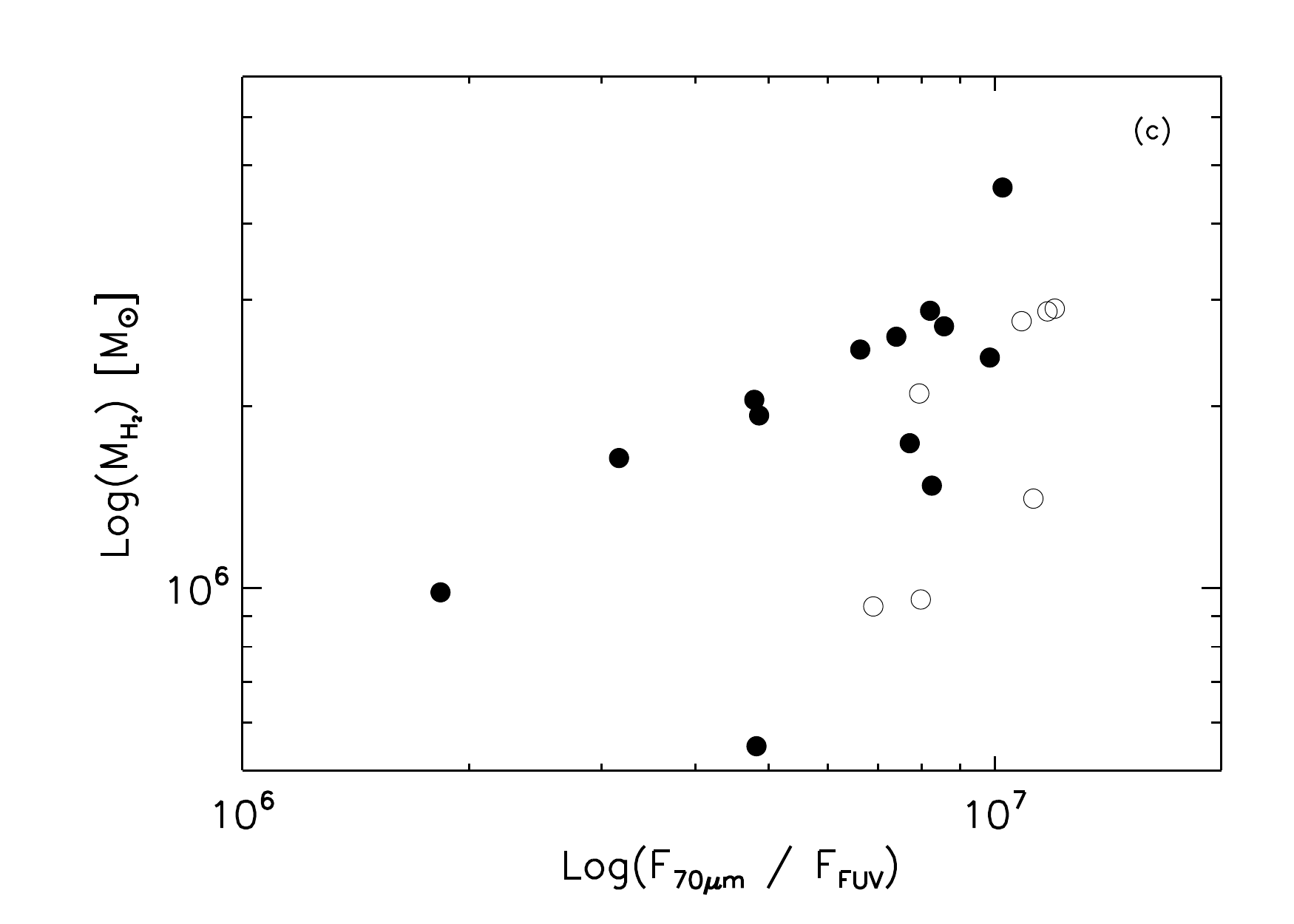}
  \hspace{-10pt}
  \includegraphics[width=7.0cm,clip=]{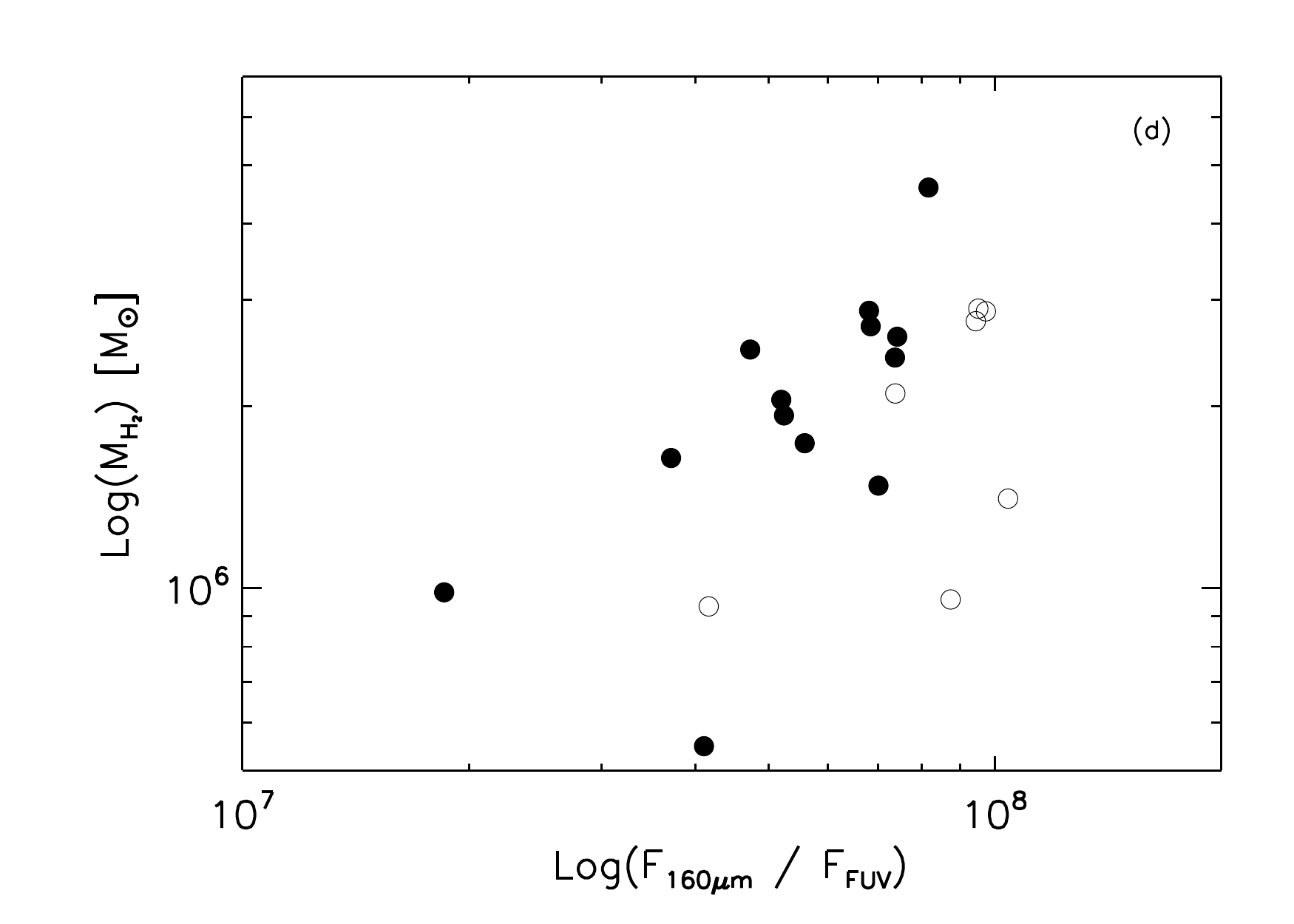}\\
  \caption{{\bf Panel $a$}: The beam-averaged total molecular gas mass as a function of distance from the galactic centre is shown. 
  {\bf Panels $b$, $c$ and $d$}: The IR-to-FUV ratio as a function of molecular gas mass is shown. The open circles represent the UV-dim positions while the filled circles represent the UV-bright positions (see the text).}
  \label{fig:mirfuvmass}
\end{figure*}

Because of its low critical density ($n_{\rm crit} \approx 10^{3}$~cm$^{-3}$), CO is generally 
thought to be a good tracer for the total gas mass in galaxies. 
As seen in Table~\ref{tab:mass}, the central region of NGC~0628 has the highest 
amount of molecular gas confined within the adopted beam of $22^{\prime\prime}$. 
The average total gas mass and gas surface density
at the UV-bright positions are $M_{\rm H_{2}} = 21.6\,\pm\,0.8\times10^{5}~M_{\sun}$ (with a median value 
of $M_{\rm H_{2}} = 20.5\,\pm\,0.5\times10^{5}~M_{\sun}$) 
and $\Sigma_{H_{2}} = 17.1\,\pm\,0.6$~$M_{\sun}$~pc$^{-2}$ 
(with a median value of $\Sigma_{H_{2}} = 16.2\,\pm\,0.7$~$M_{\sun}$~pc$^{-2}$), respectively. 
On the other hand, average molecular gas mass and gas surface density at the UV-dim positions are 
$M_{\rm H_{2}} = 19.9\,\pm\,1.3\times10^{5}~M_{\sun}$ (with a median value of $M_{\rm H_{2}} = 21.0\,\pm\,1.2\times10^{5}~M_{\sun}$) and $\Sigma_{H_{2}} = 15.8\,\pm\,1.0$~$M_{\sun}$~pc$^{-2}$ (with a median value 
of $\Sigma_{H_{2}} = 16.6\,\pm\,1.6$~$M_{\sun}$~pc$^{-2}$), respectively. 
These results indicate that the UV-bright and UV-dim positions have similar molecular gas mass and gas surface density.
This is true even after excluding the positions $0$ and $18$, where the beam-averaged total molecular 
gas mass has the highest and the lowest values, respectively. 
Panel $a$ of Figure~\ref{fig:mirfuvmass} indicates that $M_{\rm H_{2}}$ decreases as a function of galactocentric radius 
following the same pattern as seen in the integrated intensities (see panel $b$ of Figure~\ref{fig:ratios}). The same trend is seen for $\Sigma_{H_{2}}$ and $N_{\rm H_{2}}$ as a function of radius (see Table~\ref{tab:mass}). The panels $b$, $c$, and $d$ of Figure~\ref{fig:mirfuvmass} indicate that there is a linear correlation between the beam-averaged total molecular gas mass and extinction (i.e. the IR-to-FUV ratio): as the IR-to-FUV ratio increases the molecular gas mass also increases. It is also worth to note that the metallicity can strongly affect the CO-to-H$_{2}$ conversion factor used to estimate the total beam-averaged H$_{2}$ mass \citep{glo12}. Effects of both shielding and metallicity, which are not within the 
scope of this study, play an essential role in determining the molecular gas content of GMCs.

\begin{figure*}
  \includegraphics[width=7.0cm,clip=]{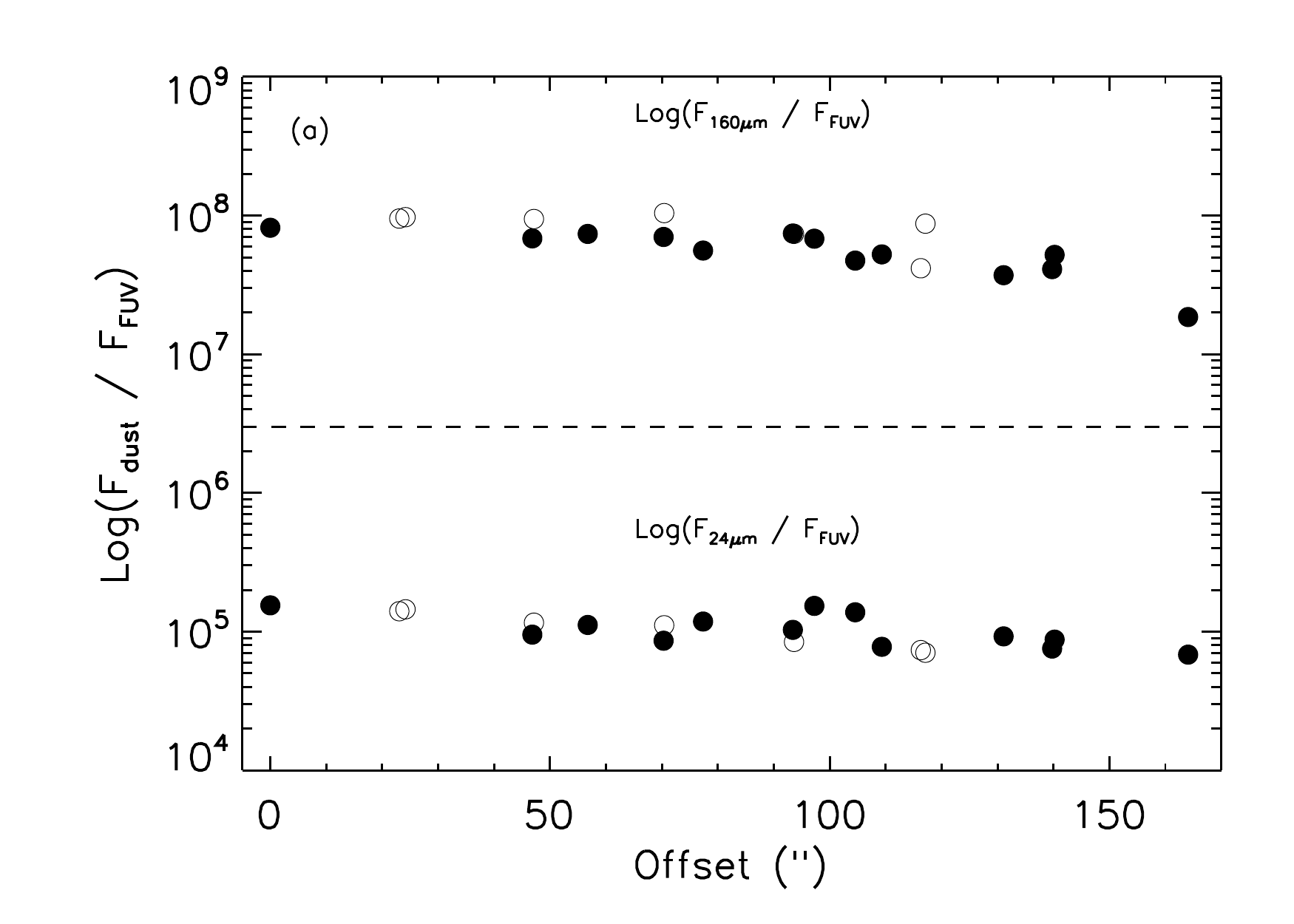}
   \hspace{-10pt}
  \includegraphics[width=7.0cm,clip=]{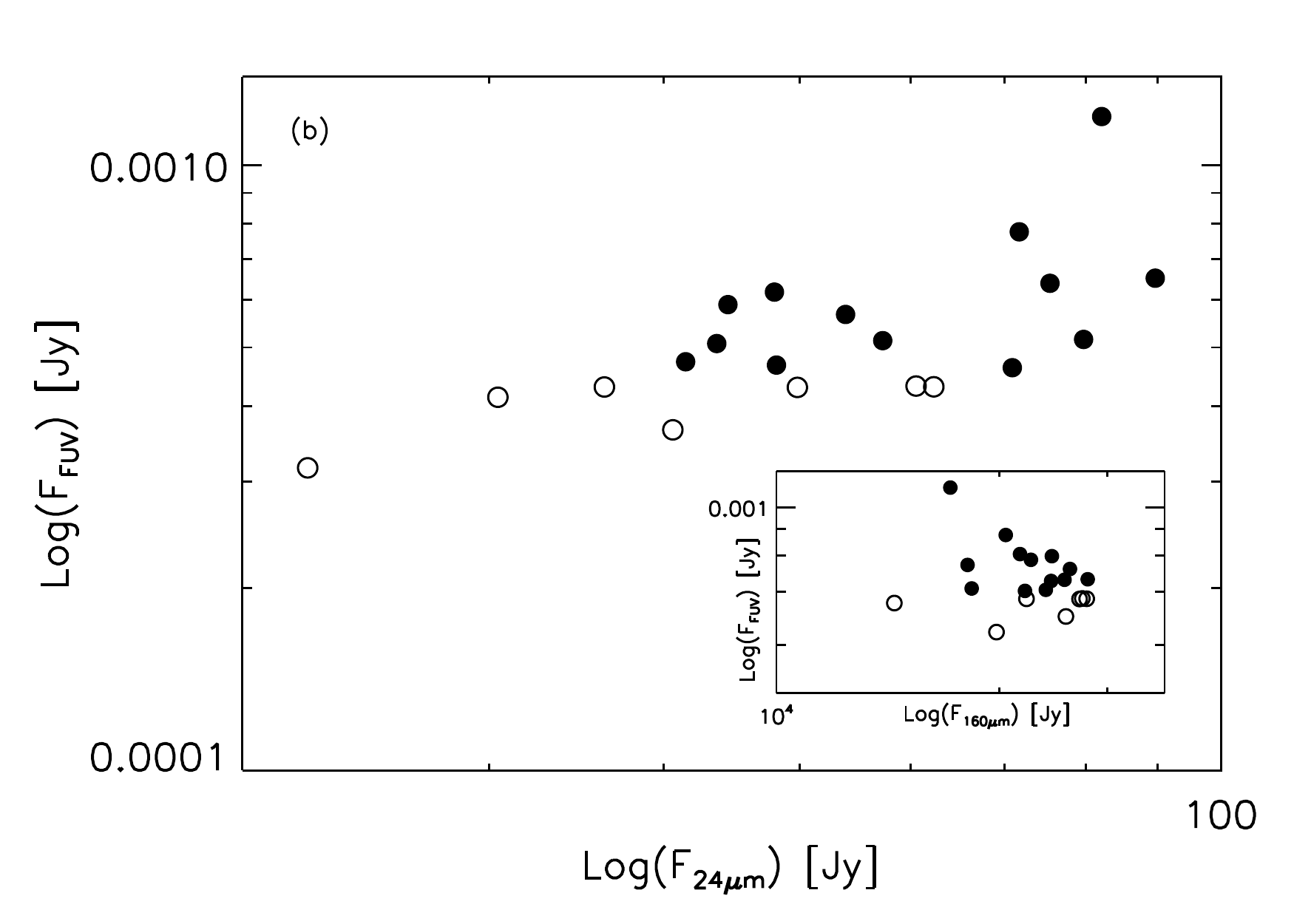} \\
  \includegraphics[width=7.0cm,clip=]{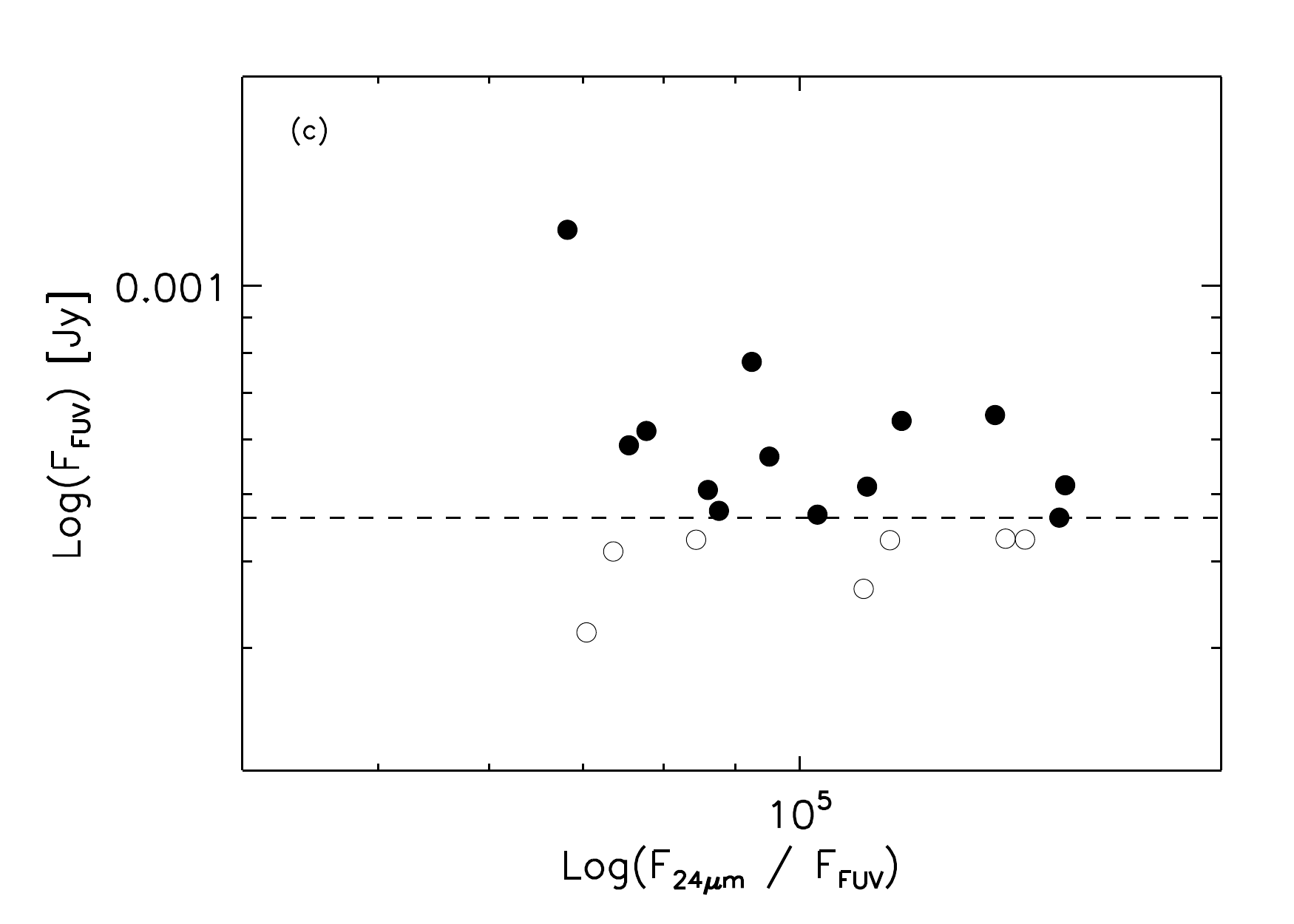}
  \hspace{-10pt}
   \includegraphics[width=7.0cm,clip=]{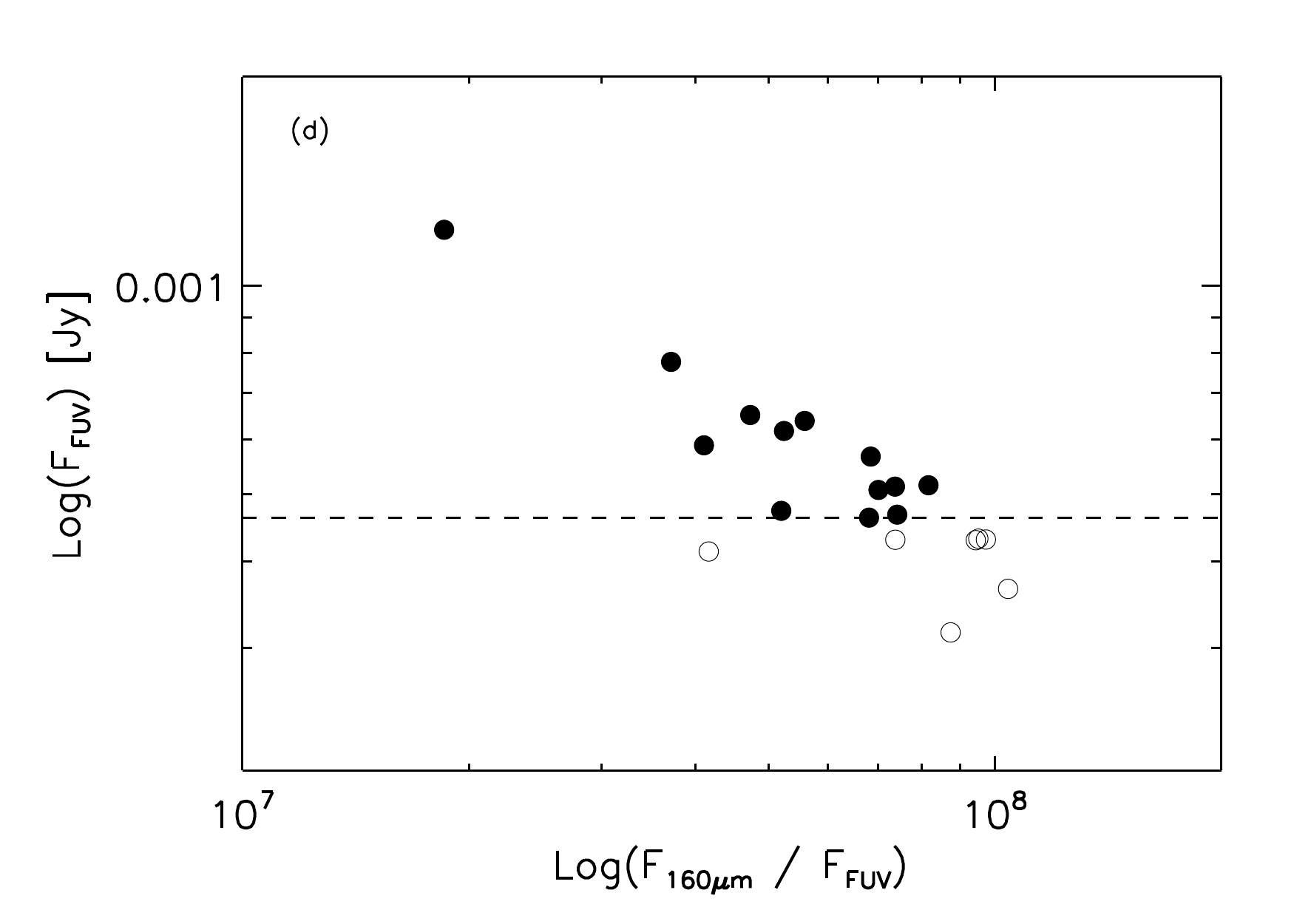}\\
  \caption{{\bf Panel $a$}: The extinction as a function of distance from the galactic centre. 
  {\bf Panel $b$}:  The beam-averaged $24\micron$ and $160\micron$ (the embedded figure) fluxes are shown as a function of FUV flux. {\bf Panels $c$ and $d$}: The extinction for $24\micron$ and $160\micron$ emission is shown as a function of FUV flux. The dashed line shows the position with the lowest FUV flux among the initially selected UV-bright positions located outside the cut (see the text). The filled symbols represent the UV-bright positions while the open symbols stand for the UV-dim positions in all panels.}
  \label{fig:firfuv}
\end{figure*}

As seen from panel $a$ of Figure~\ref{fig:firfuv}, the IR-to-FUV ratio decreases as a function of the galactocentric radius, similar to the molecular gas mass. As revealed by panel $b$ of Figure~\ref{fig:firfuv}), there is a linear correlation between UV-brightness and $24\micron$ emission, and most positions brighter at $24\micron$ emission are the UV-bright positions. However, both groups of positions share a similar range for $160\micron$ fluxes (see the embedded figure in panel $b$ of Figure~\ref{fig:firfuv}). The panels $c$ and $d$ of Figure~\ref{fig:firfuv} show that as the UV-brightness decreases, the IR-to-FUV ratio increases, considering the all positions studied. Since UV-bright positions are brighter at both UV and $24\micron$ emission, and UV-dim positions are the opposite, both groups of positions share a similar range for $F_{24\micron}/F_{\rm FUV}$ ratios (see panel $c$ of Fig.~\ref{fig:firfuv}). However, the $F_{160\micron}/F_{\rm FUV}$ ratio is higher at almost all UV-dim positions (panel $d$ of Figure~\ref{fig:firfuv}). The reason for this difference in the ratio is that the UV-bright positions have stronger FUV radiation field, so more young massive OB stars in their ISM (also indicated by stronger $24\micron$ emission and hosting more H\,{\small II} regions), compared to the UV-dim positions. A high value of $F_{160\micron}/F_{\rm FUV}$ could indicate two possible scenarios. Firstly, the extinction at the UV-dim positions could be dominant and absorb the FUV, and thus depressing the $F_{\rm FUV}$. Secondly, star formation efficiency at the UV-dim positions could be lower, causing weaker UV radiation field. The second scenario is a more plausible explanation for the UV-dim positions. $24\micron$ emission is a good tracer for warm gas with temperature approaching $100$~K (possibly heated by star formation). The weakness in $24\micron$ emission, therefore, indicates that there is not much FUV (so not much star formation) to make the dust warm enough at the UV-dim positions. This could cause the $F_{160\micron}/F_{\rm FUV}$ ratio to be higher at those positions compared to the UV-bright ones (see panel $d$ of Fig.~\ref{fig:firfuv}). On the contrary, the UV-bright positions (which are brighter at $24\micron$ emission and include much more H\,{\small II} regions) have lower $F_{160\micron}/F_{\rm FUV}$ ratios, as expected given the higher FUV fluxes they have. All this evidence lead us to the following conclusion. The reason for the difference in the UV-brightness seen between these two groups of positions, namely UV-bright and UV-dim positions, is not likely the extinction, but having a different level of star formation activity. The IR-to-UV ratio is also sensitive to metallicity, luminosity, and some other factors \citep{dale07}, but these factors are not within the scope of the present study.

\subsection{Modelling}
\label{sec:model}

The present model results are the best we can do with the 
current data. Nevertheless, we discuss the usage of a single-component model for the ISM where many phenomena are at play, 
such as magnetic fields, gas self-gravity, turbulent velocity field, 
UV radiation field, cosmic rays, dust shielding and gas self-shielding. 
First and foremost, not only higher transitions 
of CO, i.e. CO(4--3) and higher, but also multiple lines of high-density tracers, 
such as HCN, HCO$^{+}$, HNC, and HNCO are required to probe denser and warmer components of the gas complexes better \citep[e.g.][]{topal16}. 
Secondly, having multiple lines of such molecules is particularly important as some studies show 
that divergence could be present in the spectral line energy distributions (SLEDs) after $J = 4-3$, while it usually shows a similar trend up to $J = 3-2$, indicating different gas conditions and dense gas fractions \citep[e.g.][]{papa14}. 
Finding the most likely SLED representing the observed line ratios best, therefore, highly depends on the availability of high-$J$ transitions. Consequently, given the complex nature of the gas clouds, applying a single-component model with low-$J$ transitions of CO molecule using RADEX (or similar radiative transfer codes) can only give us some insights on the average physical conditions of relatively colder and less dense gas in the ISM. Although these caveats are important to point out, the present study provides valuable information on the physical nature of the relatively colder and less dense gas over the disc of NGC~0628 for future references. Keeping these caveats in mind, we now discuss the model results in the following.

$13$ out of $20$ positions were investigated by applying 
the LVG modelling. $8$ out of $13$ positions are UV-bright positions 
(i.e. the positions $0$, $2$, $7$, $8$, $14$, $15$, $16$ and $17$), 
while the rest are UV-dim positions located over the cut (i.e. the positions $1$, $4$, $5$, $6$, and $9$). 
Our LVG model results, based on both $\chi^{2}$ and likelihood methods, 
are listed in Table~\ref{tab:modelr}, $\Delta\chi^{2}$ contour maps are shown in Figure~\ref{fig:chi}, and 
the probability distribution functions (PDFs) for each model parameter are shown in Figure~\ref{fig:like}.

As seen from Figure~\ref{fig:chi}, the $\Delta\chi^{2}_{\rm r,min}$ contour maps reveal 
the classical banana-shaped degeneracy, i.e. inverse $n$(H$_2$)-$T_{\rm K}$ 
relationship \citep[e.g.][]{van07, topal14, papa14, topal16}. The most PDFs reveal either 
a half or a full single-peaked Gaussian profile, 
resulting in a good agreement between the peak and the median of the likelihood (see Figure~\ref{fig:like}). 
$T_{\rm K}$ and $n$(H$_2$) mostly show a half-Gaussian PDF, 
i.e. a peak at low values followed by a gradual decrease in the PDF as the values increase, 
while $N$(CO) mostly shows a complete Gaussian PDF (see Figure~\ref{fig:like}). 
As a result, the best-fitting model (in a $\chi^{2}$ sense) is mostly contained within the $68$ per cent ($1\sigma$) confidence levels for all three parameters regardless of the shape of the PDFs, except a few positions (see Table~\ref{tab:modelr} and Figure~\ref{fig:like}).

The range for the $T_{\rm K}$ is large at some positions across the disc of NGC~0628. 
This is partially due to the classical banana-shaped degeneracy seen in the $\chi^{2}$ contour maps and flatness seen in the PDFs 
(see Figure~\ref{fig:chi} and Figure~\ref{fig:like}), causing large uncertainties in the model results at 
some positions (see Table~\ref{tab:modelr}). However, the limit values can still provide us with valuable information on possible differences in temperature from one position to another. The results from both $\chi^{2}$ and likelihood methods indicate that 
$T_{\rm K}$ at position $14$ is not higher than $30$~K, and it could get even lower at position $17$. 
The temperature at these UV-bright positions could be as low as $5-6$~K, making them 
the coldest among the positions studied. However, the temperature at UV-dim positions $4$, $5$, and $9$, 
and UV-bright position $7$ is always higher than $30$~K, making them the hottest compared to the other positions studied. 
The UV-bright positions $0$ (the centre) and $15$ reveal a similar range for the $T_{\rm K}$; 
ranging from $20$ to $100$~K, while the range for UV-dim positions 
$1$ and $9$ could be wider; $20$ to $200$~K. The temperature at the remaining 
UV-bright positions $2$, $8$ and $16$, and the UV-dim position $6$, 
where the results from both methods do not overlap, 
indicates an even wider range from about $10$ to $200$~K, 
making these positions more uncertain in terms of temperature. Even though the model 
results suggest larger uncertainties in the temperature at some positions, the empirical 
(see Section~\ref{sec:ratioco}) and model results are mostly in agreement; 
the UV-bright positions mostly have lower temperature 
compared to that of the UV-dim positions. However, more positions in the disc of the galaxy should 
be studied to make a firmer conclusion.

$n$(H$_2$) and $N$(CO) also show large variations throughout the disc of the galaxy. 
The best model and most likely model results do not overlap at positions $1$, $4$, $7$ and $9$ for $n$(H$_2$), 
and positions $2$, $8$ and $16$ for $N$(CO). Except for these positions, the best model results (in a $\chi^{2}$ sense) 
are within $1\sigma$ uncertainty of the median of the likelihood for both parameters, i.e. $n$(H$_2$) and $N$(CO) at the remaining positions (see Table~\ref{tab:modelr}). As seen from Table~\ref{tab:modelr}, the UV-bright positions $0$ and $15$, and the UV-dim positions $4$, $5$ and $6$ have similar $n$(H$_2$) and $N$(CO) values. However, considering the results from both methods, 
namely $\chi^{2}$ and likelihood, most UV-bright positions (i.e. the positions $8$, $14$, $16$, and $17$) have higher 
$n$(H$_2$) and $N$(CO) compared to the UV-dim positions.

Overall, based on both $\chi^{2}$ and most likely model results, while the UV-dim positions 
mostly have warmer ISM with relatively lower $N$(CO) and $n$(H$_2$), the UV-bright positions indicate an opposite 
trend, i.e. colder ISM with higher $N$(CO) and $n$(H$_2$). These model results agree with the results based on 
empirical line ratios.

%
%
\begin{table*}
\setlength{\tabcolsep}{3pt}
 \caption{Model results for the CO gas in the disc of NGC~0628.}
 \begin{tabular}{cclllcclll}
 
   \hline
   Position & Offset  & Parameter & $\chi^{2}$ & Likelihood & Position & Offset & Parameter & $\chi^{2}$ & Likelihood\\
    &($\Delta\alpha^{\prime\prime}, \Delta\delta^{\prime\prime}$)&&&&&($\Delta\alpha^{\prime\prime}, \Delta\delta^{\prime\prime}$)&&\\
   \hline\\
  $0$ & $(0,0)$ & $T_{\rm K}$ & $30$~K & $38.4^{+38.7}_{-18.6}$~K &$8$ & $(-66.9,21.6)$ & $T_{\rm K}$ & $8$~K & $100.7^{+96.6}_{-73.9}$~K  \\
  & & $\log$($n$(H$_2$)) & $2.2$~cm$^{-3}$ & $2.5^{+0.7}_{-0.4}$~cm$^{-3}$ &  & & $\log$($n$(H$_2$)) & $5.2$& $3.6^{+2.4}_{-1.2}$~cm$^{-3}$ \\
  & & $\log$($N$(CO)) & $17.7$~cm$^{-2}$ & $17.7^{+0.2}_{-0.3}$~cm$^{-2}$ & & & $\log$($N$(CO)) & $17.2$~cm$^{-2}$ & $18.4^{+1.6}_{-0.9}$~cm$^{-2}$ \\
  & & [$^{12}$C]/[$^{13}$C]& $20$ && & & [$^{12}$C]/[$^{13}$C]& $20$ & \\\\
  
   $1$ & $(108,-43.1)$ & $T_{\rm K}$ & $25$~K & $86.9^{+98.9}_{-62.5}$~K & $9$ & $(-88.7,30.1)$ & $T_{\rm K}$ & $185$~K & $117.9^{+88.5}_{-86.1}$~K\\
  & & $\log$($n$(H$_2$)) & $2.0$~cm$^{-3}$ & $4.2^{+1.9}_{-1.7}$~cm$^{-3}$ && & $\log$($n$(H$_2$)) & $2.0$~cm$^{-3}$ & $2.7^{+2.5}_{-0.5}$~cm$^{-3}$\\
  & & $\log$($N$(CO))& $18.5$~cm$^{-2}$ & $19.2^{+1.2}_{-1.0}$~cm$^{-2}$ && & $\log$($N$(CO)) & $17.5$~cm$^{-2}$ & $17.6^{+1.5}_{-1.0}$~cm$^{-2}$\\
   & & [$^{12}$C]/[$^{13}$C]& $40$ && & & [$^{12}$C]/[$^{13}$C]& $20$ & \\\\
  
   $2$ & $(87,-34)$ & $T_{\rm K}$ & $7$~K & $50.8^{+81.8}_{-34.7}$~K &$14$ &$(-35.7,-44.1)$& $T_{\rm K}$ & $6$~K & $7.4^{+24.8}_{-1.6}$~K\\
  & & $\log$($n$(H$_2$)) & $3.7$~cm$^{-3}$ & $2.7^{+1.8}_{-0.5}$~cm$^{-3}$ & & & $\log$($n$(H$_2$)) & $6.2$~cm$^{-3}$ & $5.0^{+1.3}_{-1.9}$~cm$^{-3}$\\
  & & $\log$($N$(CO)) & $17.2$~cm$^{-2}$ & $17.7^{+0.3}_{-0.4}$~cm$^{-2}$ & & & $\log$($N$(CO)) & $17.7$~cm$^{-2}$ & $18.0^{+1.1}_{-0.3}$~cm$^{-2}$\\
  & & [$^{12}$C]/[$^{13}$C]& $30$ && & & [$^{12}$C]/[$^{13}$C]& $90$ & \\\\
  
   $4$ & $(43.9,-17)$ & $T_{\rm K}$ & $135$~K & $128.4^{+78.0}_{-69.4}$~K&$15$ &$(-73.1,64.2)$& $T_{\rm K}$ & $25$~K & $43.5^{+62}_{-25.3}$~K   \\ 
  & & $\log$($n$(H$_2$)) & $2.0$~cm$^{-3}$ & $2.3^{+0.4}_{-0.2}$~cm$^{-3}$ & & & $\log$($n$(H$_2$)) & $2.2$~cm$^{-3}$ & $2.5^{+1.0}_{-0.4}$~cm$^{-3}$  \\
  & & $\log$($N$(CO)) & $17.5$~cm$^{-2}$ & $17.4^{+0.3}_{-0.4}$~cm$^{-2}$  & & & $\log$($N$(CO)) & $17.7$~cm$^{-2}$ & $17.6^{+0.3}_{-0.3}$~cm$^{-2}$  \\
    & & [$^{12}$C]/[$^{13}$C]& $20$ && & & [$^{12}$C]/[$^{13}$C]& $30$ & \\\\
  
   $5$ & $(22.4,-9.2)$ & $T_{\rm K}$ & $65$~K & $125.4^{+81.9}_{-76.7}$~K &$16$ &$(9.3,76.8)$& $T_{\rm K}$ & $10$~K & $40.9^{+100.3}_{-28.1}$~K  \\ 
  & & $\log$($n$(H$_2$)) & $2.5$~cm$^{-3}$ & $2.4^{+1.0}_{-0.3}$~cm$^{-3}$  & & & $\log$($n$(H$_2$)) & $4.0$~cm$^{-3}$ & $4.5^{+1.7}_{-1.8}$~cm$^{-3}$ \\
  & & $\log$($N$(CO)) & $17.5$~cm$^{-2}$ & $17.6^{+0.3}_{-0.3}$~cm$^{-2}$ & & & $\log$($N$(CO)) & $18.2$~cm$^{-2}$ & $19.2^{+1.1}_{-0.8}$~cm$^{-2}$ \\
   & & [$^{12}$C]/[$^{13}$C]& $20$ && & & [$^{12}$C]/[$^{13}$C]& $60$ & \\\\
 
   $6$ & $(-21.9, 7.2)$ & $T_{\rm K}$ & $13$~K & $91.7^{+97.9}_{-66.2}$~K &$17$ &$(-13.4,103.7)$& $T_{\rm K}$ & $5^{\ast}$~K & $8.0^{+8.5}_{-2.4}$~K  \\ 
  & & $\log$($n$(H$_2$)) & $3.2$~cm$^{-3}$ & $2.6^{+1.6}_{-0.5}$~cm$^{-3}$ & & & $\log$($n$(H$_2$)) & $4.2$~cm$^{-3}$ & $3.3^{+2.2}_{-1.0}$~cm$^{-3}$   \\
  & & $\log$($N$(CO)) & $17.2$~cm$^{-2}$ & $17.4^{+0.4}_{-0.4}$~cm$^{-2}$ & & & $\log$($N$(CO)) & $17.5$~cm$^{-2}$ & $18.0^{+0.6}_{-0.5}$~cm$^{-2}$   \\
  & & [$^{12}$C]/[$^{13}$C]& $20$ && & & [$^{12}$C]/[$^{13}$C]& $90$ & \\\\
  
   $7$ & $(-44.4,14.9)$ & $T_{\rm K}$ & $160$~K & $126.5^{+81.7}_{-72.8}$~K& &&&&   \\ 
  & & $\log$($n$(H$_2$)) & $2.0$~cm$^{-3}$ & $2.4^{+0.4}_{-0.3}$~cm$^{-3}$ & &&&& \\
  & & $\log$($N$(CO)) & $17.2$~cm$^{-2}$ & $17.1^{+0.5}_{-1.5}$~cm$^{-2}$ & &&&& \\
  & & [$^{12}$C]/[$^{13}$C]& $20$ && & &&& \\\\

    \hline\\
 \end{tabular}
\parbox[t]{\textwidth}{\textit{Notes:} Likelihood results list the median values 
and $68$ per cent (1$\sigma$) confidence level. A star ($\ast$) indicates 
a physical parameter lying at the edge of the model grid.}
 \label{tab:modelr}
\end{table*}
%
%
%
%
\section{Conclusions}
\label{sec:conc}
We studied $20$ positions across the disc of a spiral galaxy NGC~0628. $11$ out of $20$ positions 
were selected over the SE-NW cut, while the rest of the positions 
were selected in the southern and northern arms of the galaxy. 
$13$ out of $20$ positions are brighter in the UV and hosting 
more H\,{\small II} regions compared to the UV-dim positions with opposite 
characteristics. This enables us to probe; i) radial variations in the physical conditions of the gas; 
ii) differences in the physical conditions between the UV-bright and UV-dim positions; iii) correlations between 
strong star formation activity indicators (such as UV, IR radiation and H\,{\small II} regions) and molecular gas if any. 
Our main conclusions are summarised below.

\begin{enumerate}

\item Our beam-averaged velocity-integrated CO line intensities indicate that the central 
region of NGC~0628 is brighter in CO compared to the positions in the arms 
and inter-arms of the galaxy. The CO intensity decreases steadily with some fluctuations 
as a function of radius up to $5.8$~kpc from the centre. 
The decrease is more clearly seen on both sides of the SE-NW cut.
\item The range for the $R_{13}$ ratios is wider than that for $R_{12}$ ratios in NGC~0628, 
and the $R_{12}$ ratios show a smoother distribution with smaller uncertainties across the disc of the galaxy. 
The $R_{12}$ ratios range from $1$ to $2$, with an average value of $1.51\,\pm\,0.02$, while the $R_{13}$ 
ratios range from $2$ to $10$, with an average value of $4.3\,\pm\,0.1$ across the disc of NGC~0628. 
The UV-dim positions have lower $R_{12}$ and $R_{13}$ ratios (relatively warmer gas) 
whereas the UV-bright positions have higher ratios (relatively colder gas).
\item The $R_{11}$ ratios, as a tracer for diffused gas, show more fluctuations compared to that of 
$R_{12}$ and $R_{13}$ ratios in NGC~0628. The $R_{11}$ ratio increases up to $50^{\prime\prime}$ 
(or equally $1.7$~kpc) and then starts to fluctuate with a tendency to decrease. 
In contrast with the increase in the $R_{11}$ ratios, the $R_{13}$ ratios decrease, 
indicating that the gas gets warmer and thinner up to $1.7$~kpc. 
Outside the central $1.7$~kpc, CO(1--0) integrated intensity at some UV-bright positions 
shows an increase causing sudden decrease and increase in the ratios. 
\item The range of $R_{12}$ and $R_{13}$ ratios along the disc of NGC~0628 
is similar to that found in the centre of spirals, like our own Milky Way, indicating similar gas temperatures. 
The range of $R_{13}$ ratios is wider than that found in the centre of $61$ galaxies from different Hubble types. 
The $R_{13}$ ratios in the disc of NGC~0628 are higher than that found at the centre of starbursts, indicating 
that NGC~0628 has relatively colder ISM. The range of $R_{11}$ ratios in NGC~0628 is $5 \le R_{11} \le 19$ given the error bars, 
similar to that found in the central regions of spirals and Seyferts. While the $R_{11}$ ratios 
at some positions in the arms and inter-arms of NGC~0628 are similar to that found 
at the centre of starbursts, it is lower in the centre of the galaxy.
\item Beam-averaged molecular gas mass and the gas surface density at the centre of the galaxy is the highest. 
As expected from the decrease seen in the integrated line intensities, there is also a decrease in 
$M_{\rm H_{2}}$, $\Sigma_{H_{2}}$ and $N_{\rm H_{2}}$ as a function of the radius 
from the centre of the galaxy. We found that the UV-bright and UV-dim positions have 
similar beam-averaged total molecular gas mass and gas surface densities on average. We also 
found that there is a linear correlation between the molecular gas mass and the extinction (i.e. IR-to-FUV ratio). 
\item The extinction decreases as a function of galactocentric radius similar to the molecular gas mass. 
UV-bright and UV-dim positions have similar $F_{24\micron}/F_{\rm FUV}$ ratios, while the $F_{160\micron}/F_{\rm FUV}$ is higher at the UV-dim positions. This is due to the fact that UV-bright positions are brighter at both $24\micron$ and UV, while UV-dim positions are dimmer at both wavelengths, but they have similar $F_{160\micron}$ fluxes. This indicates that the reason for the difference in the UV-brightness between these two groups of positions is not the extinction, but having a different level of star formation activity.
\item The UV-bright positions have more diffused gas, indicated by a slightly higher $R_{11}$ ratio found at those positions. 
Since the UV-bright positions are also hosting many H\,{\small II} regions and brighter at $24\micron$ emission, it is natural to expect to have a higher level of star formation activity at those positions, which in turn could make the gas more diffused, compared to the UV-dim positions. 
\item Both $\chi^{2}$ and most likely model results indicate that the UV-dim positions 
mostly have warmer molecular gas with relatively lower $N$(CO) and $n$(H$_2$), while the UV-bright positions indicate an opposite 
trend, i.e. colder molecular gas with higher $N$(CO) and $n$(H$_2$). Additionally, our best model results indicate that the UV-dim positions have abundance ratio ranges from $20$ to $40$, while the abundance ratio at the UV-bright positions is mostly higher, i.e. ranges from $30$ to $90$.

\end{enumerate}
%
%
\section*{Acknowledgements}
ST would like to thank the anonymous referee for his/her
insightful comments and suggestions. ST thanks Timothy A. Davis for very useful discussions.
ST thanks Irene San Jose Garcia and Ferhat F. Ozeren for 
conducting the observation and also thanks to the staff at IRAM $30$m 
telescope for their help with the observations. ST was supported by the Republic of Turkey, Ministry of National
Education, The Philip Wetton Graduate Scholarship at Christ
Church. ST acknowledge the usage of the HyperLeda database (http://leda.univ-lyon1.fr). 
This research has made use of the NASA/ IPAC Infrared Science Archive, 
which is operated by the Jet Propulsion Laboratory, California Institute of 
Technology, under contract with the National Aeronautics and Space Administration.
\bibliographystyle{mn2e}
\bibliography{reference}
%
%
\appendix

\section{Spectra}
\label{sec:Ap1}
%
%
%
%
\begin{figure*}
  \includegraphics[width=7.0cm,clip=]{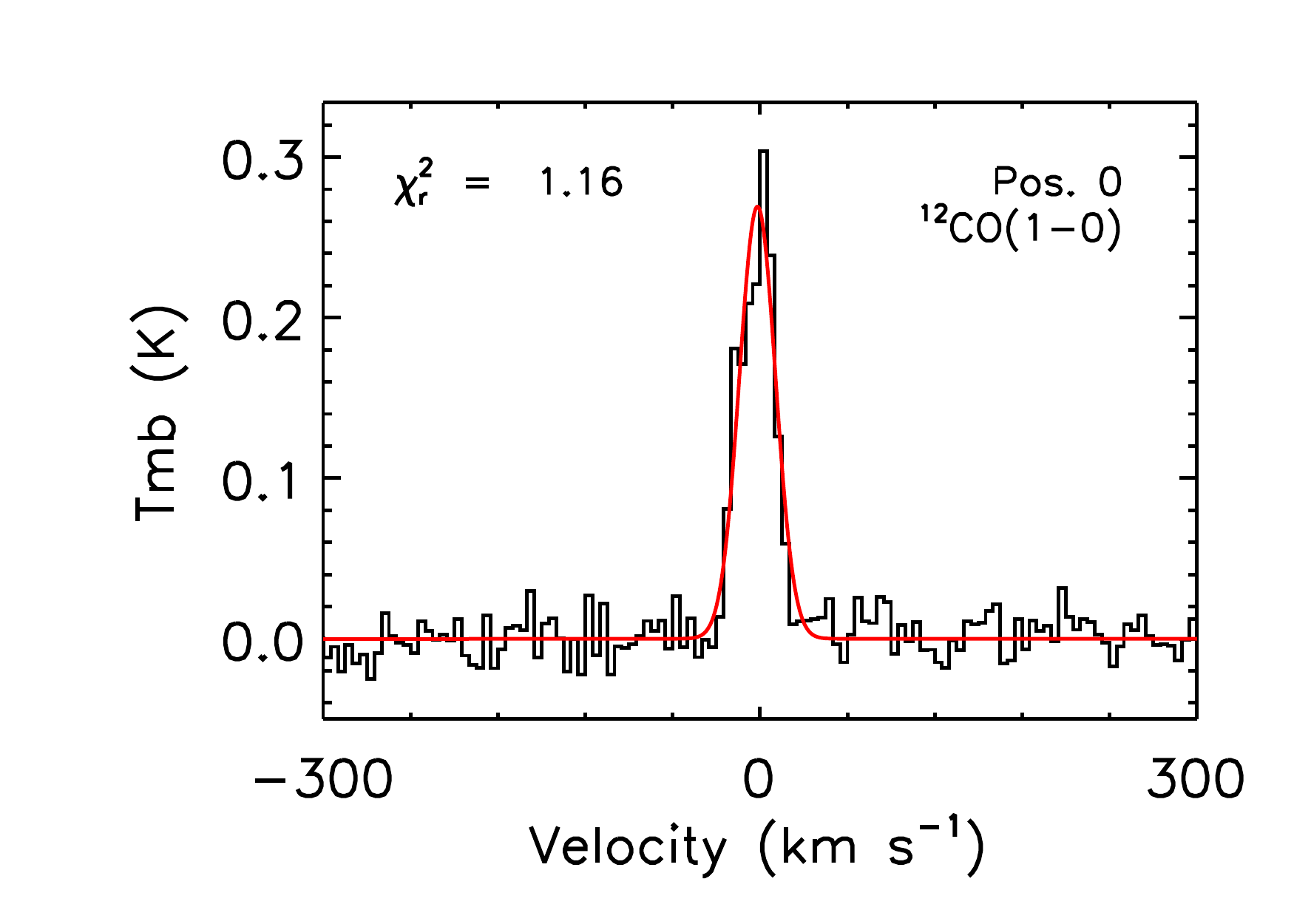}
   \includegraphics[width=7.0cm,clip=]{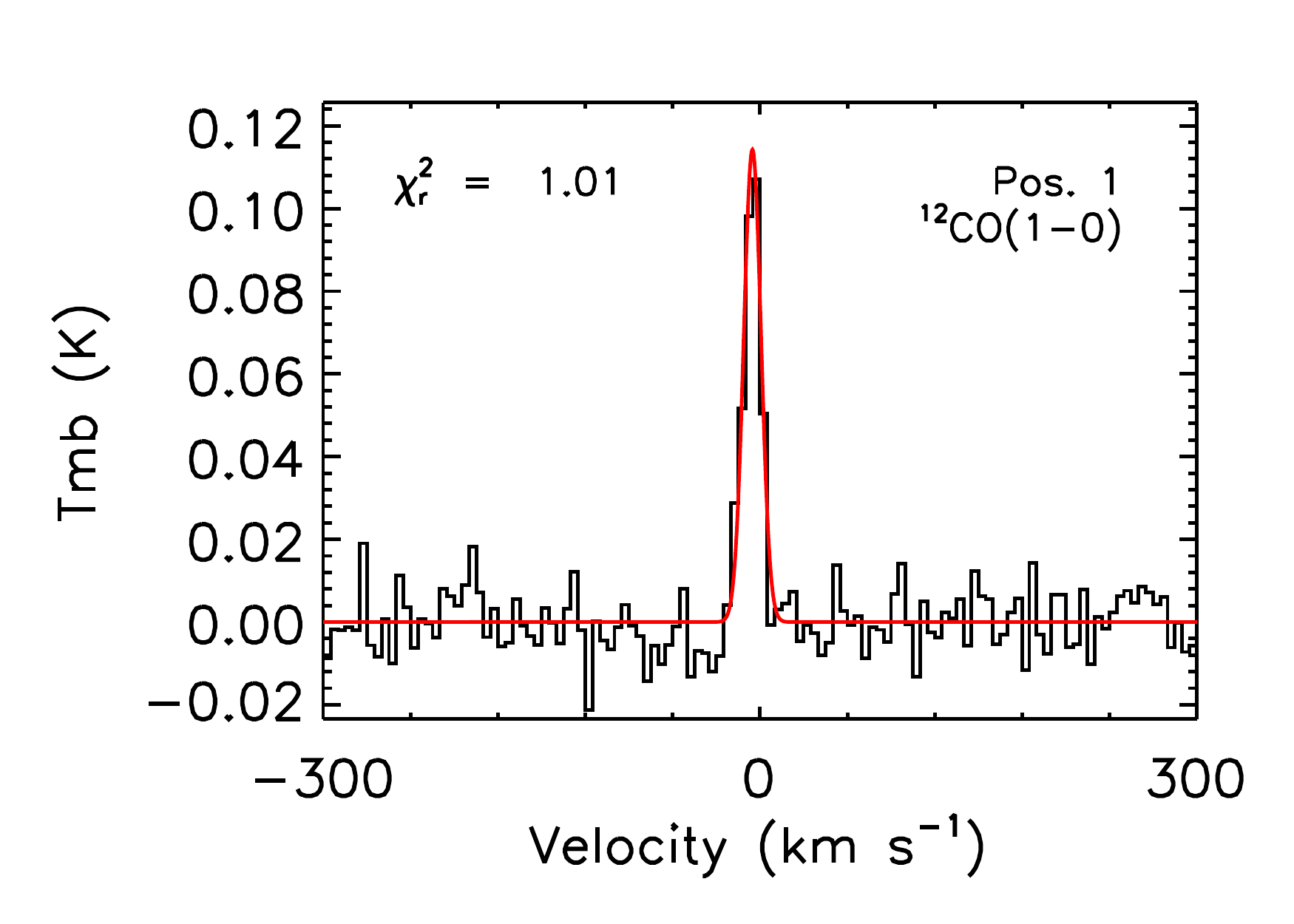}\\
  \includegraphics[width=7.0cm,clip=]{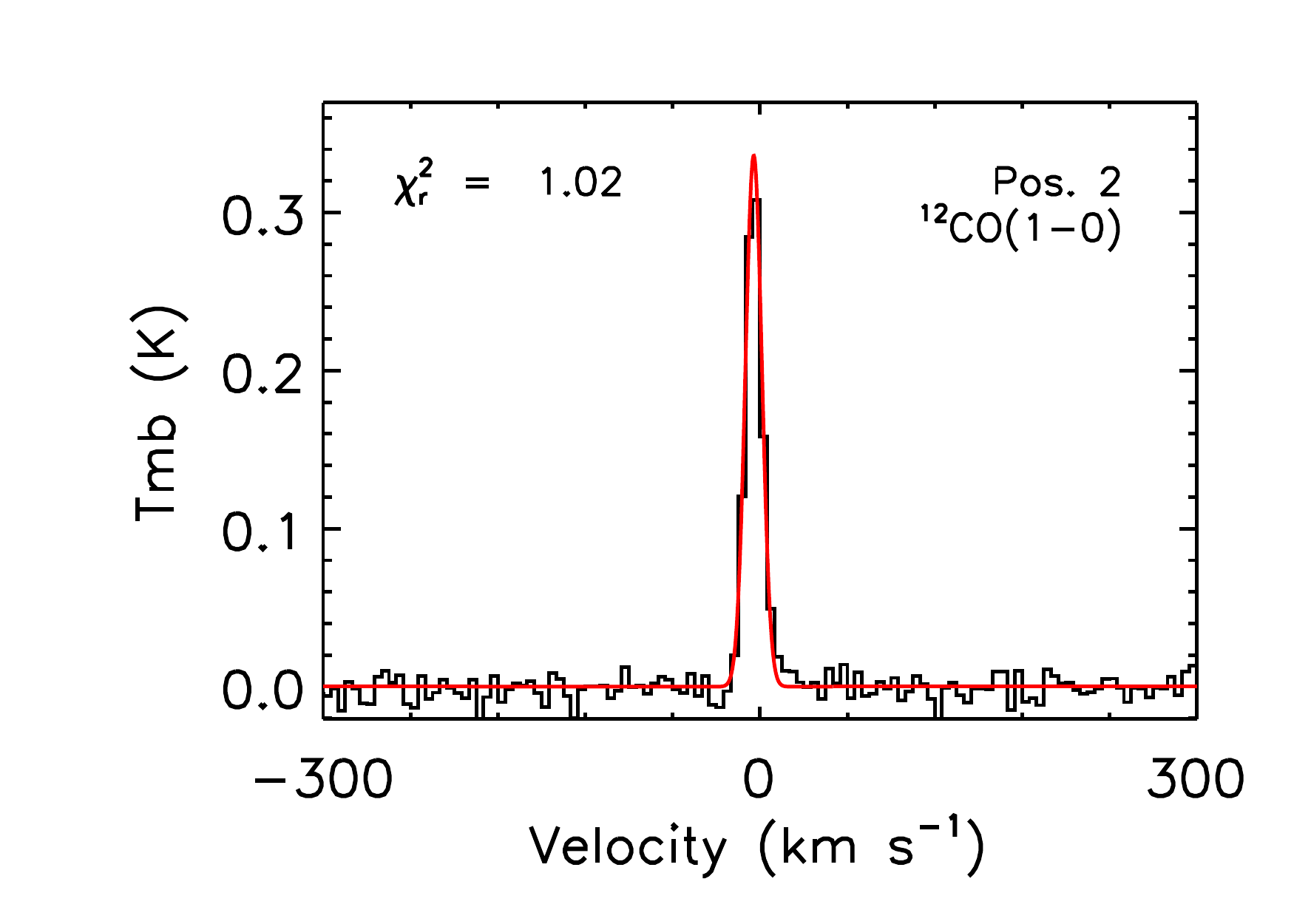}
   \includegraphics[width=7.0cm,clip=]{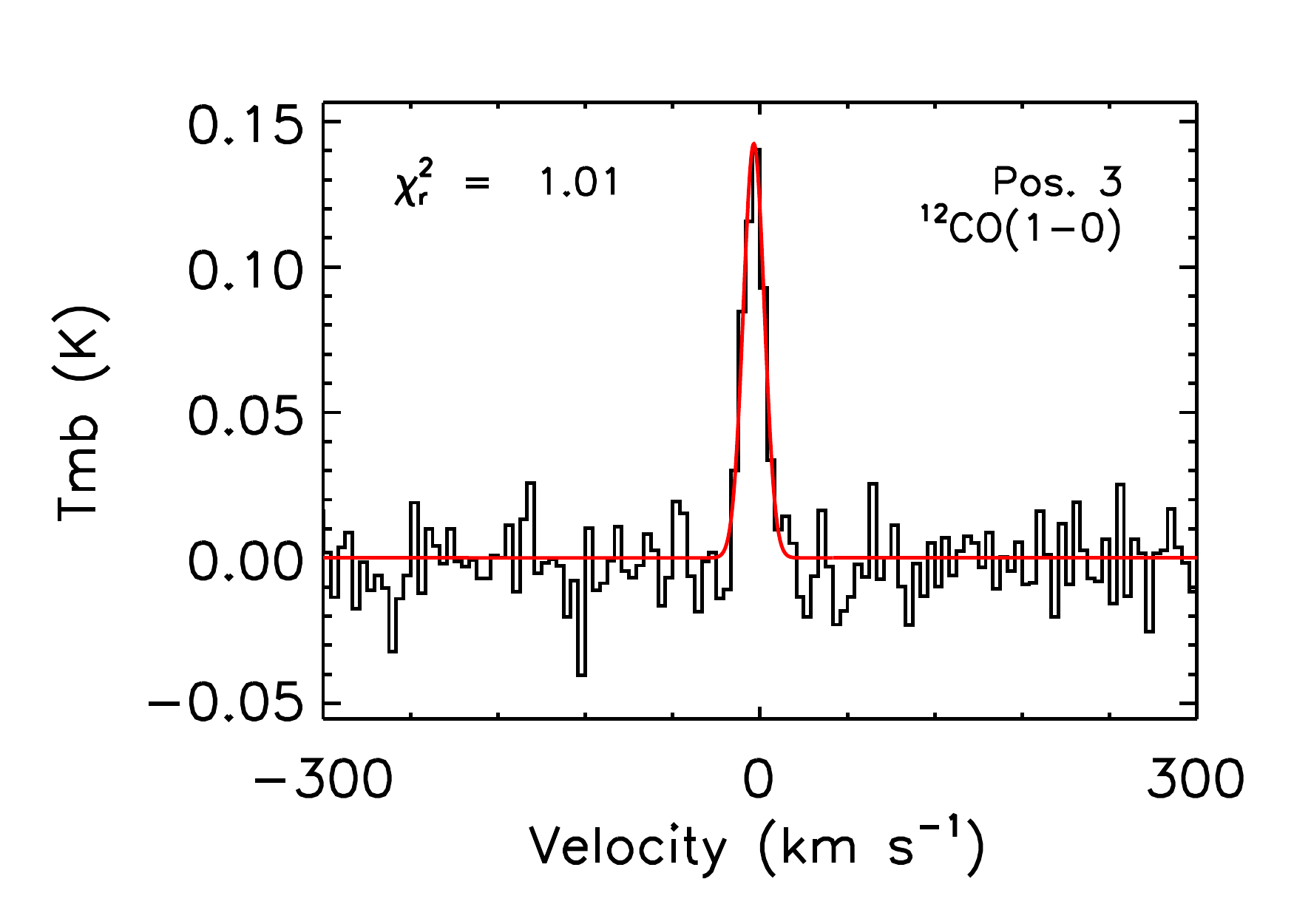}\\
    \includegraphics[width=7.0cm,clip=]{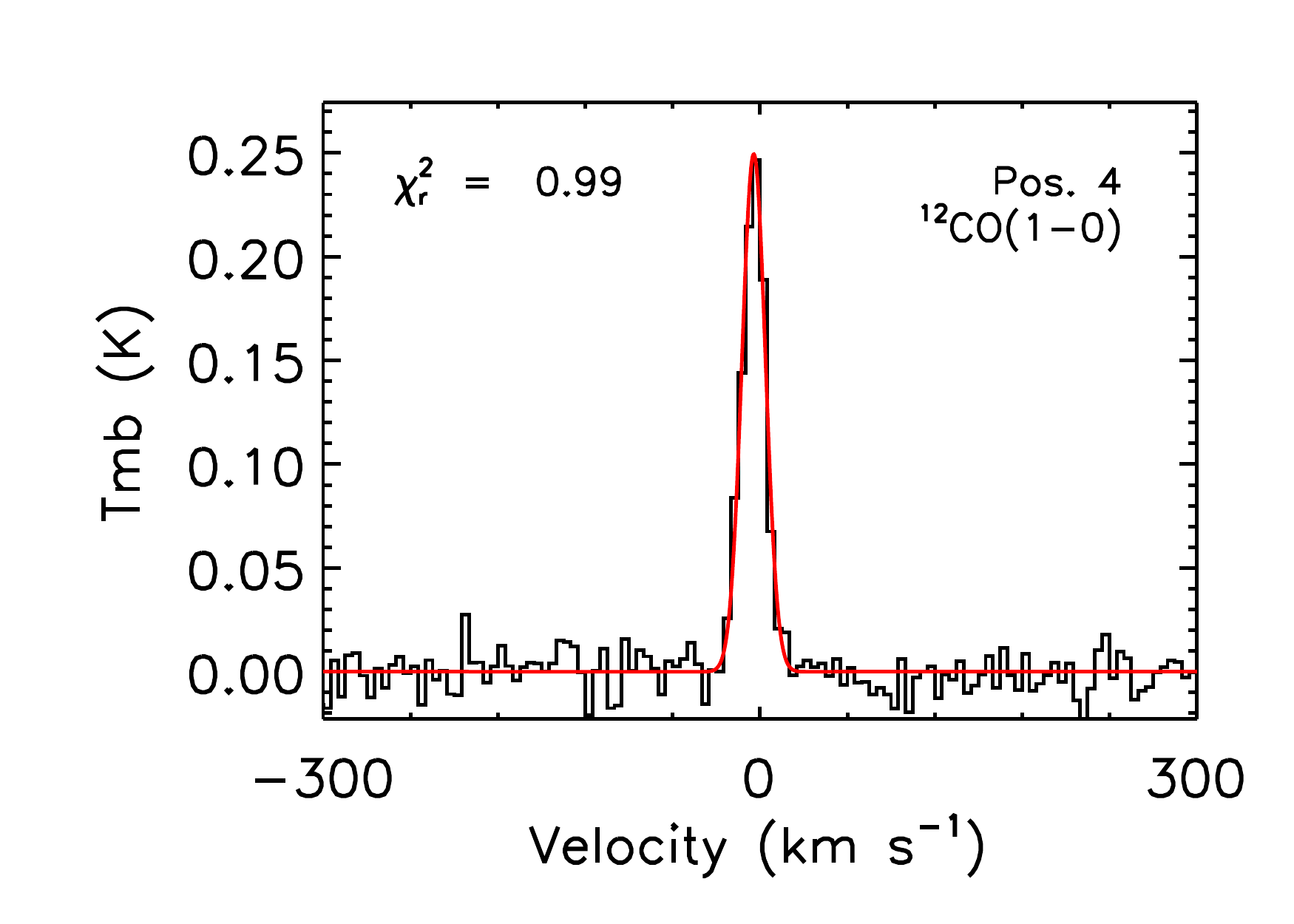}
   \includegraphics[width=7.0cm,clip=]{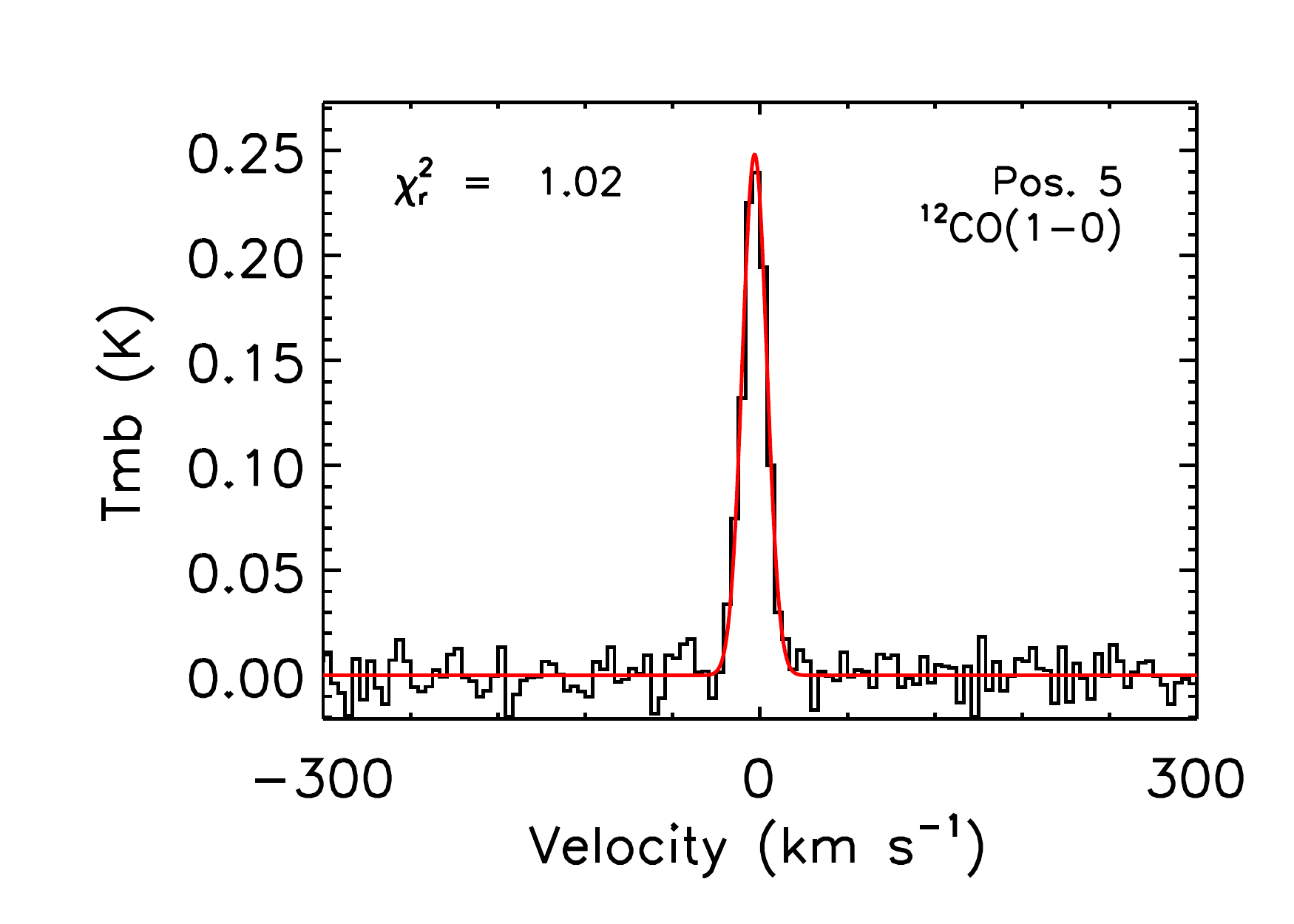}\\
   \includegraphics[width=7.0cm,clip=]{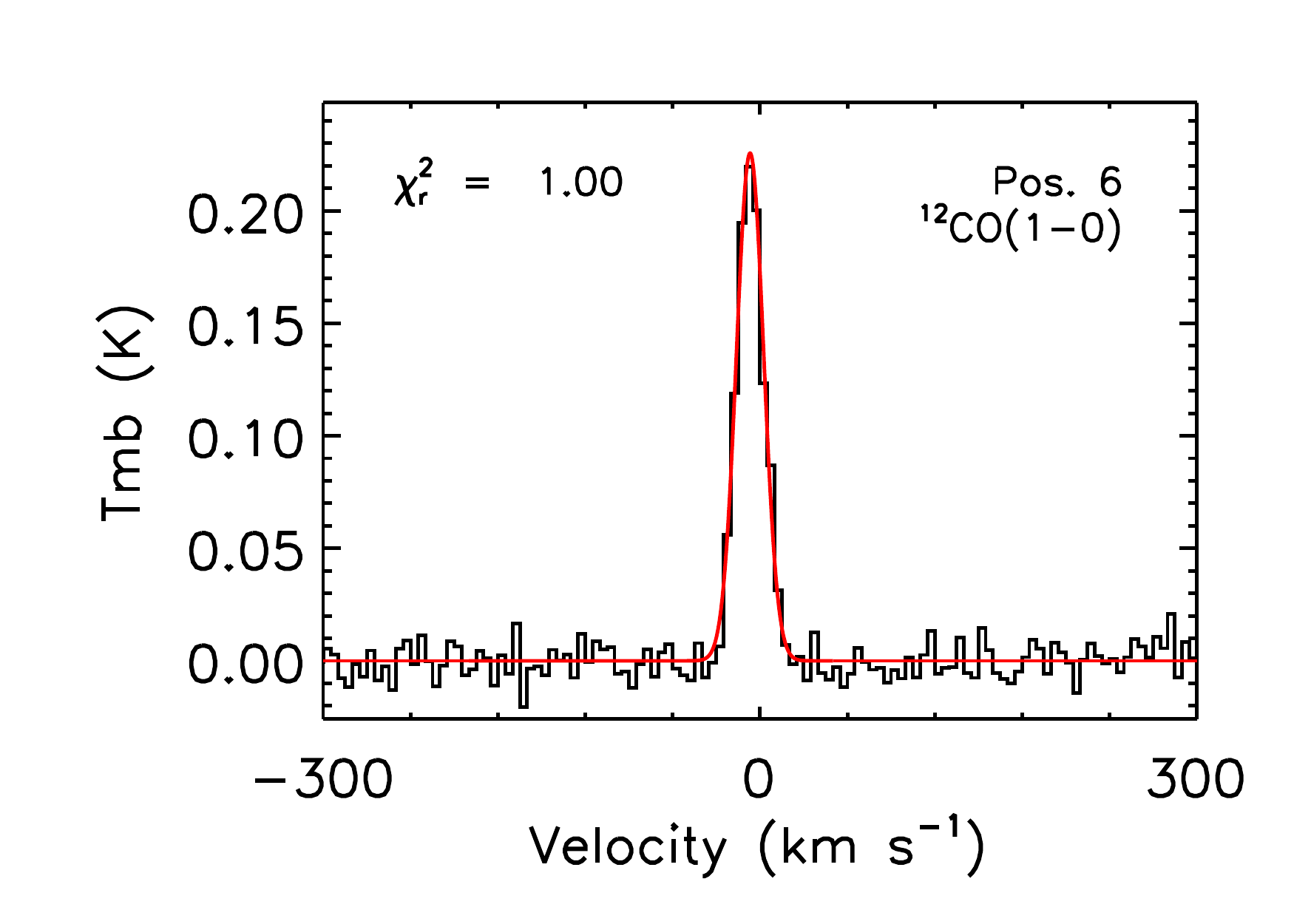}
   \includegraphics[width=7.0cm,clip=]{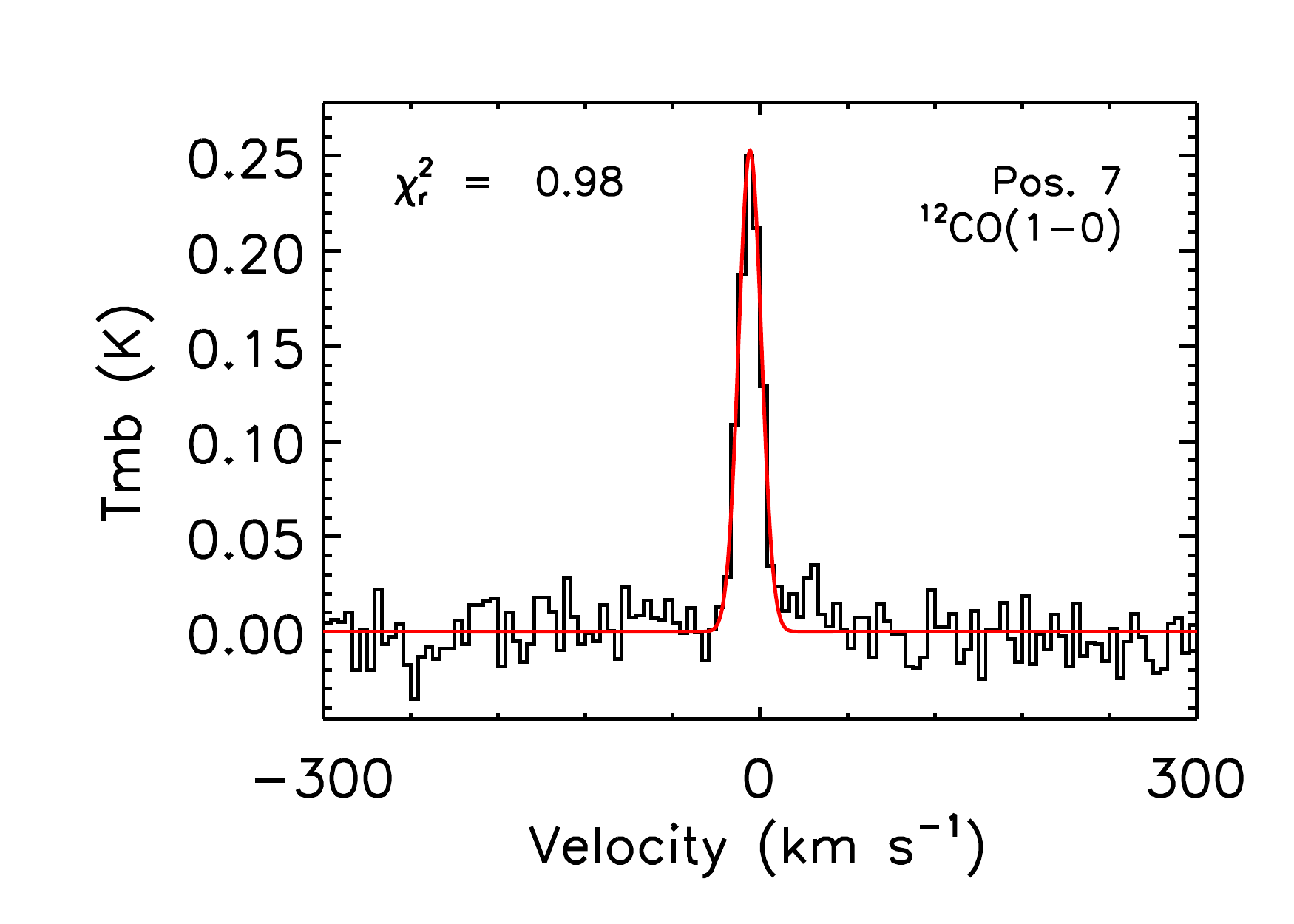}\\
  \caption{Integrated spectra of CO(1--0) emission observed in the disc of NGC~0628. Gaussian fits are overlaid. 
  The value of $\chi_{\rm r}^2$ is also shown in each panel.}
  \label{fig:spec1}
\end{figure*}

%
%
\addtocounter{figure}{-1}
\begin{figure*}
  \includegraphics[width=7.0cm,clip=]{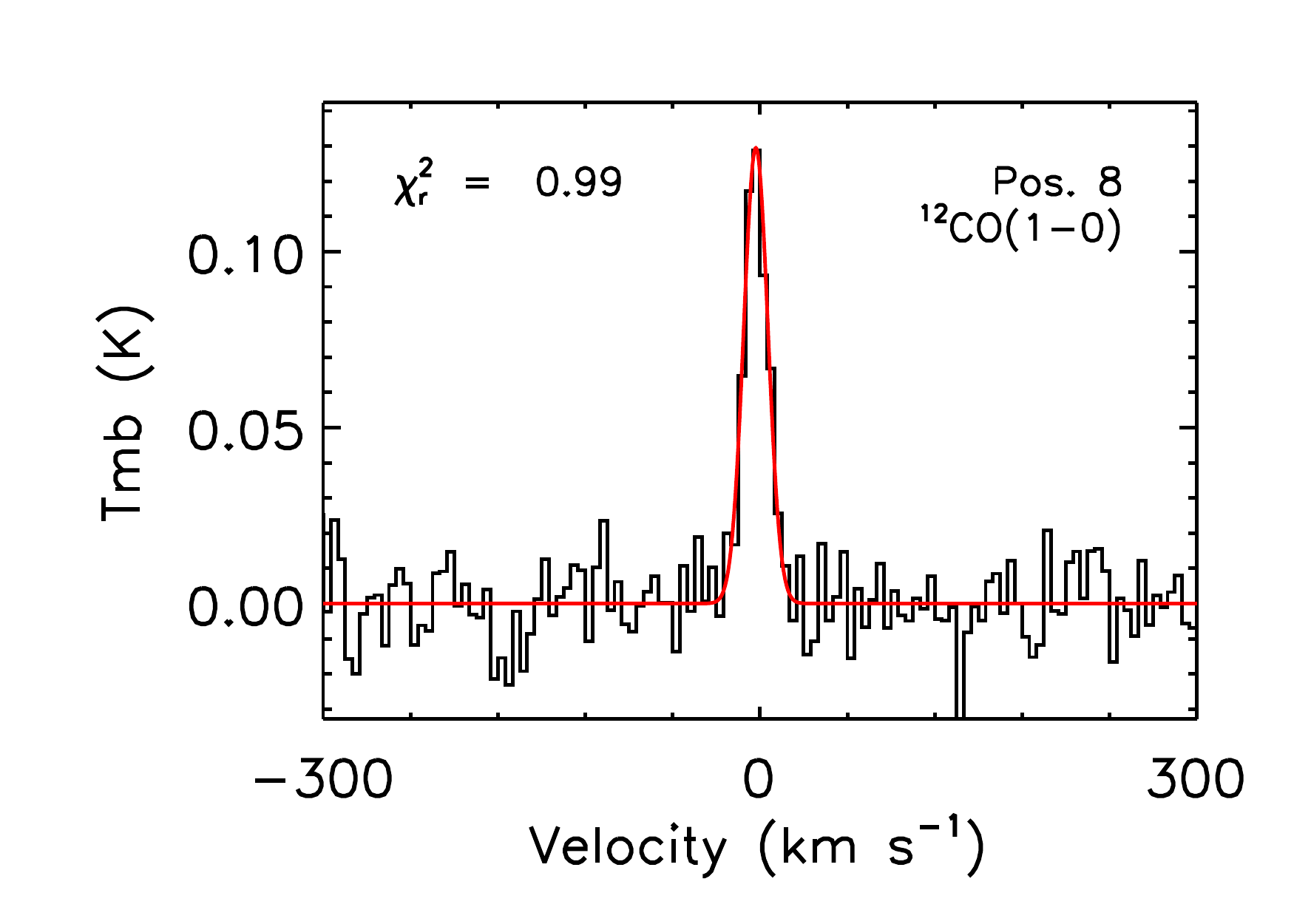}
   \includegraphics[width=7.0cm,clip=]{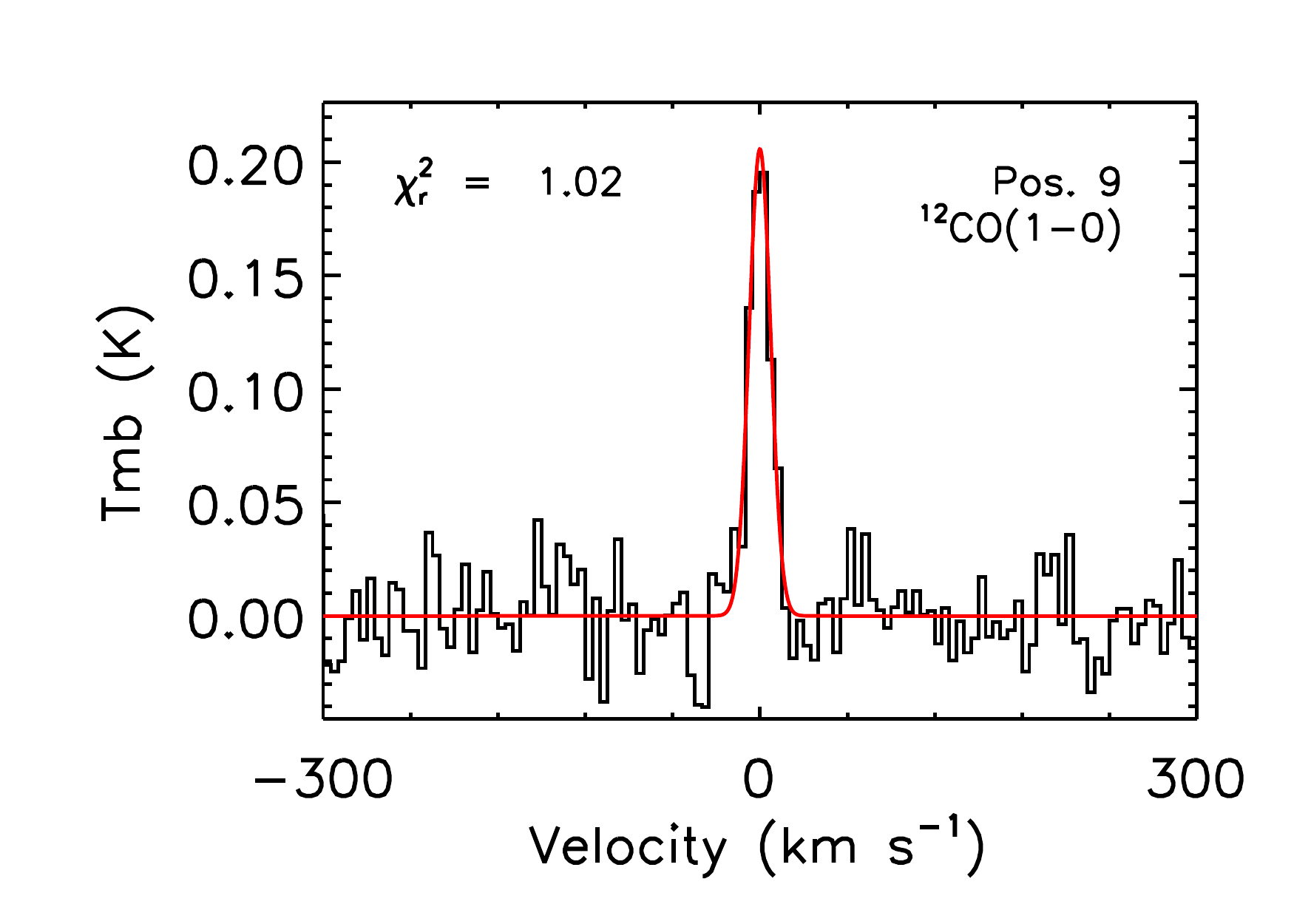}\\
  \includegraphics[width=7.0cm,clip=]{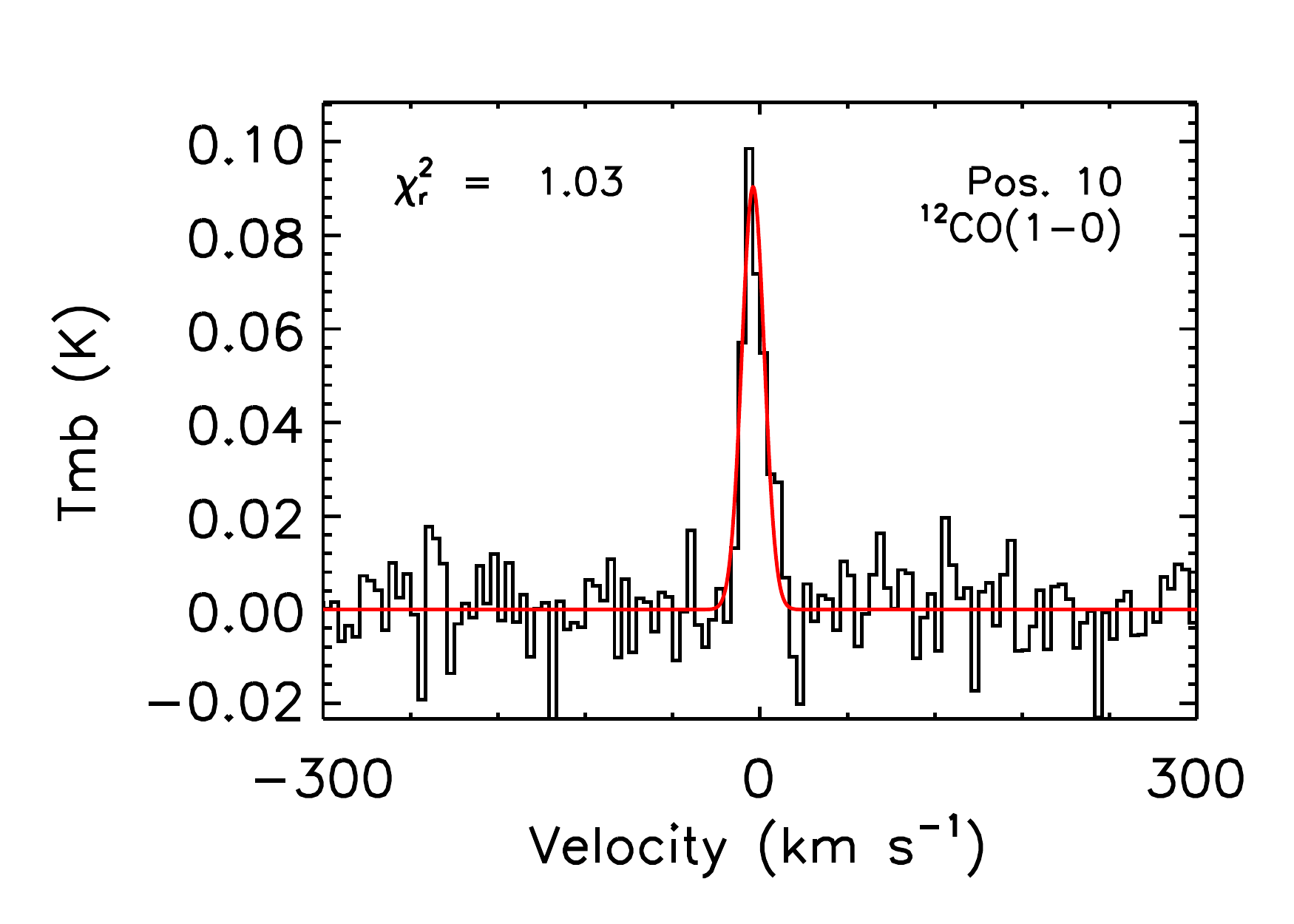}
   \includegraphics[width=7.0cm,clip=]{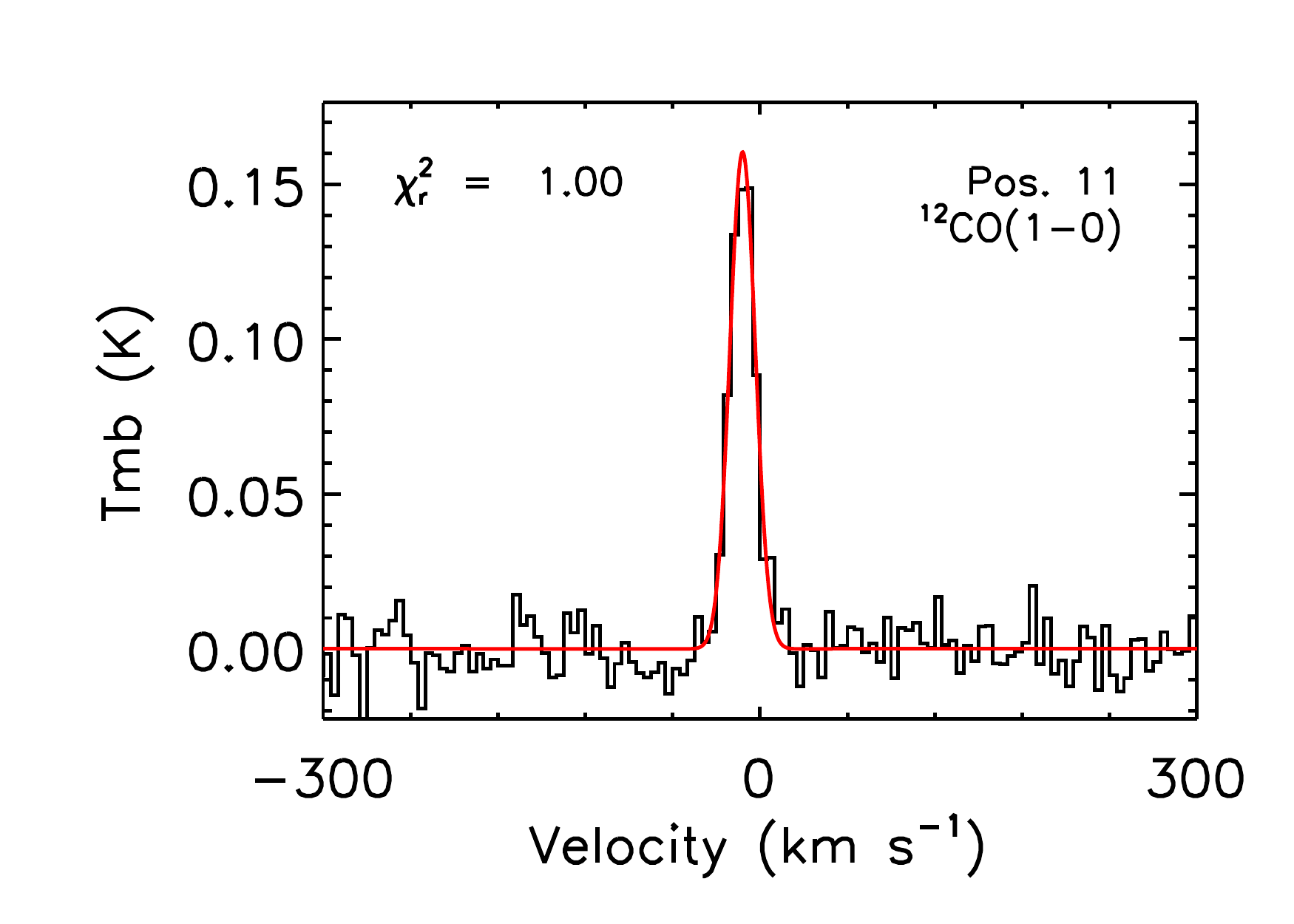}\\
    \includegraphics[width=7.0cm,clip=]{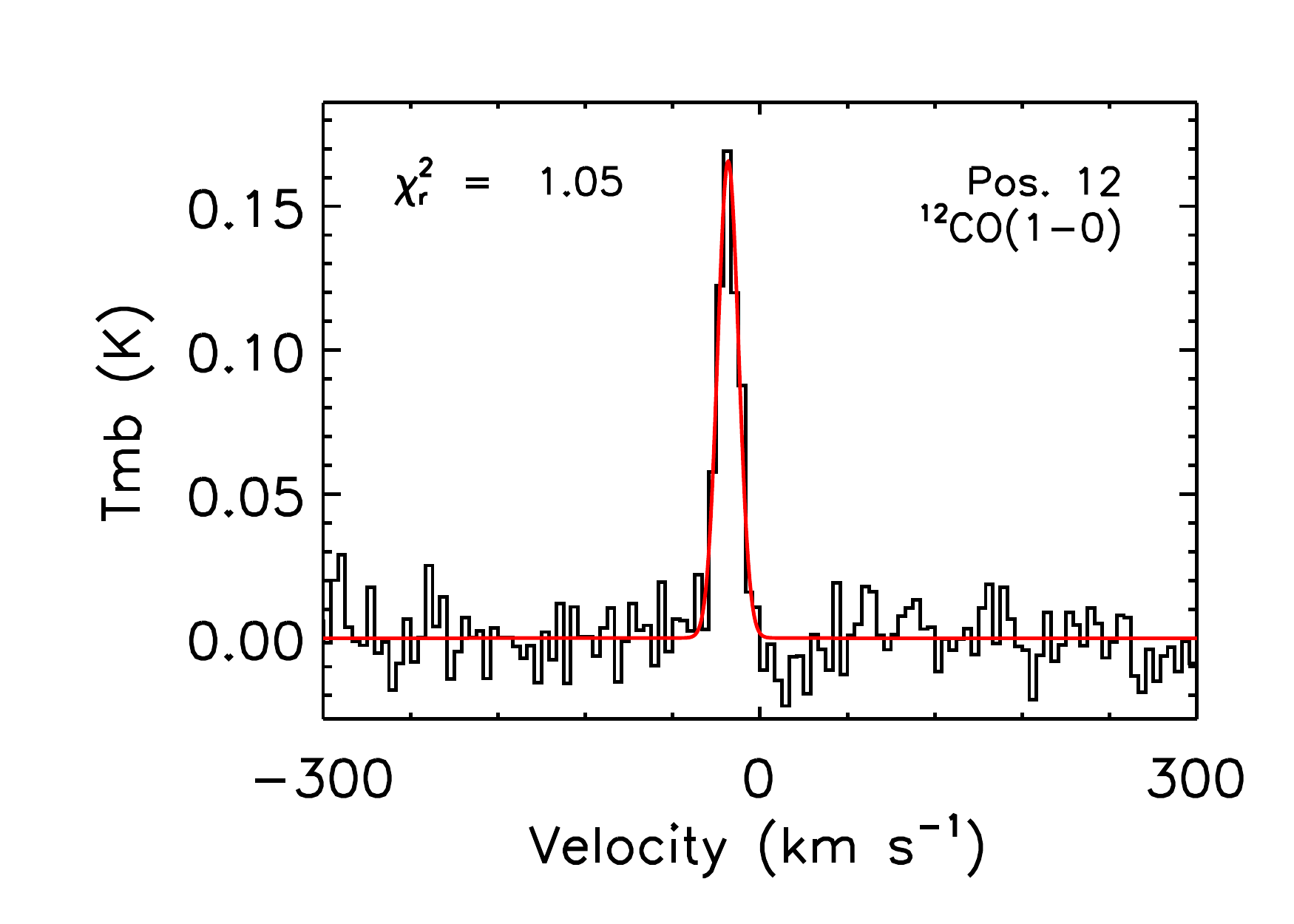}
   \includegraphics[width=7.0cm,clip=]{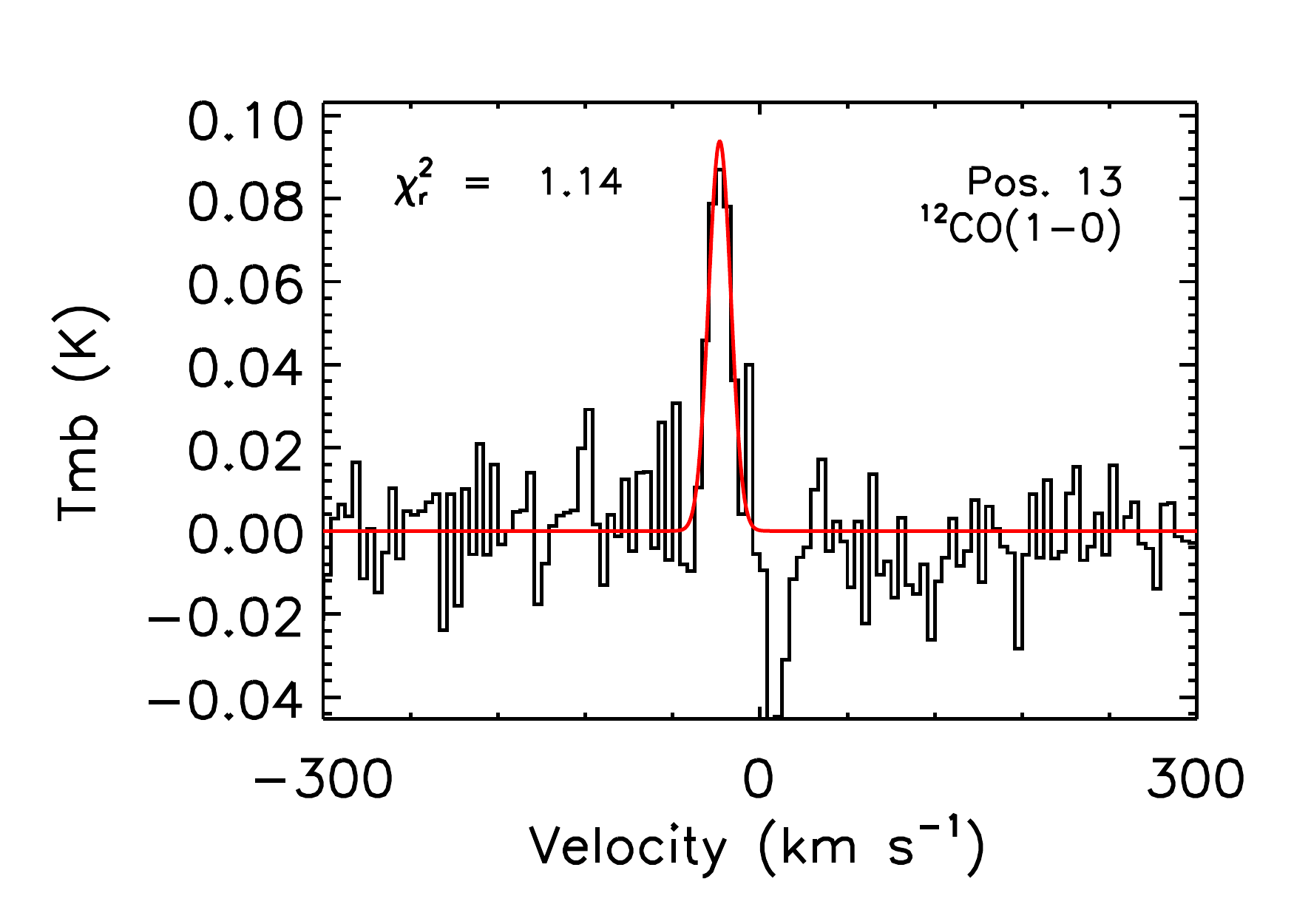}\\
   \includegraphics[width=7.0cm,clip=]{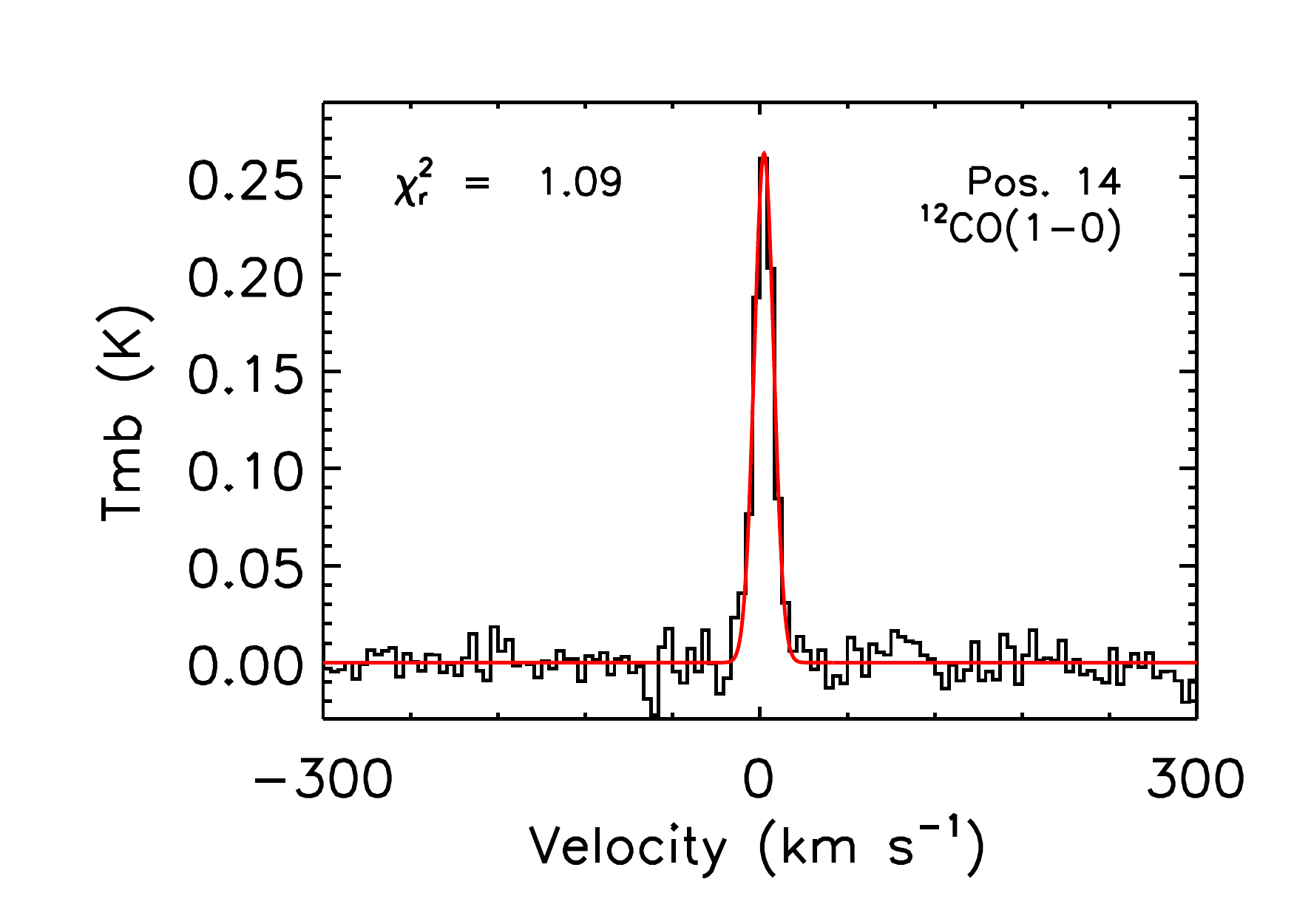}
   \includegraphics[width=7.0cm,clip=]{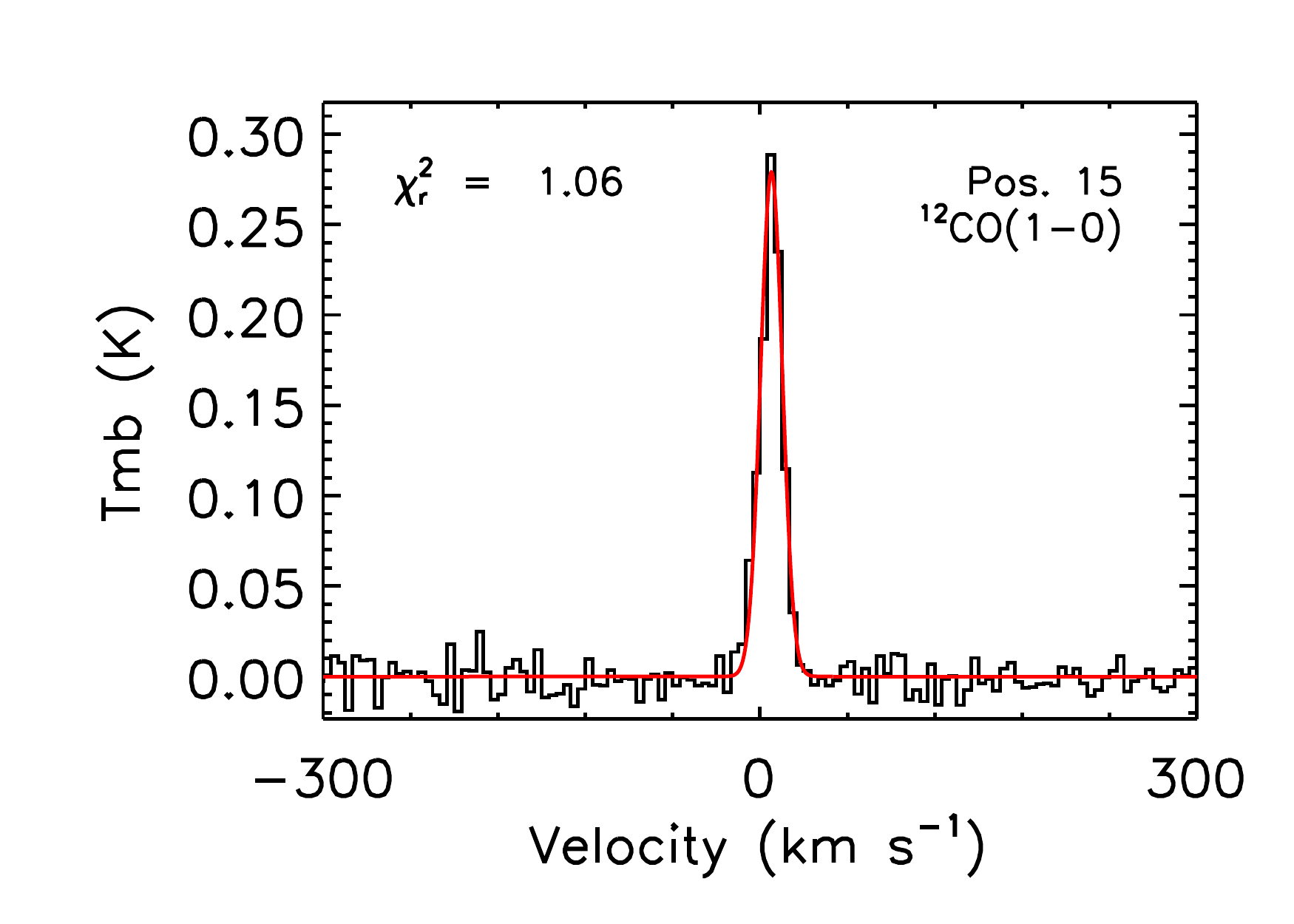}\\
  \caption{Continued.}
  \label{fig:spec1}
\end{figure*}

%
%
\addtocounter{figure}{-1}
\begin{figure*}
  \includegraphics[width=7.0cm,clip=]{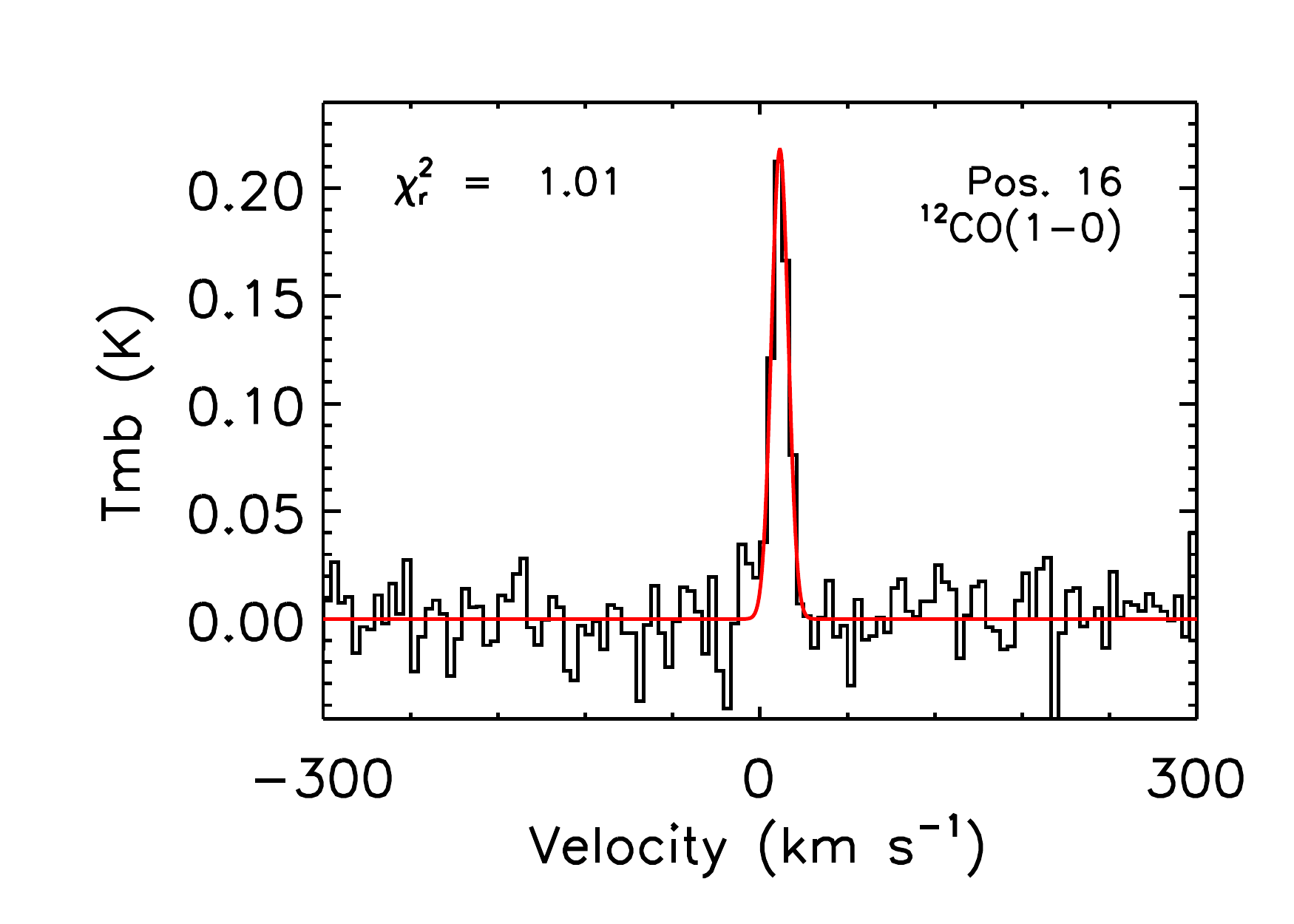}
   \includegraphics[width=7.0cm,clip=]{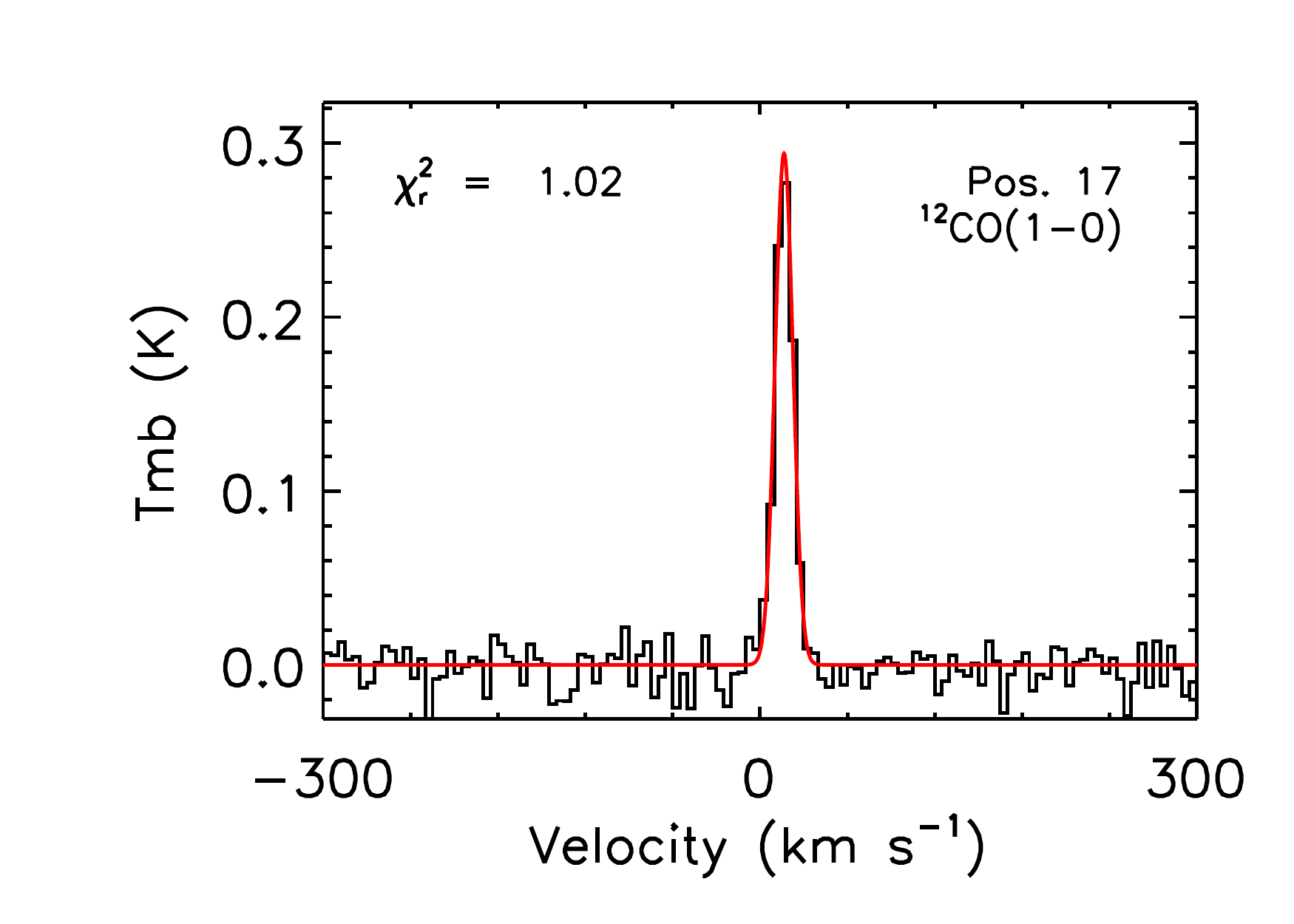}\\
  \includegraphics[width=7.0cm,clip=]{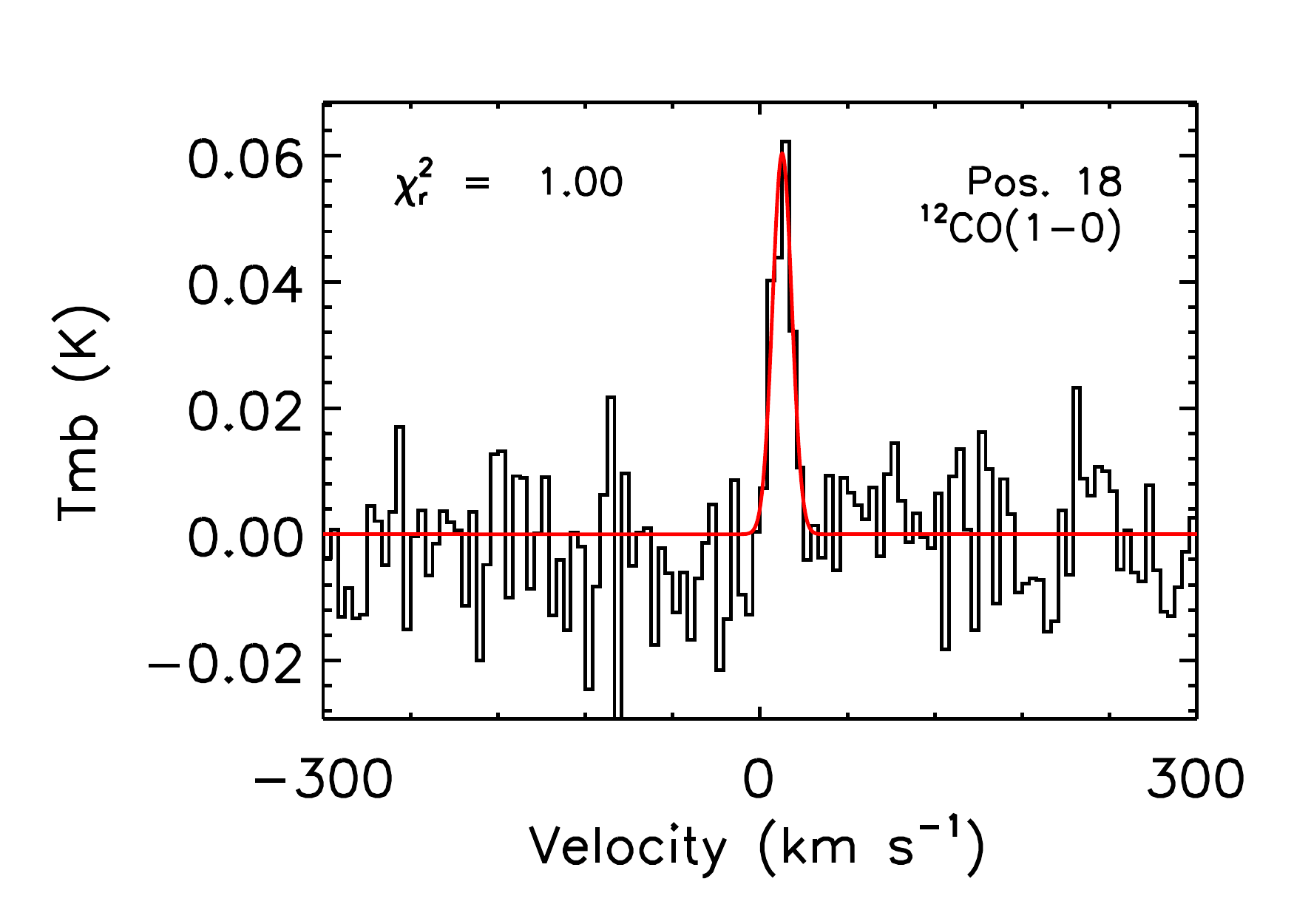}
   \includegraphics[width=7.0cm,clip=]{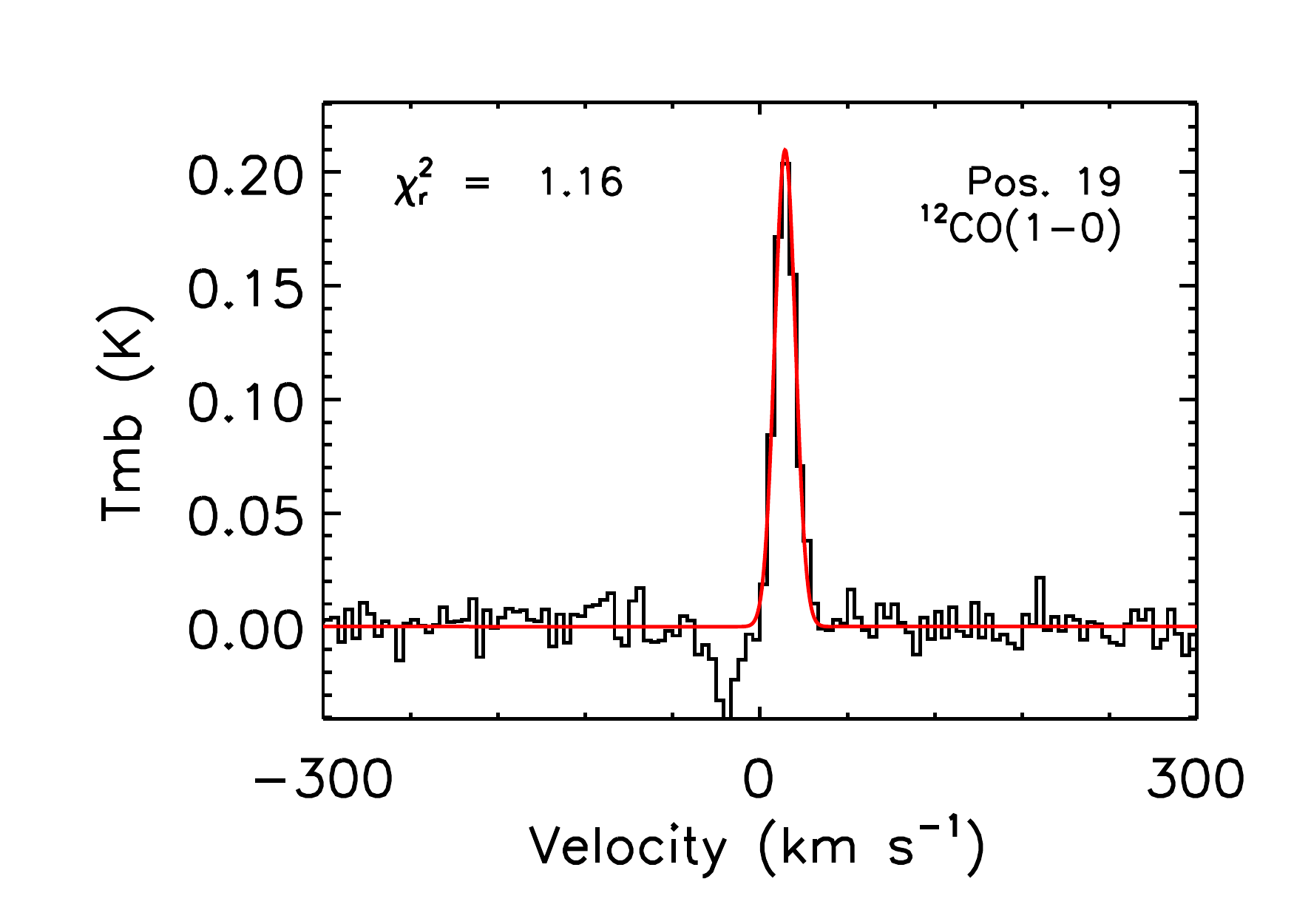}\\
  \caption{Continued.}
  \label{fig:spec1}
\end{figure*}

\begin{figure*}
  \includegraphics[width=7.0cm,clip=]{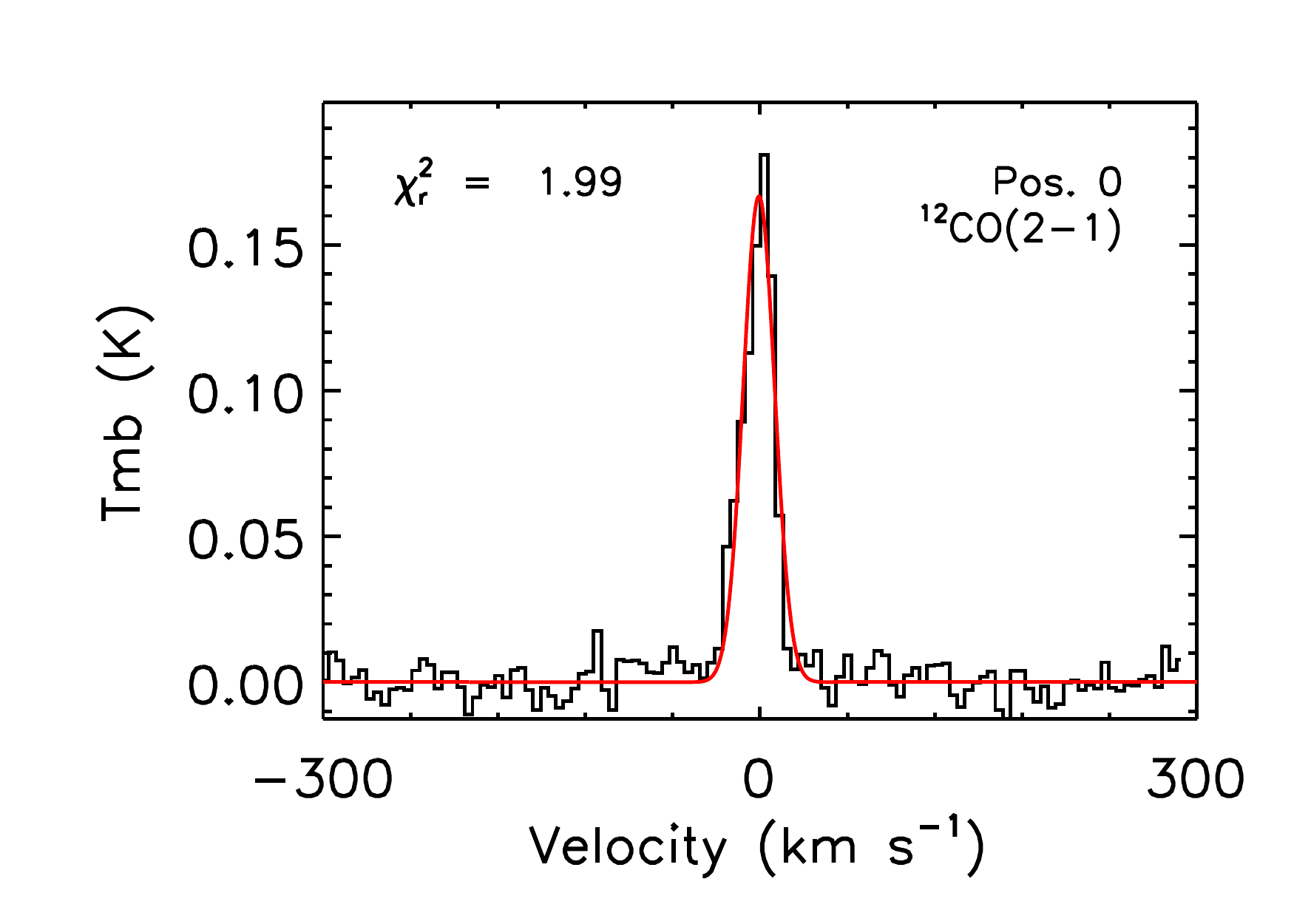}
   \includegraphics[width=7.0cm,clip=]{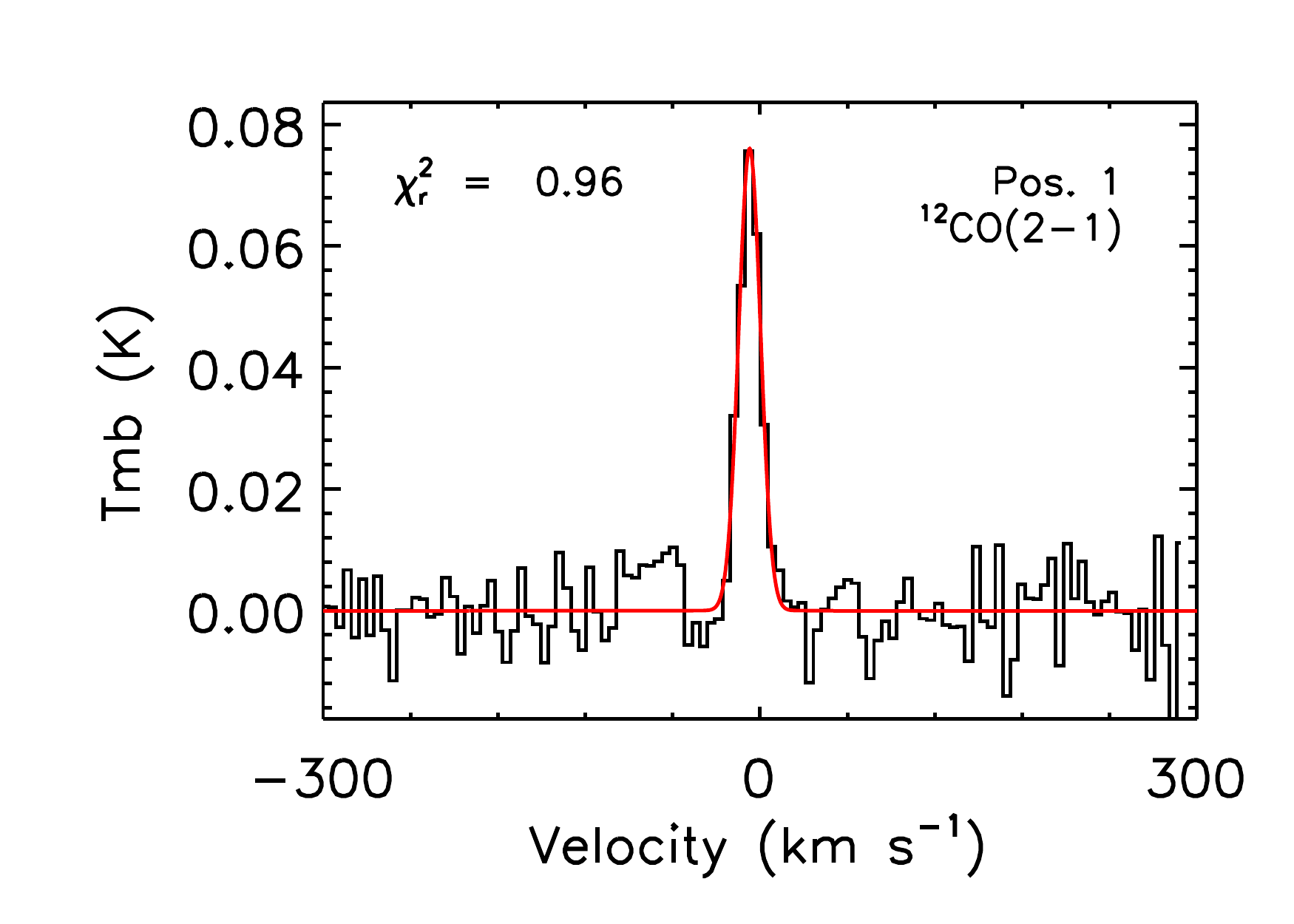}\\
  \includegraphics[width=7.0cm,clip=]{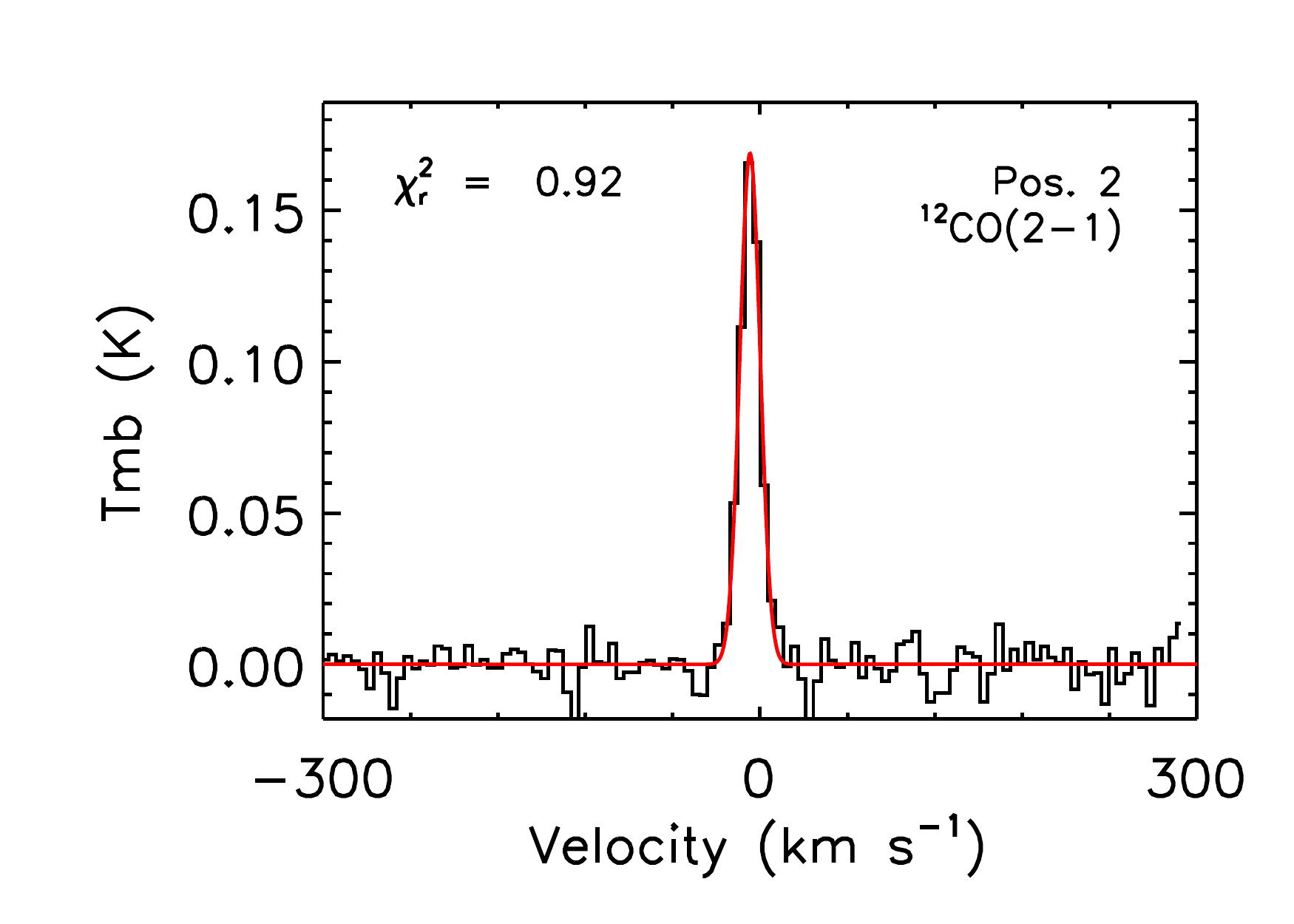}
   \includegraphics[width=7.0cm,clip=]{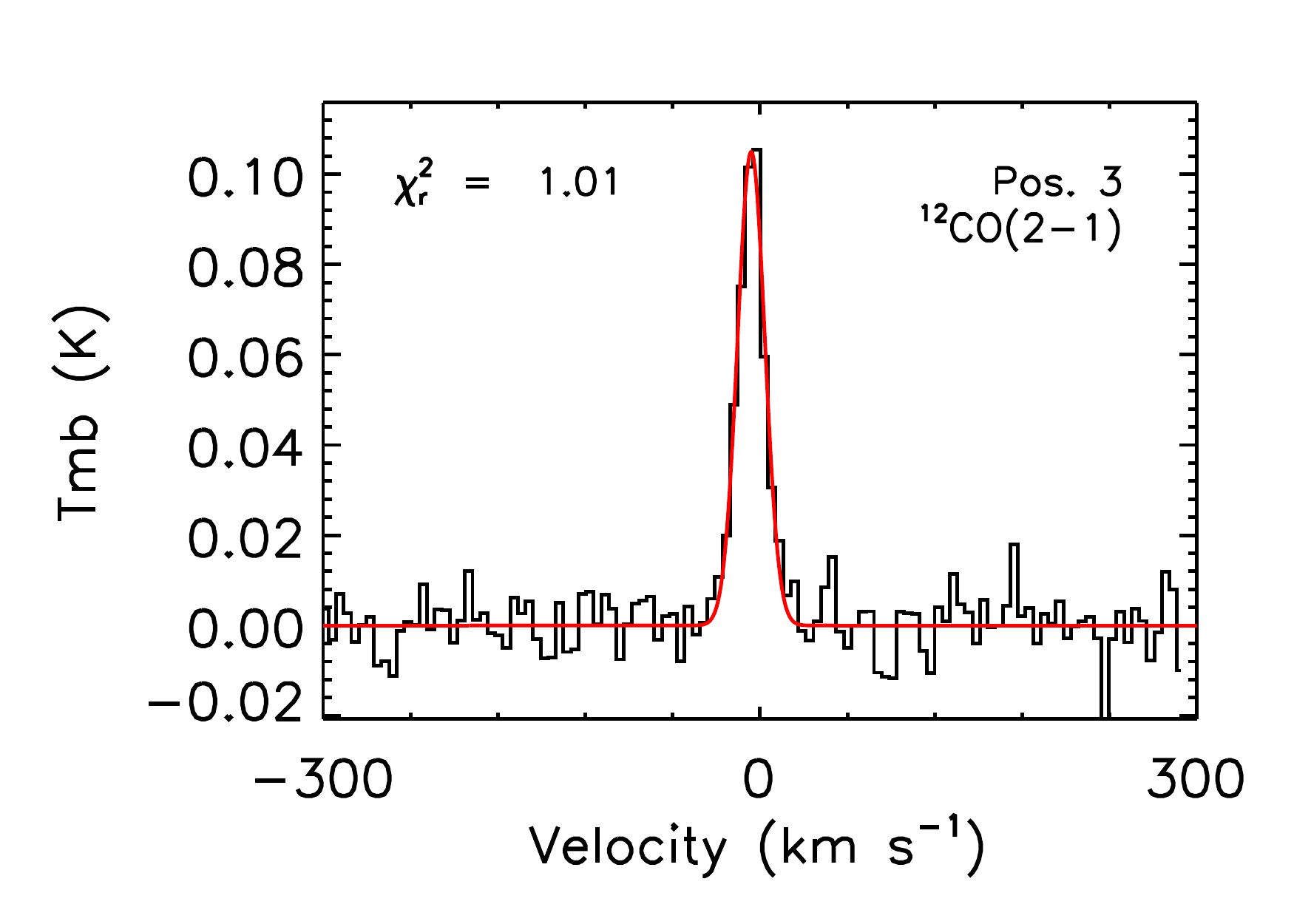}\\
    \includegraphics[width=7.0cm,clip=]{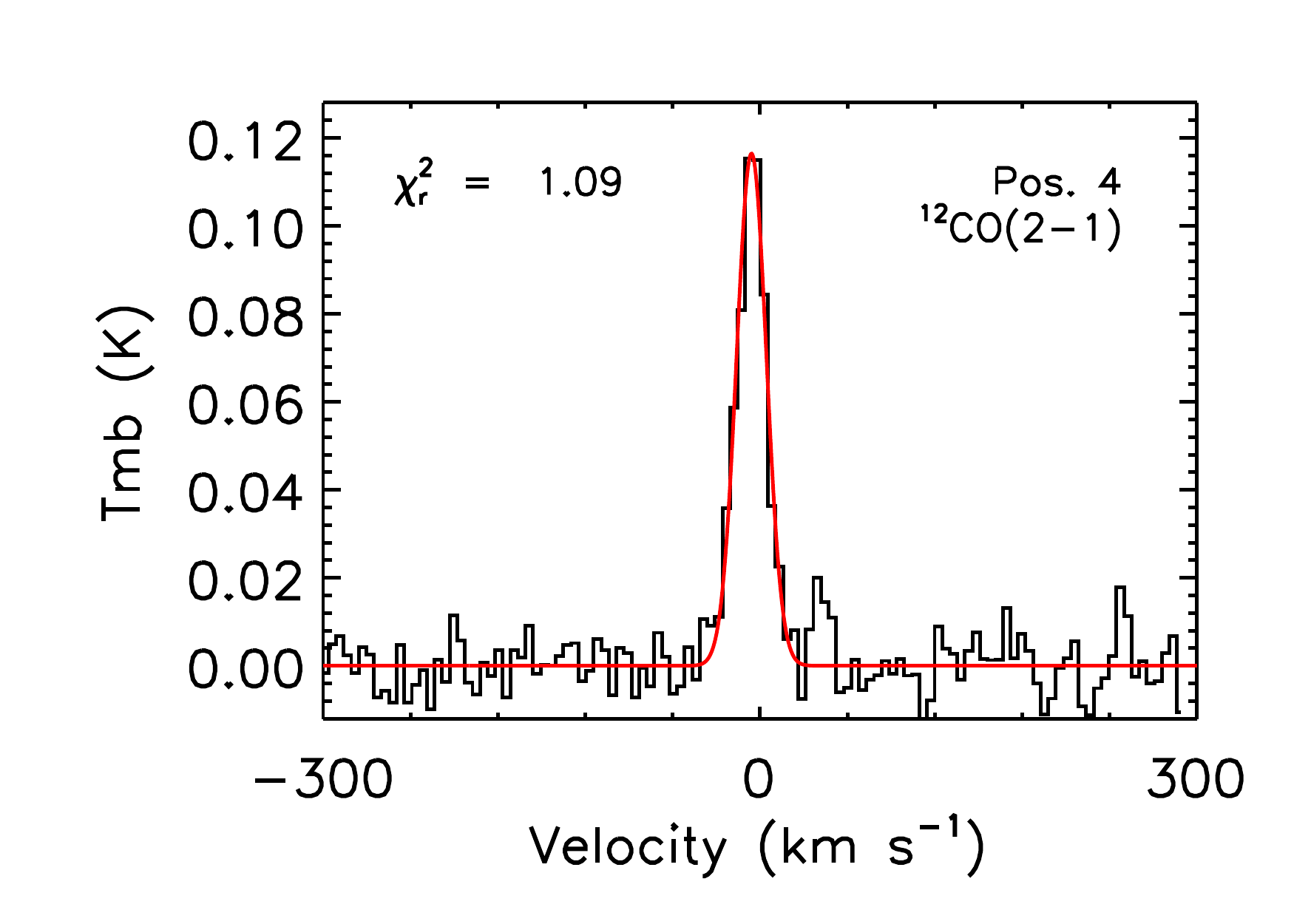}
   \includegraphics[width=7.0cm,clip=]{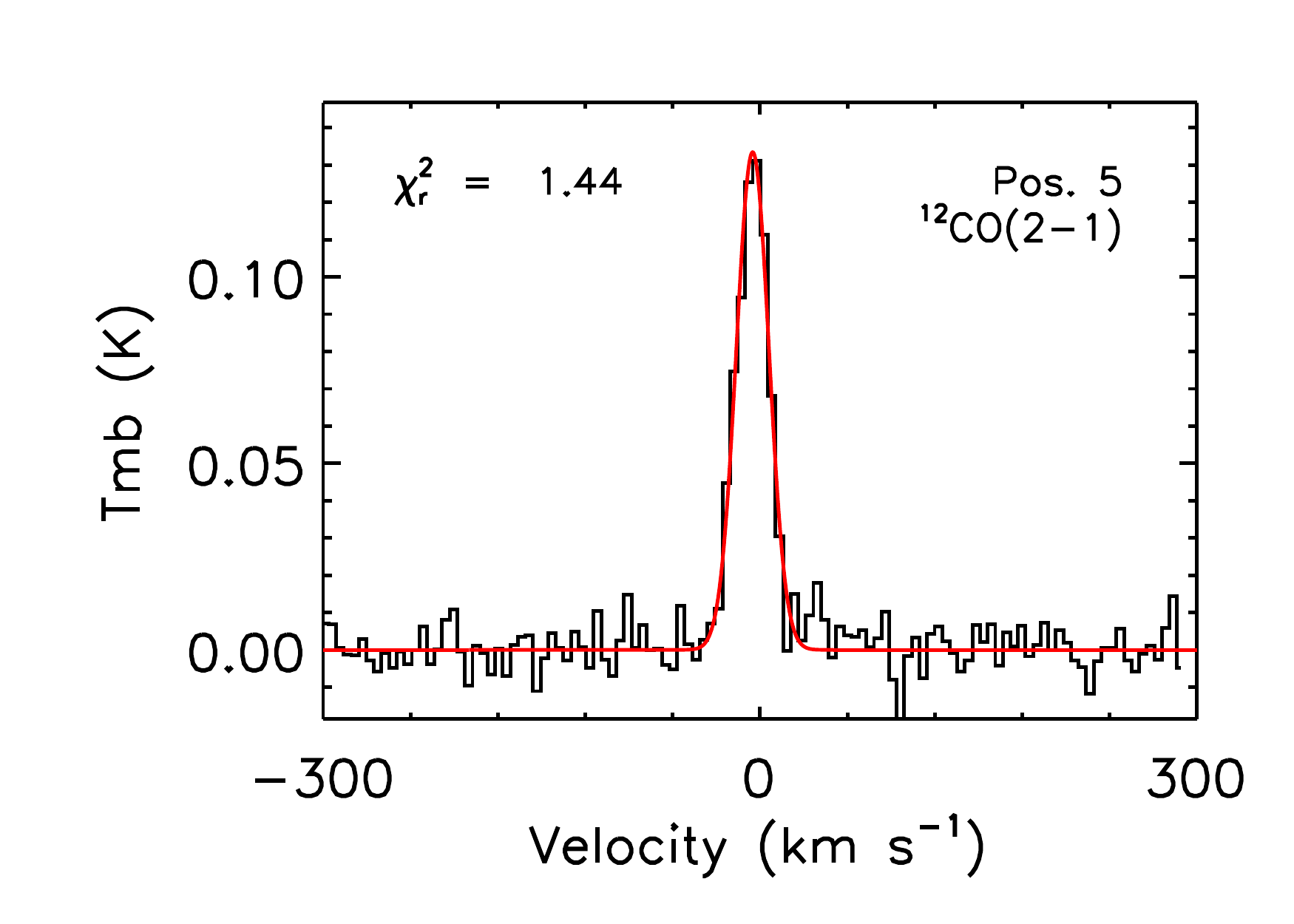}\\
   \includegraphics[width=7.0cm,clip=]{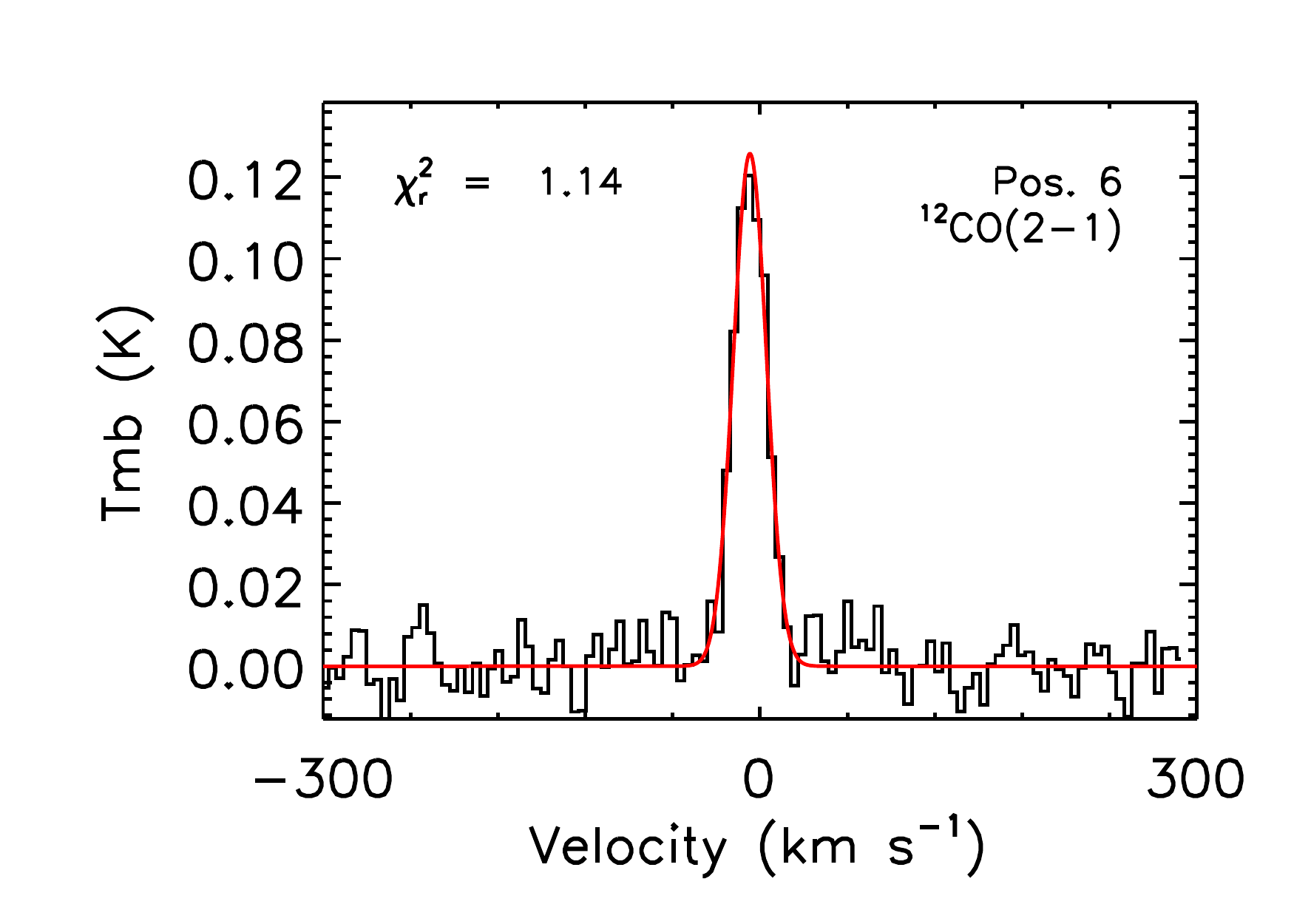}
   \includegraphics[width=7.0cm,clip=]{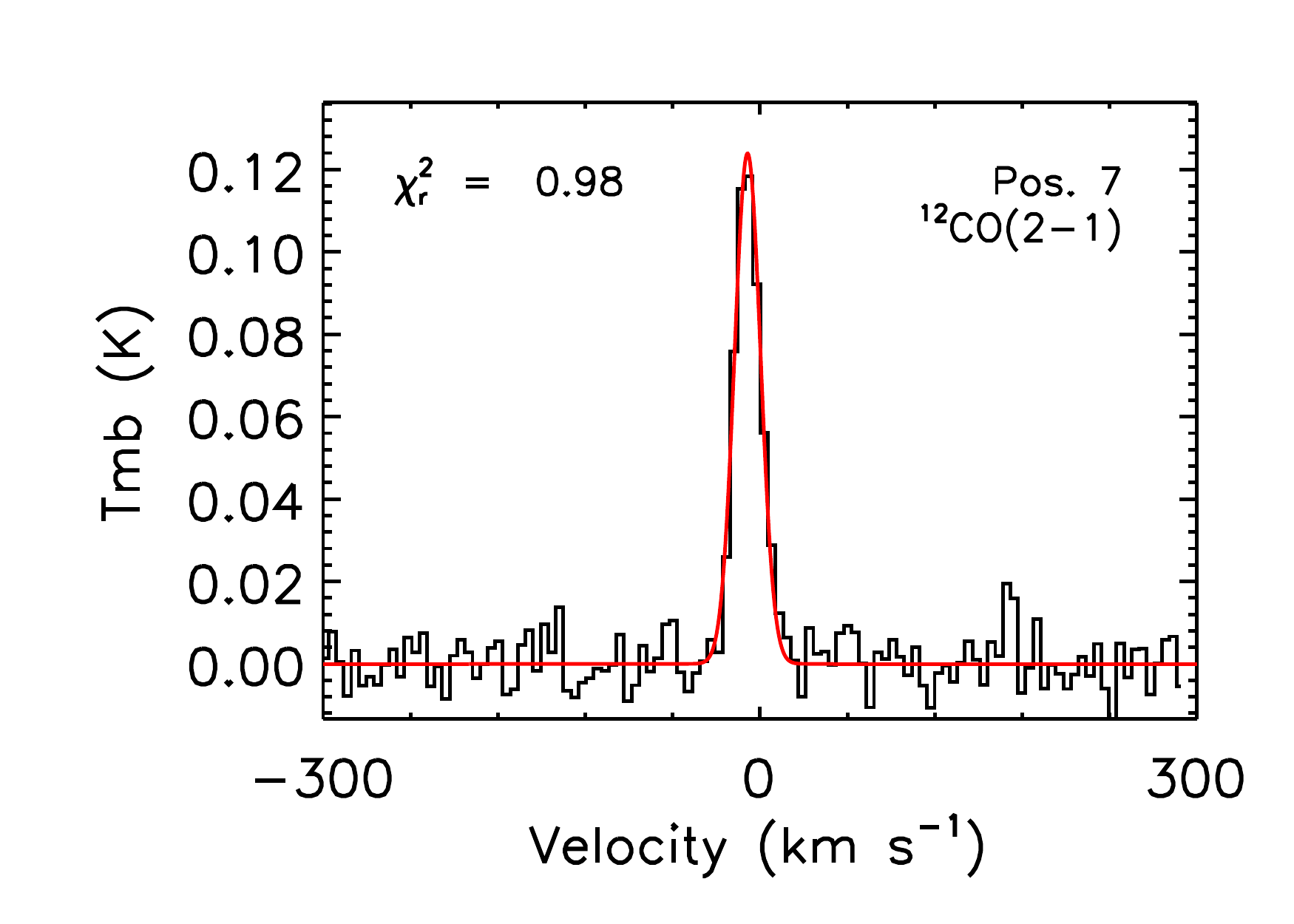}\\
  \caption{Integrated spectra of the literature CO(2--1) data extracted from the cubes are shown. Gaussian fits are overlaid. 
  The value of $\chi_{\rm r}^2$ is also shown in each panel.}
  \label{fig:spec2}
\end{figure*}

%
%
\addtocounter{figure}{-1}
\begin{figure*}
  \includegraphics[width=7.0cm,clip=]{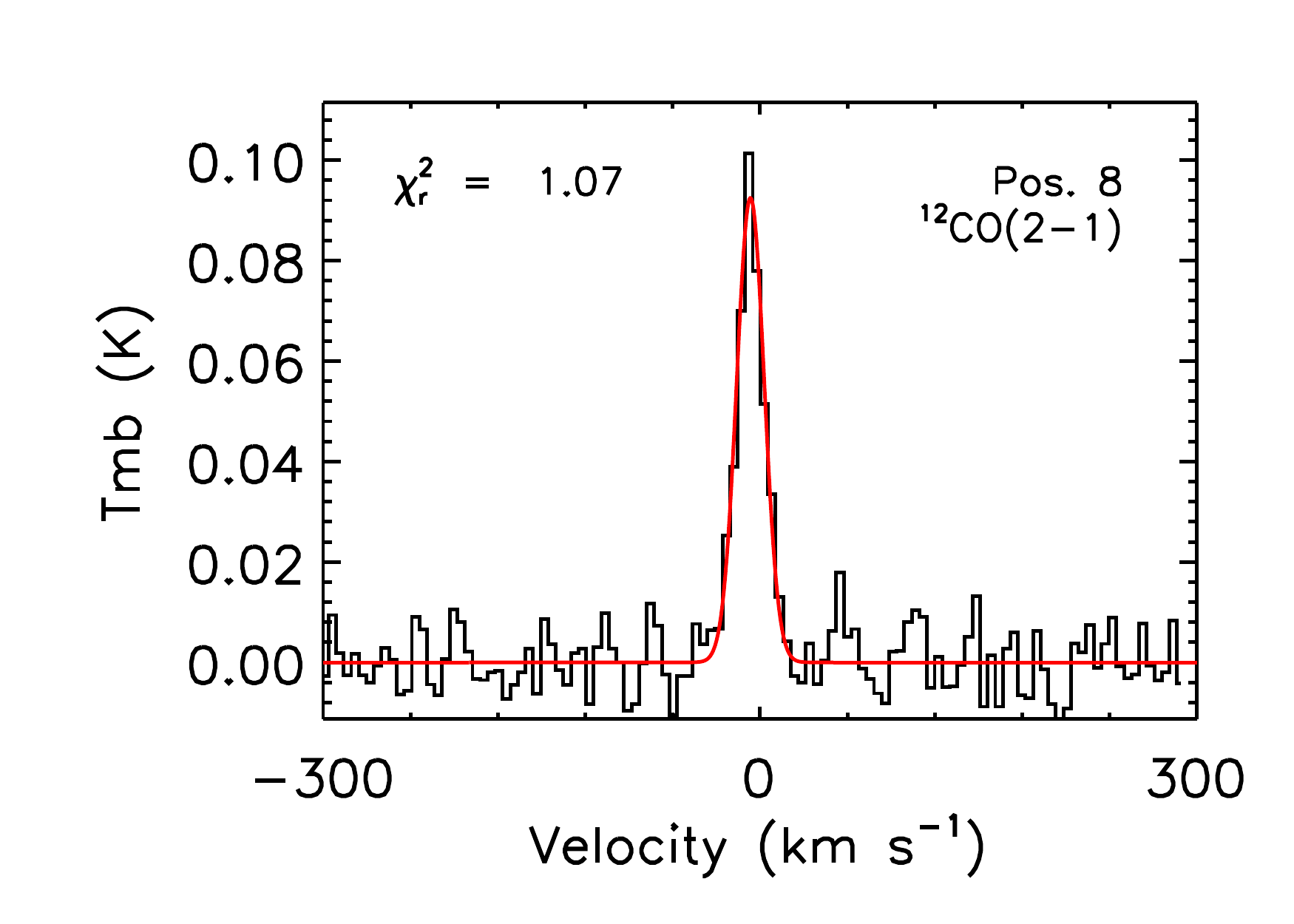}
   \includegraphics[width=7.0cm,clip=]{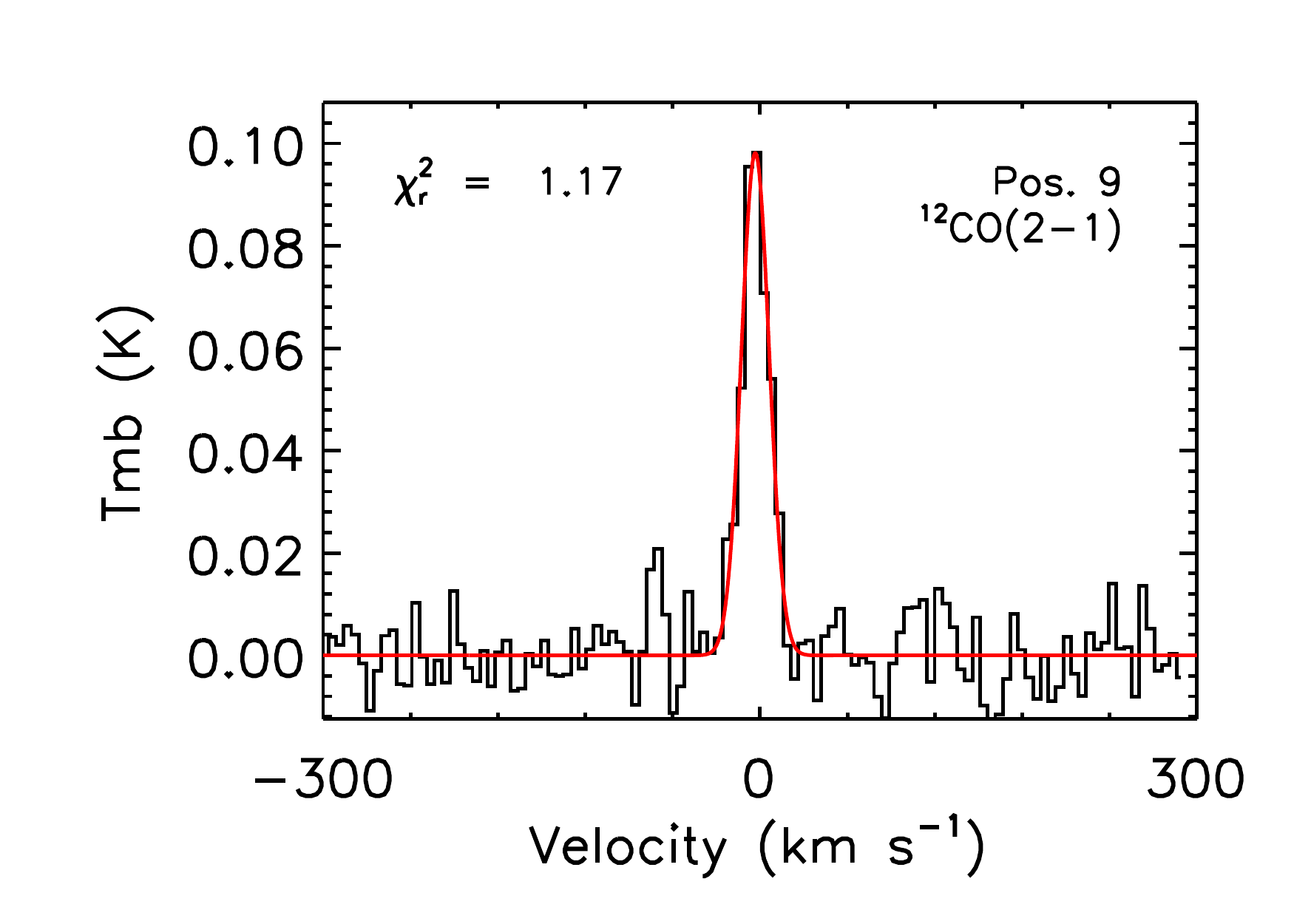}\\
  \includegraphics[width=7.0cm,clip=]{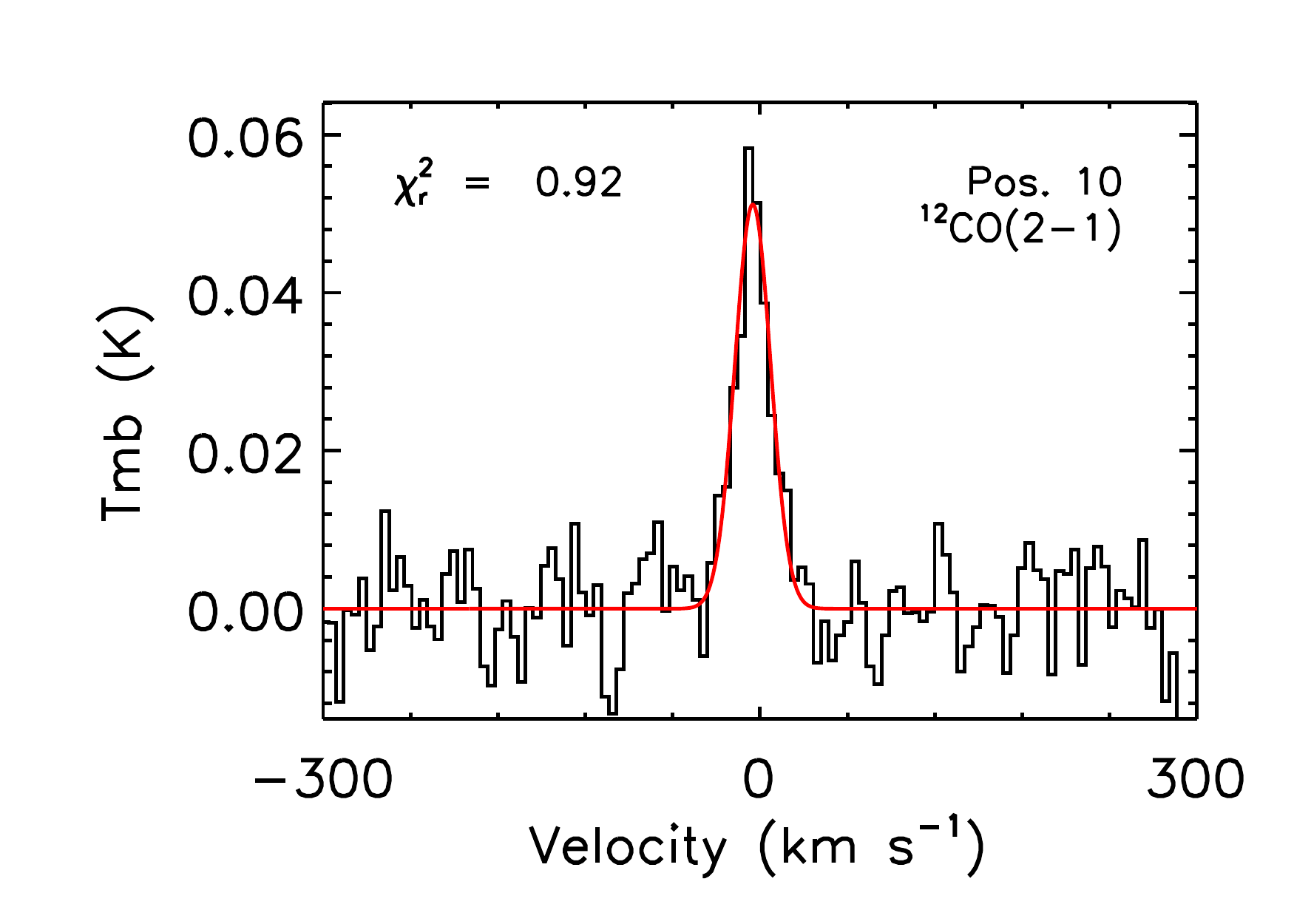}
   \includegraphics[width=7.0cm,clip=]{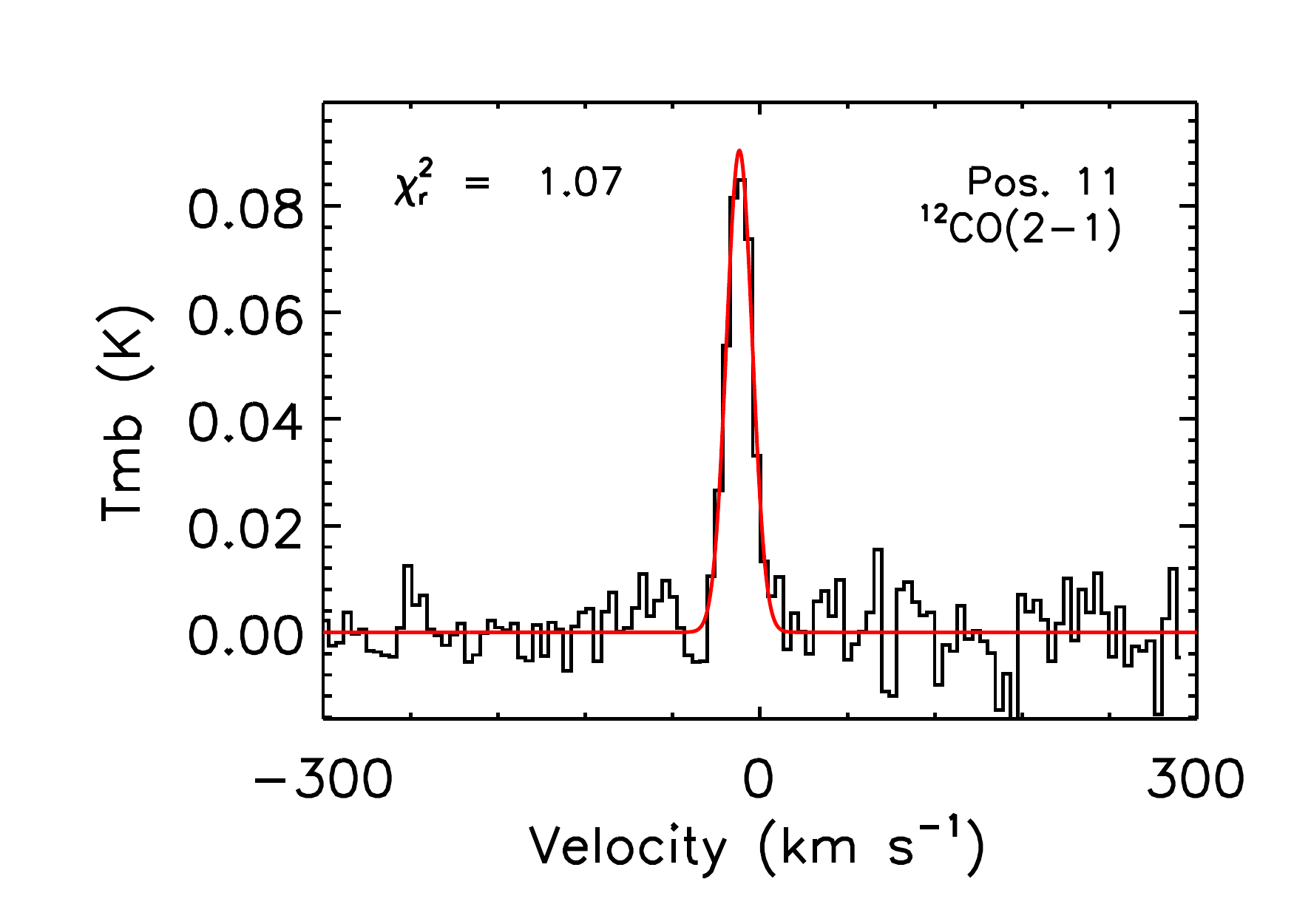}\\
    \includegraphics[width=7.0cm,clip=]{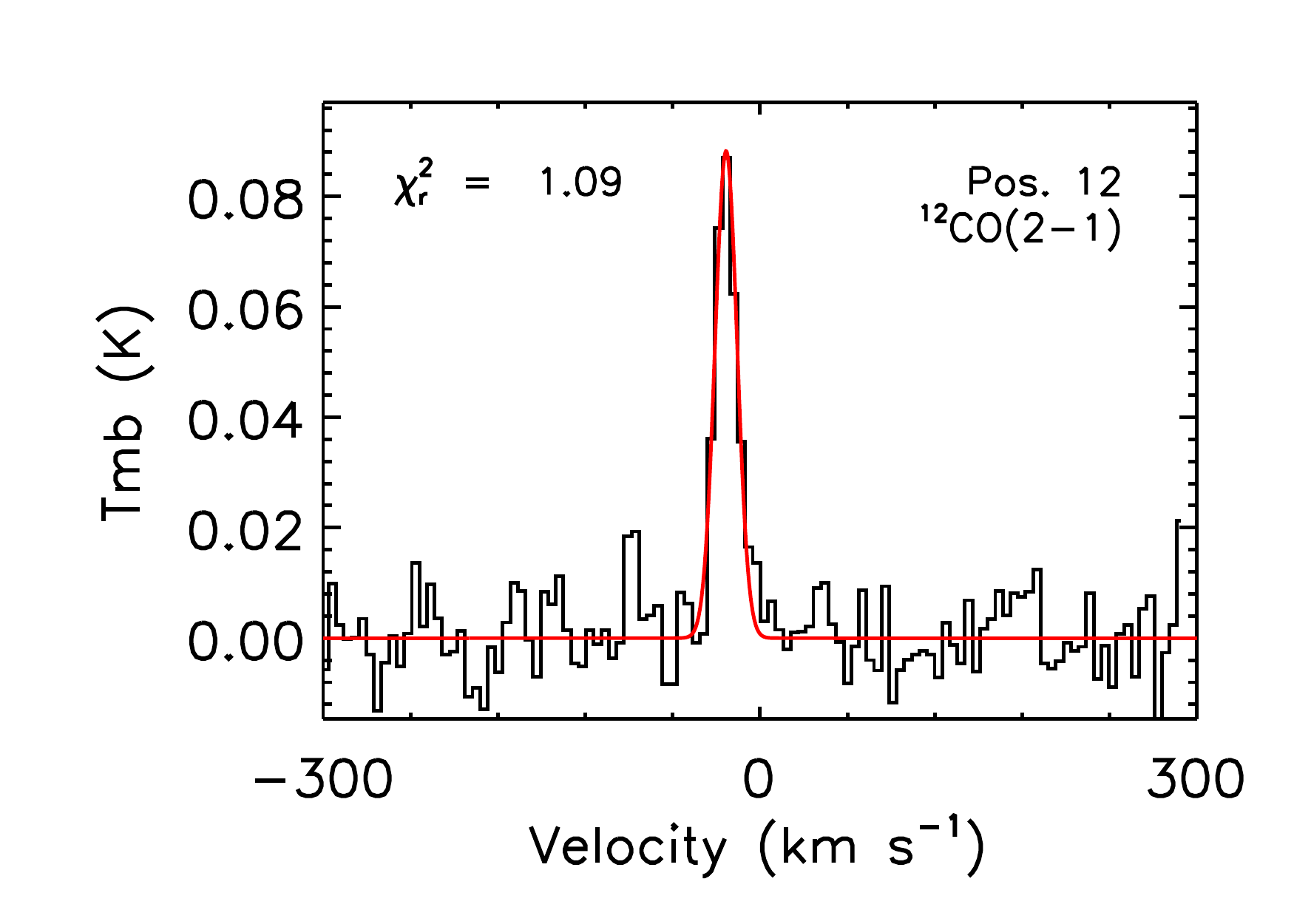}
   \includegraphics[width=7.0cm,clip=]{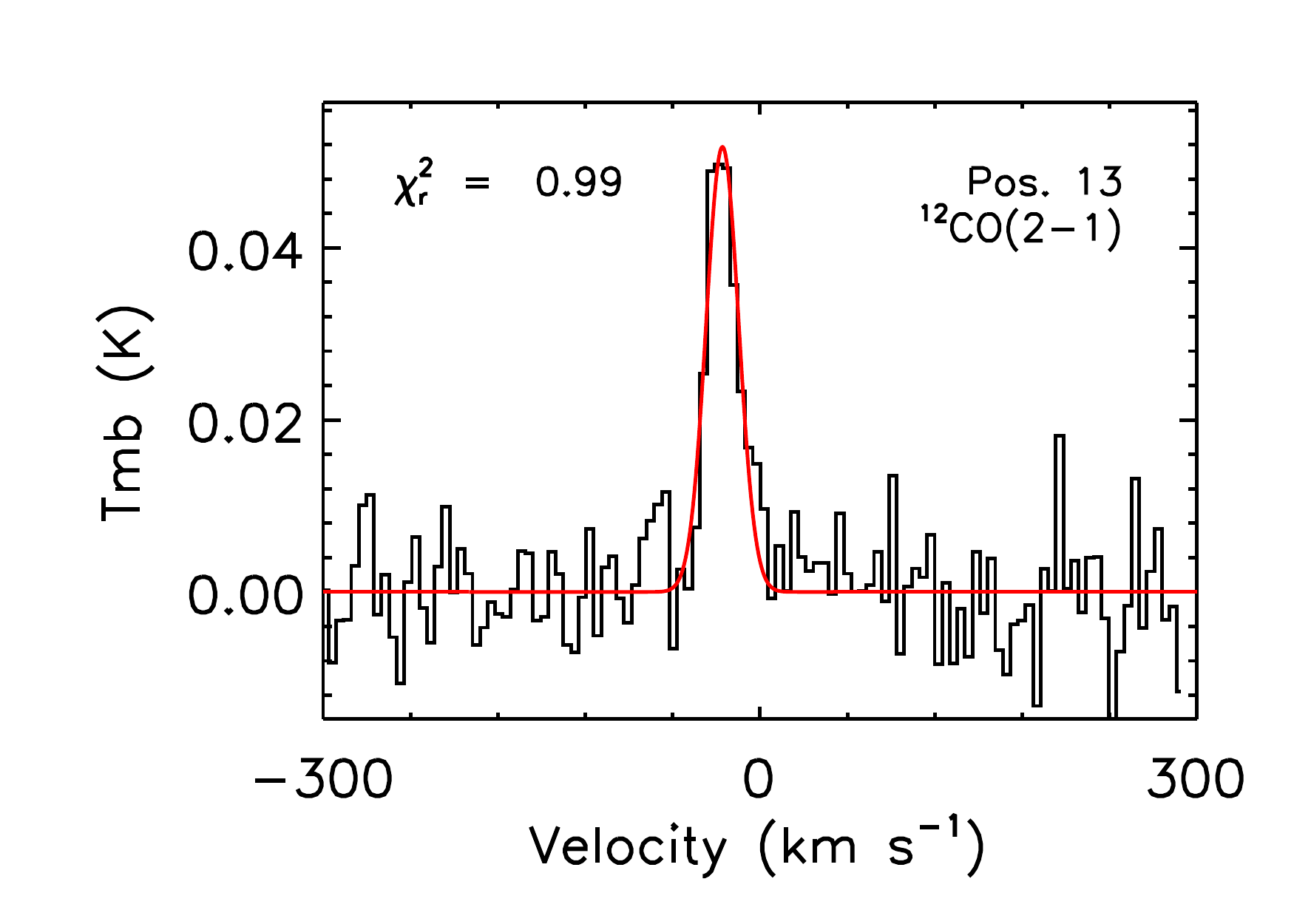}\\
   \includegraphics[width=7.0cm,clip=]{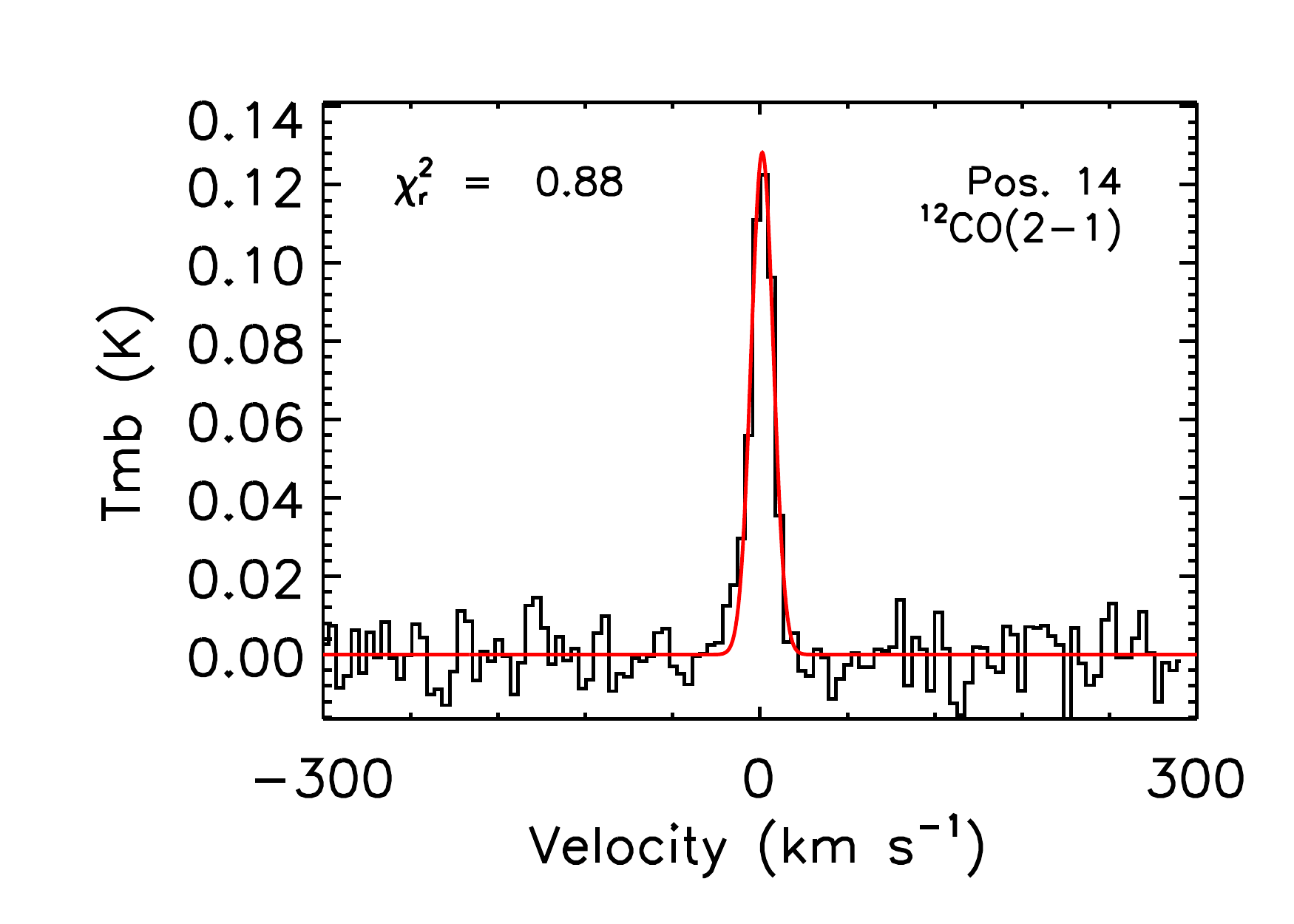}
   \includegraphics[width=7.0cm,clip=]{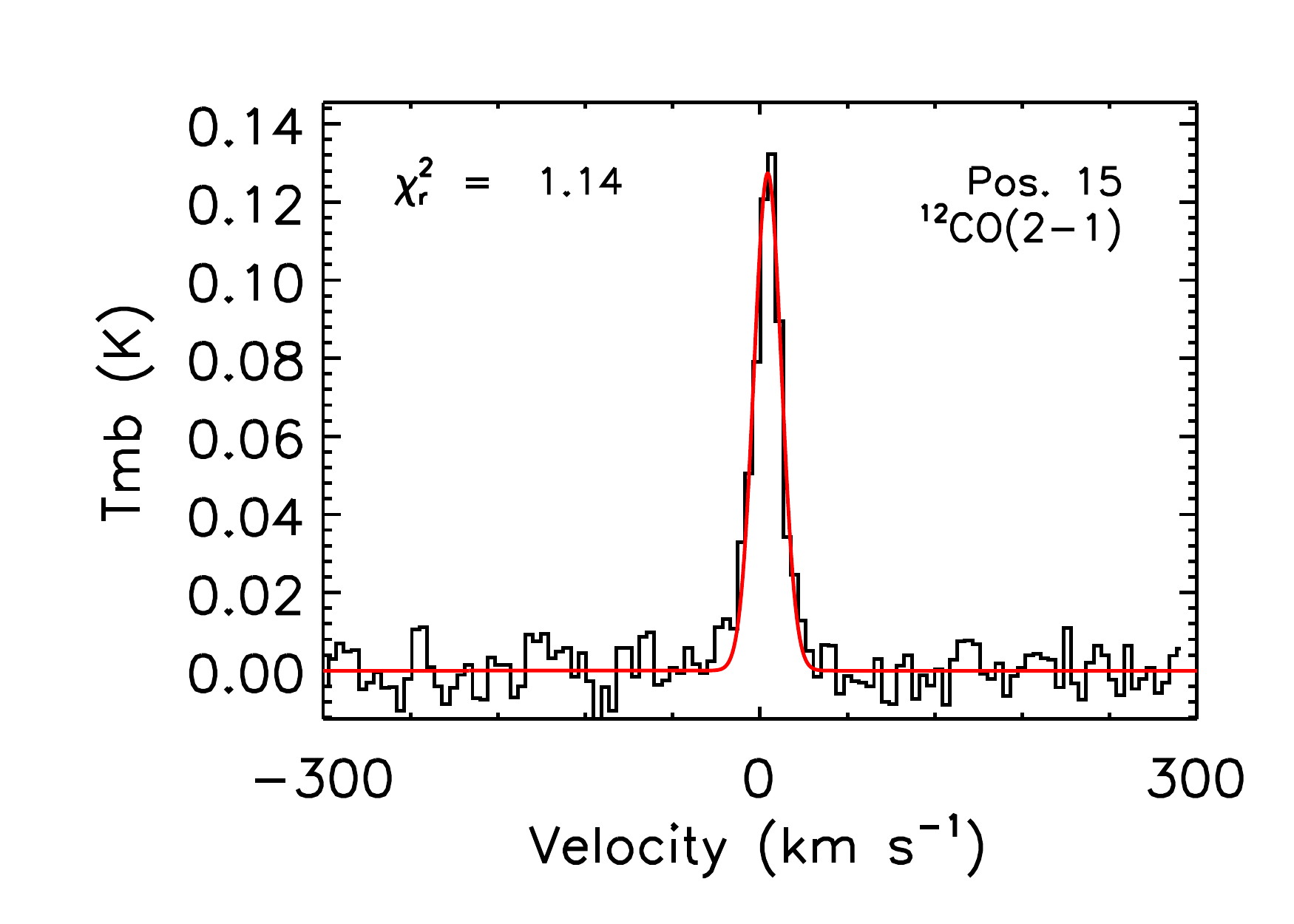}\\
  \caption{Continued.}
  \label{fig:spec2}
\end{figure*}

%
%
\addtocounter{figure}{-1}
\begin{figure*}
  \includegraphics[width=7.0cm,clip=]{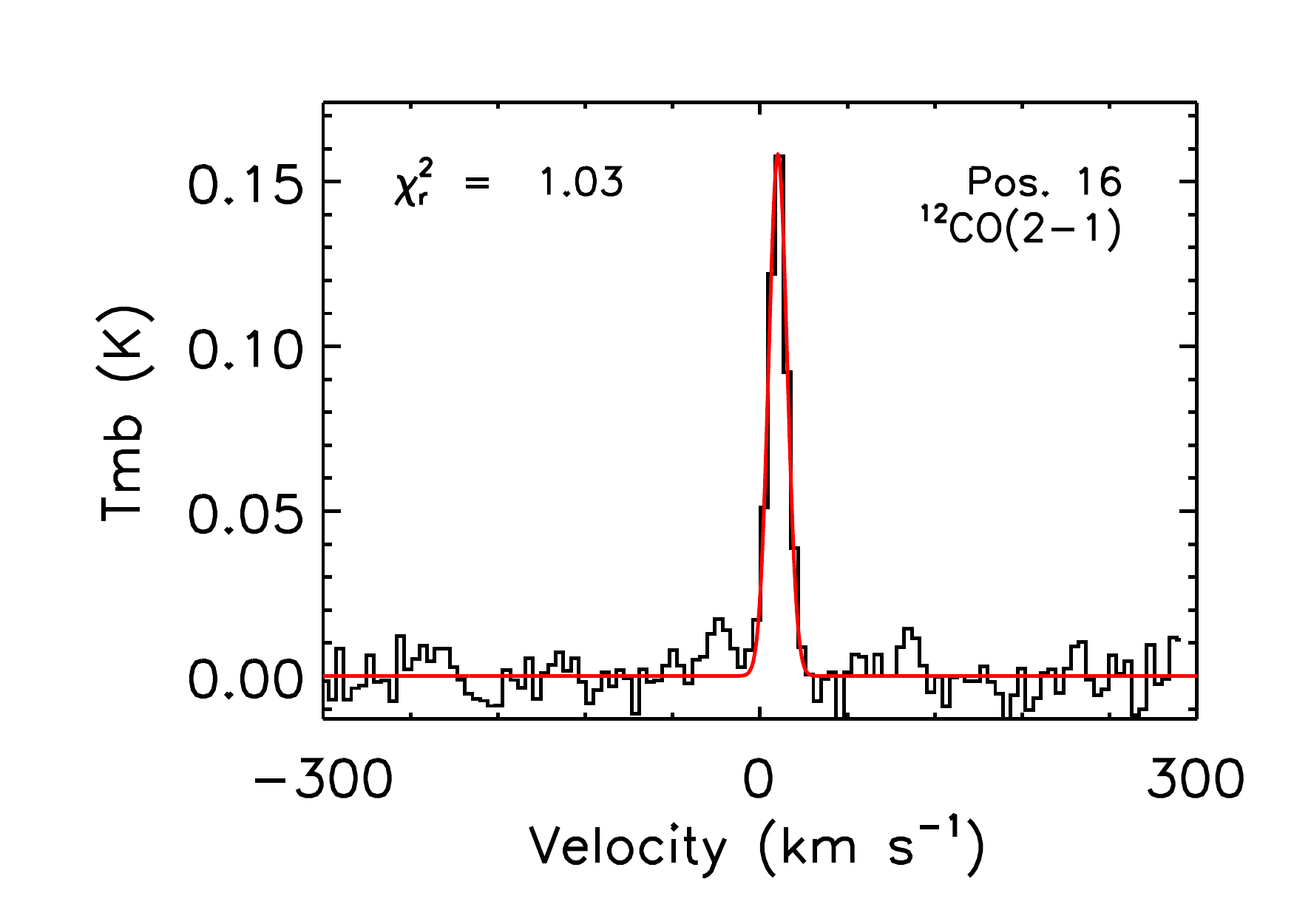}
   \includegraphics[width=7.0cm,clip=]{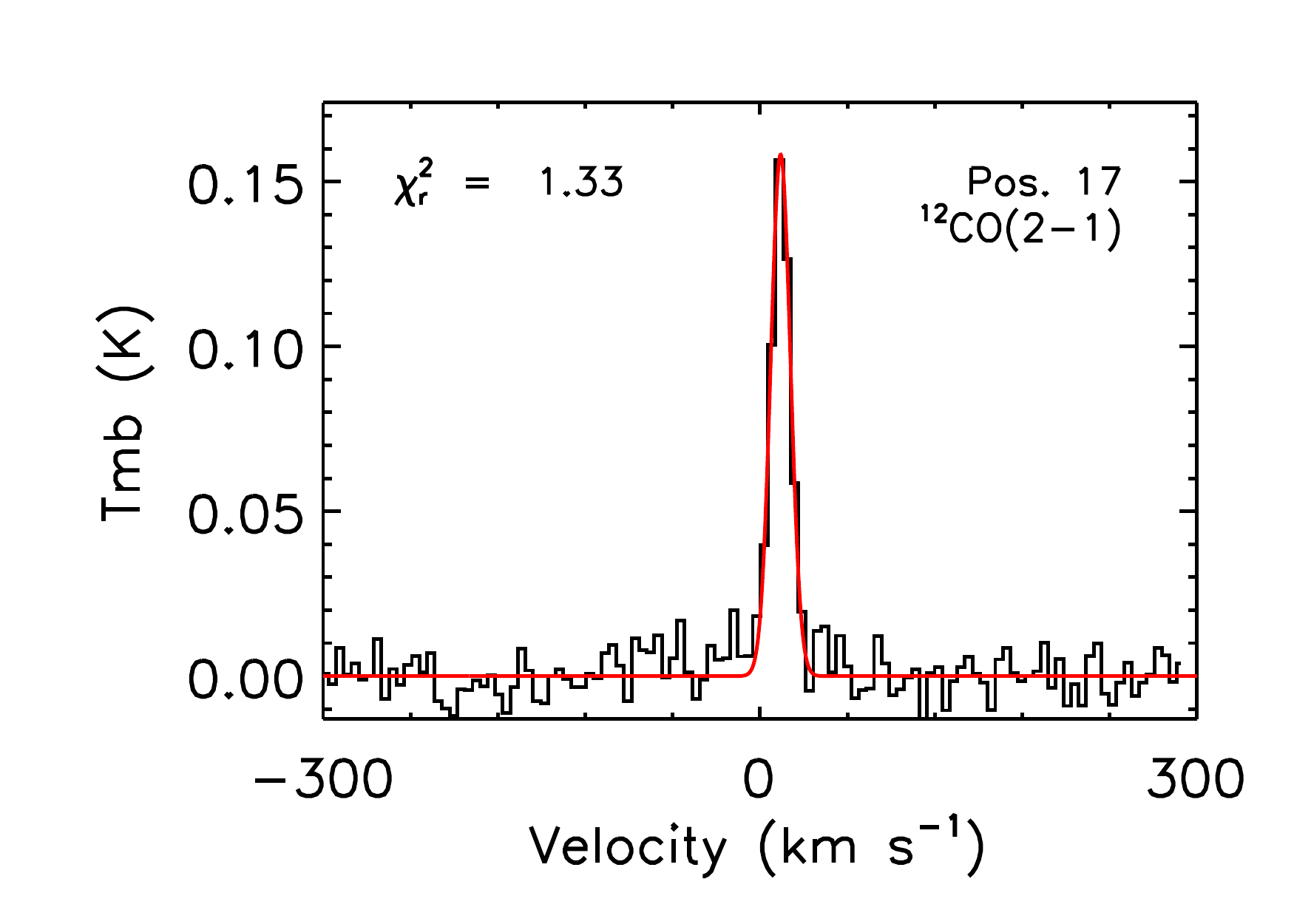}\\
  \includegraphics[width=7.0cm,clip=]{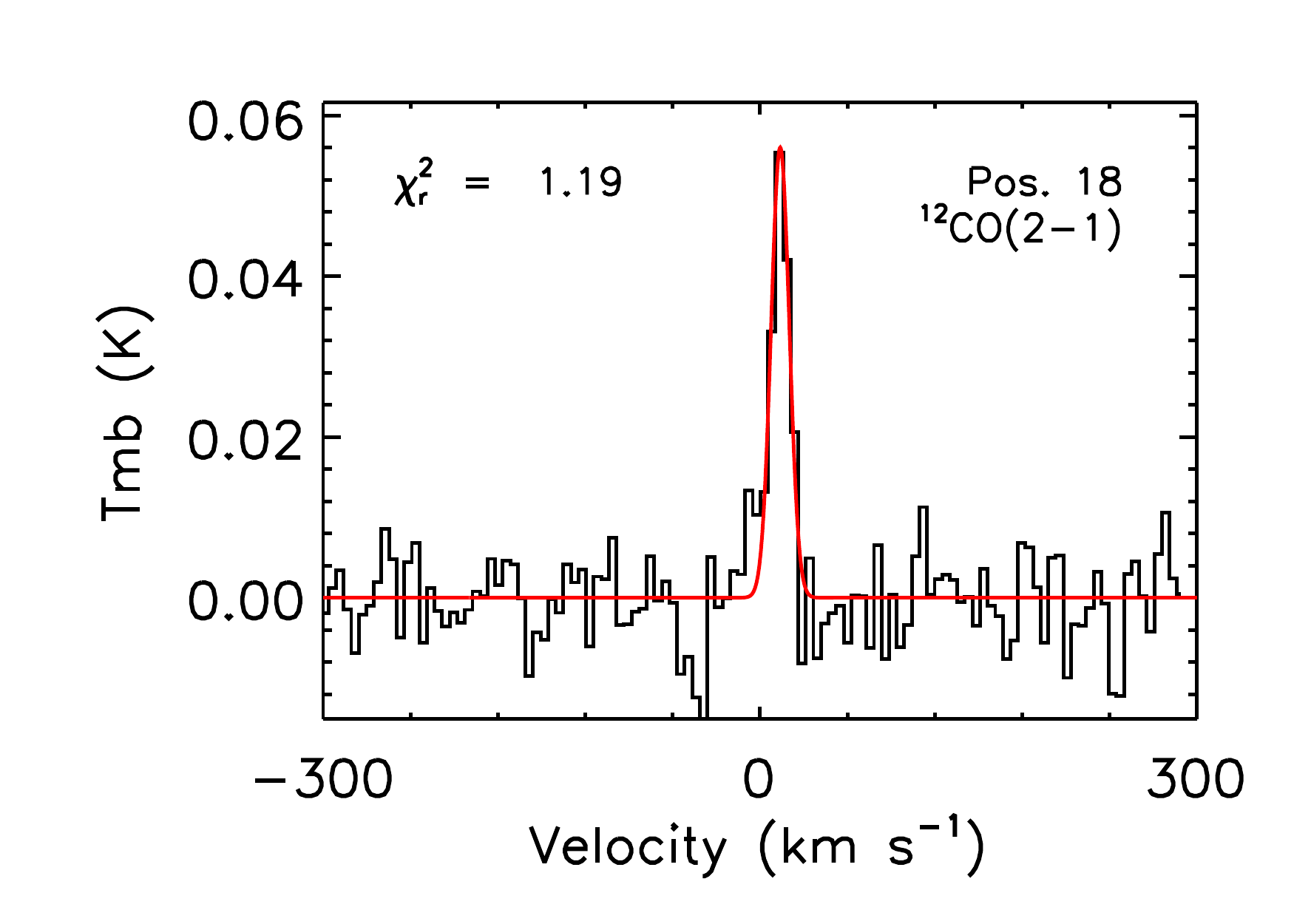}
   \includegraphics[width=7.0cm,clip=]{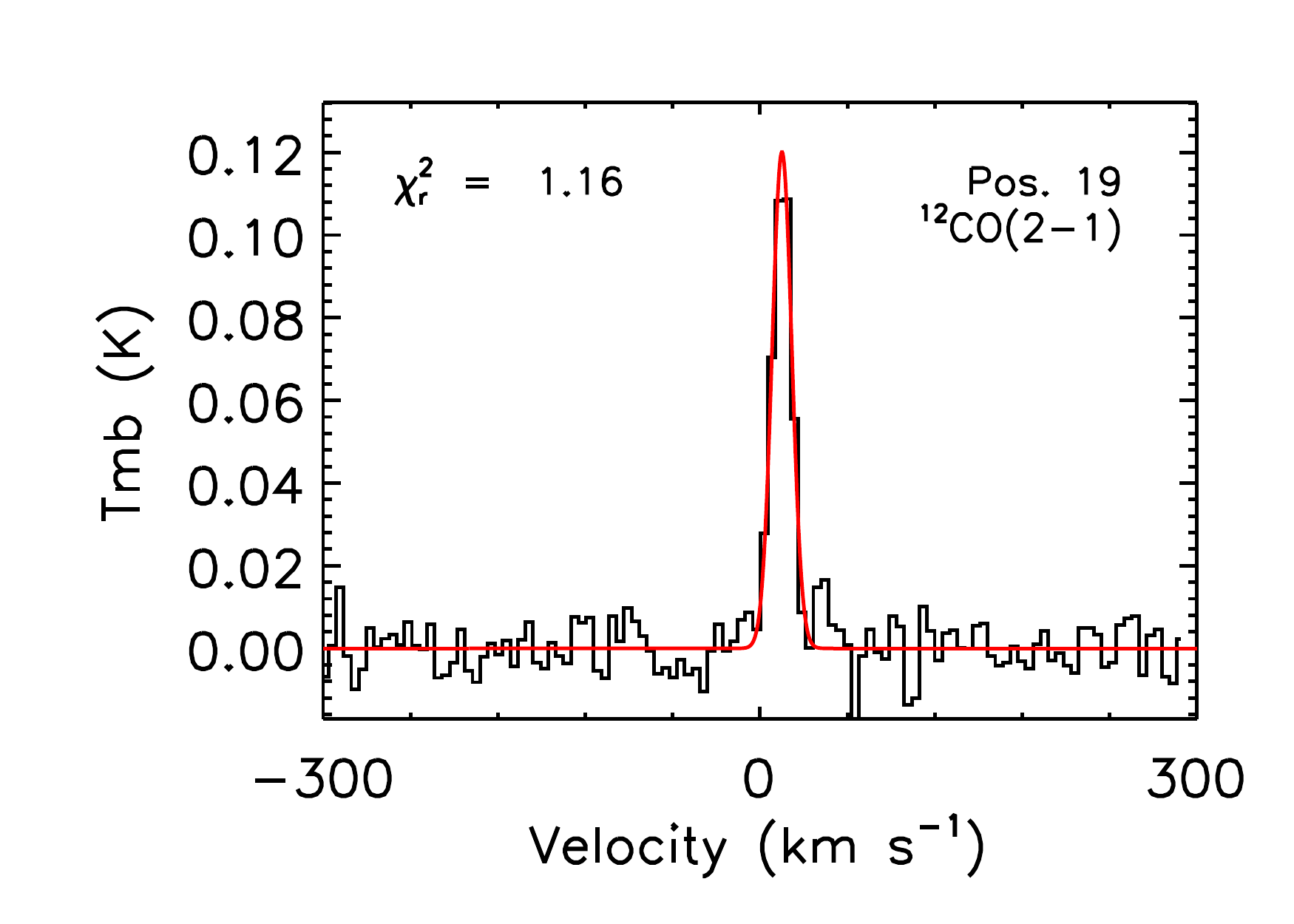}\\
  \caption{Continued.}
  \label{fig:spec2}
\end{figure*}

\clearpage

\begin{figure*}
  \includegraphics[width=7.0cm,clip=]{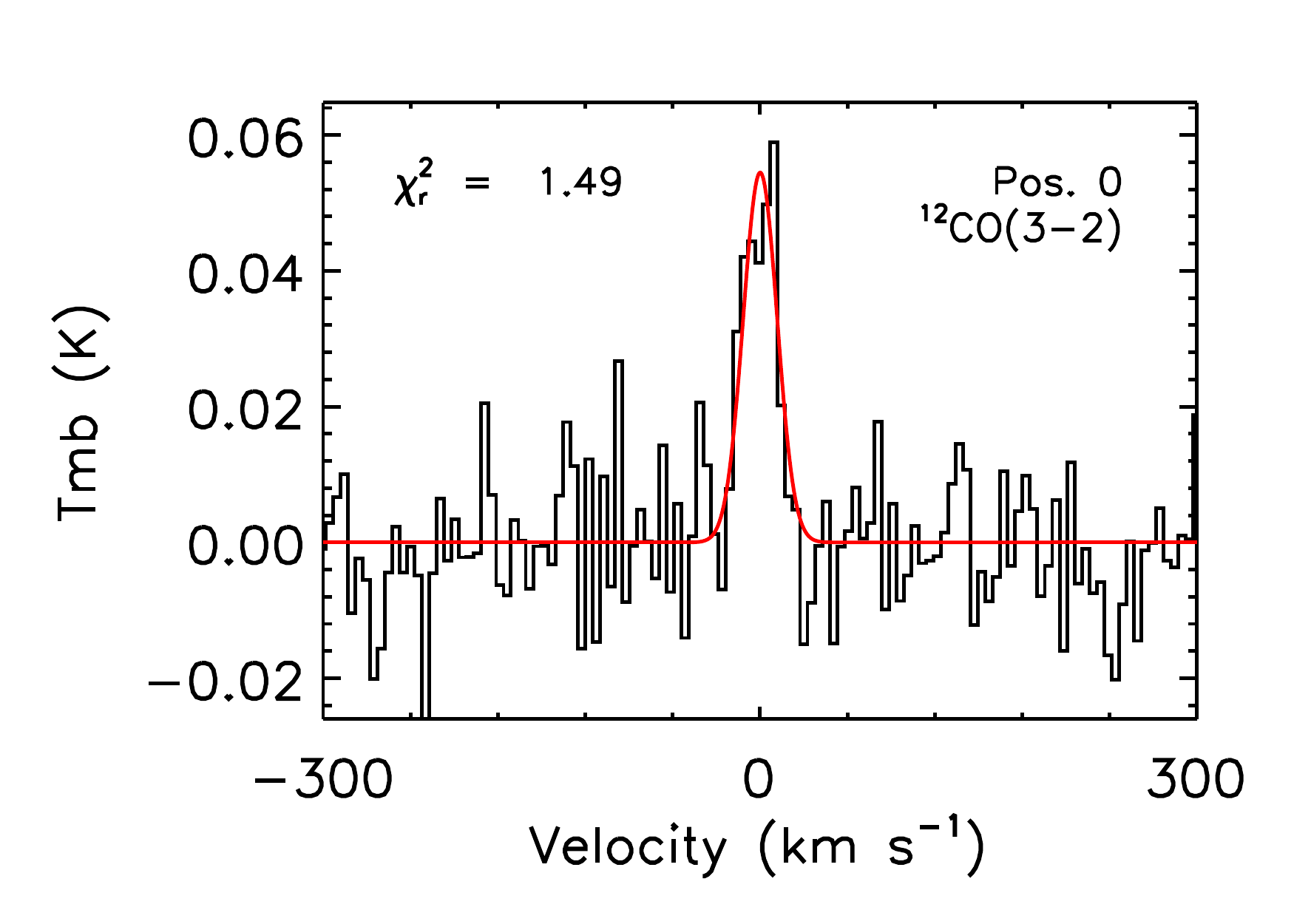}
   \includegraphics[width=7.0cm,clip=]{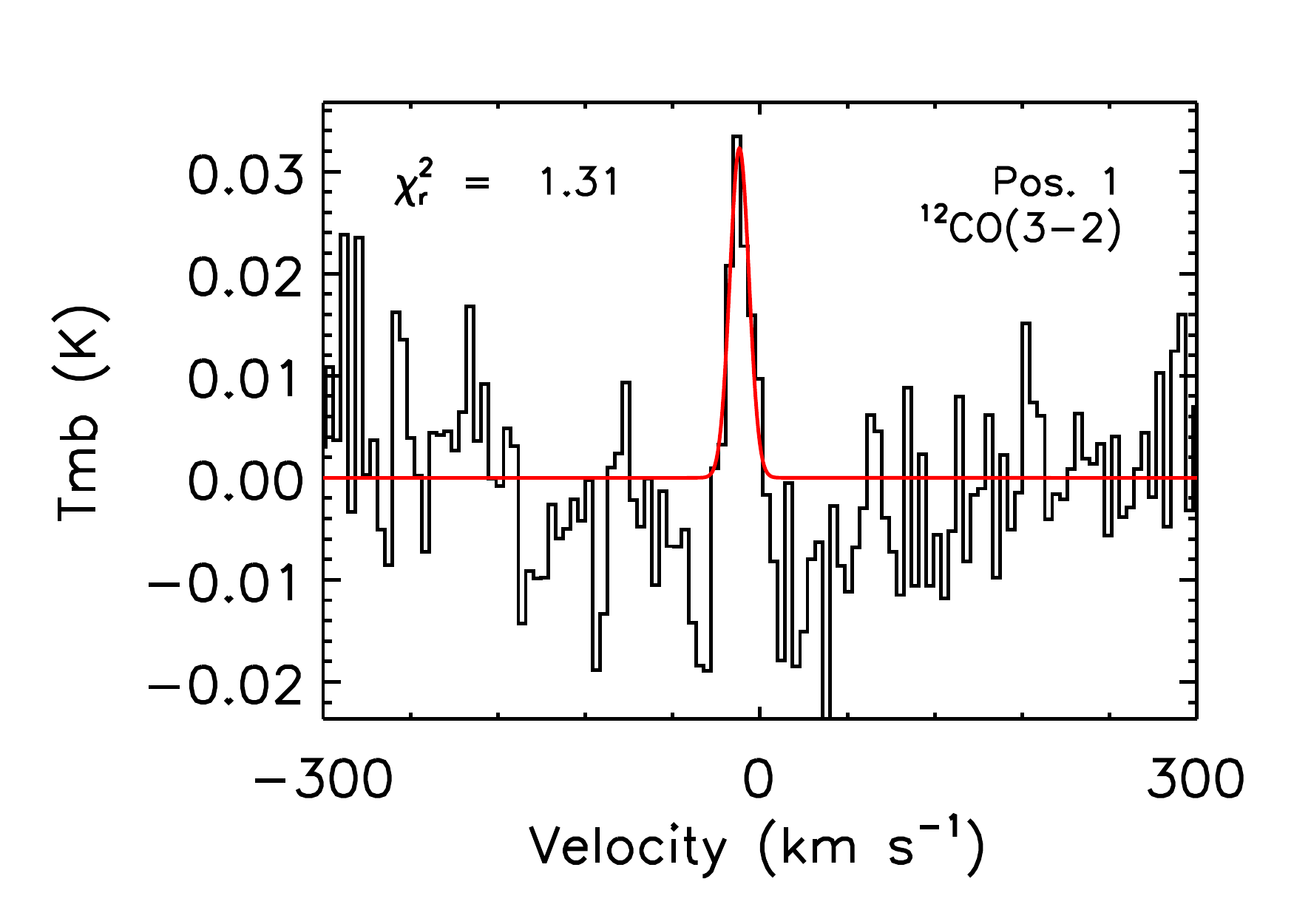}\\
  \includegraphics[width=7.0cm,clip=]{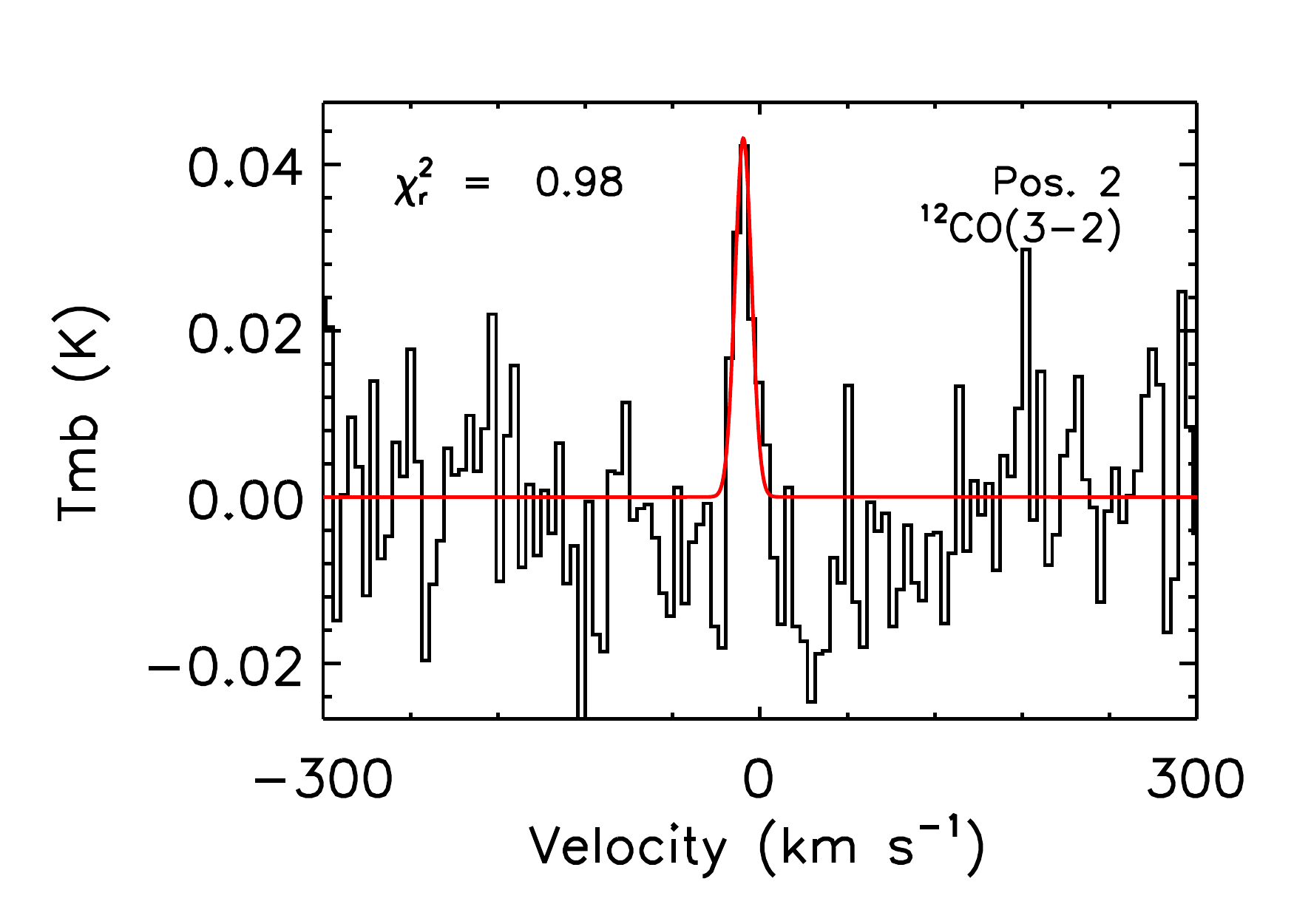}
   \includegraphics[width=7.0cm,clip=]{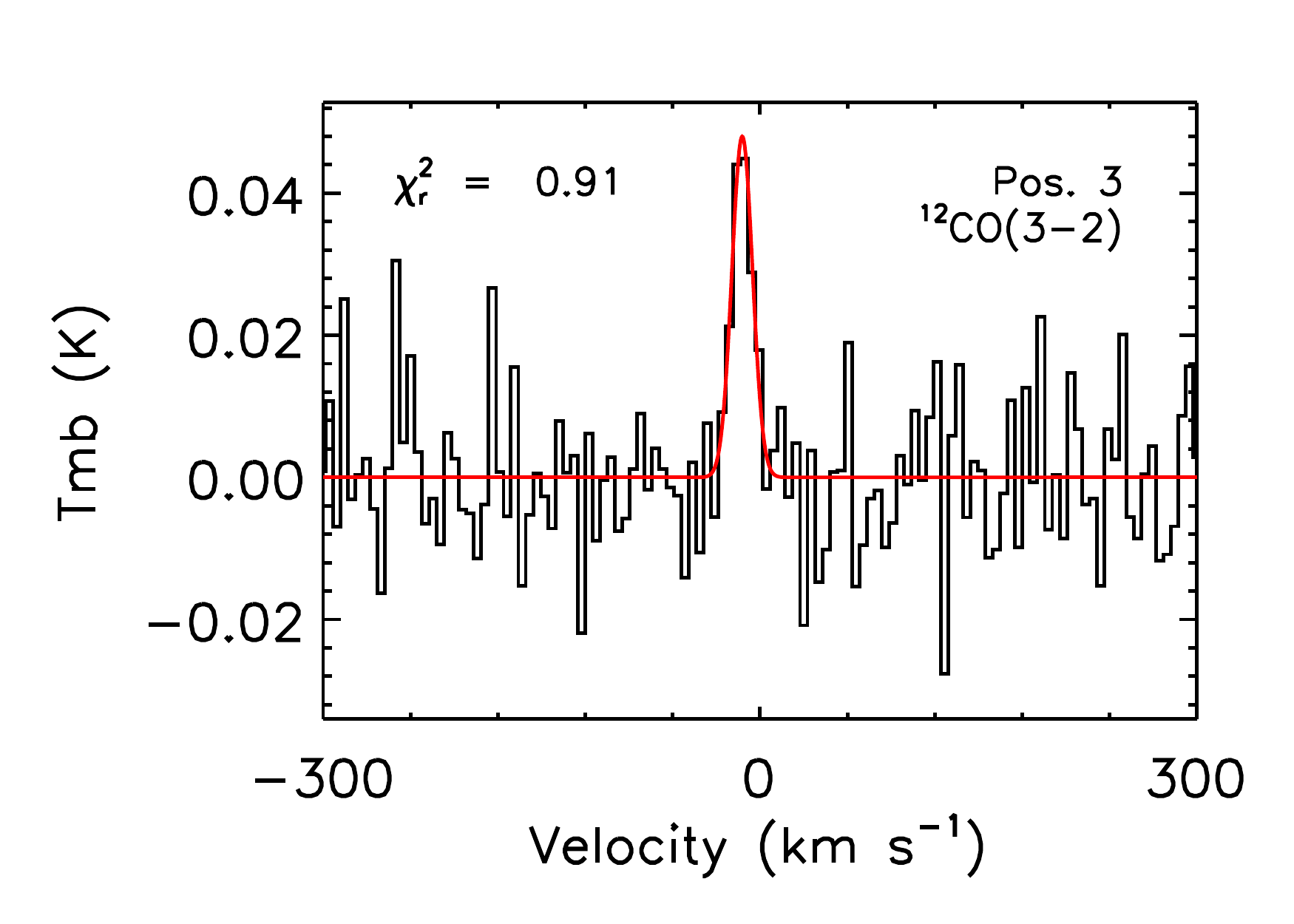}\\
    \includegraphics[width=7.0cm,clip=]{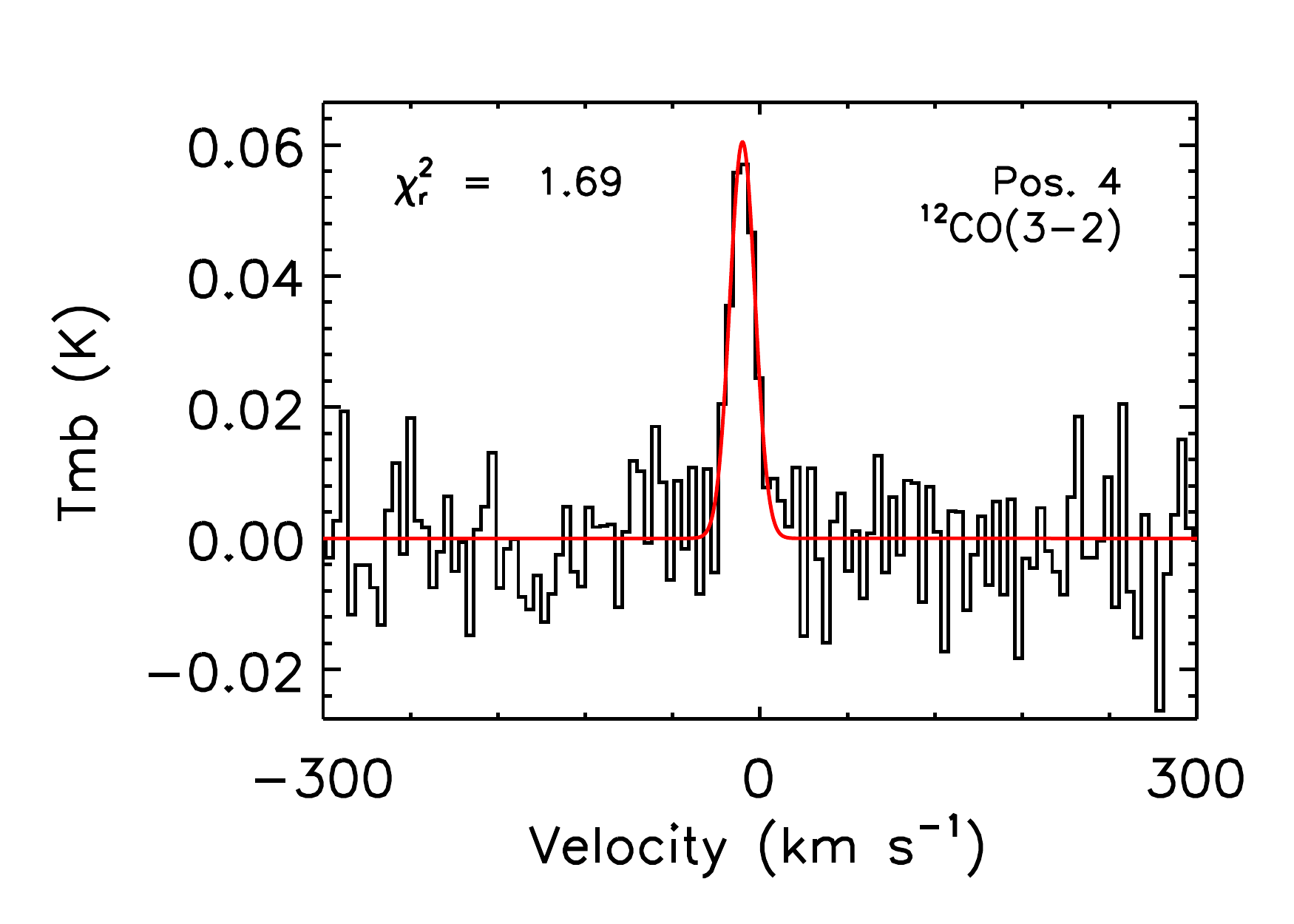}
   \includegraphics[width=7.0cm,clip=]{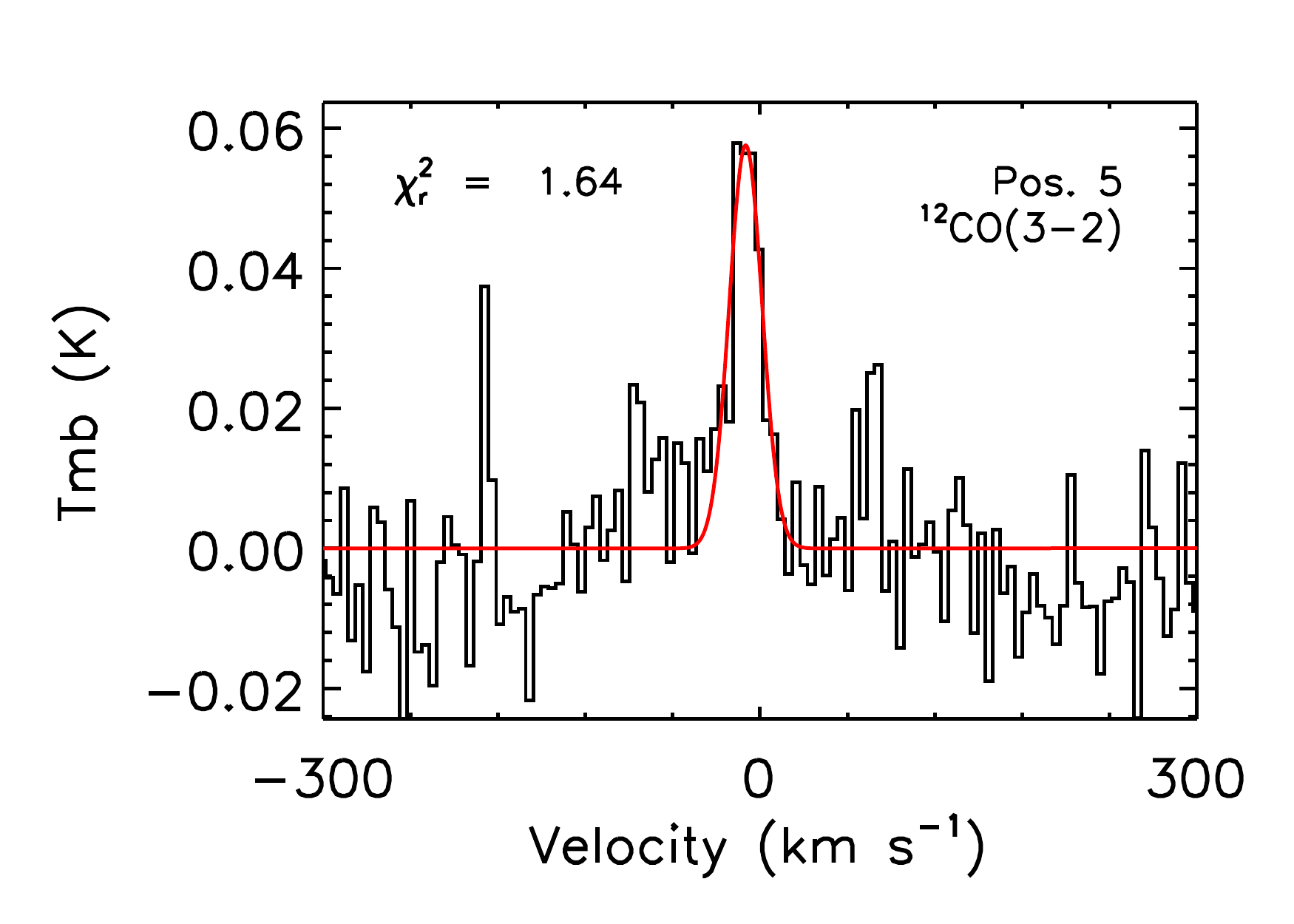}\\
   \includegraphics[width=7.0cm,clip=]{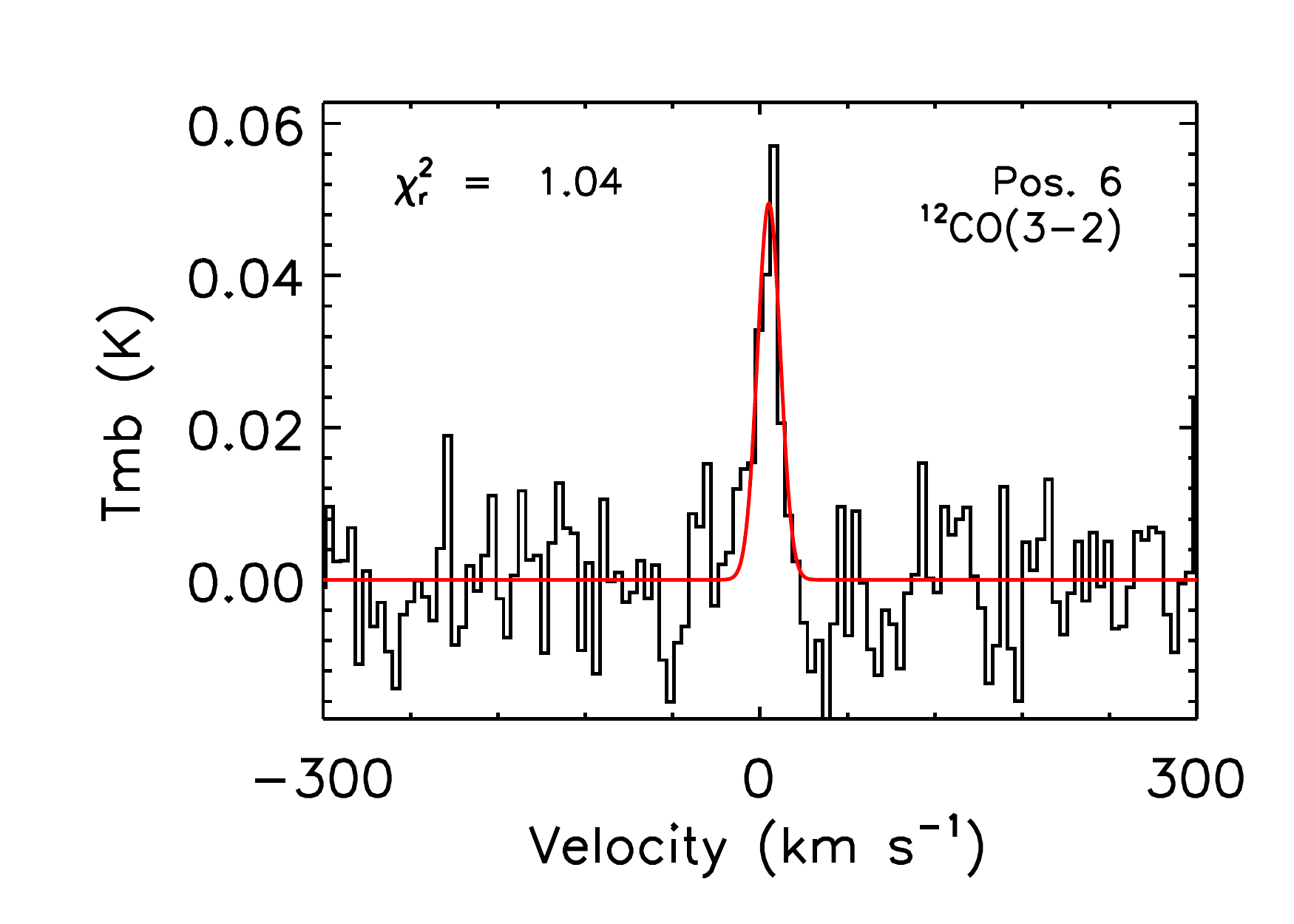}
   \includegraphics[width=7.0cm,clip=]{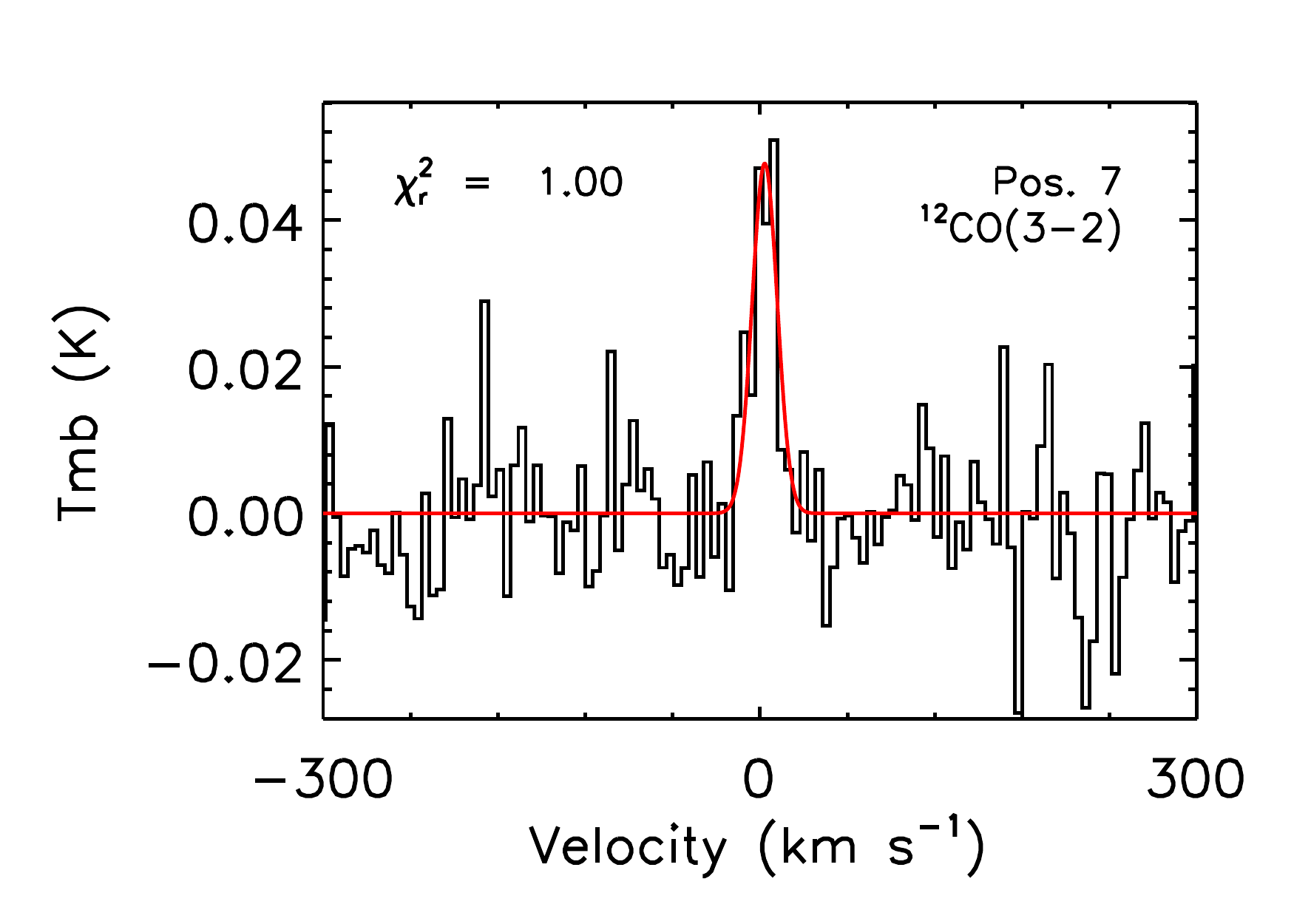}\\
  \caption{Integrated spectra of the literature CO(3--2) data extracted from the cubes are shown. Gaussian fits are overlaid. 
  The value of $\chi_{\rm r}^2$ is also shown in each panel.}
  \label{fig:spec3}
\end{figure*}

\addtocounter{figure}{-1}
\begin{figure*}
  \includegraphics[width=7.0cm,clip=]{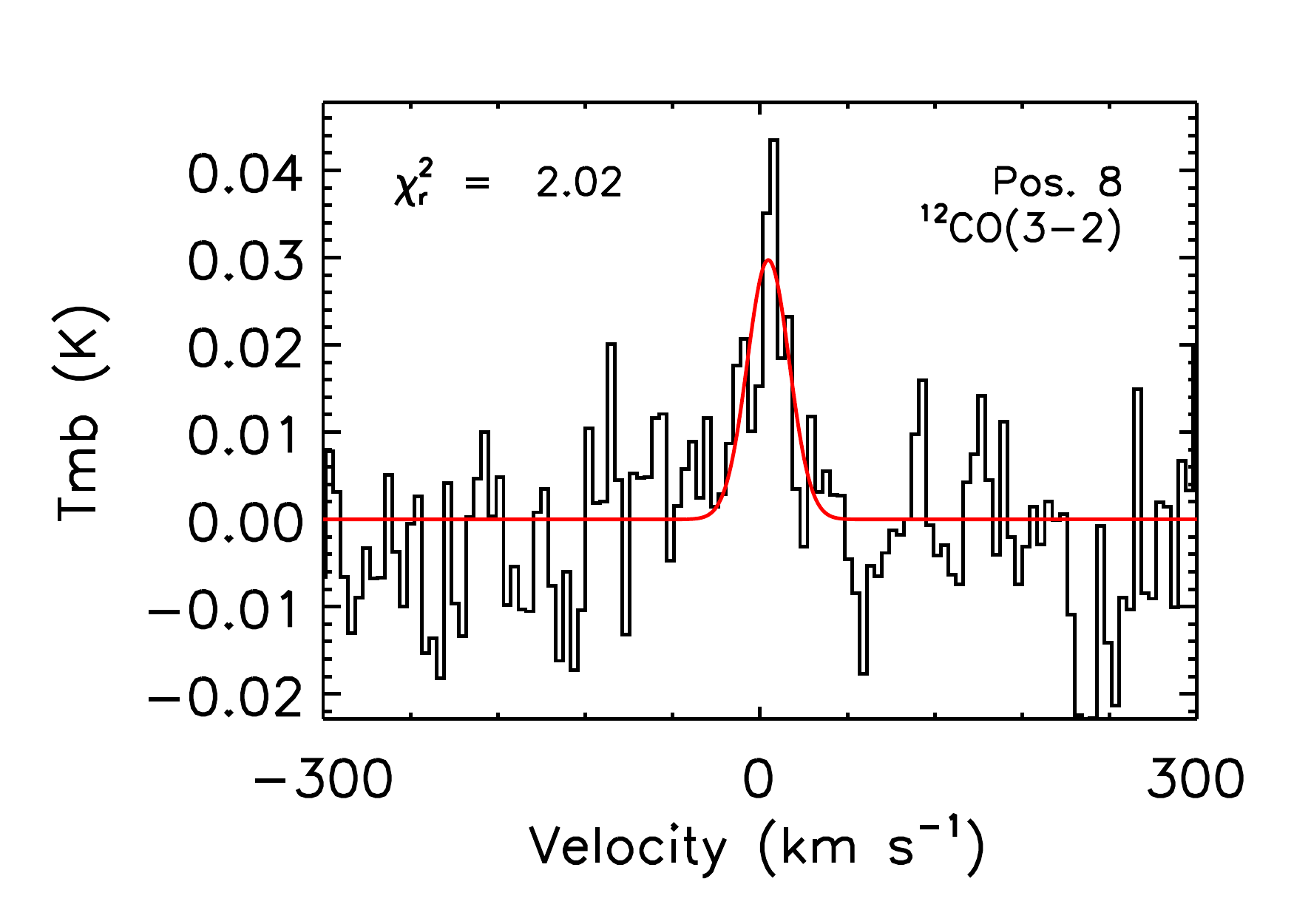}
   \includegraphics[width=7.0cm,clip=]{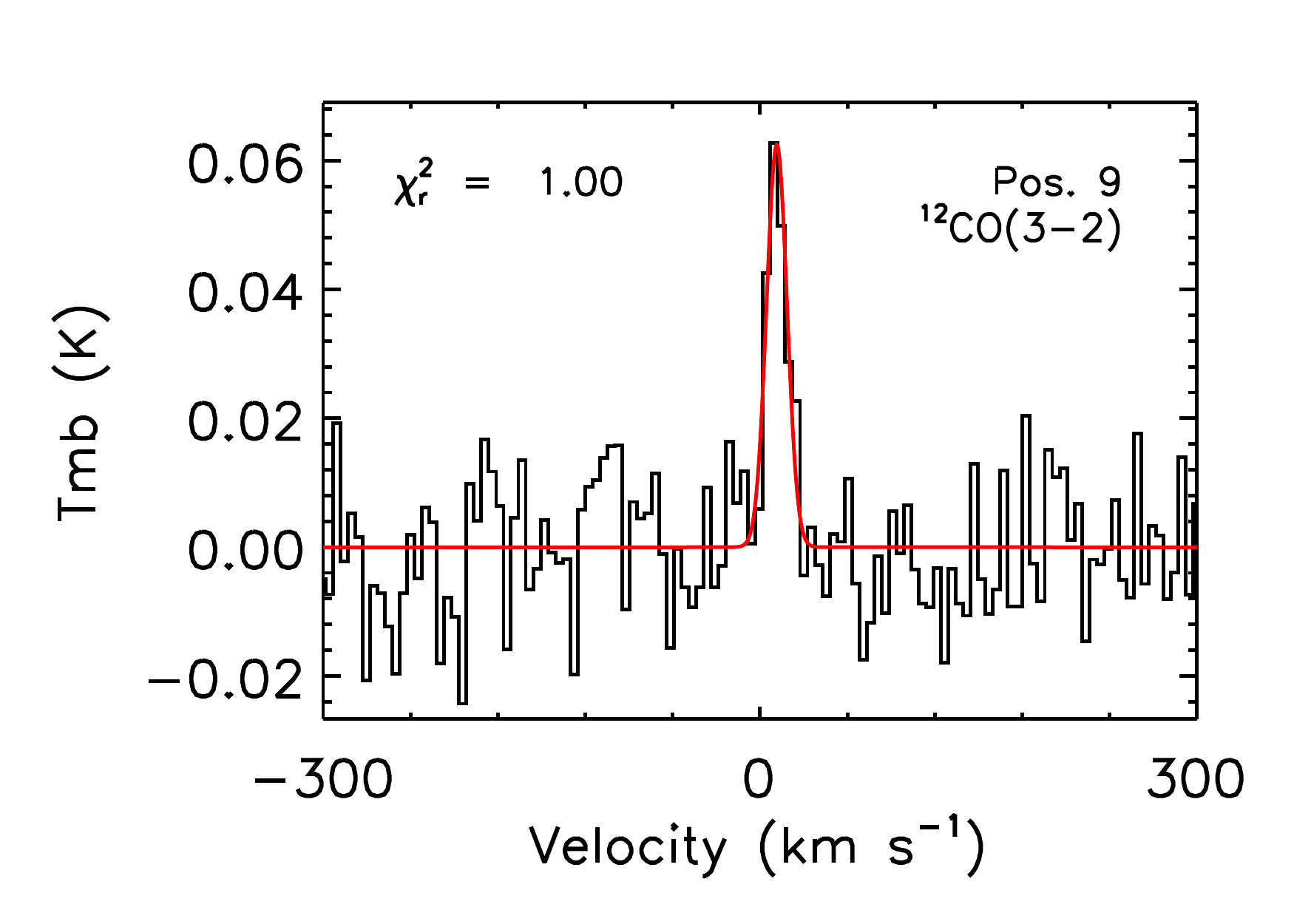}\\
  \includegraphics[width=7.0cm,clip=]{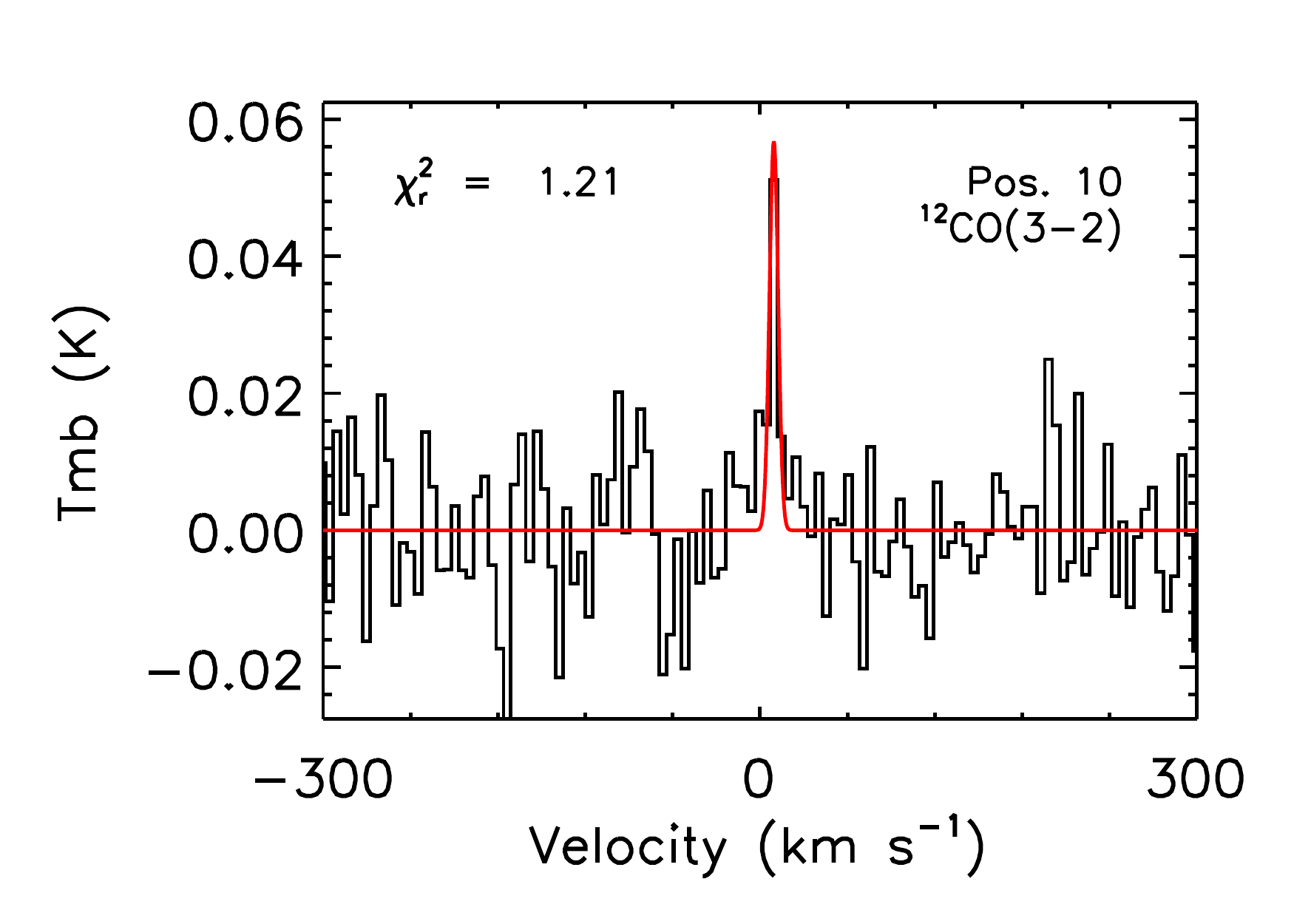}
   \includegraphics[width=7.0cm,clip=]{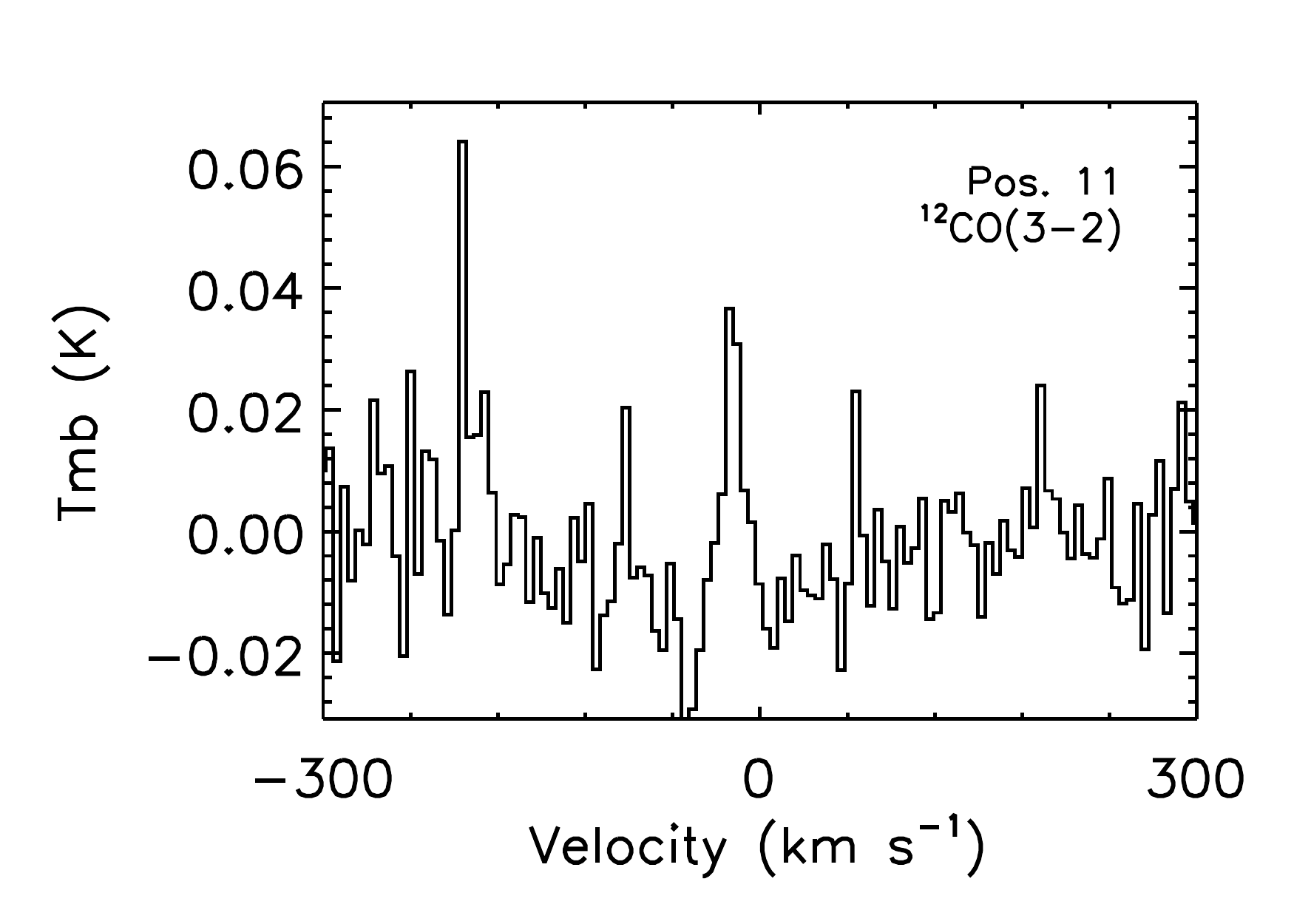}\\
    \includegraphics[width=7.0cm,clip=]{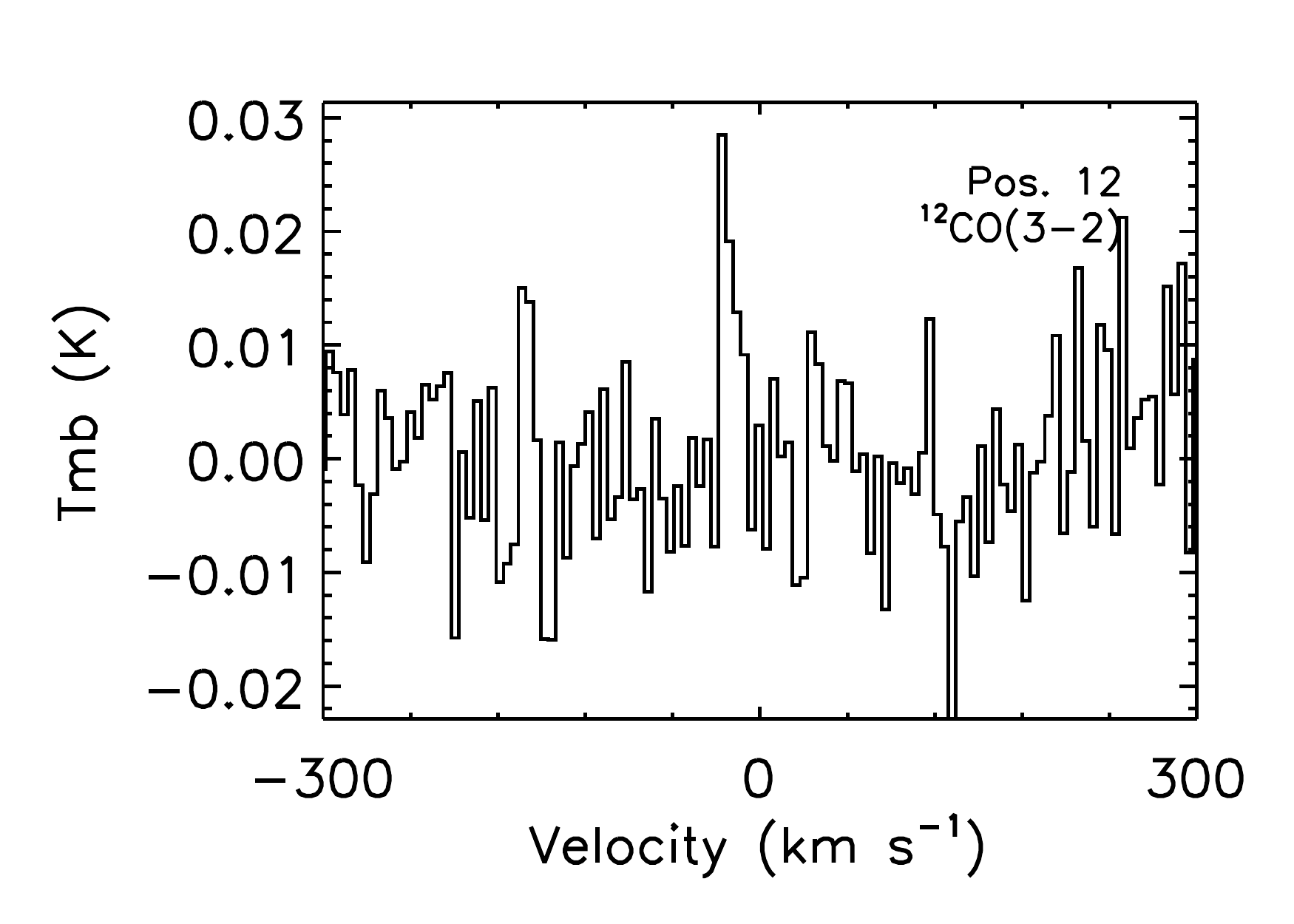}
   \includegraphics[width=7.0cm,clip=]{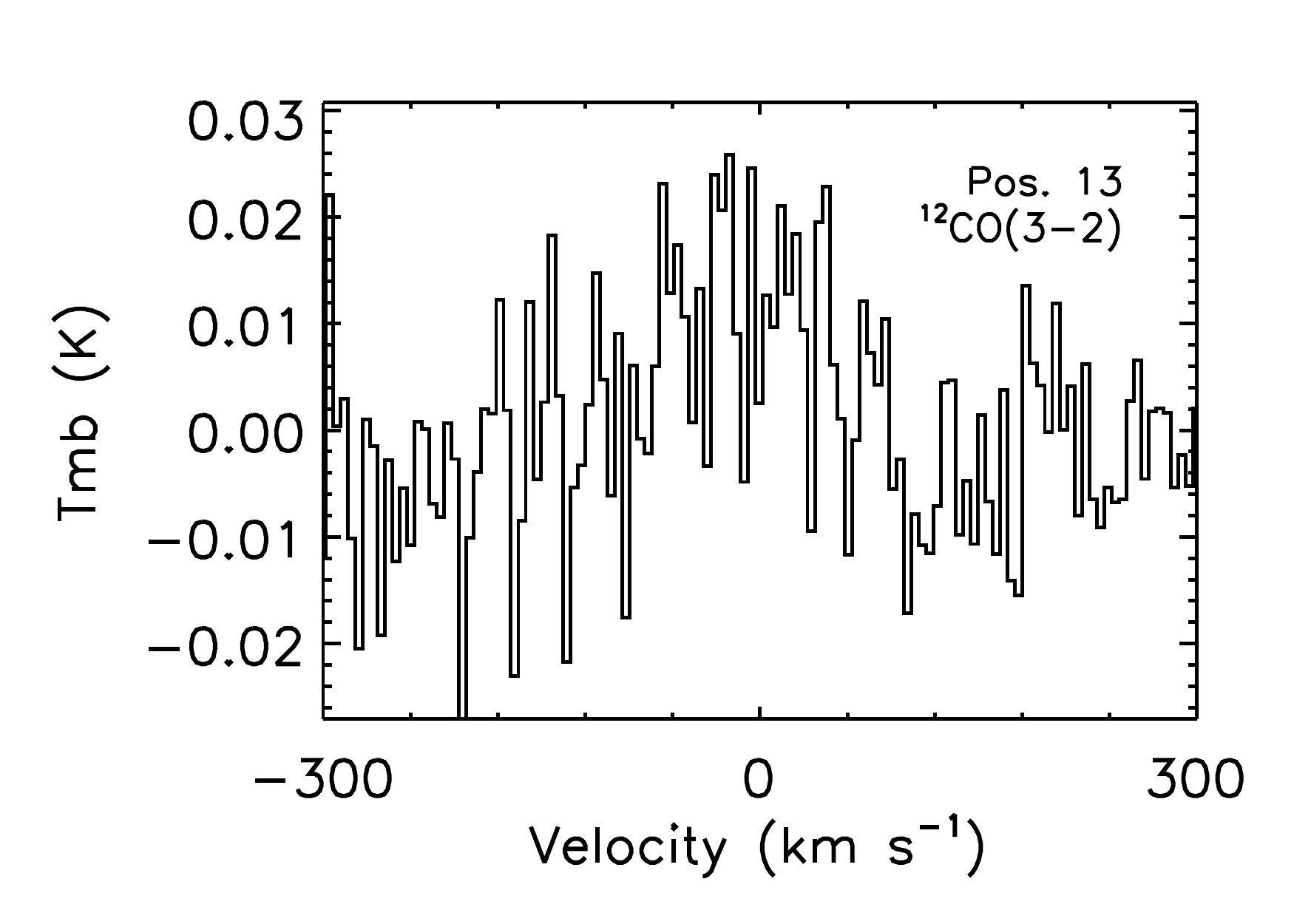}\\
   \includegraphics[width=7.0cm,clip=]{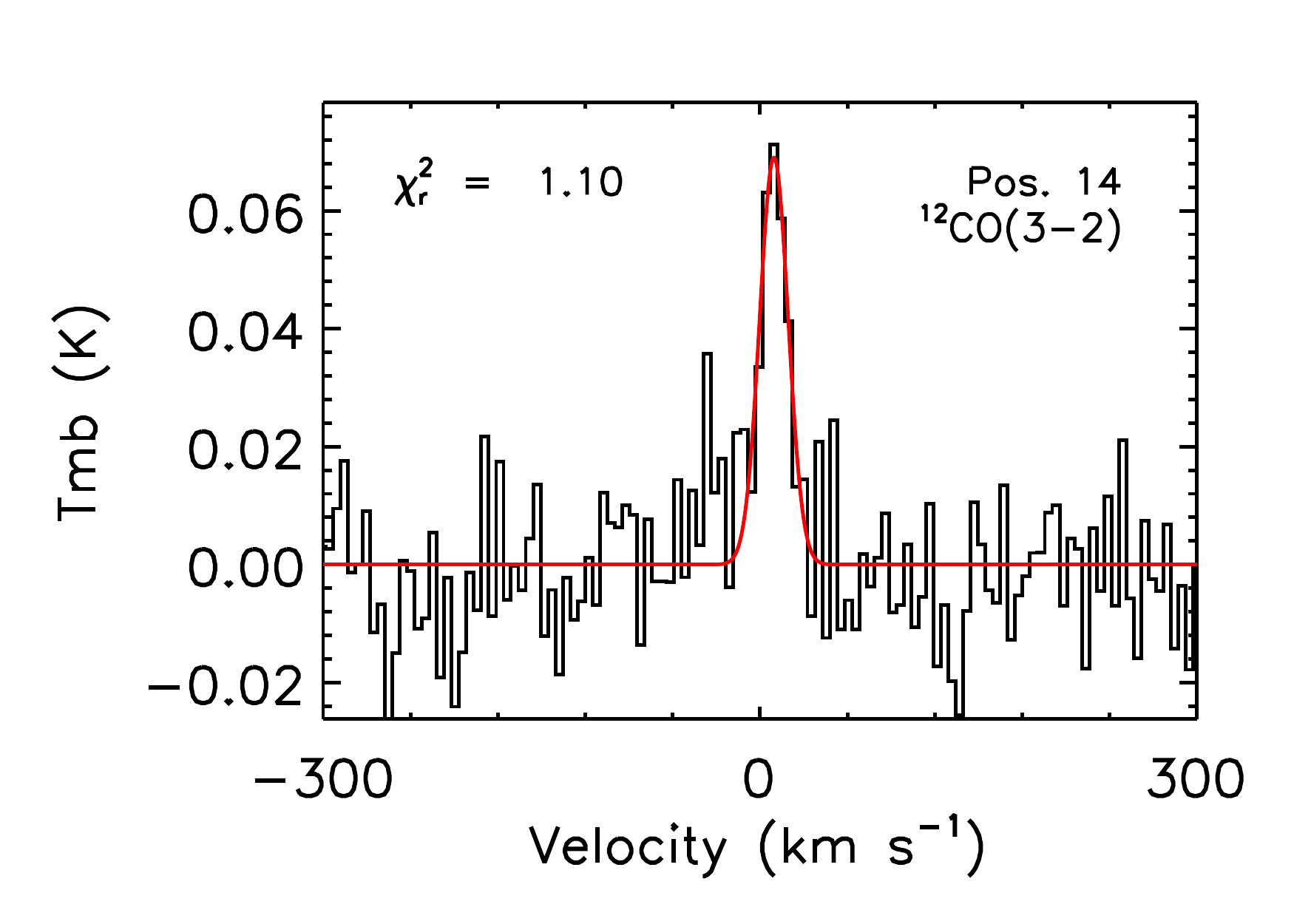}
   \includegraphics[width=7.0cm,clip=]{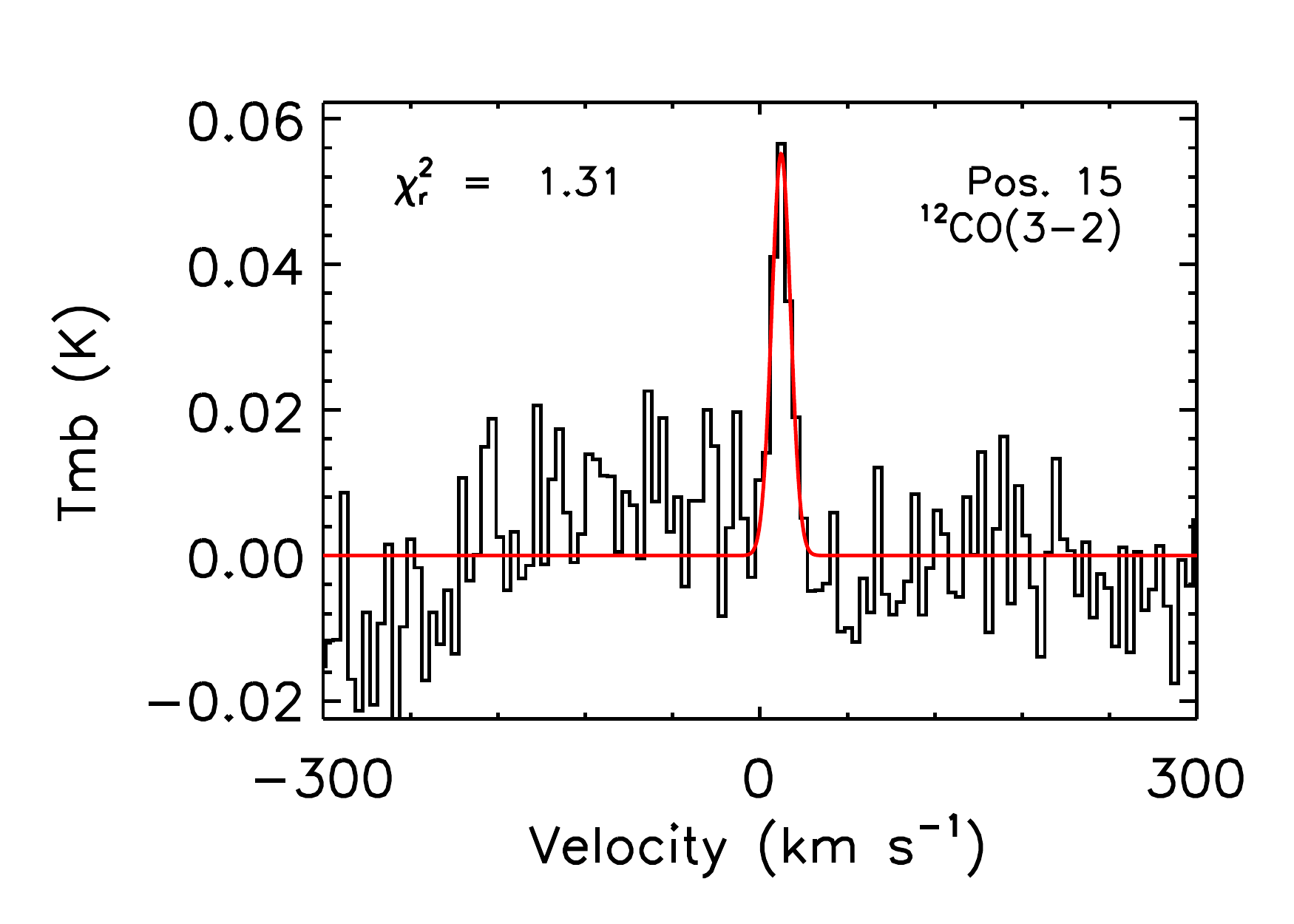}\\
  \caption{Continued.}
  \label{fig:spec3}
\end{figure*}

\addtocounter{figure}{-1}
\begin{figure*}
  \includegraphics[width=7.0cm,clip=]{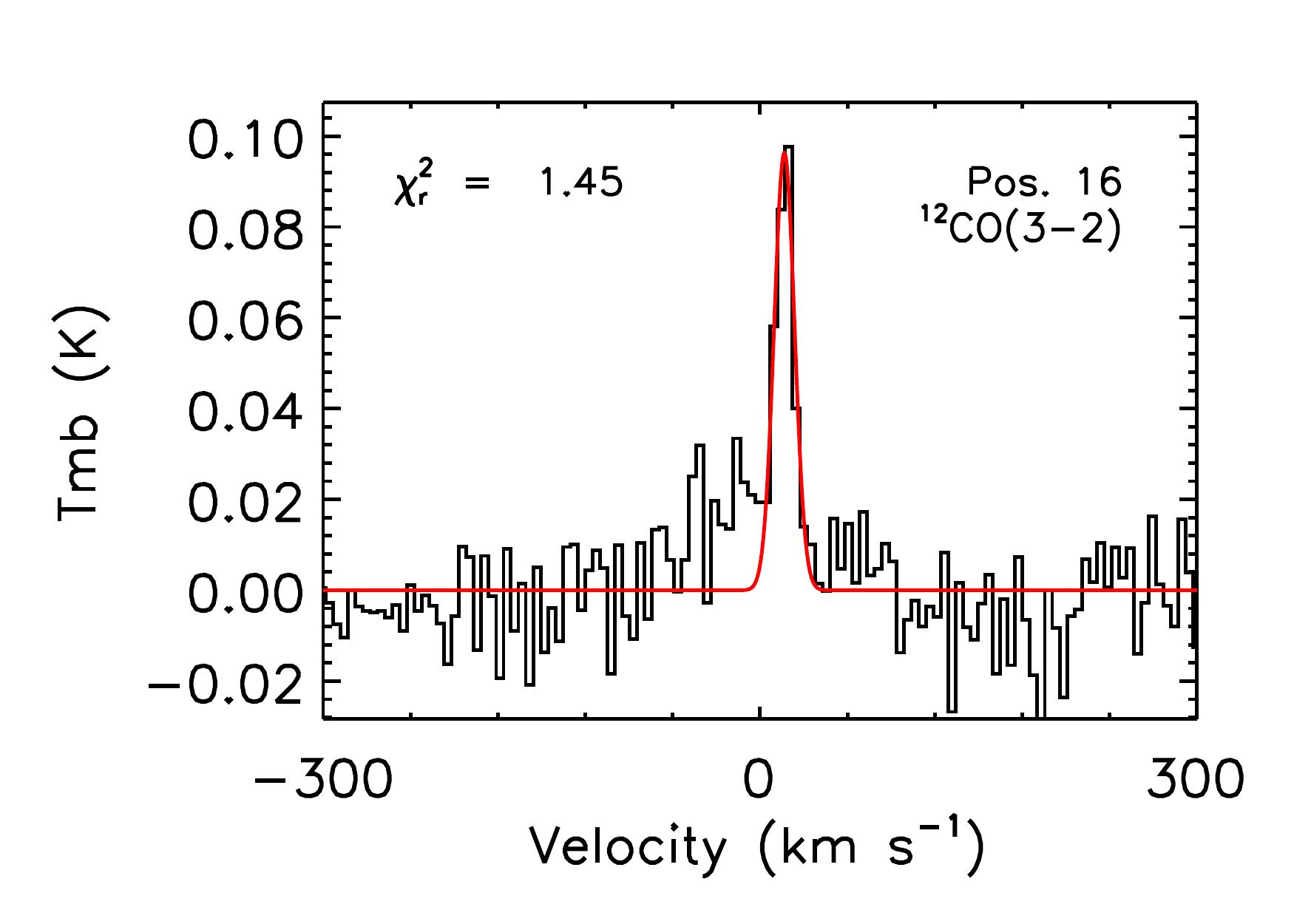}
   \includegraphics[width=7.0cm,clip=]{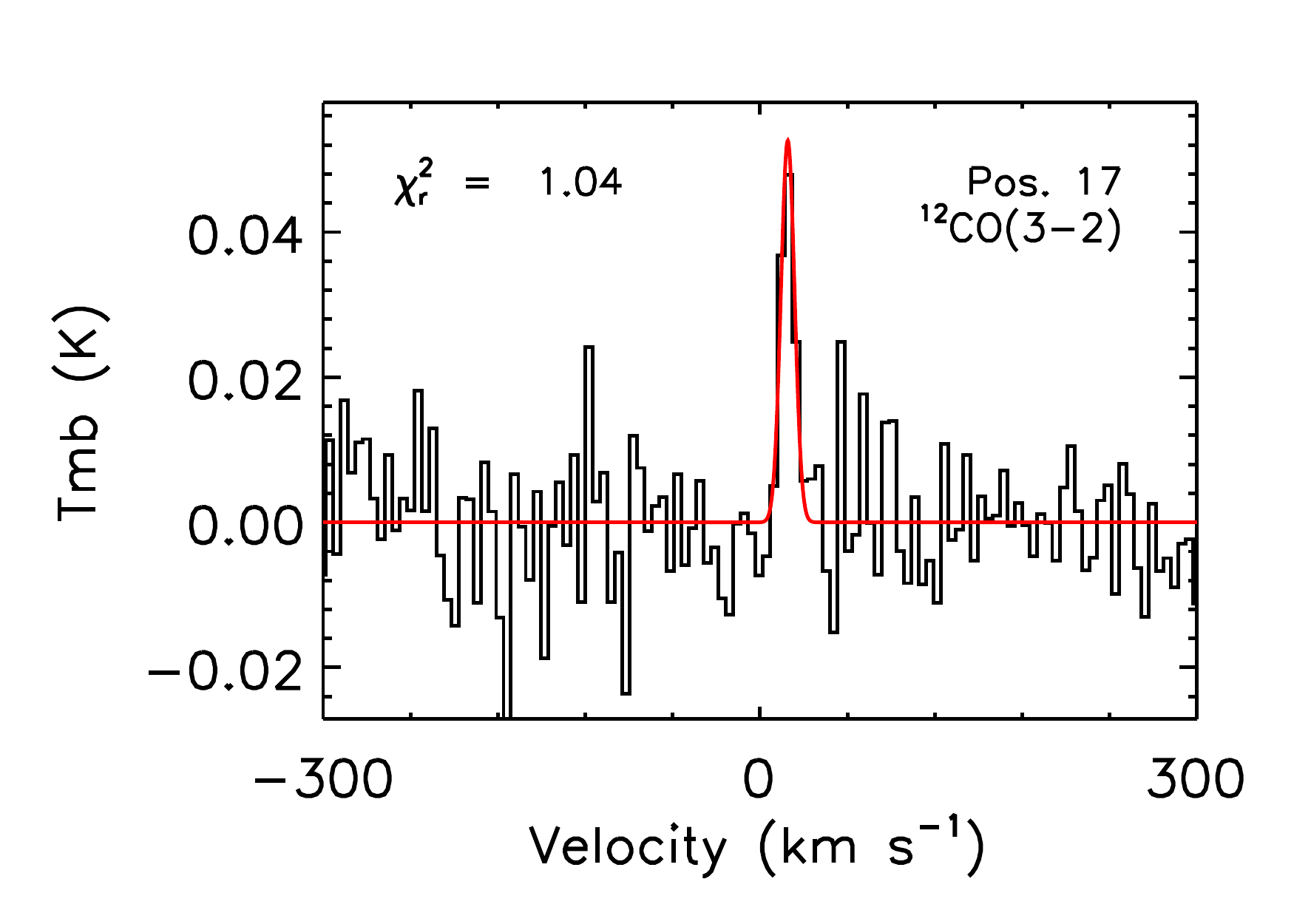}\\
  \includegraphics[width=7.0cm,clip=]{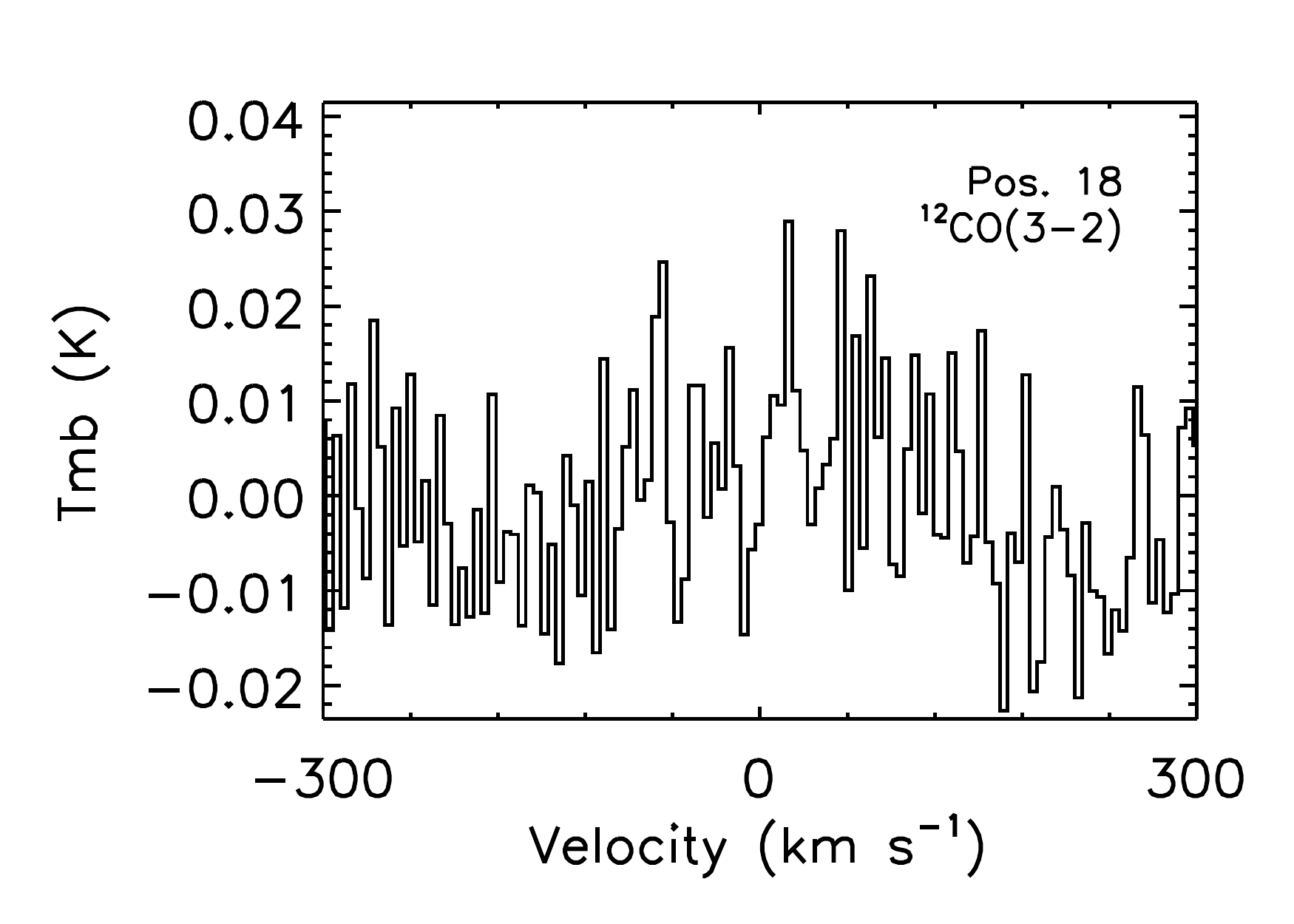}
   \includegraphics[width=7.0cm,clip=]{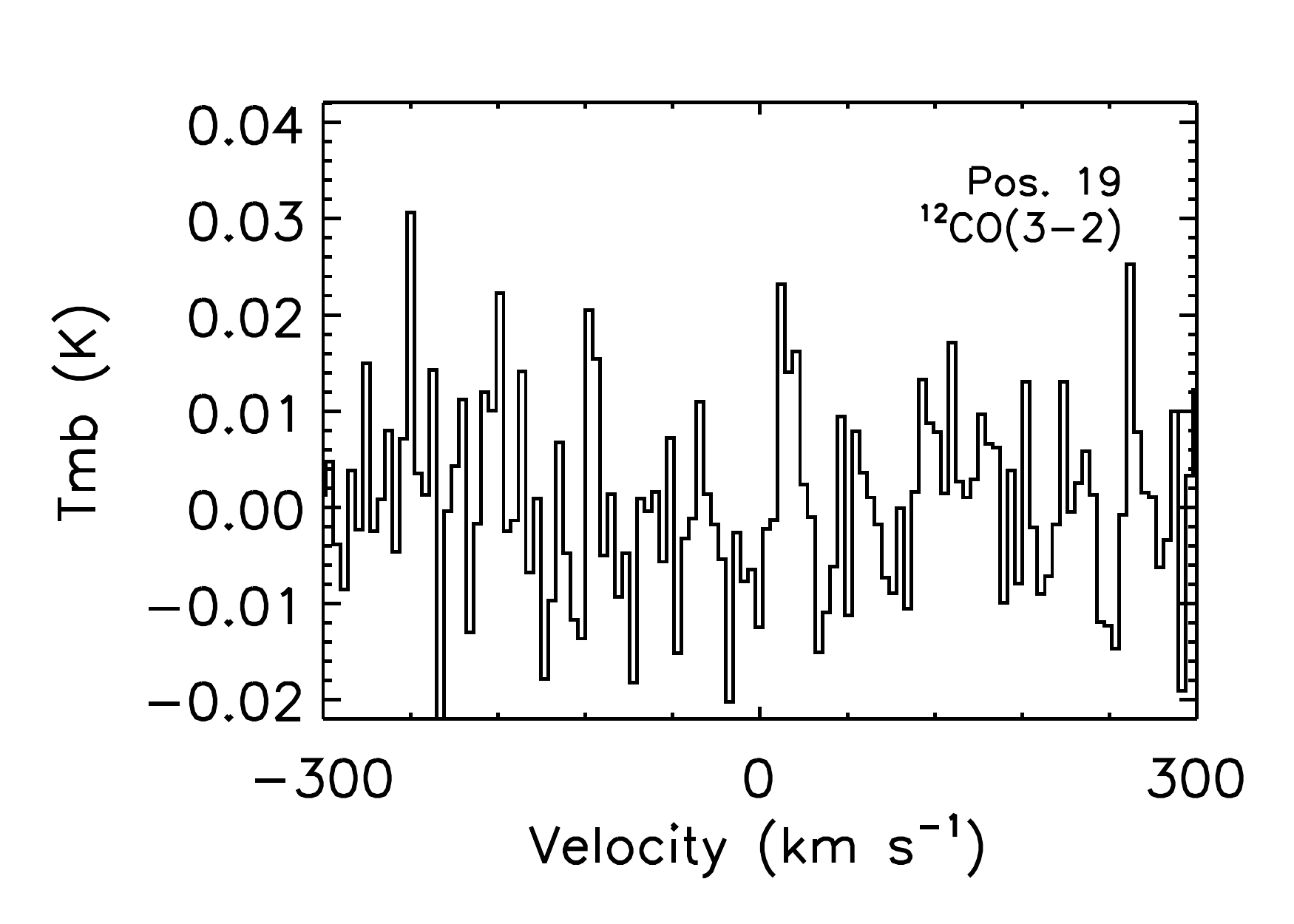}\\
  \caption{Continued.}
  \label{fig:spec3}
\end{figure*}

\clearpage

\begin{figure*}
  \includegraphics[width=7.0cm,clip=]{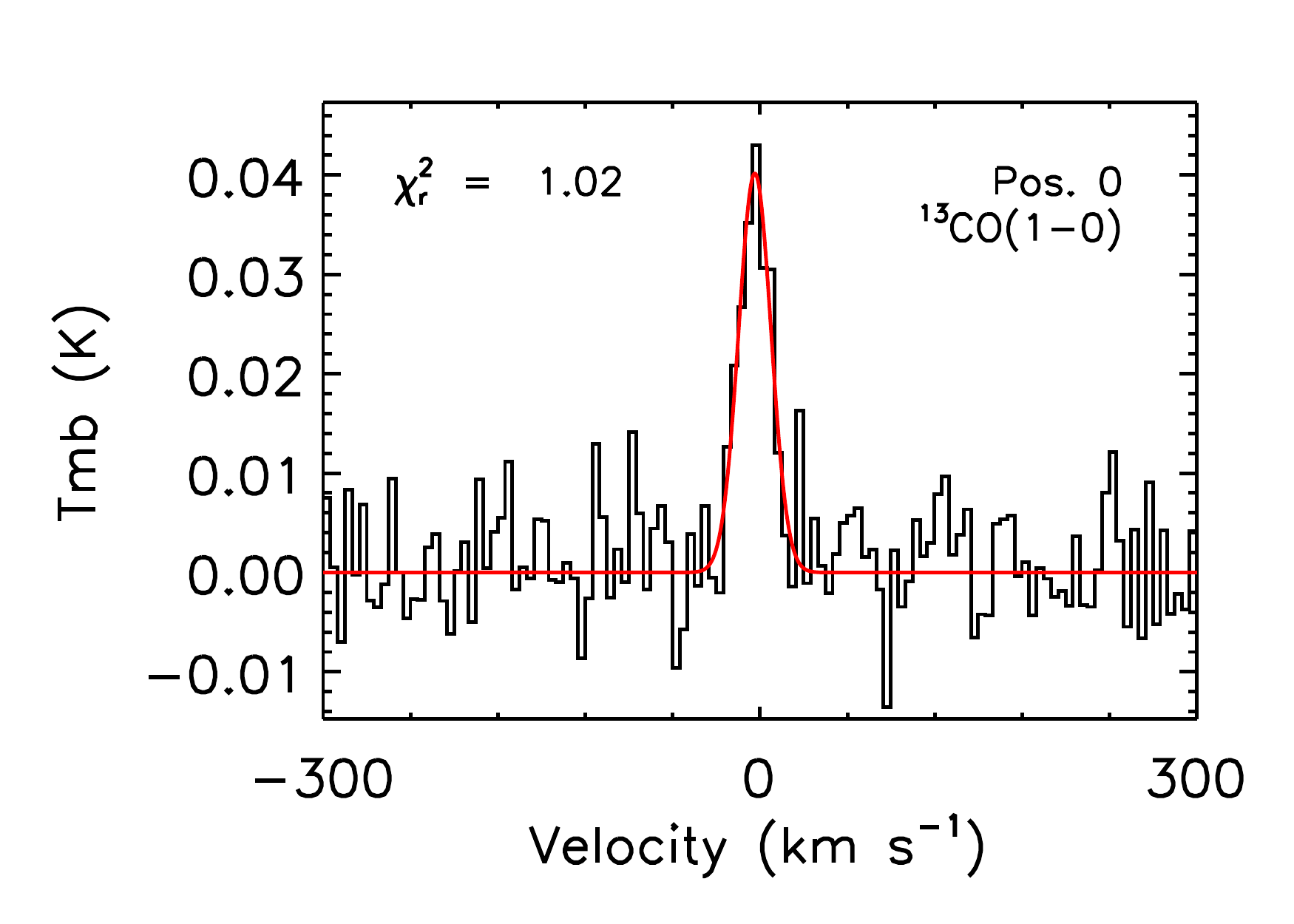}
   \includegraphics[width=7.0cm,clip=]{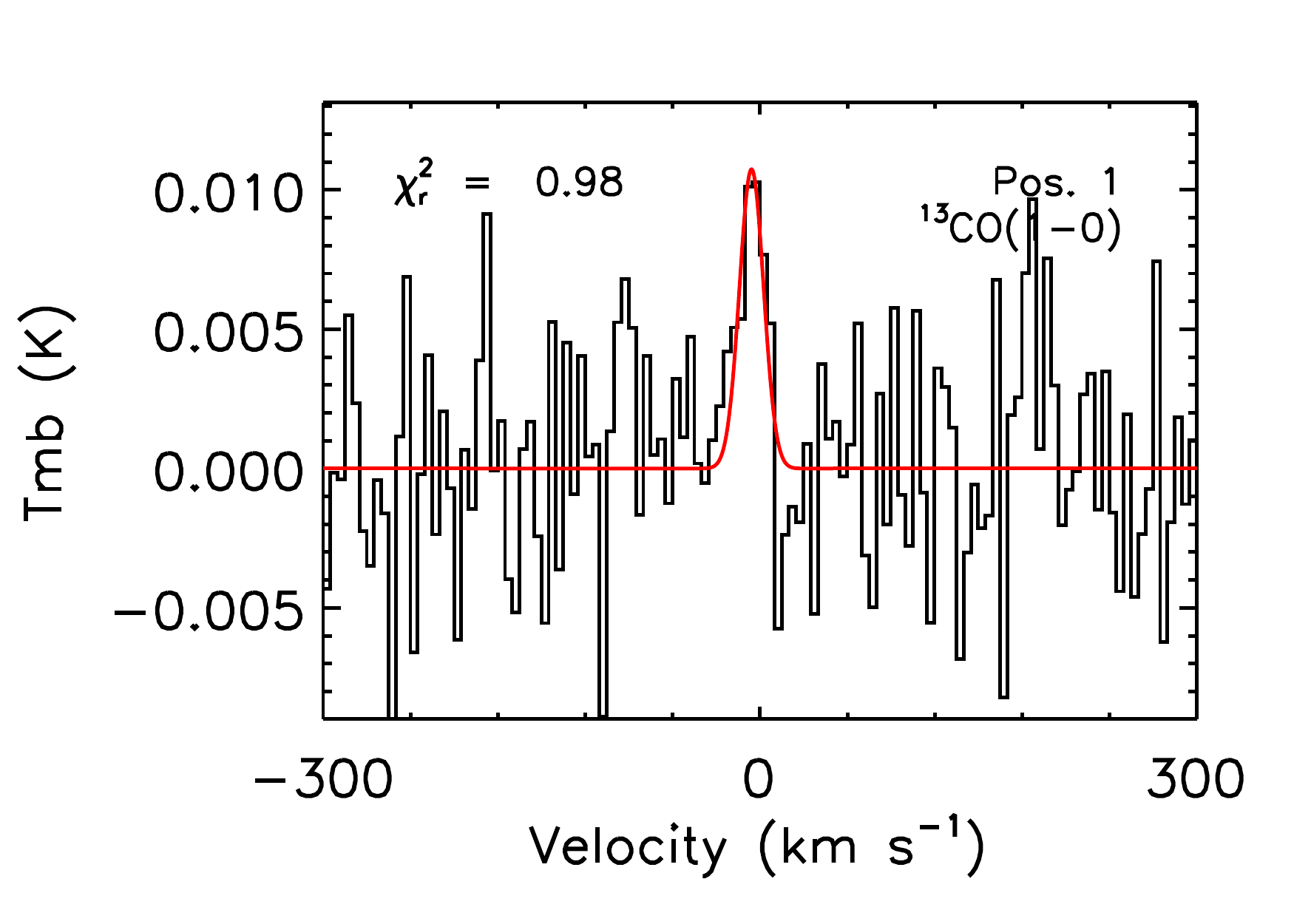}\\
  \includegraphics[width=7.0cm,clip=]{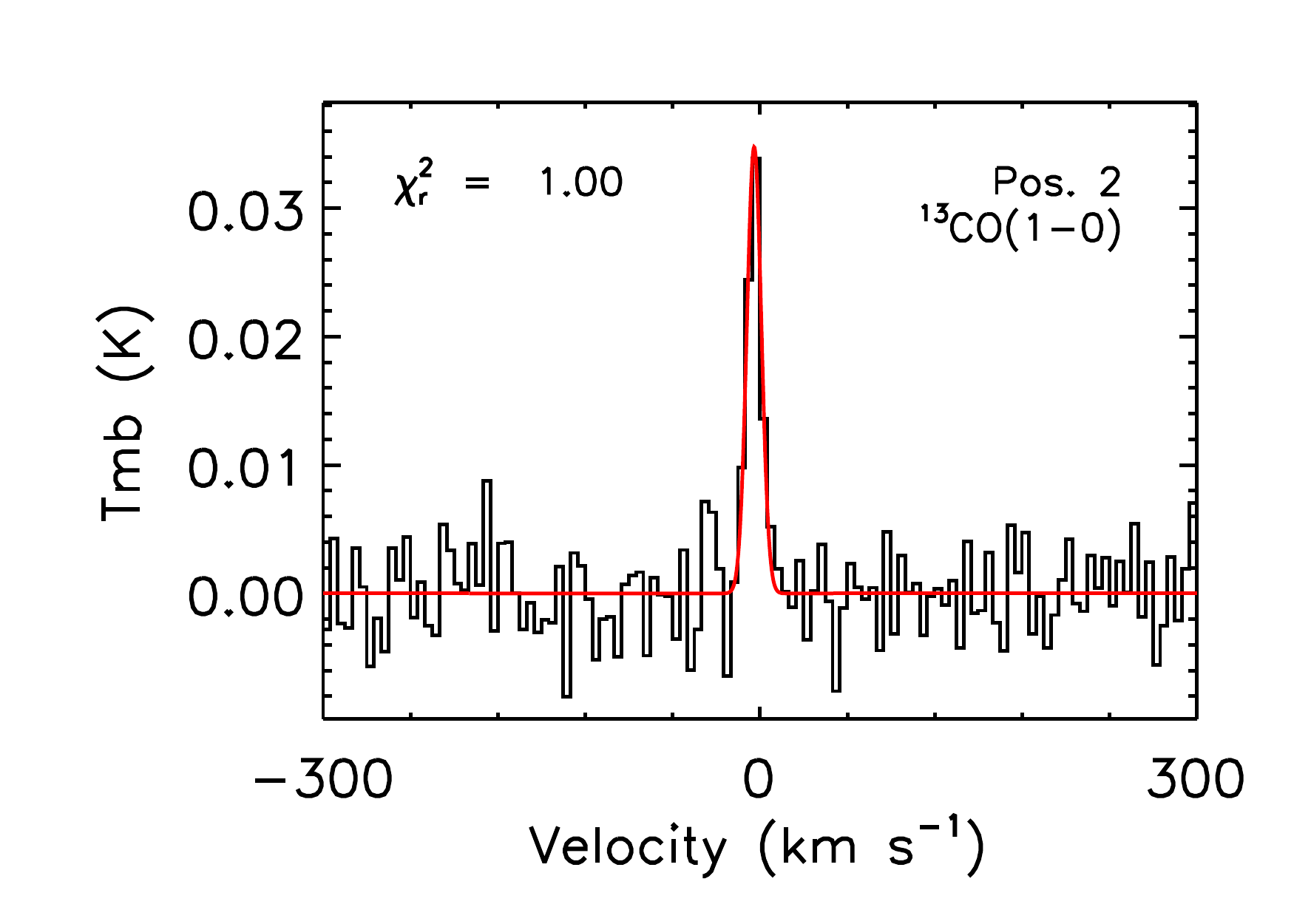}
   \includegraphics[width=7.0cm,clip=]{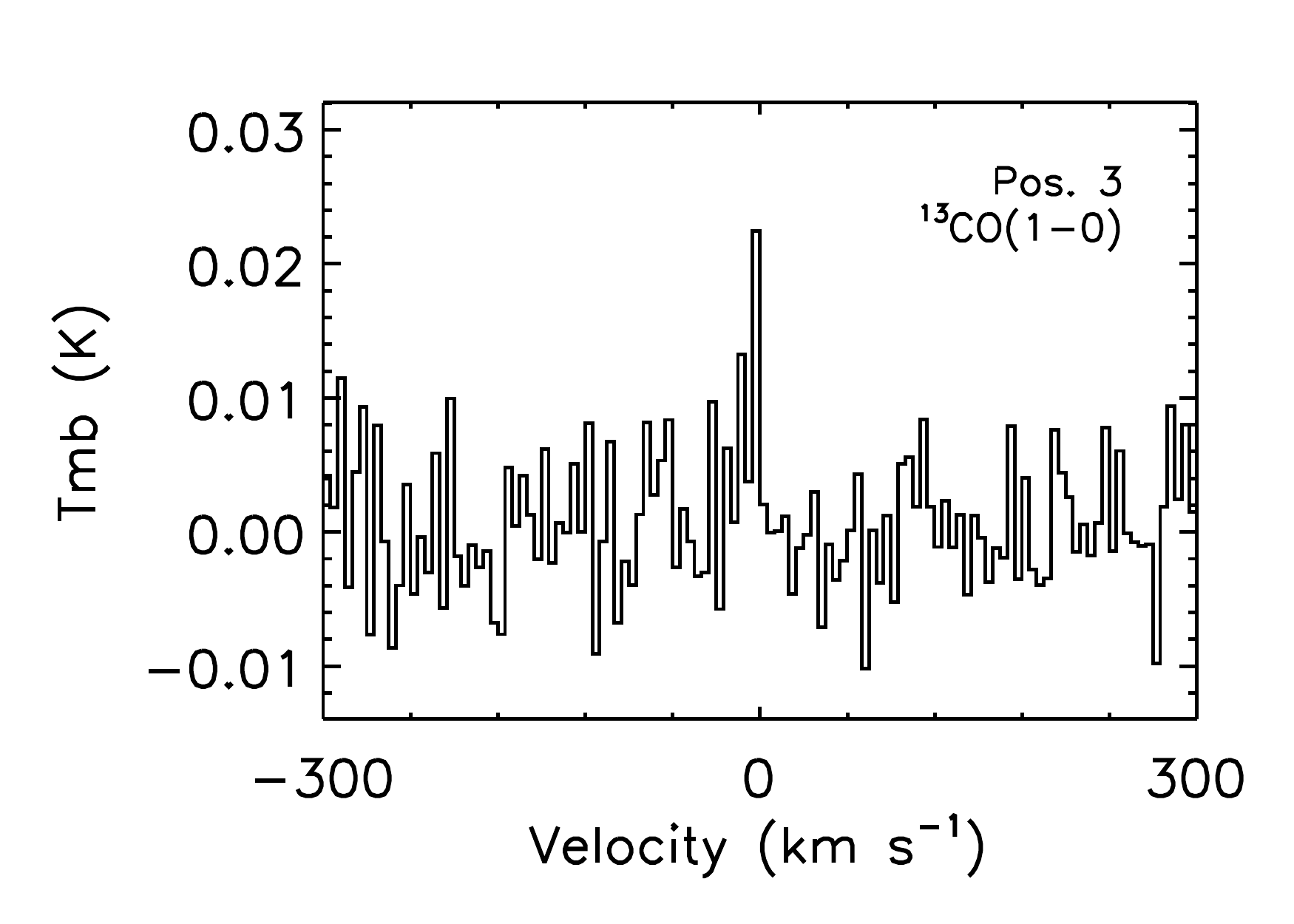}\\
    \includegraphics[width=7.0cm,clip=]{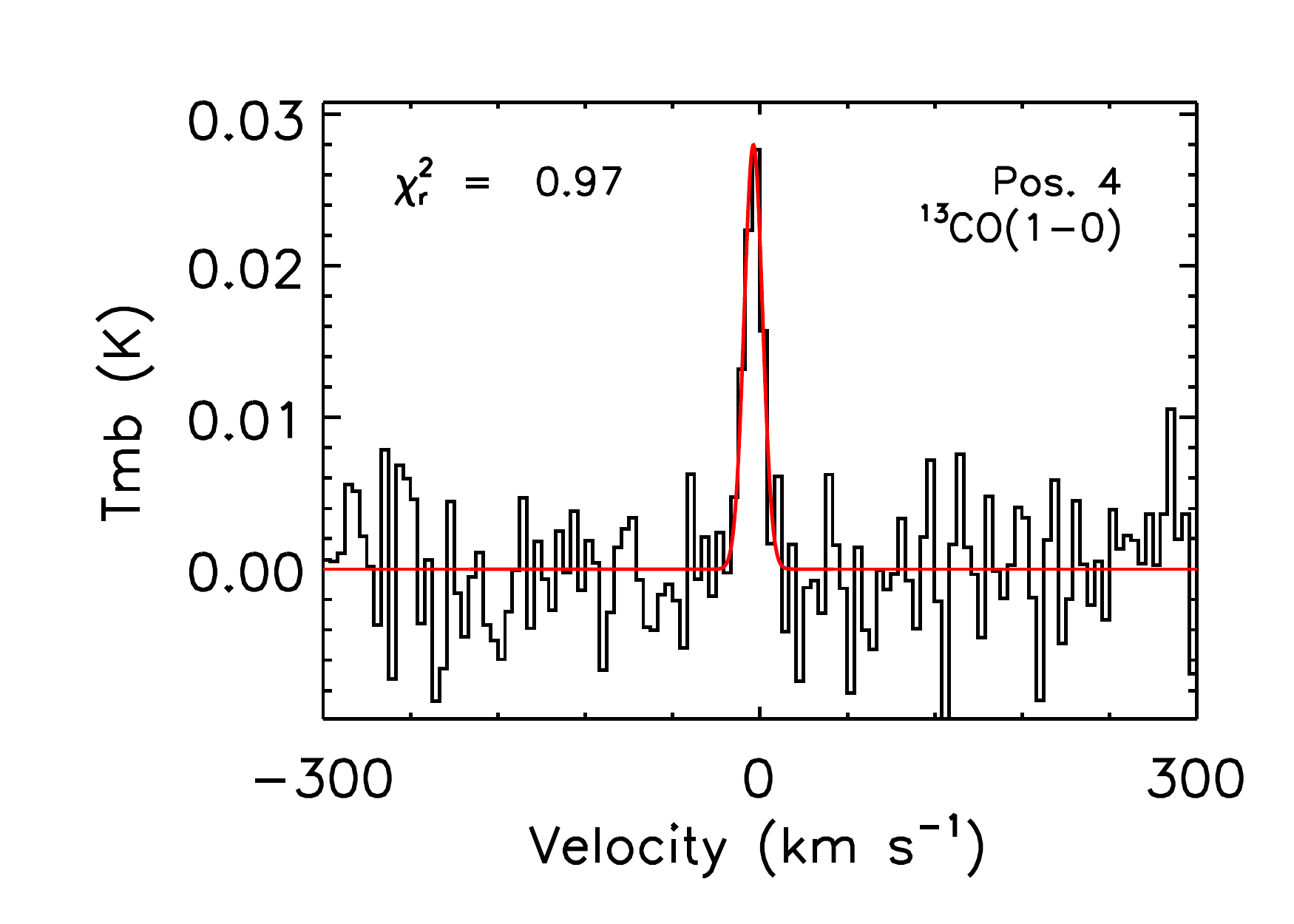}
   \includegraphics[width=7.0cm,clip=]{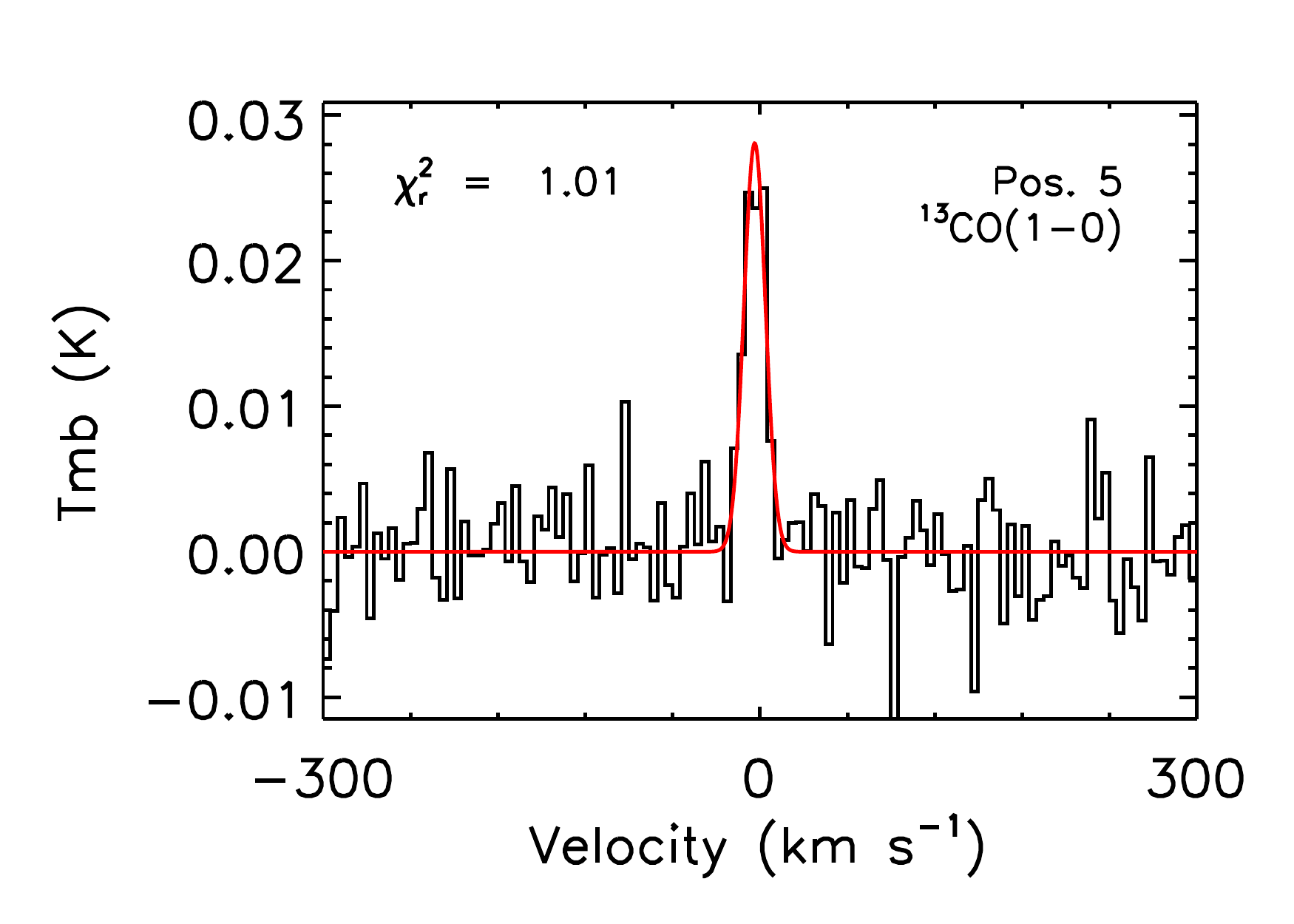}\\
   \includegraphics[width=7.0cm,clip=]{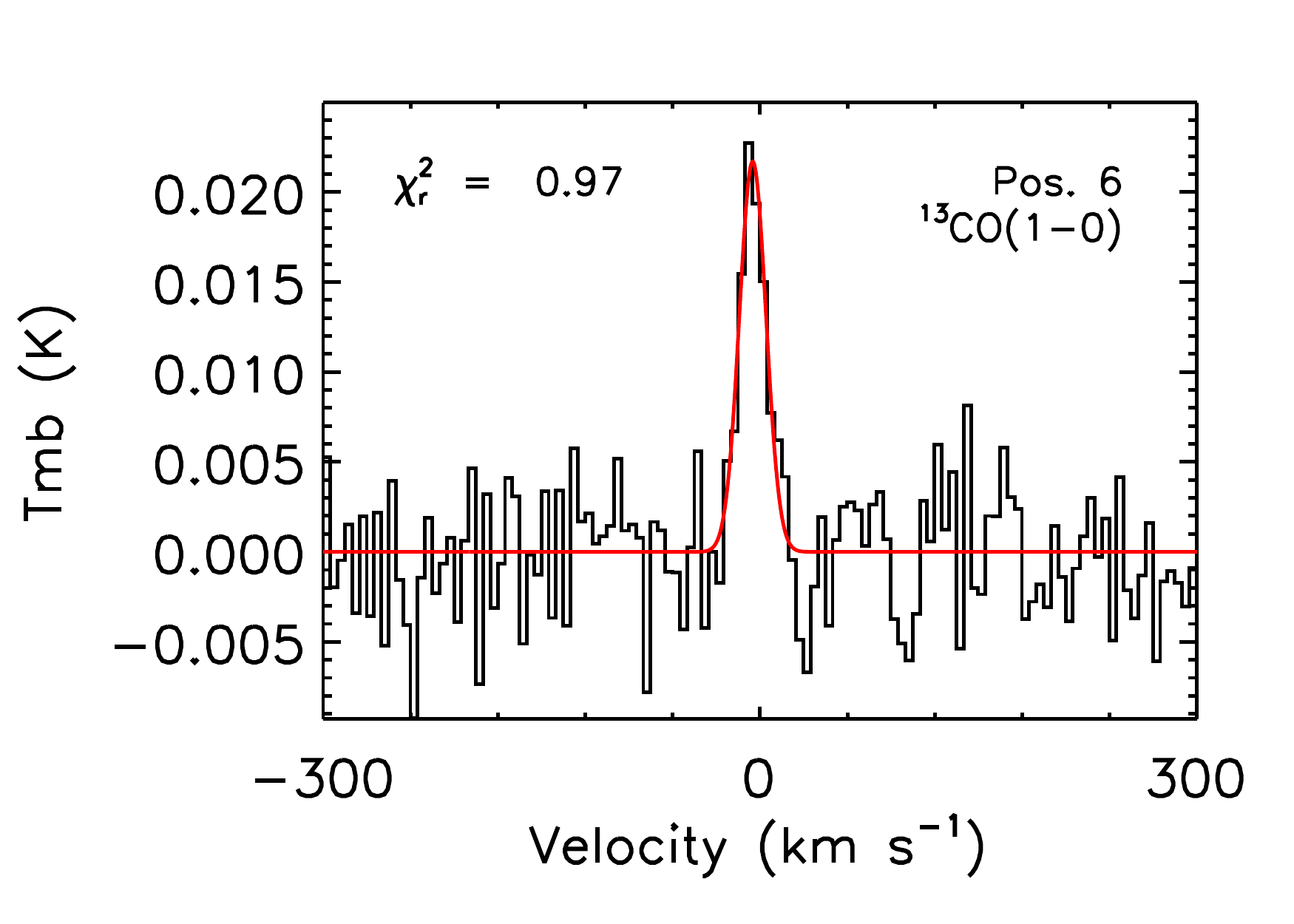}
   \includegraphics[width=7.0cm,clip=]{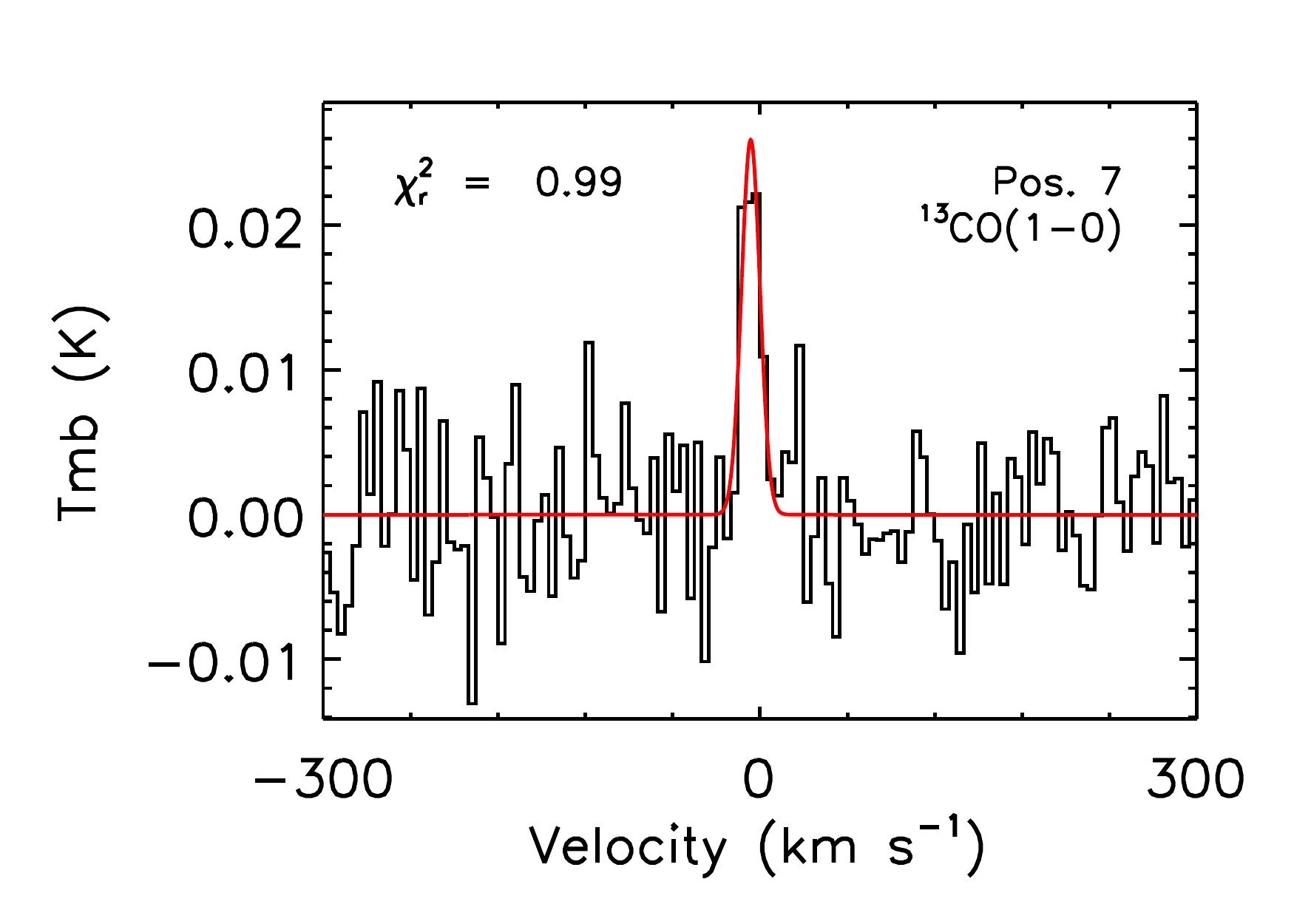}\\
  \caption{Integrated spectra of $^{13}$CO(1--0) emission observed in the disc of NGC~0628. Gaussian fits are overlaid. 
  The value of $\chi_{\rm r}^2$ is also shown in each panel.}
  \label{fig:spec4}
\end{figure*}

\addtocounter{figure}{-1}
\begin{figure*}
  \includegraphics[width=7.0cm,clip=]{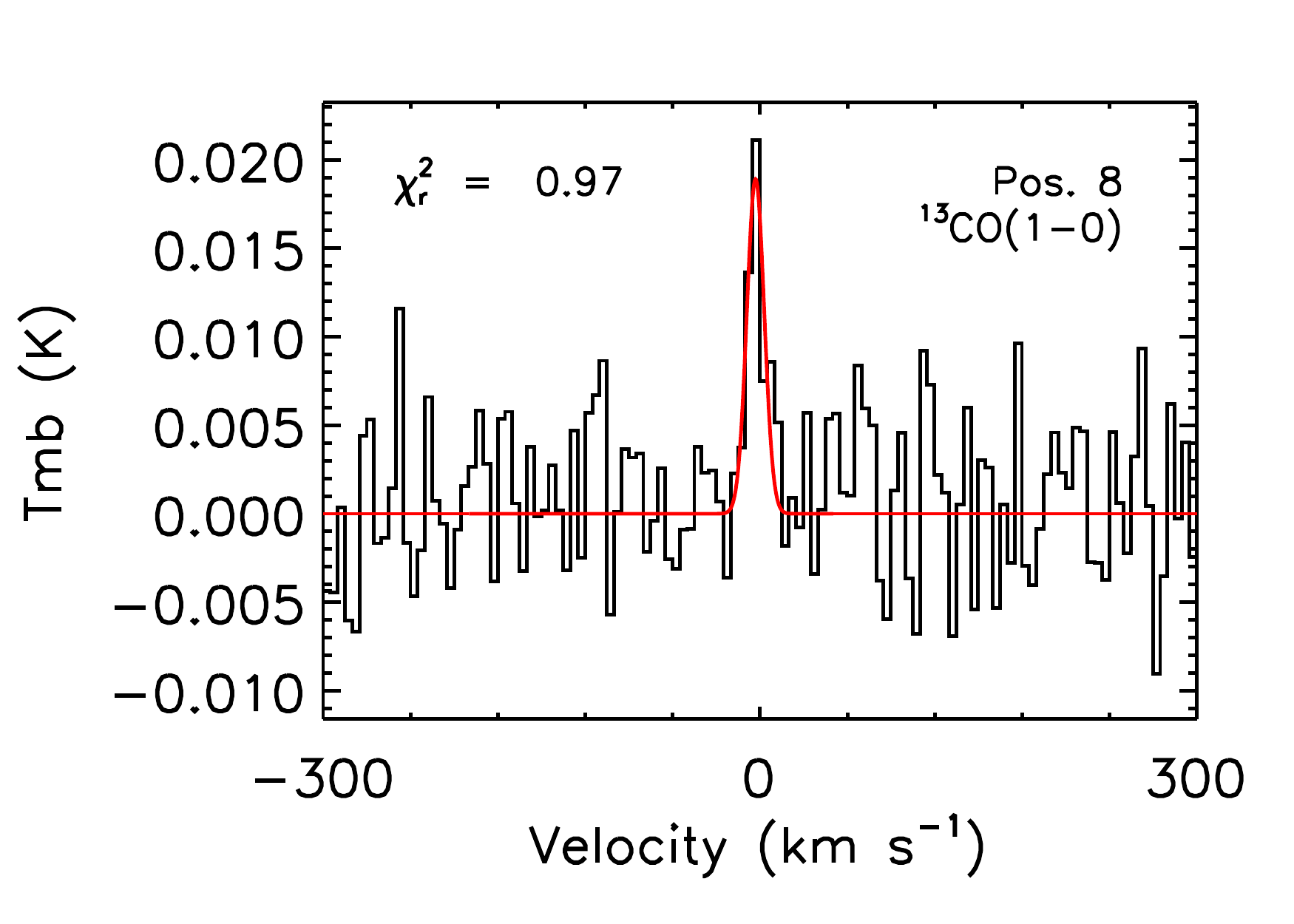}
   \includegraphics[width=7.0cm,clip=]{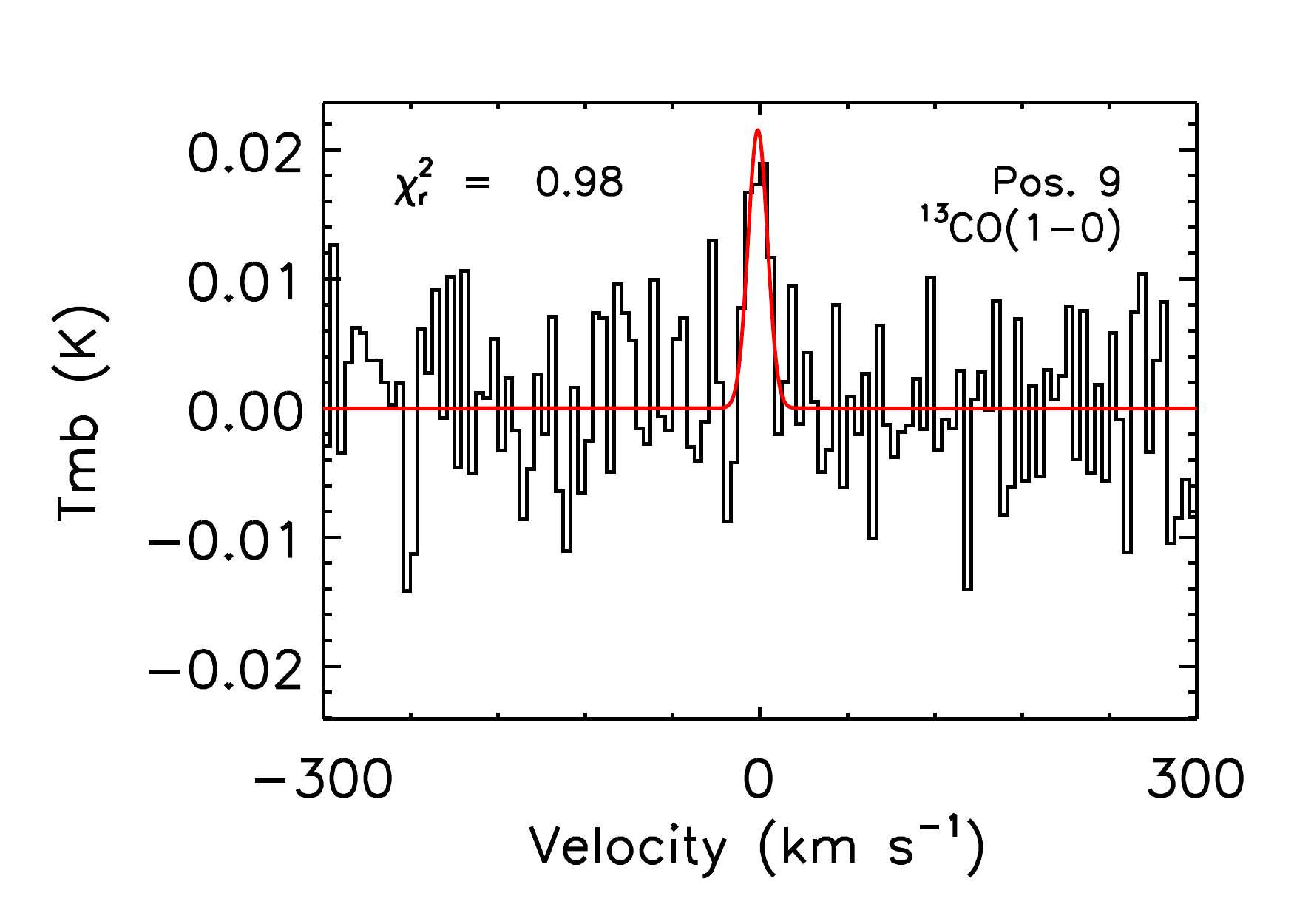}\\
  \includegraphics[width=7.0cm,clip=]{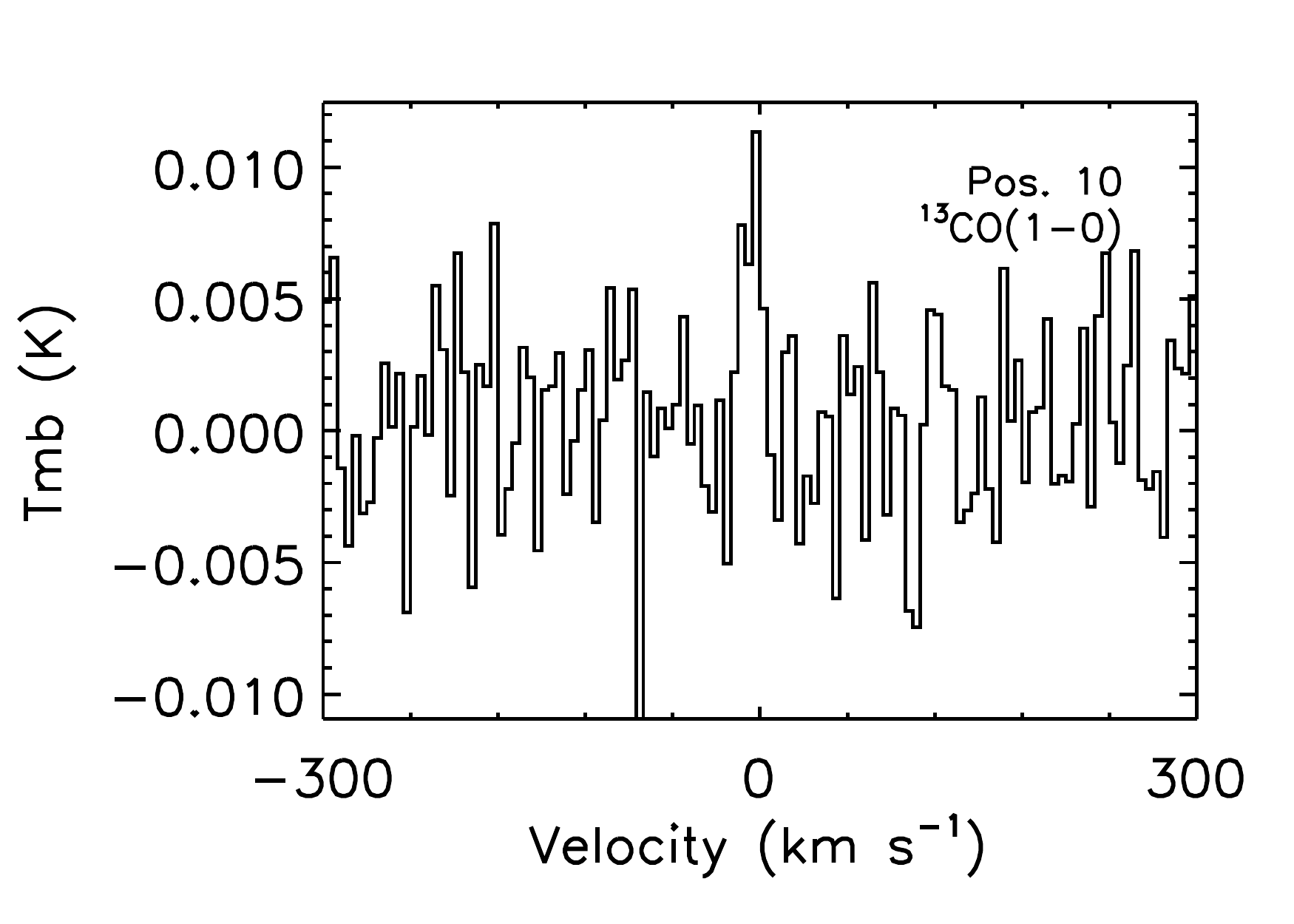}
   \includegraphics[width=7.0cm,clip=]{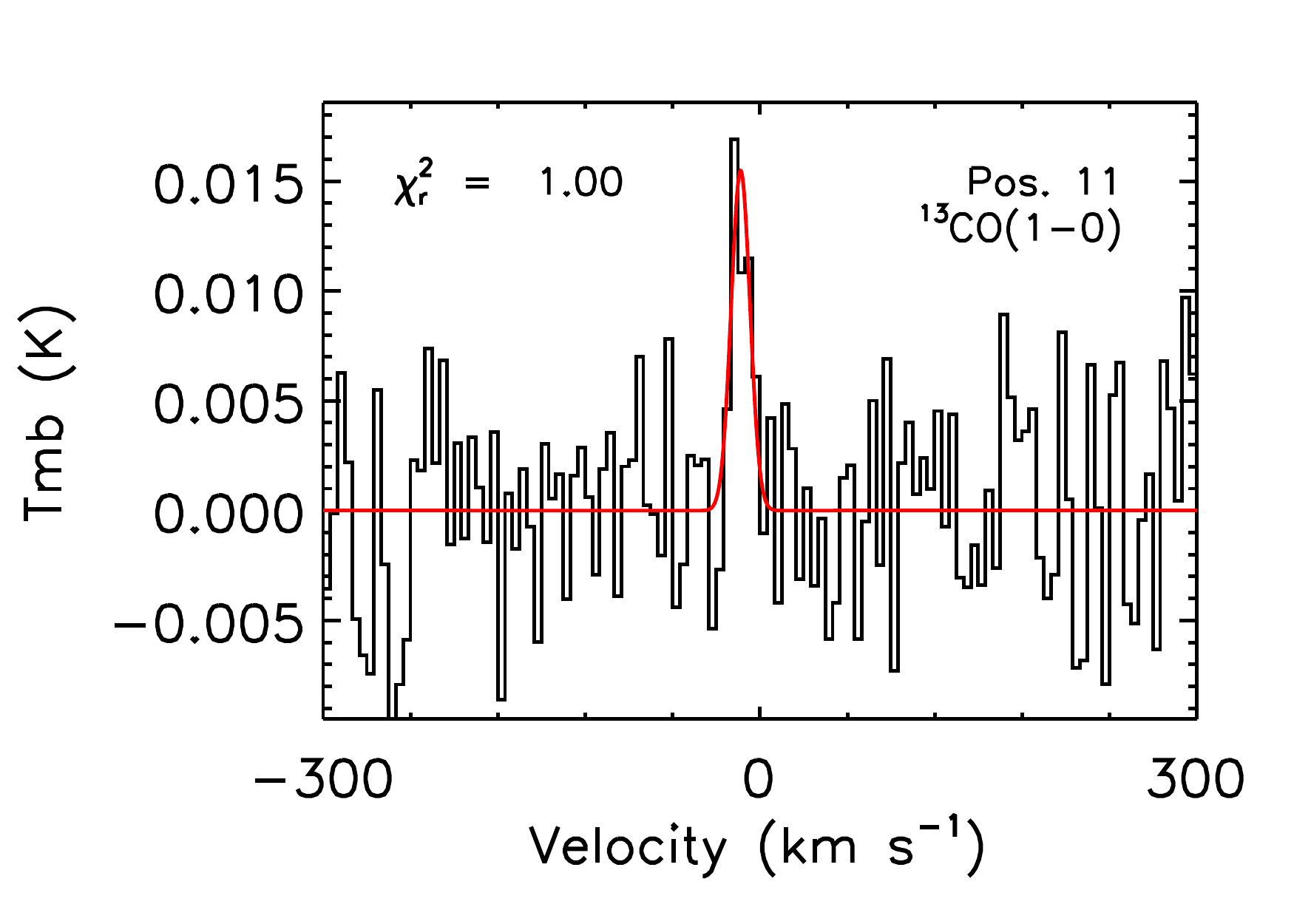}\\
    \includegraphics[width=7.0cm,clip=]{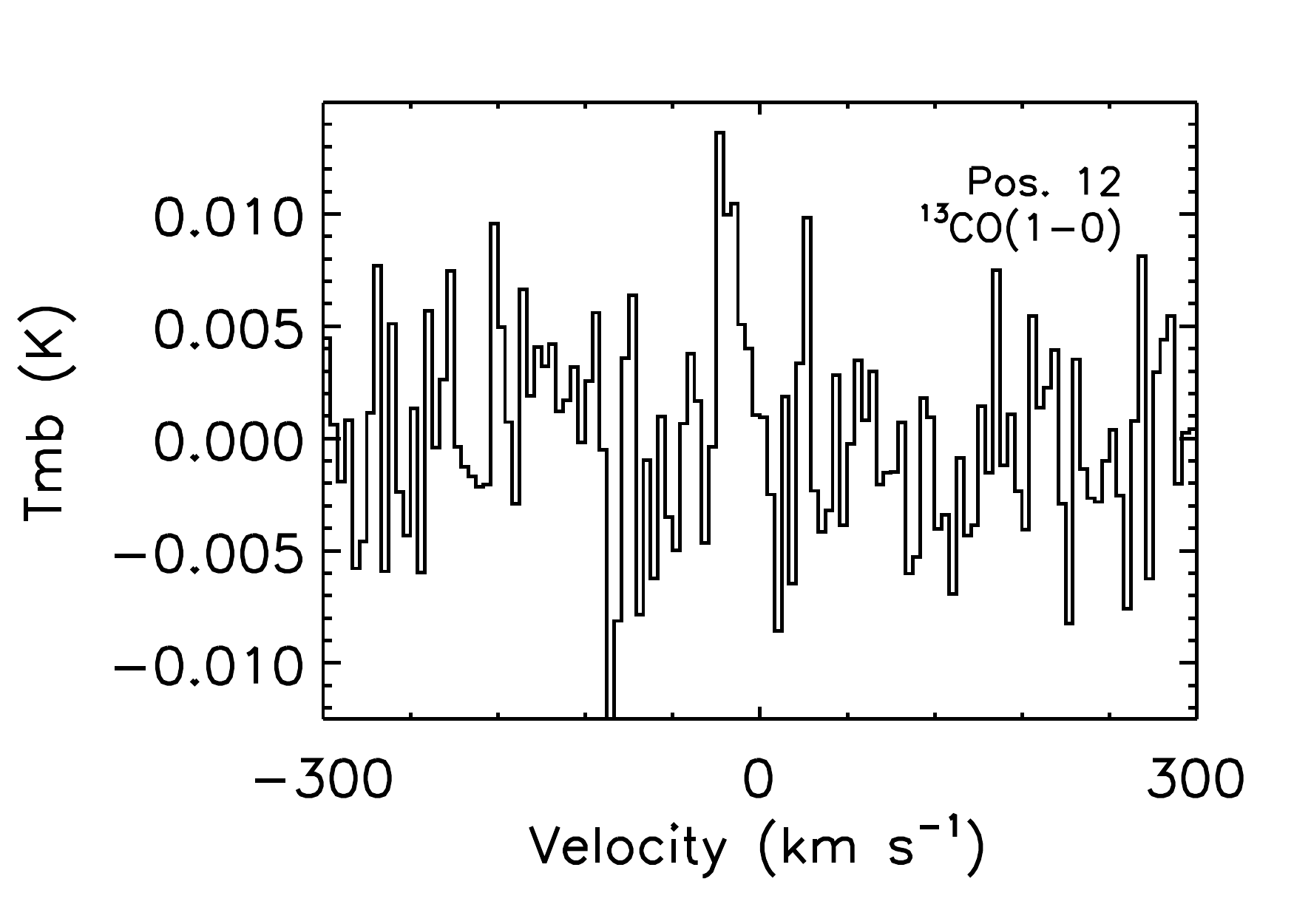}
   \includegraphics[width=7.0cm,clip=]{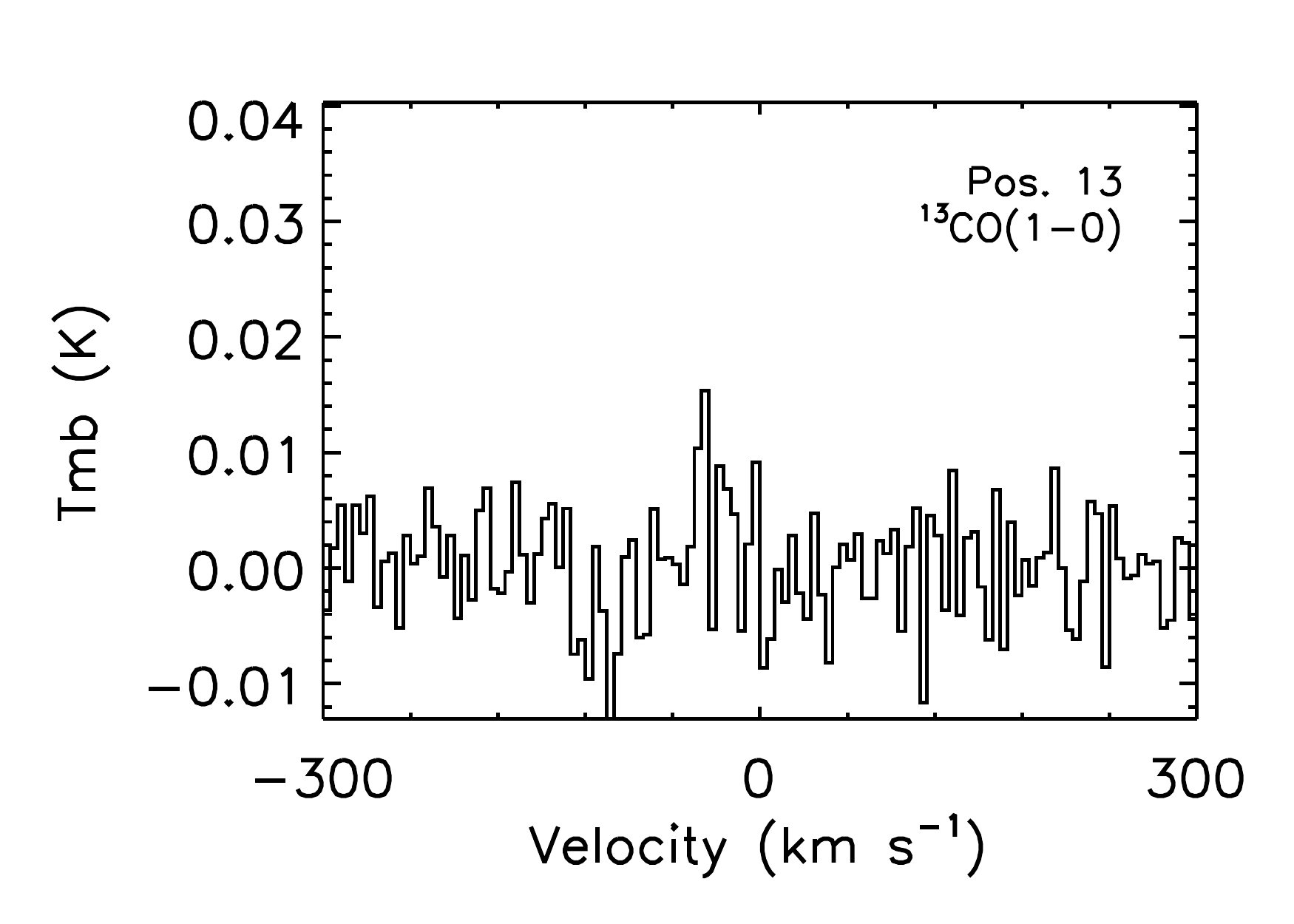}\\
   \includegraphics[width=7.0cm,clip=]{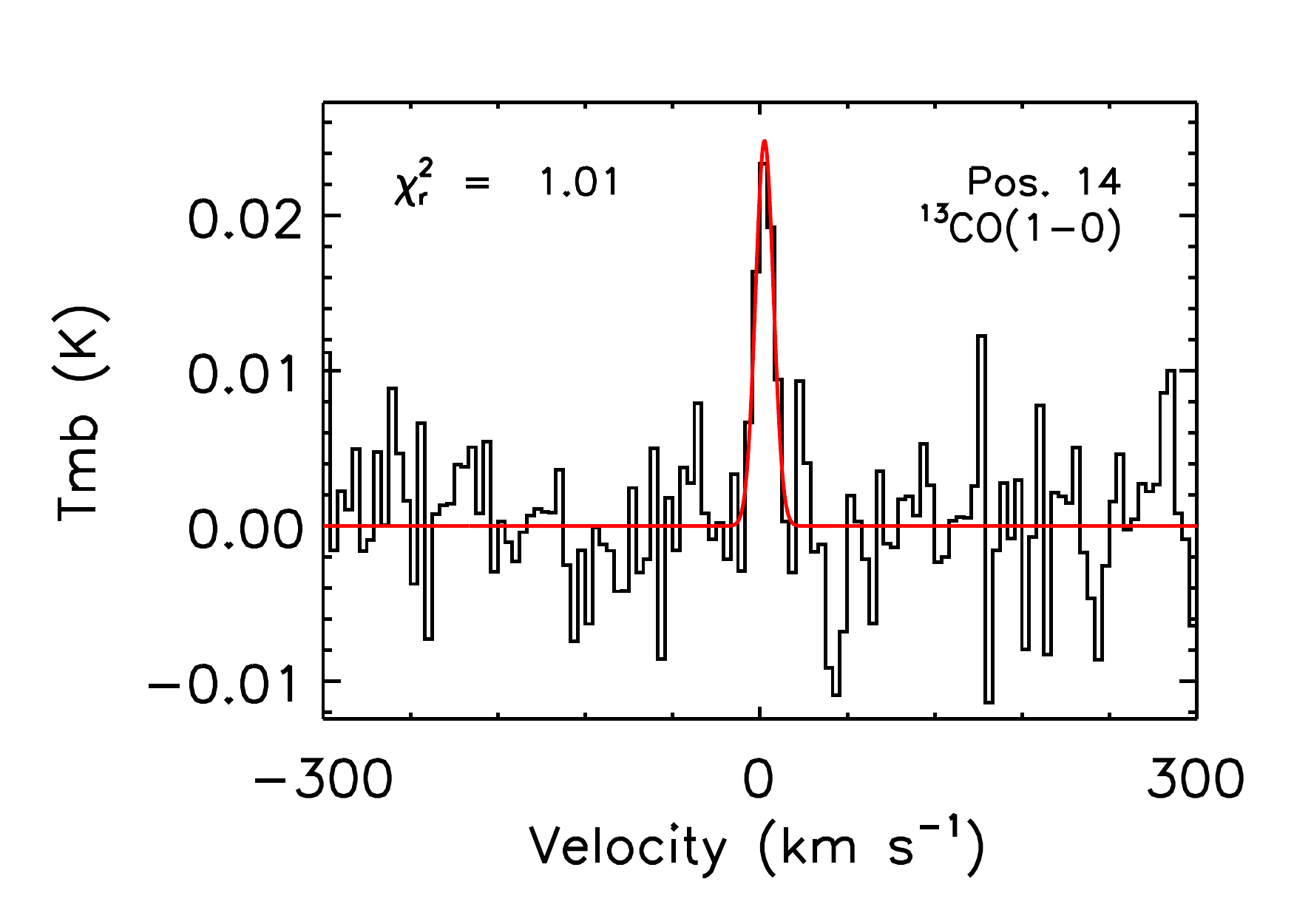}
   \includegraphics[width=7.0cm,clip=]{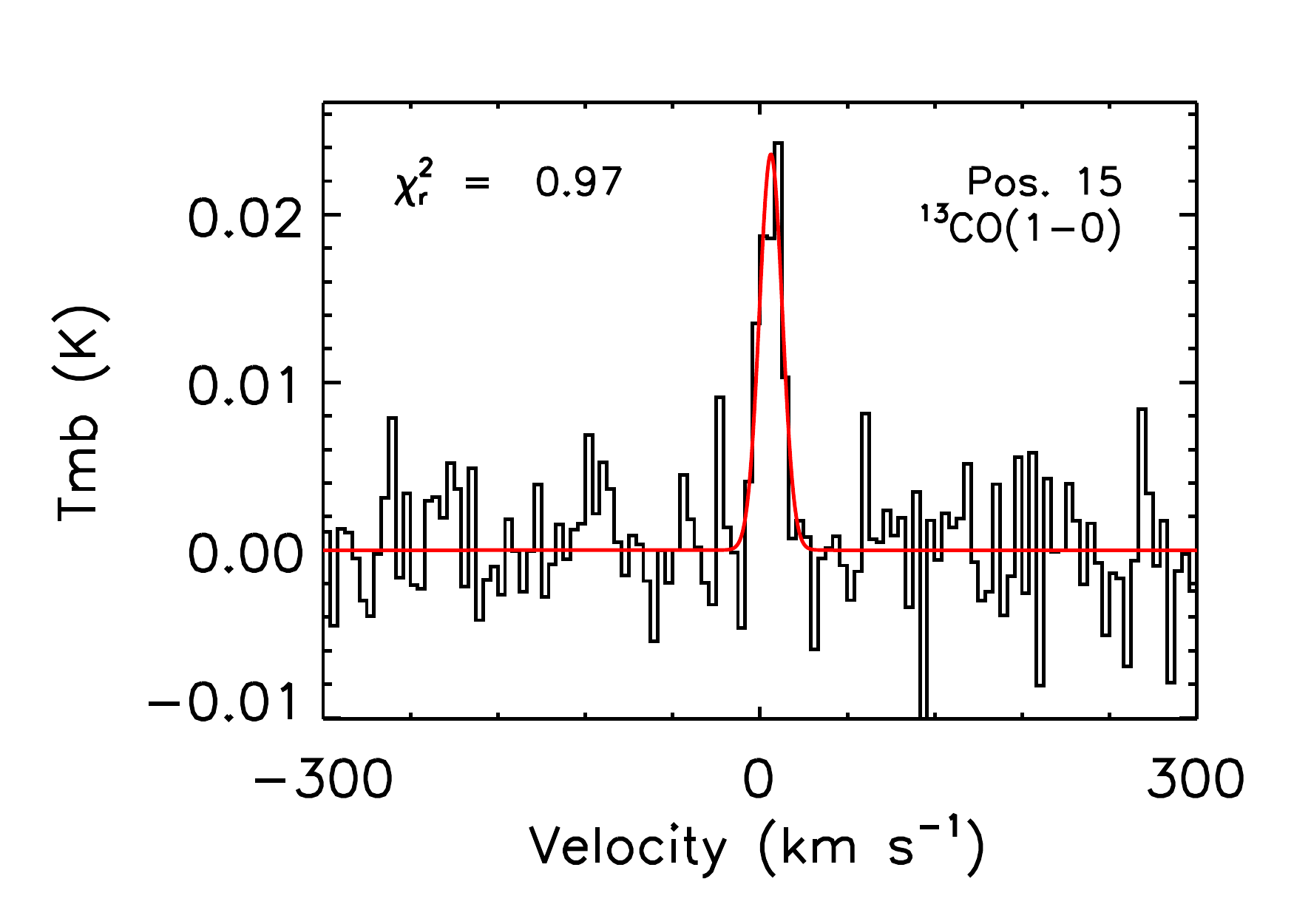}\\
  \caption{Continued.}
  \label{fig:spec4}
\end{figure*}

\addtocounter{figure}{-1}
\begin{figure*}
  \includegraphics[width=7.0cm,clip=]{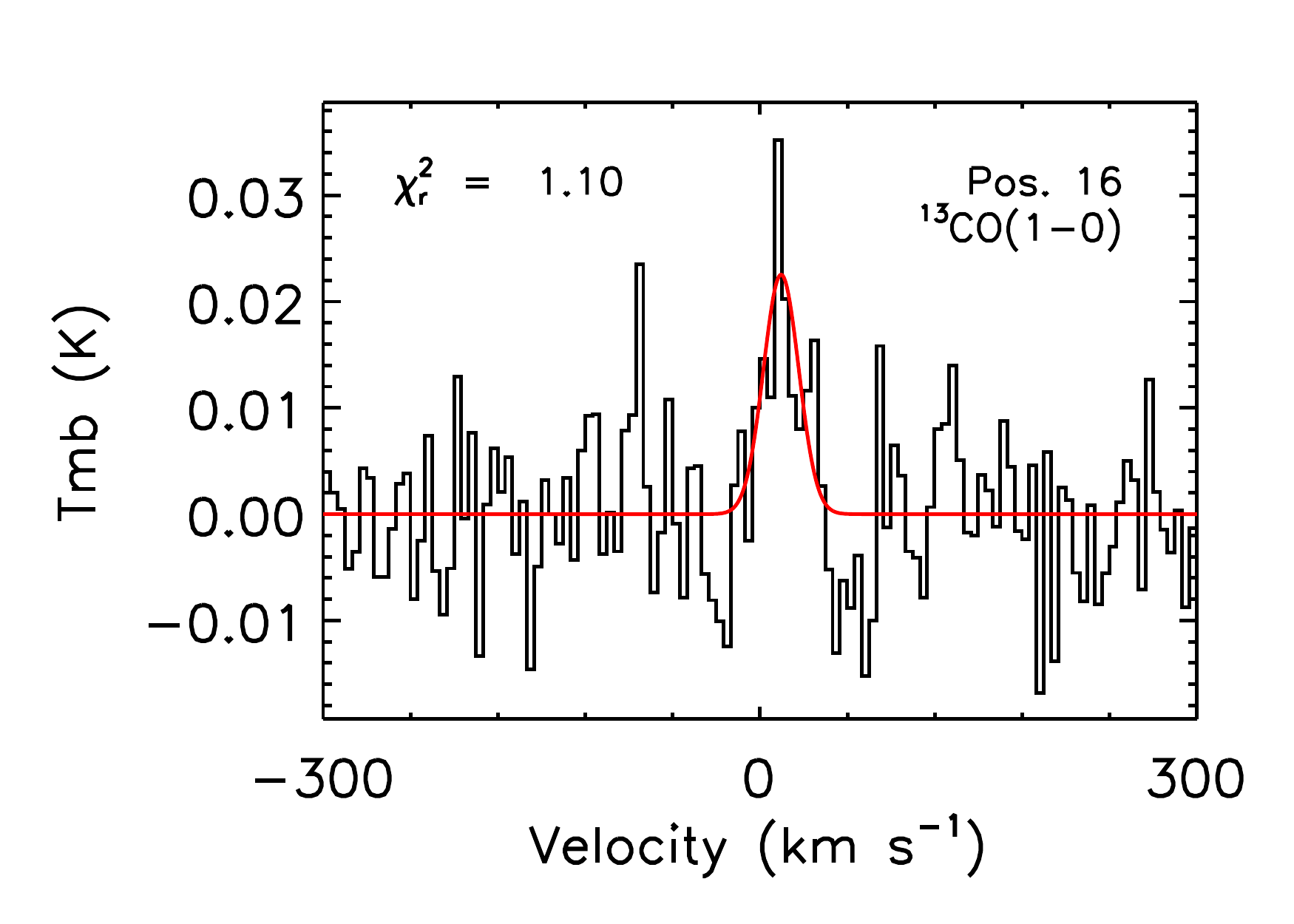}
   \includegraphics[width=7.0cm,clip=]{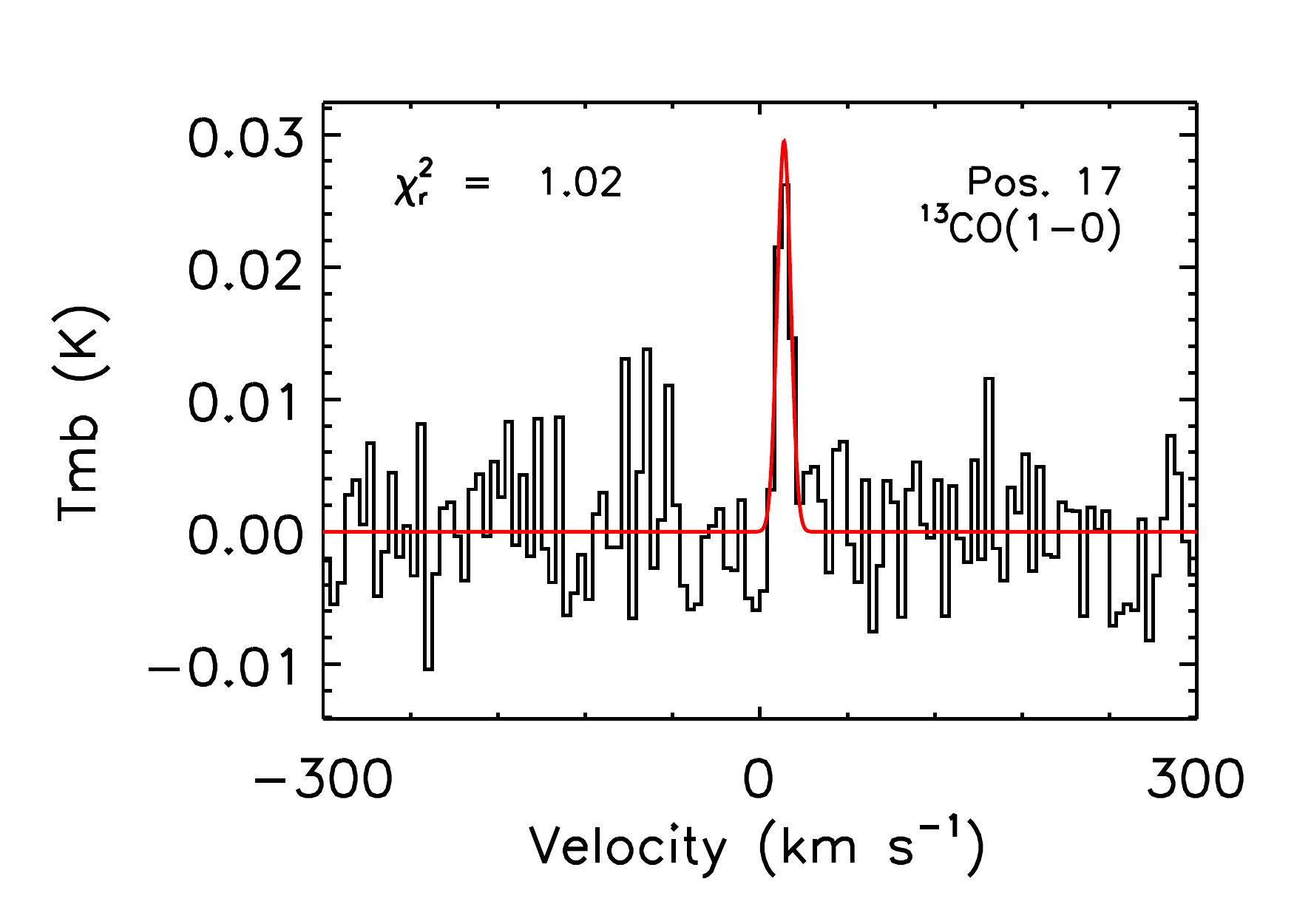}\\
  \includegraphics[width=7.0cm,clip=]{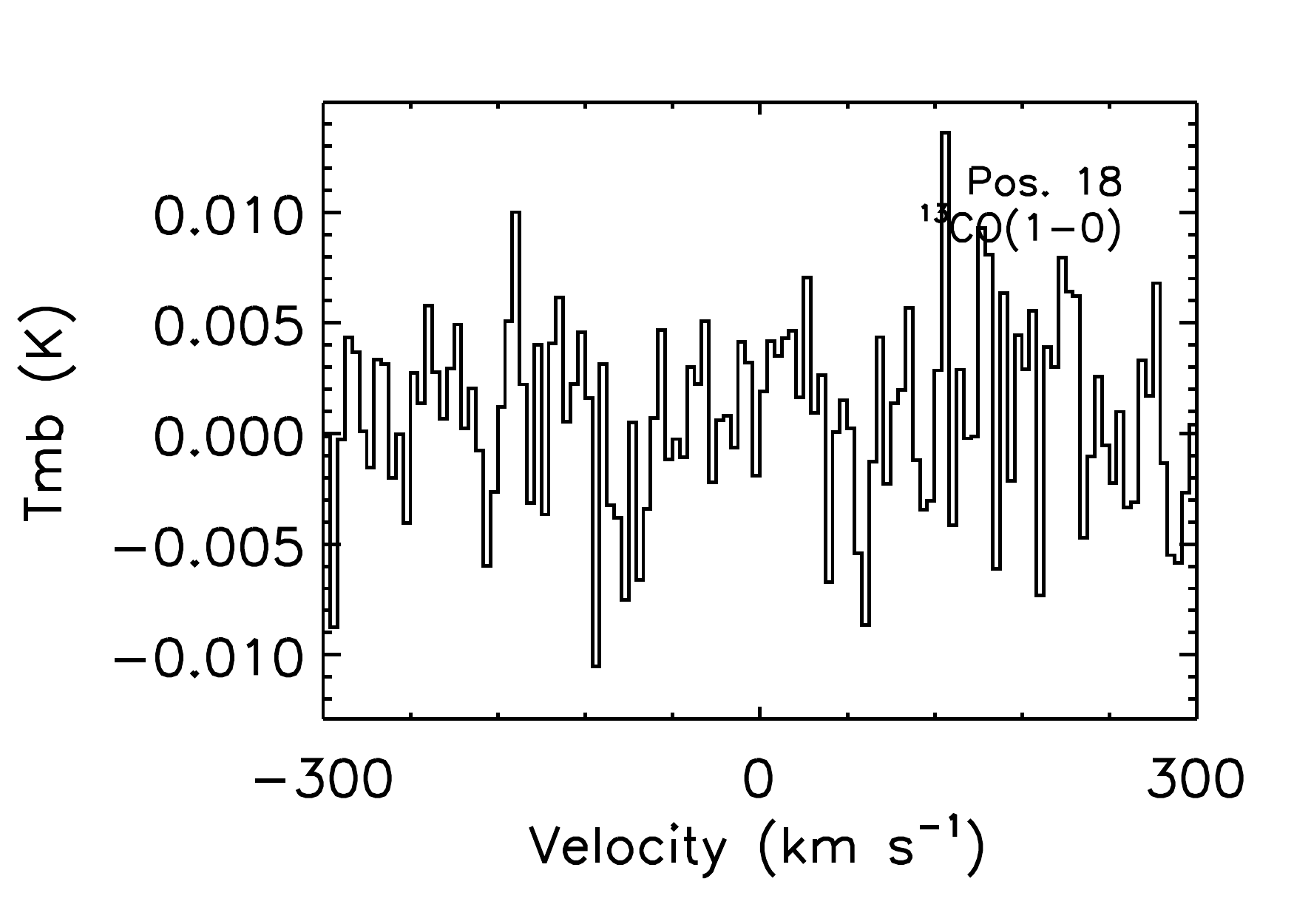}
   \includegraphics[width=7.0cm,clip=]{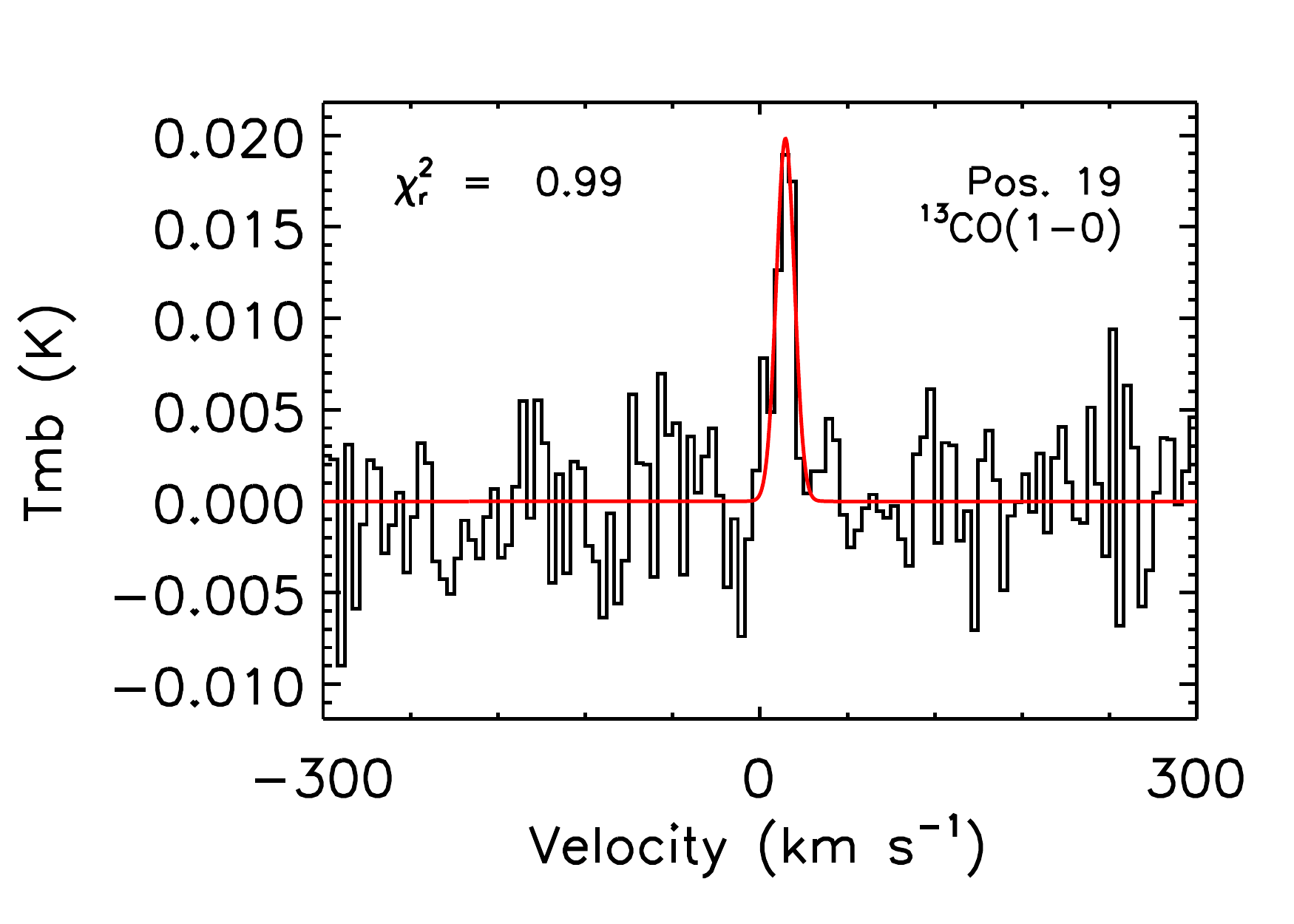}\\
  \caption{Continued.}
  \label{fig:spec4}
\end{figure*}

\clearpage

%

%
%
\begin{figure*}
  \includegraphics[width=7.0cm,clip=]{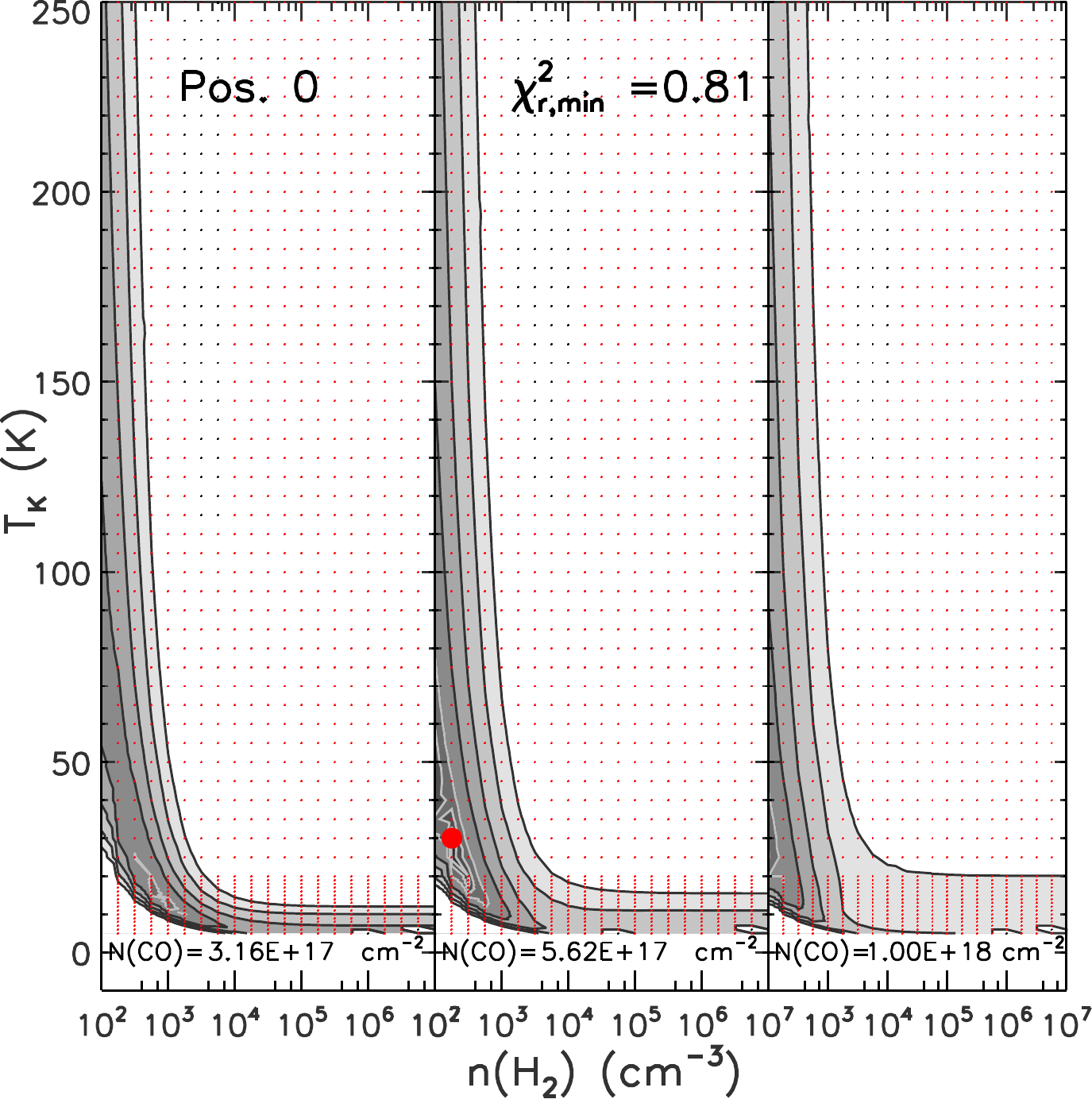}
   \includegraphics[width=7.0cm,clip=]{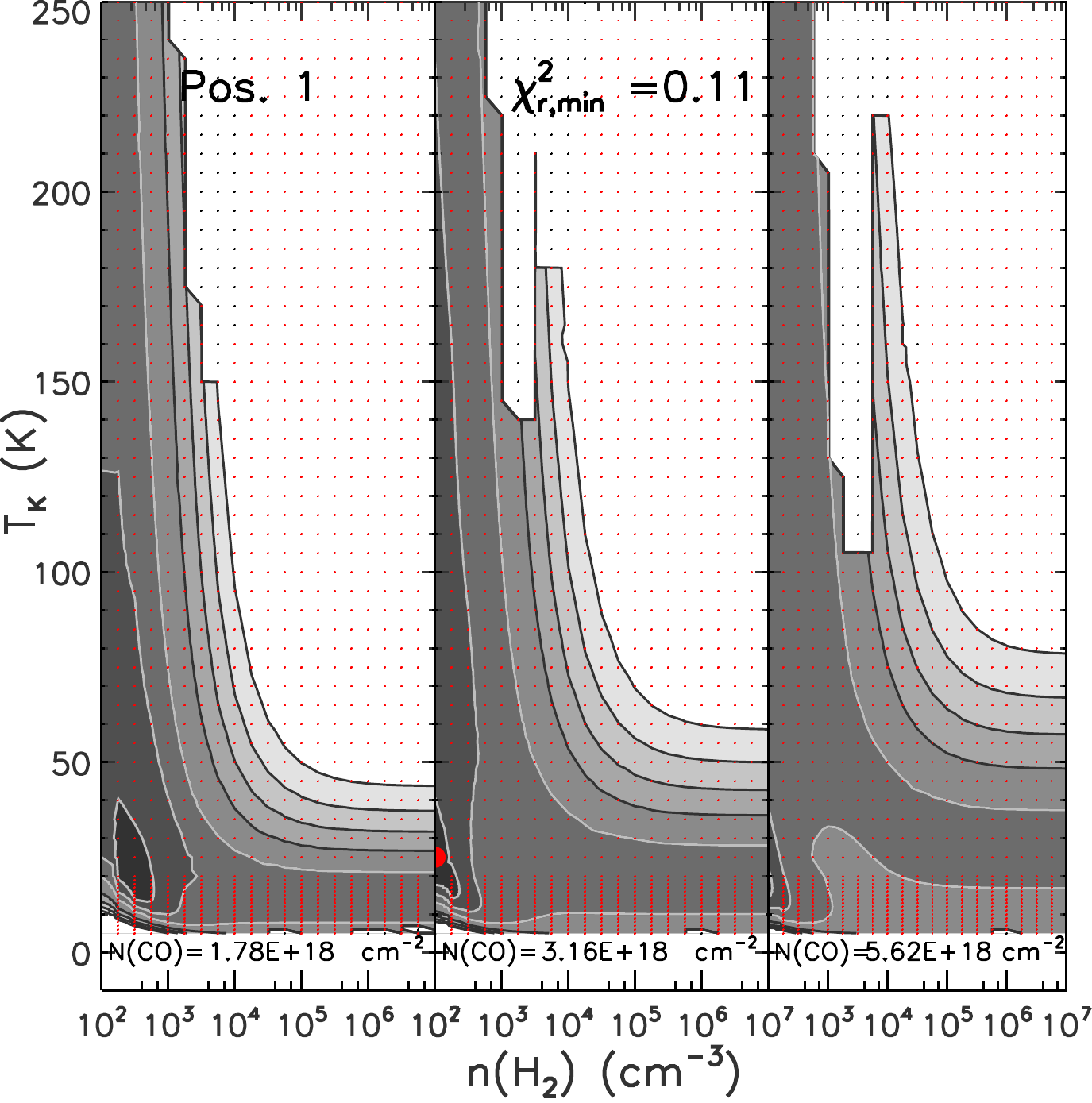}\\
  \includegraphics[width=7.0cm,clip=]{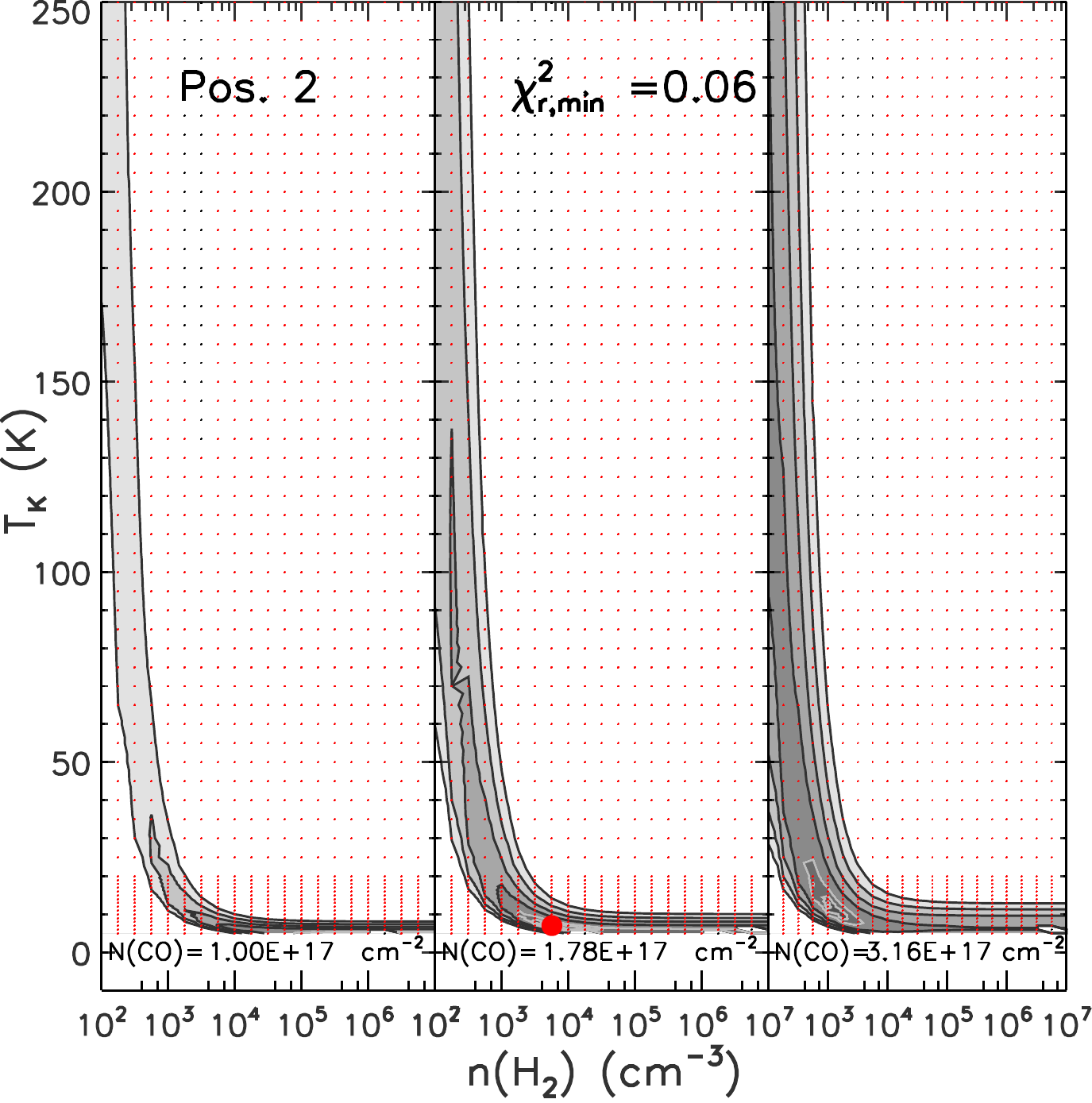}
   \includegraphics[width=7.0cm,clip=]{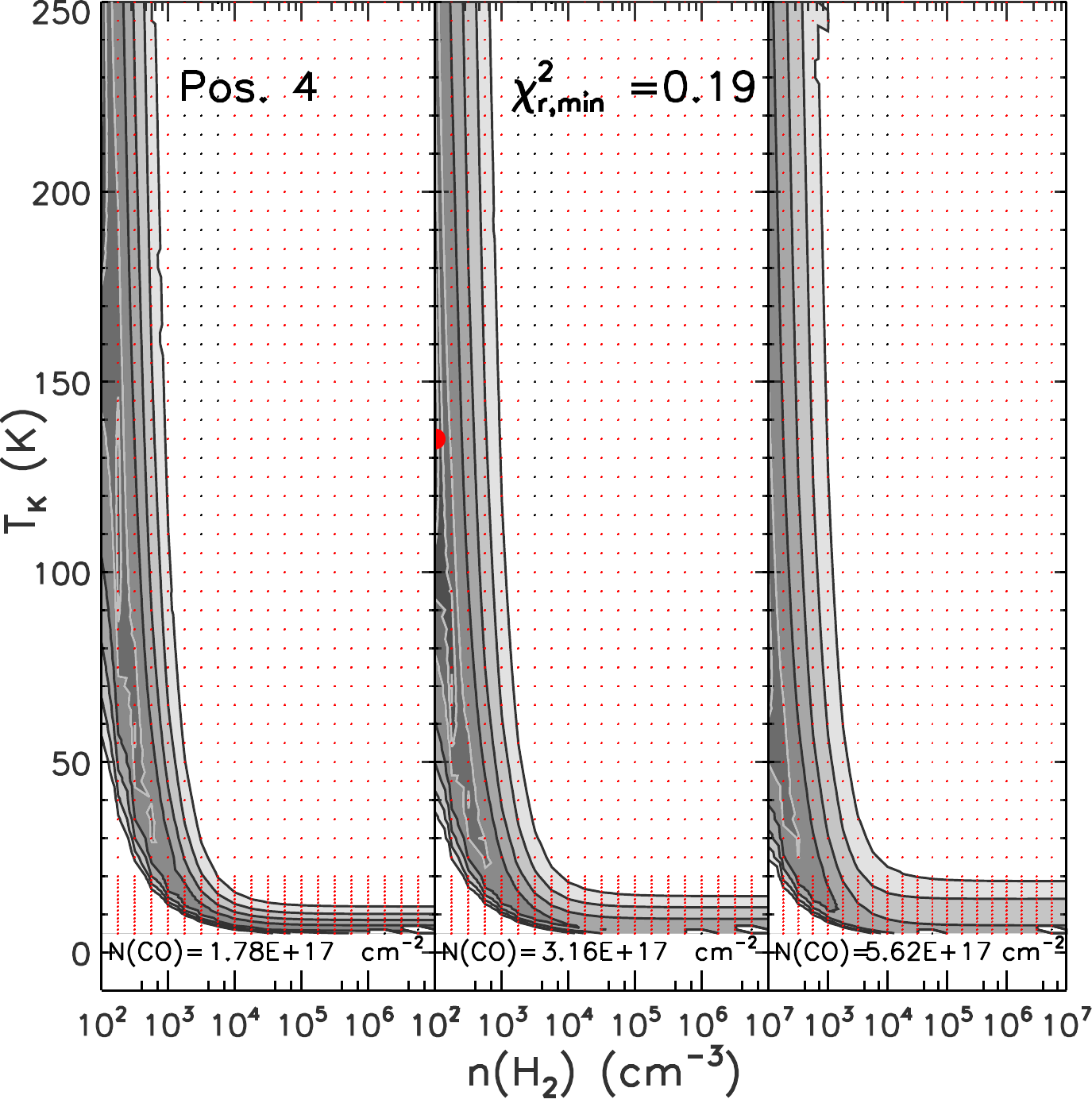}\\
    \includegraphics[width=7.0cm,clip=]{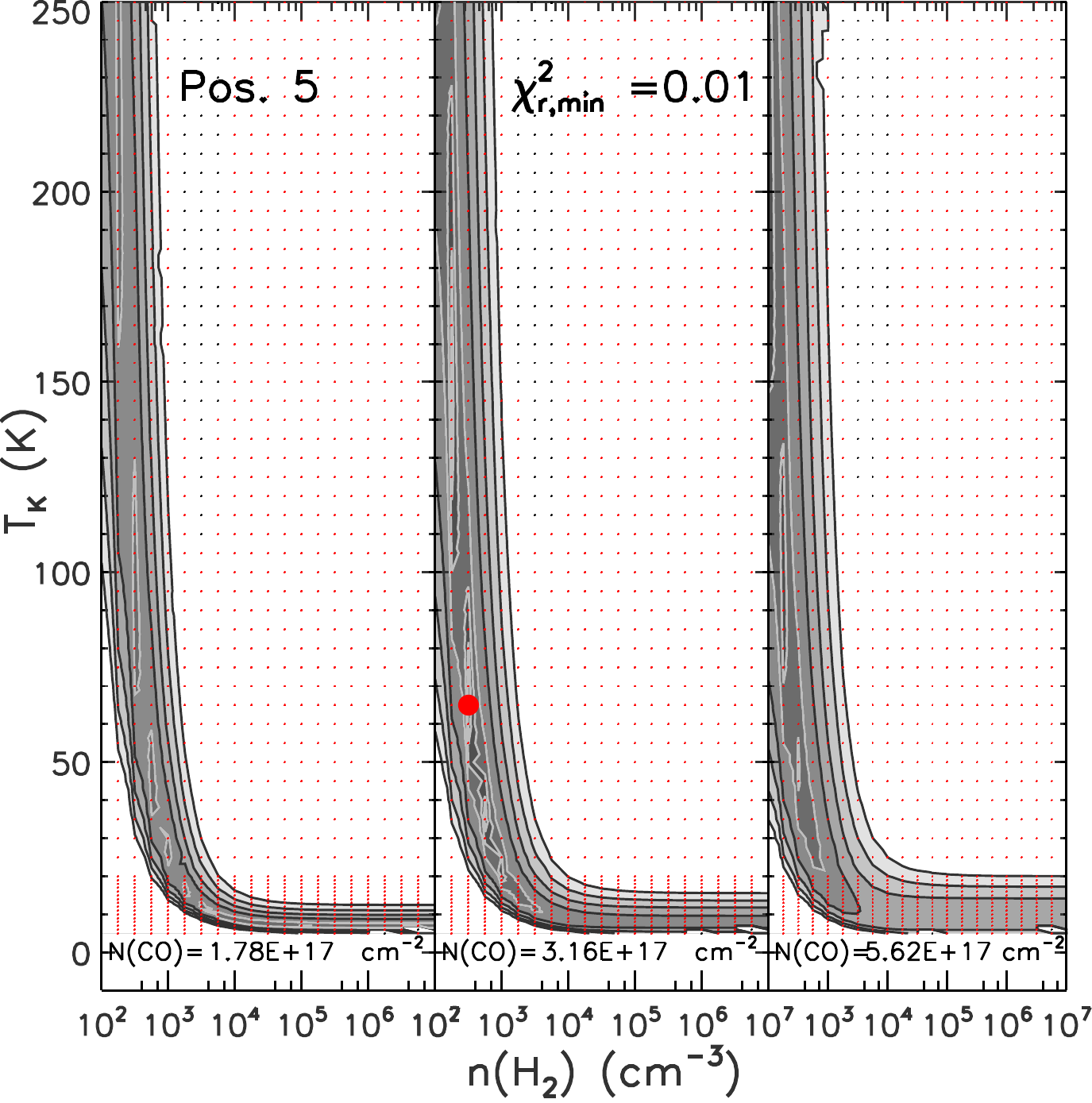}
   \includegraphics[width=7.0cm,clip=]{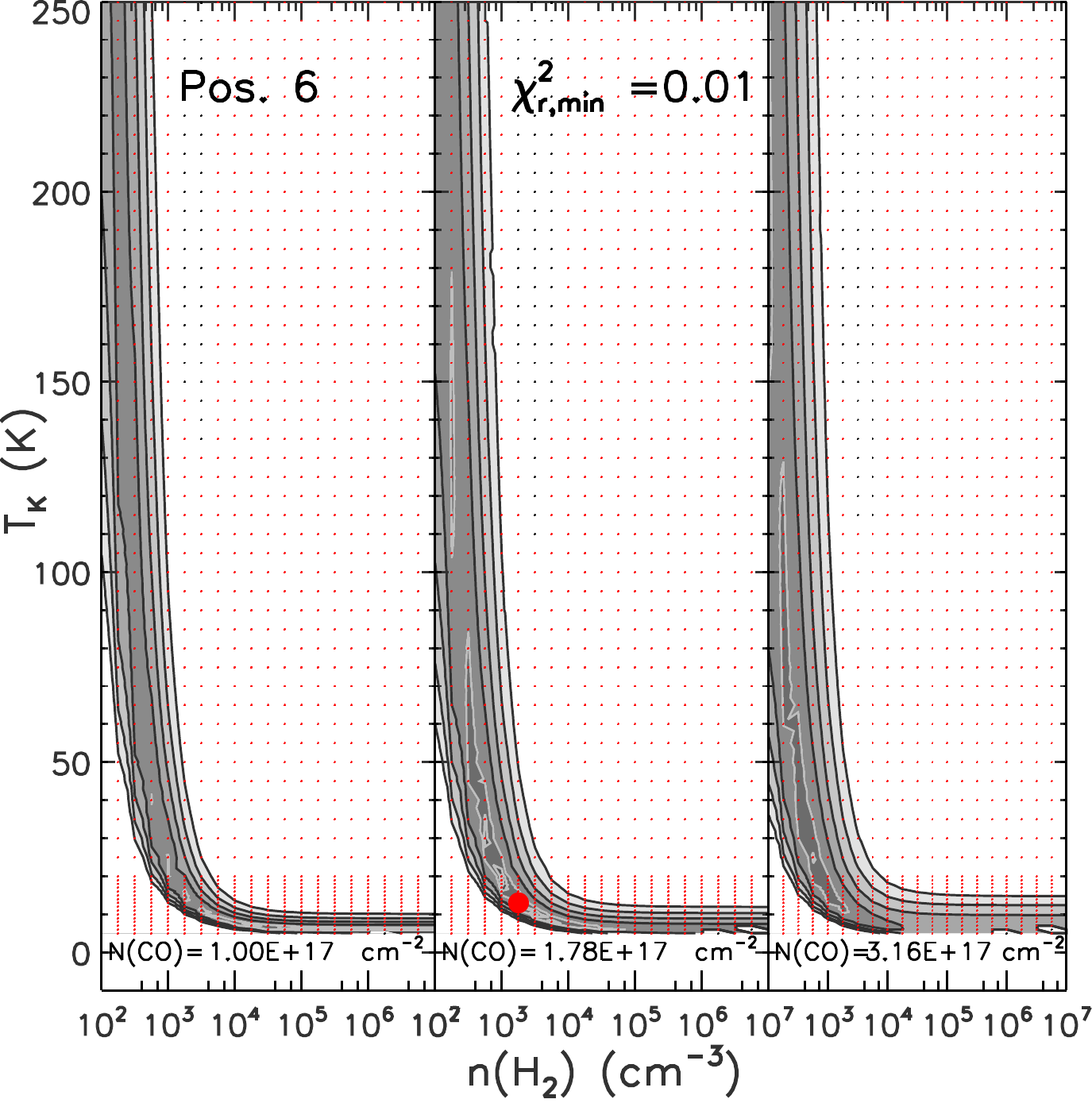}\\
  \caption{$\Delta\chi^{2}_{\rm r,min} = \chi^{2}_{\rm r} - \chi^{2}_{\rm r,min}$ maps for CO gas in NGC~0628 (positions $0$, $1$, $2$, $4$, $5$, $6$). For each position $\chi^{2}_{\rm r}$ is shown as a function of T$_{\rm K}$ and n(H$_{2}$) for three values of CO column density $N$(CO) centred around the best fit. The values of $N$(CO) are indicated at the bottom of each panel. As listed in Table~\ref{tab:modelr}, the values of $\Delta\chi^{2}_{\rm r,min}$ and position numbers are also shown at the top of each panel. The model outputs are indicated by red dots while the best-fitting model are represented by a red filled circle. Black dots represent bad models (e.g., too low opacity; see \citealt{van07}). The contour confidence levels are, $0.2\sigma$ ($16$\% probability that the best model is enclosed: the darkest greyscale), $0.5\sigma$ ($38$\%), $1\sigma$ ($68$\%), $2\sigma$, $3\sigma$, $4\sigma$ and $5\sigma$ (the lightest greyscale), for $2$ degrees of freedom (three line ratios). The values for the level from $0.2\sigma$ to $5\sigma$ are $0.35$, $0.97$, $2.30$, $6.18$, $11.83$, $19.33$ and $28.74$ respectively. The confidence levels from $0.2\sigma$ to $1\sigma$ are separated by grey lines and those from $2\sigma$ to $5\sigma$ by black lines.}
  \label{fig:chi}
\end{figure*}

%
\addtocounter{figure}{-1}
\begin{figure*}
  \includegraphics[width=7.0cm,clip=]{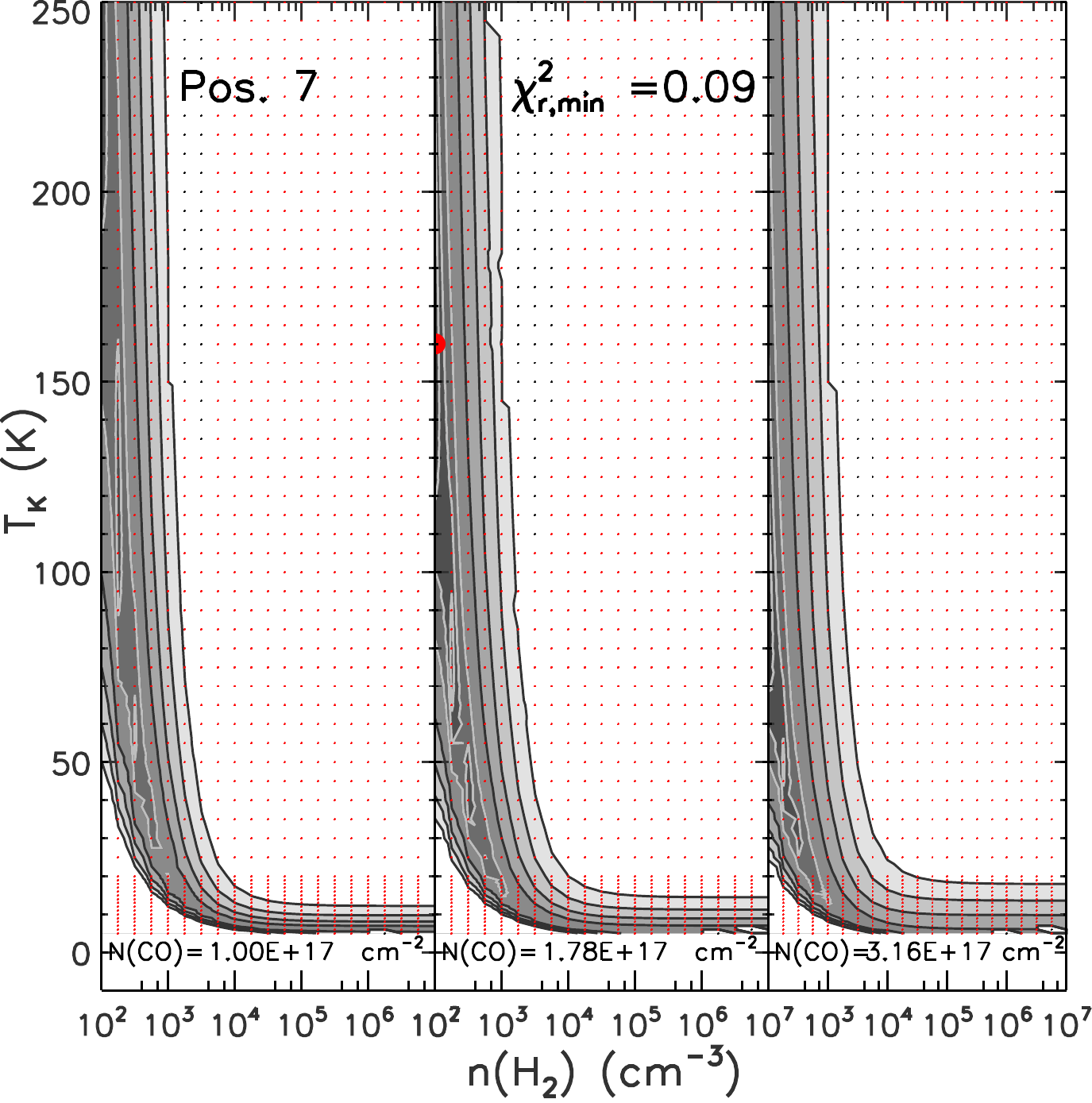}
   \includegraphics[width=7.0cm,clip=]{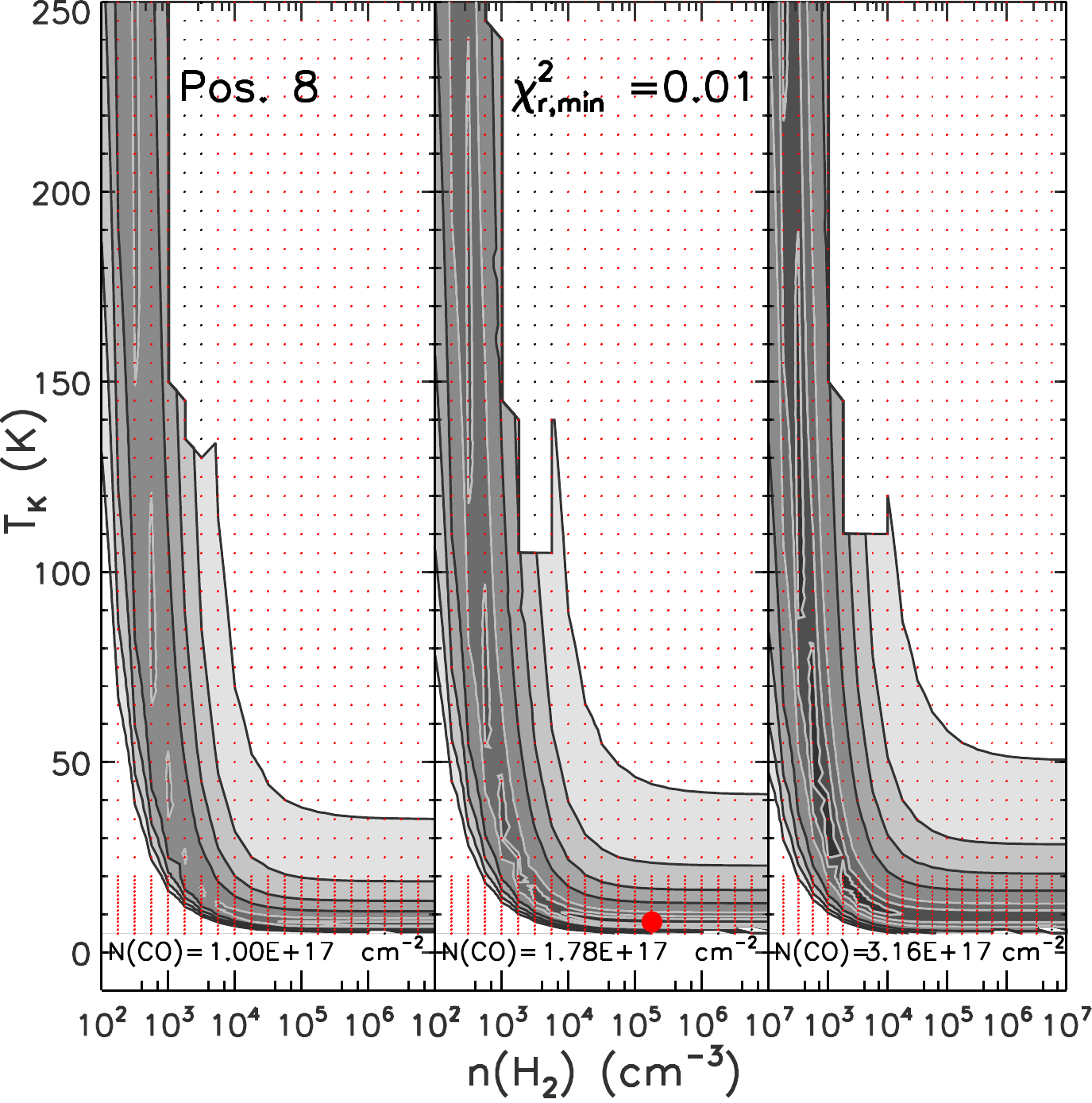}\\
  \includegraphics[width=7.0cm,clip=]{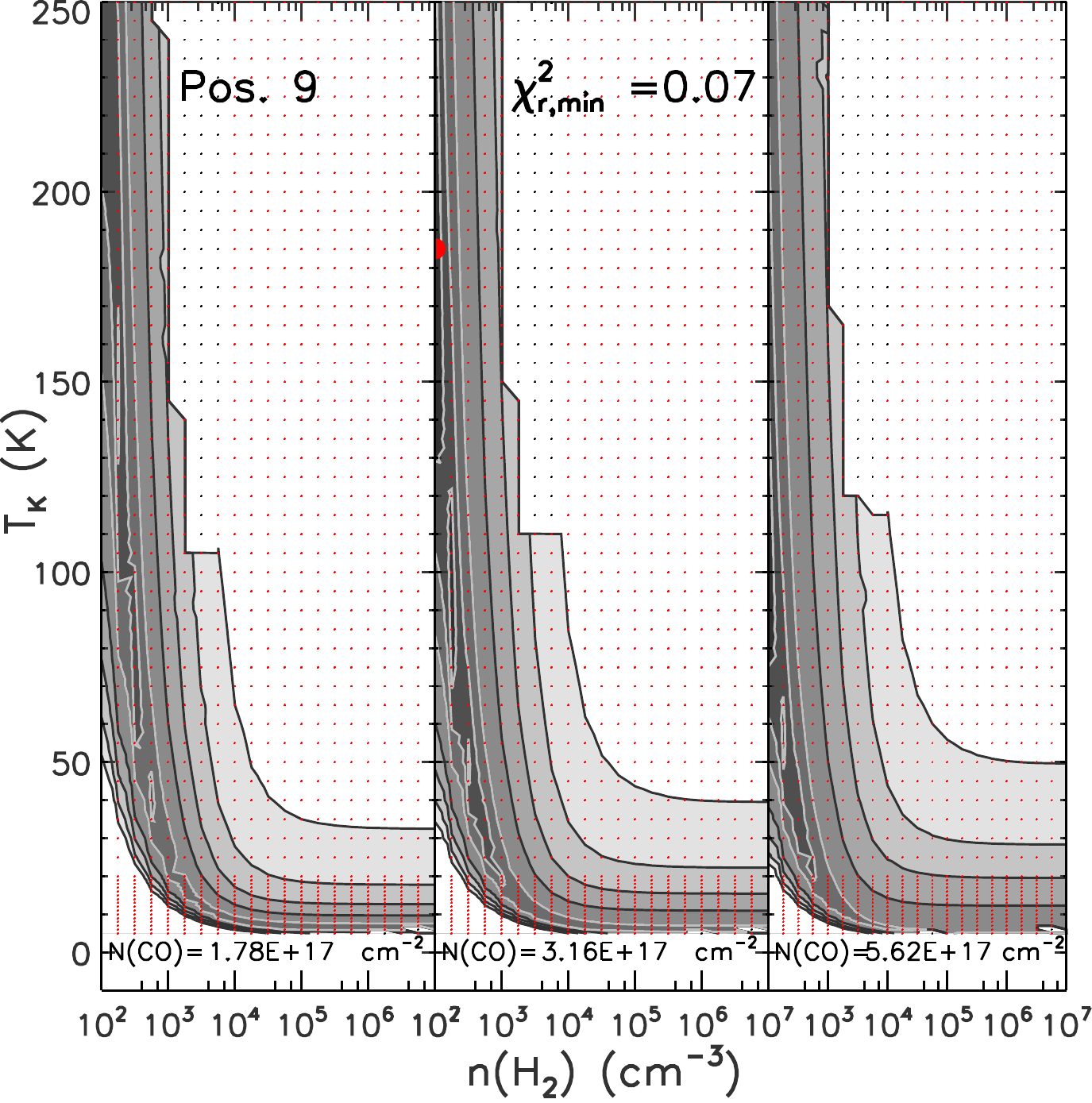}
   \includegraphics[width=7.0cm,clip=]{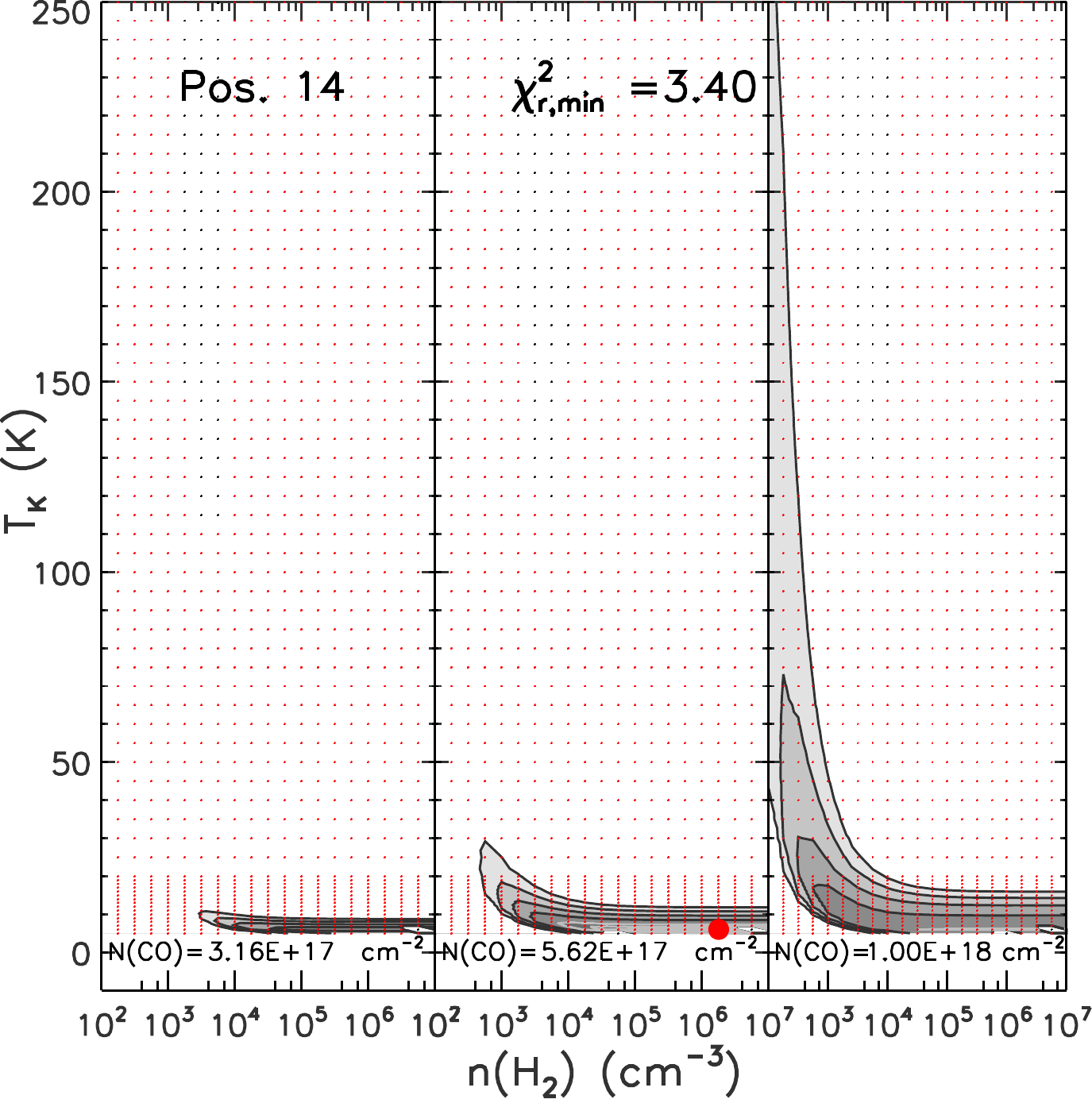}\\
    \includegraphics[width=7.0cm,clip=]{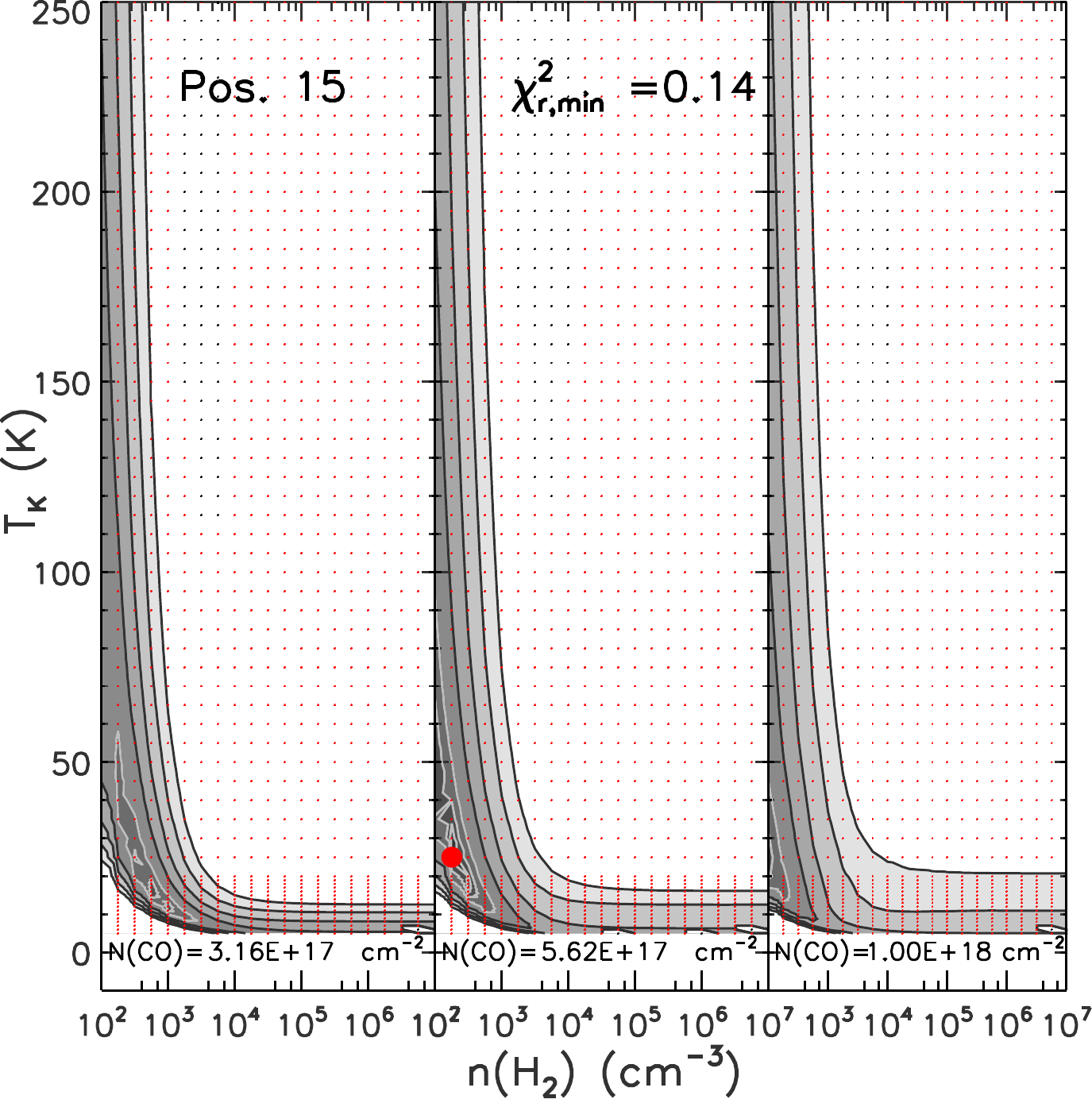}
   \includegraphics[width=7.0cm,clip=]{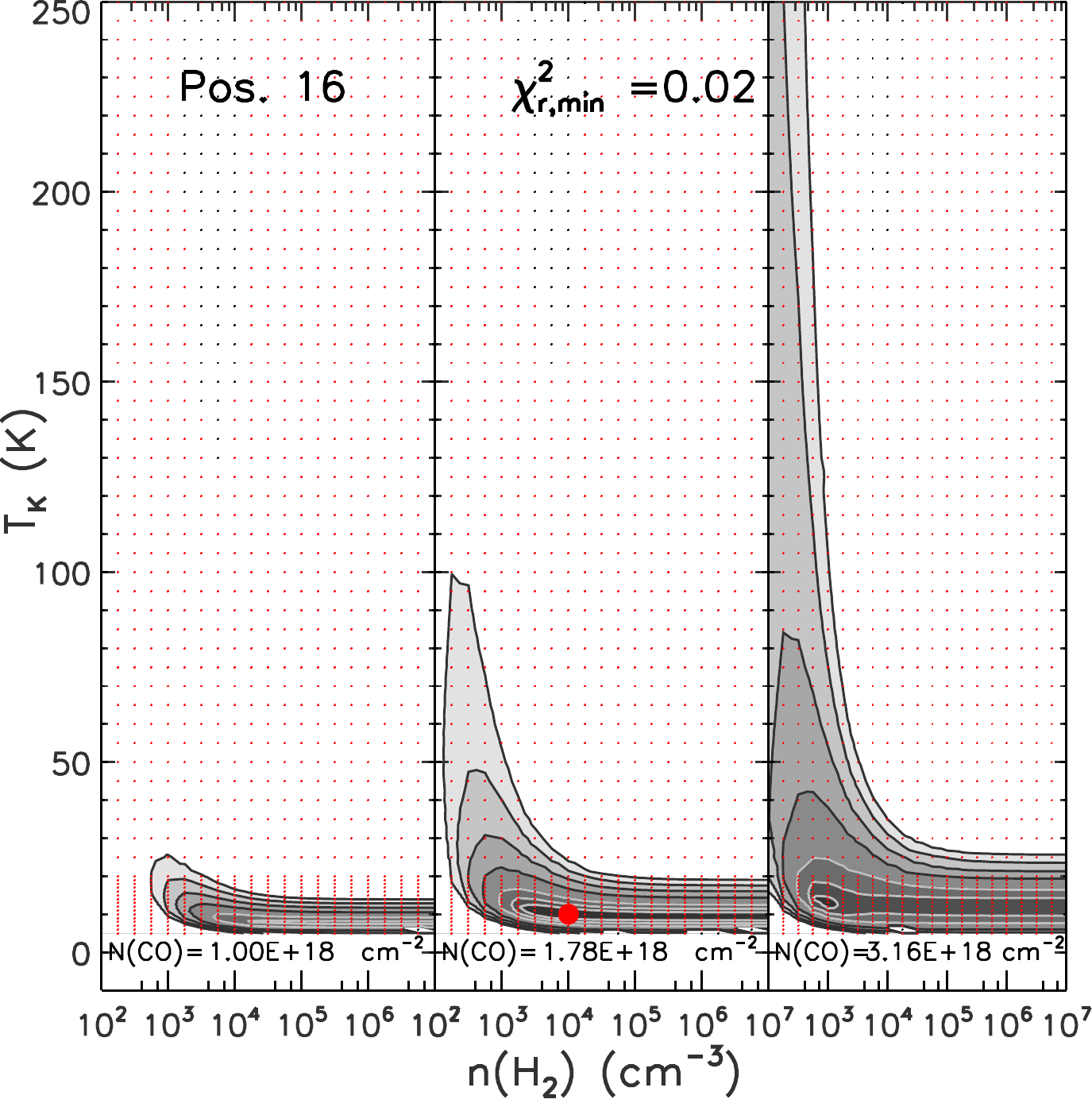}\\
  \caption{{\bf Continued.} Same plots as Fig.\ref{fig:chi} but for positions $7$, $8$, $9$, $14$, $15$, and $16$.}
  \label{fig:chi}
\end{figure*}

%
\addtocounter{figure}{-1}
\begin{figure*}
   \includegraphics[width=7.0cm,clip=]{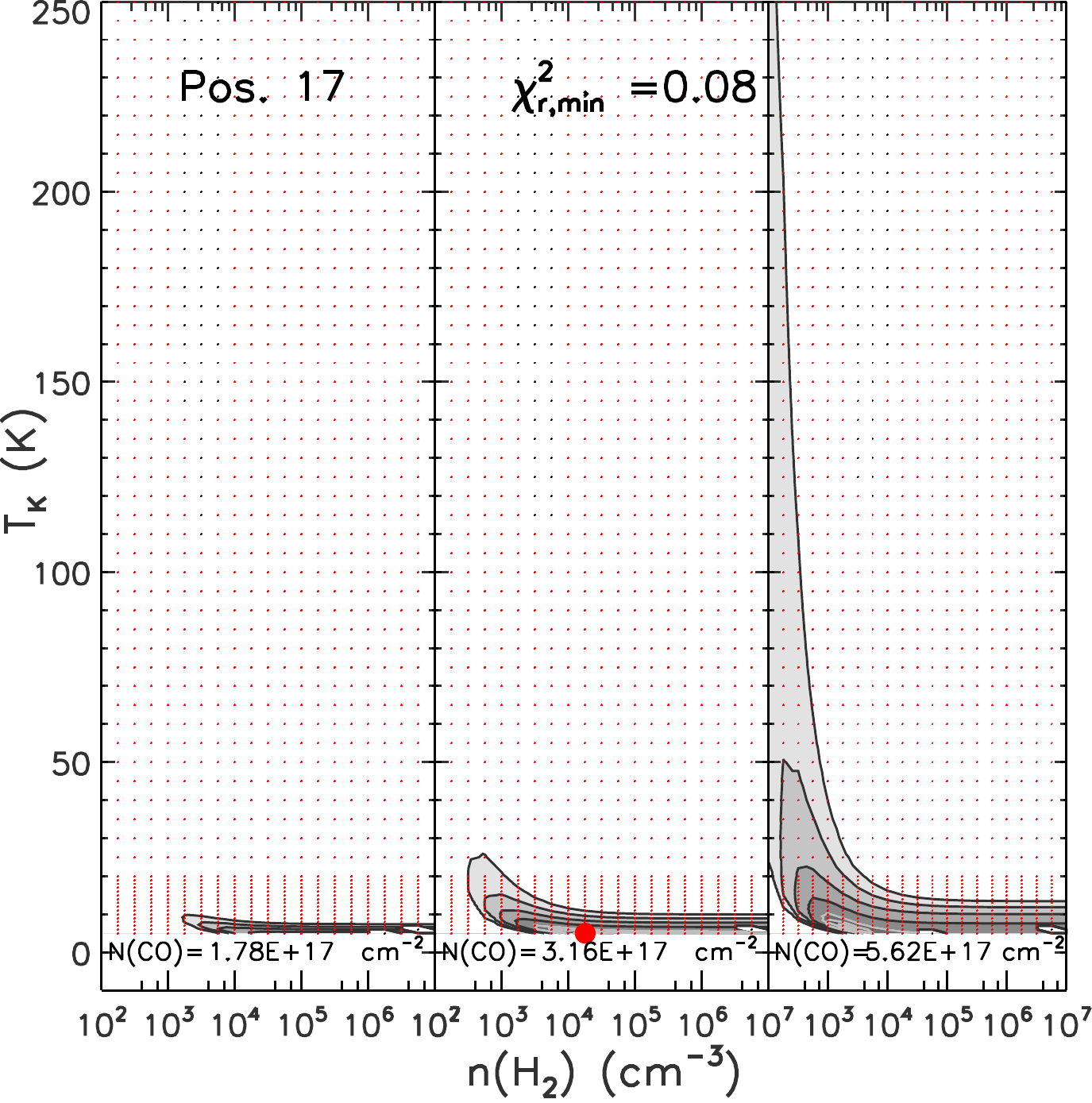}\\
  \caption{{\bf Continued.} Same plots as Fig.\ref{fig:chi} but for position $17$.}
  \label{fig:chi}
\end{figure*}

\clearpage

%
%
\begin{figure*}
  \includegraphics[width=7.0cm,clip=]{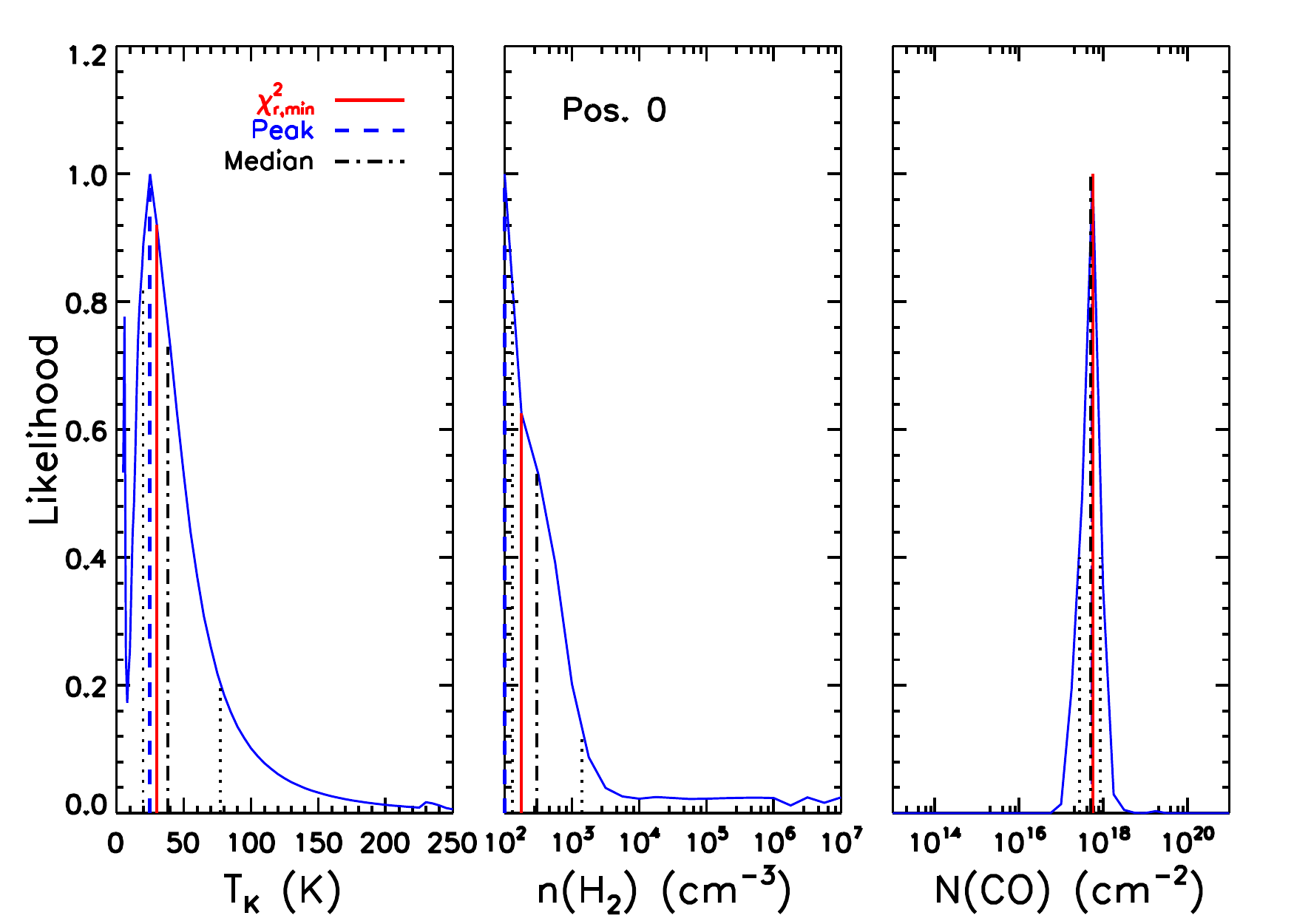}
   \includegraphics[width=7.0cm,clip=]{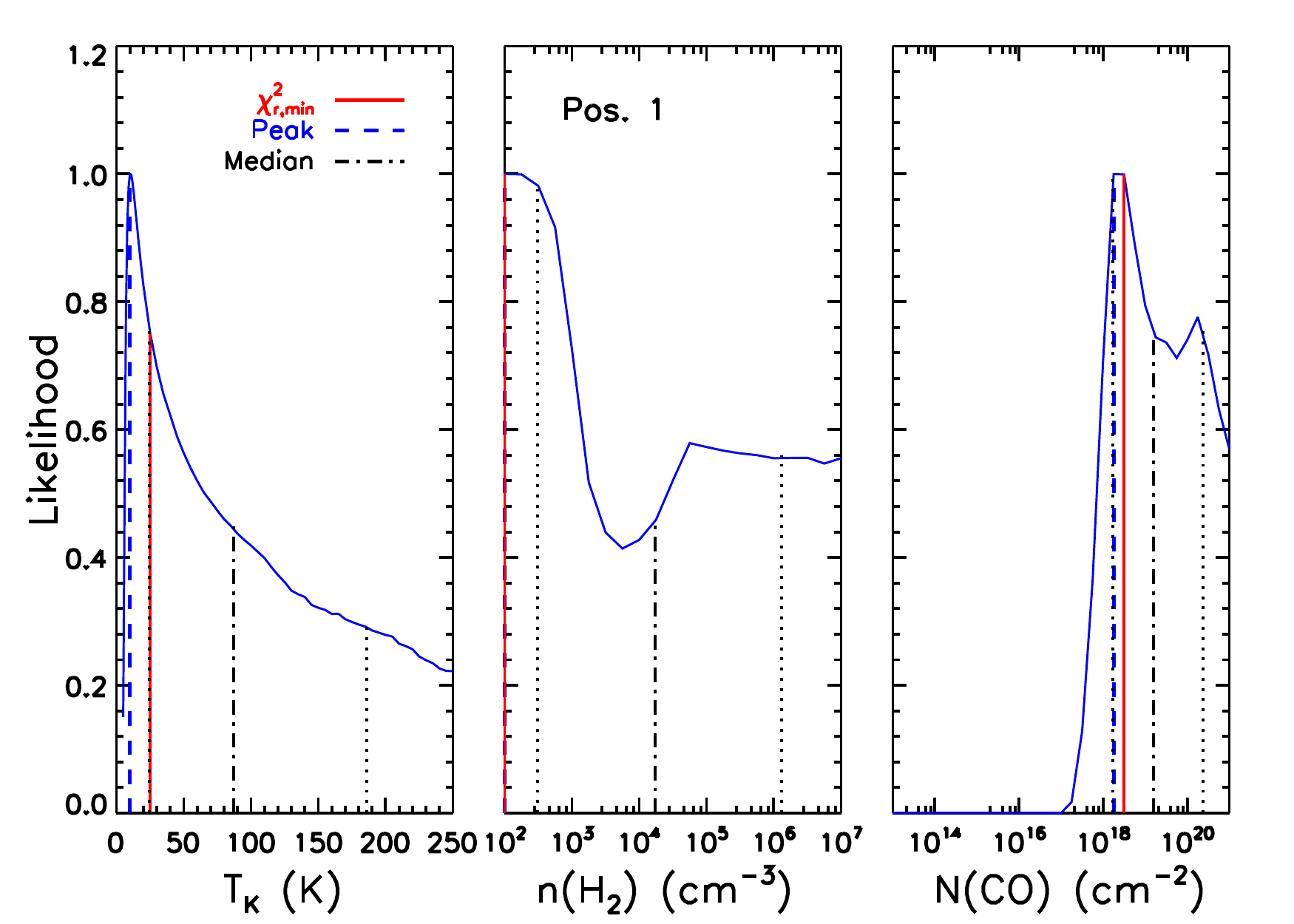}\\
  \includegraphics[width=7.0cm,clip=]{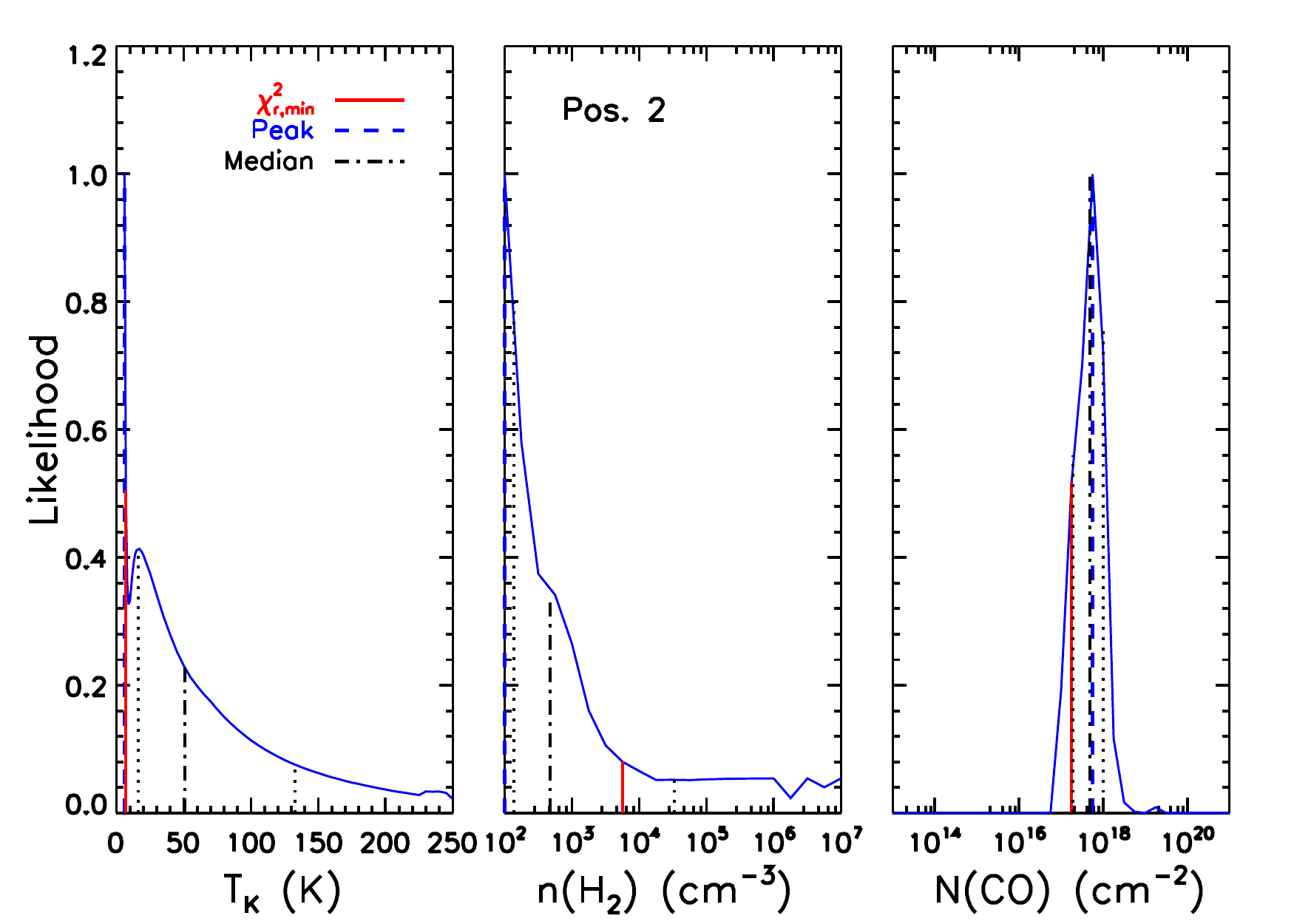}
   \includegraphics[width=7.0cm,clip=]{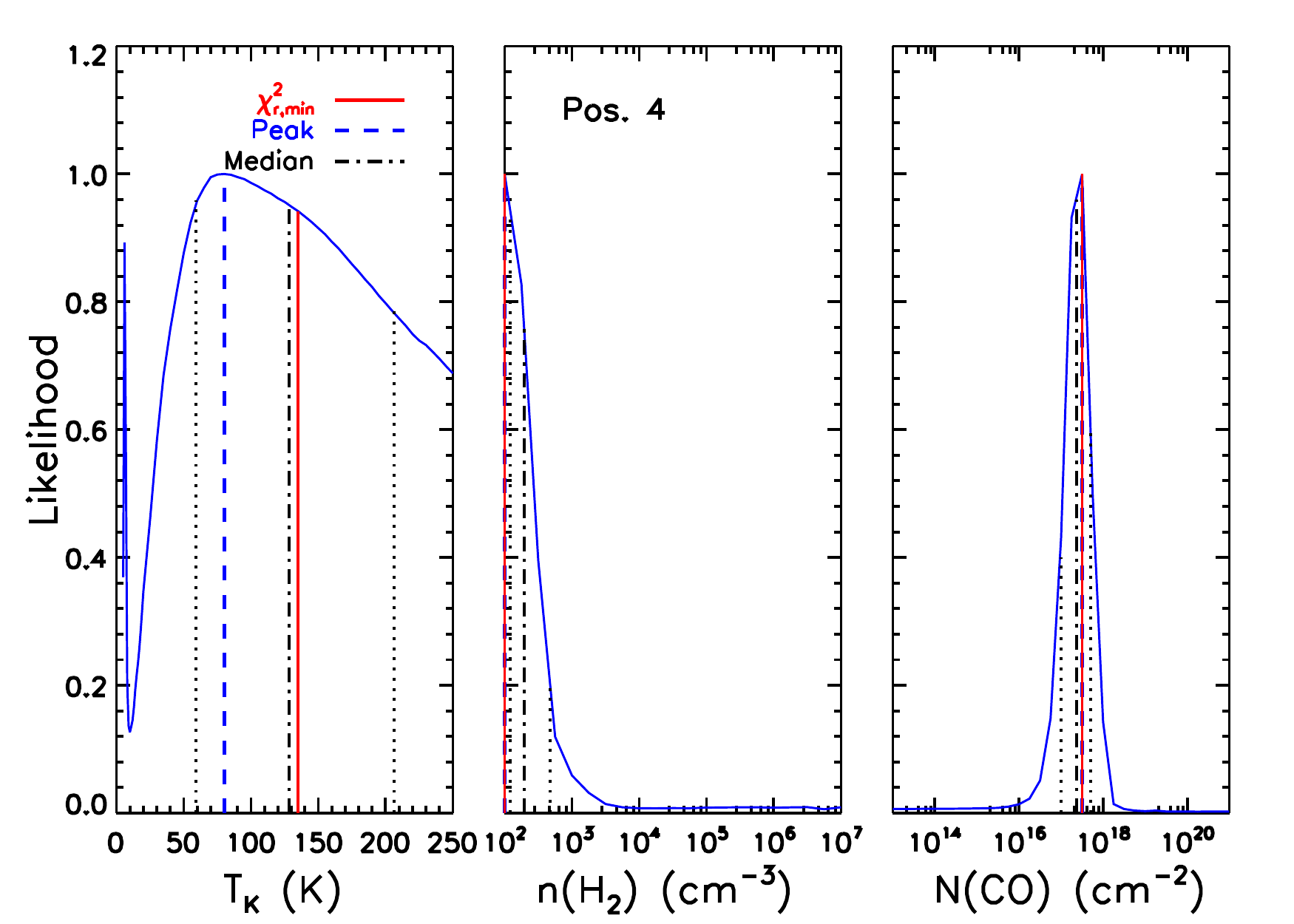}\\
    \includegraphics[width=7.0cm,clip=]{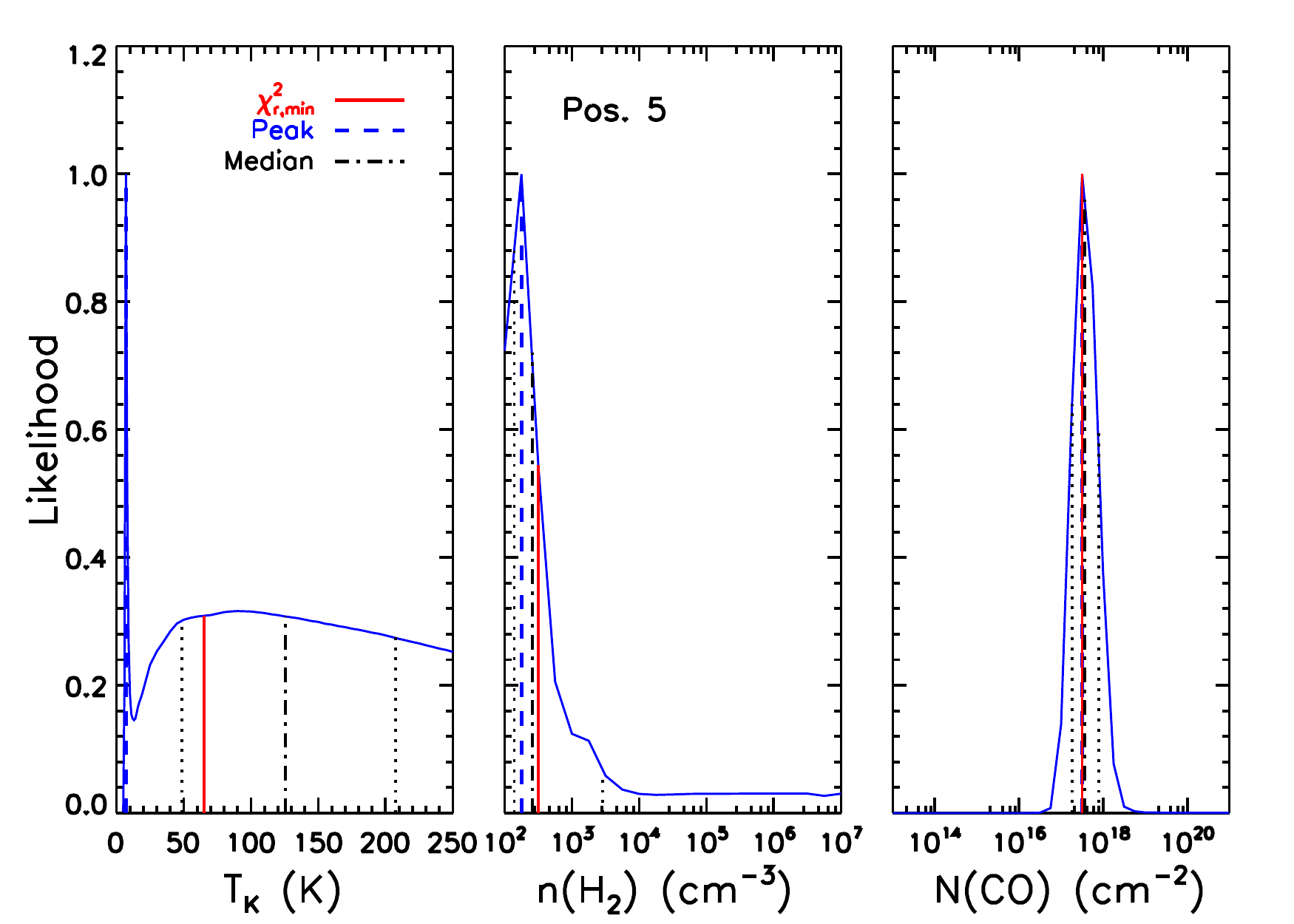}
   \includegraphics[width=7.0cm,clip=]{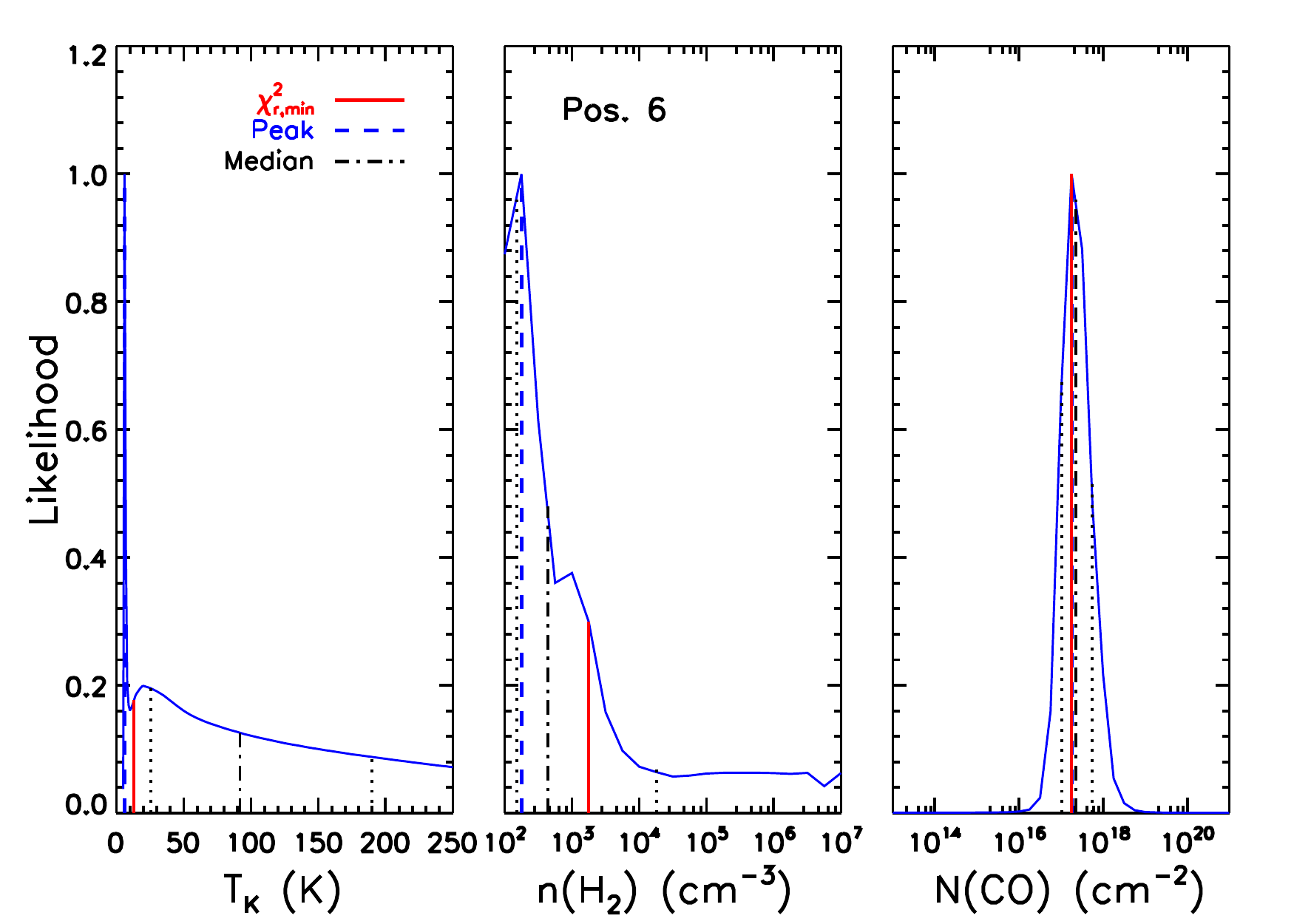}\\
   \includegraphics[width=7.0cm,clip=]{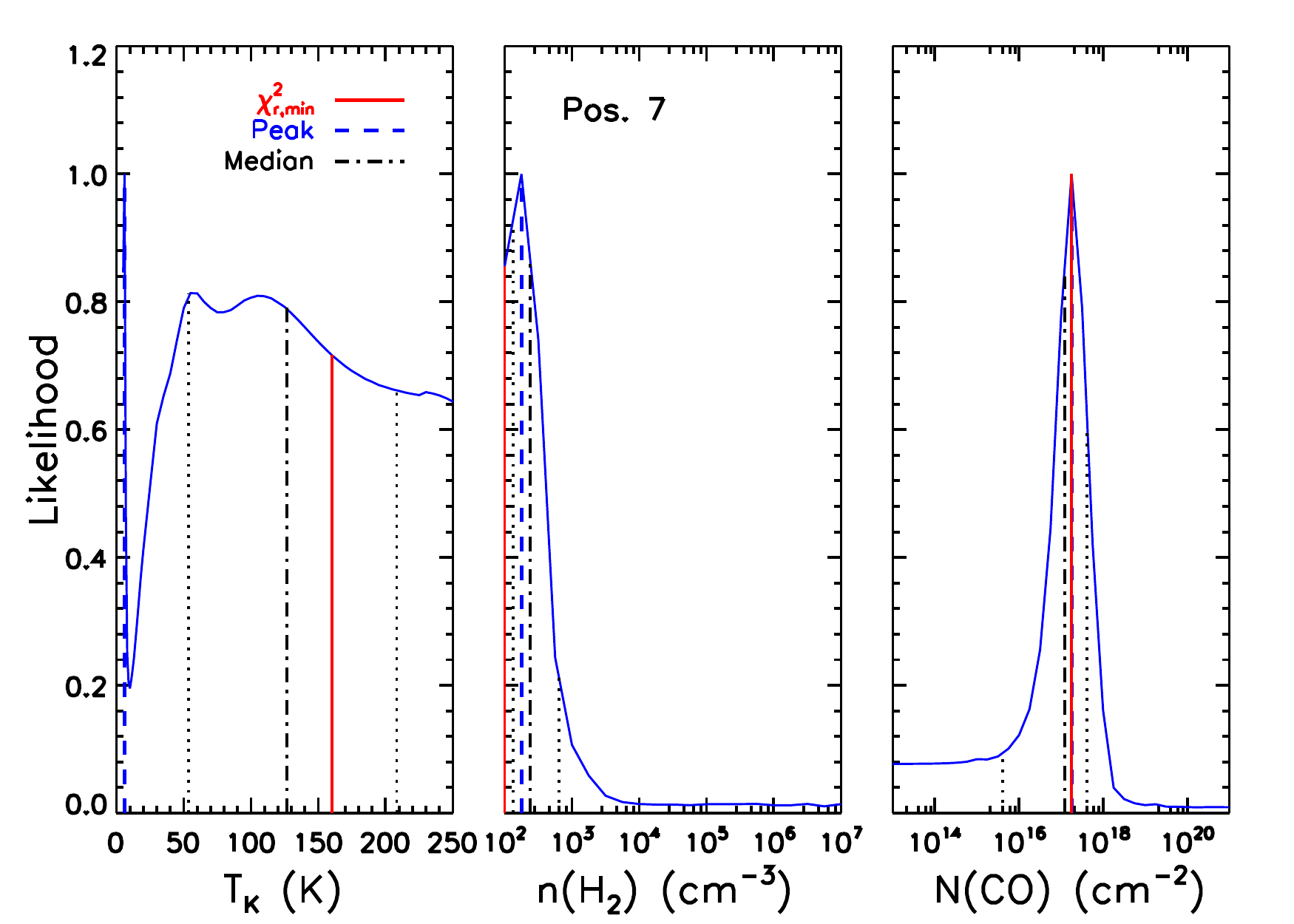}
   \includegraphics[width=7.0cm,clip=]{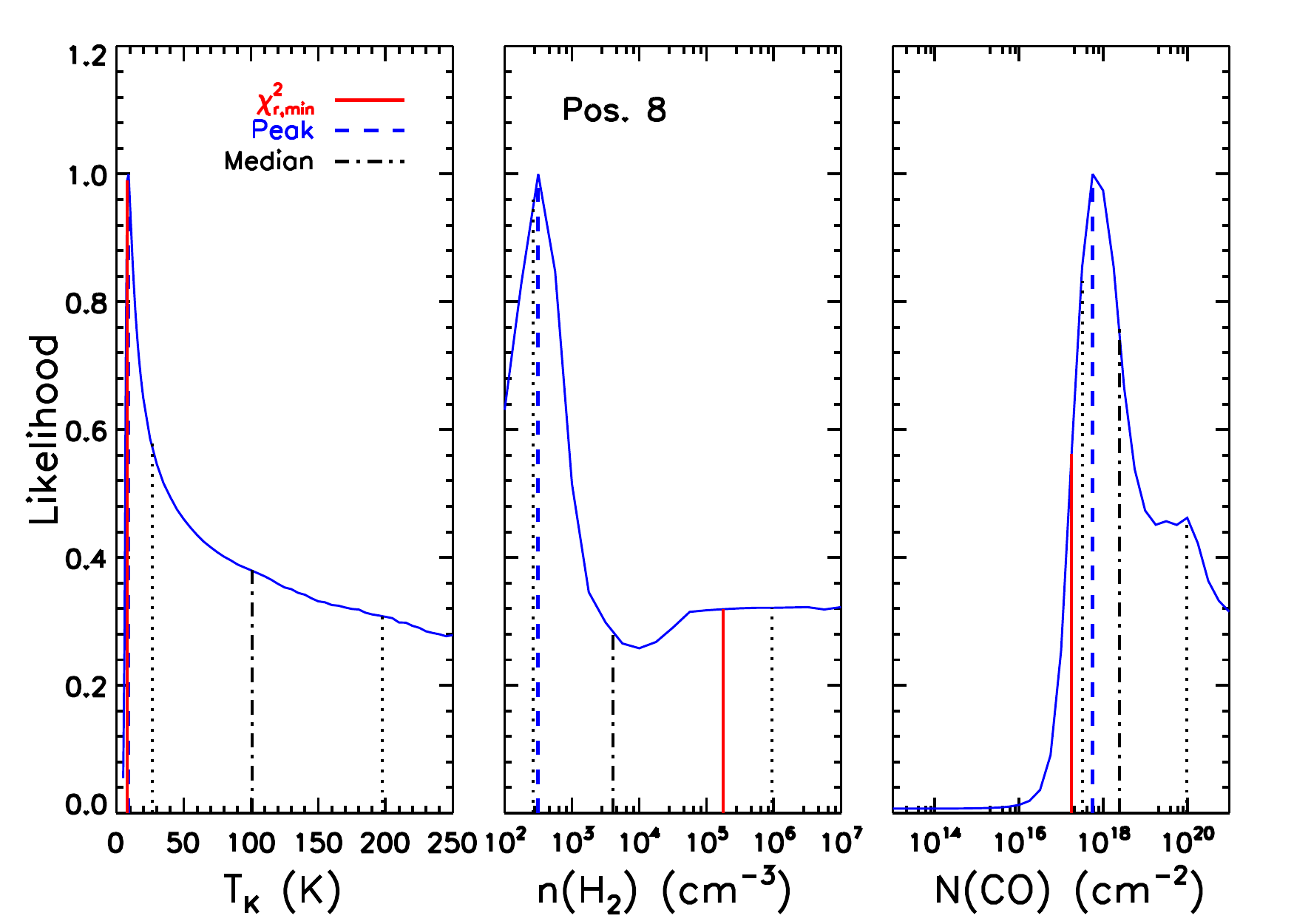}\\
  \caption{PDF of each model parameter marginalized over the other two, for CO gas in NGC~0628 
  (positions $0$, $1$, $2$, $4$, $5$, $6$, $7$ and $8$). In each PDF, the peak (most likely) and median value within 
  the model grid are identified with a dashed blue and dash-dotted black line, respectively.
The $68$ per cent ($1\sigma$) confidence level around the median is indicated by dotted black lines. 
  The best-fitting model in a $\chi^{2}$ sense is indicated by a solid red line.}
  \label{fig:like}
\end{figure*}
%
%
%
\addtocounter{figure}{-1}
\begin{figure*}
  \includegraphics[width=7.0cm,clip=]{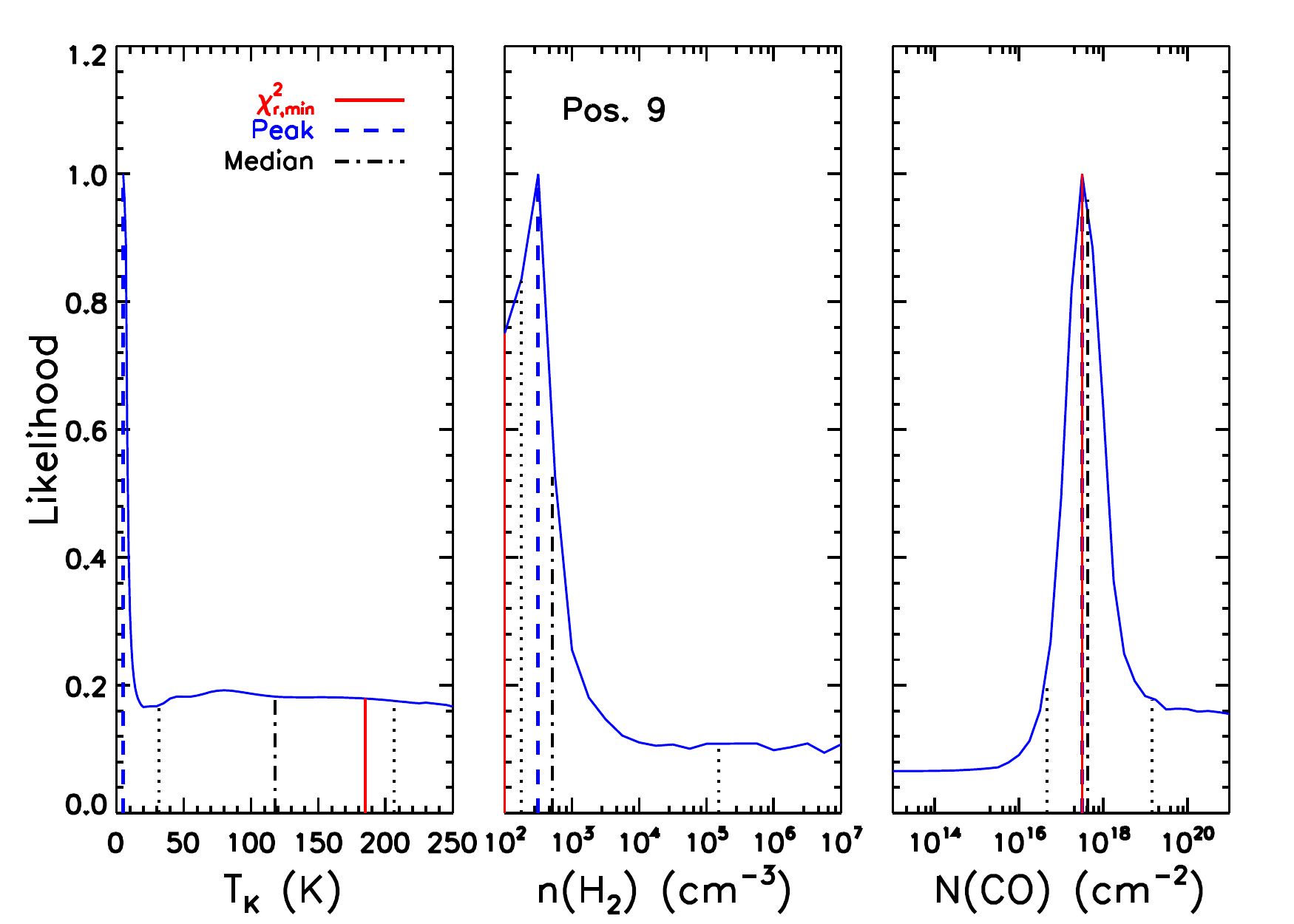}
   \includegraphics[width=7.0cm,clip=]{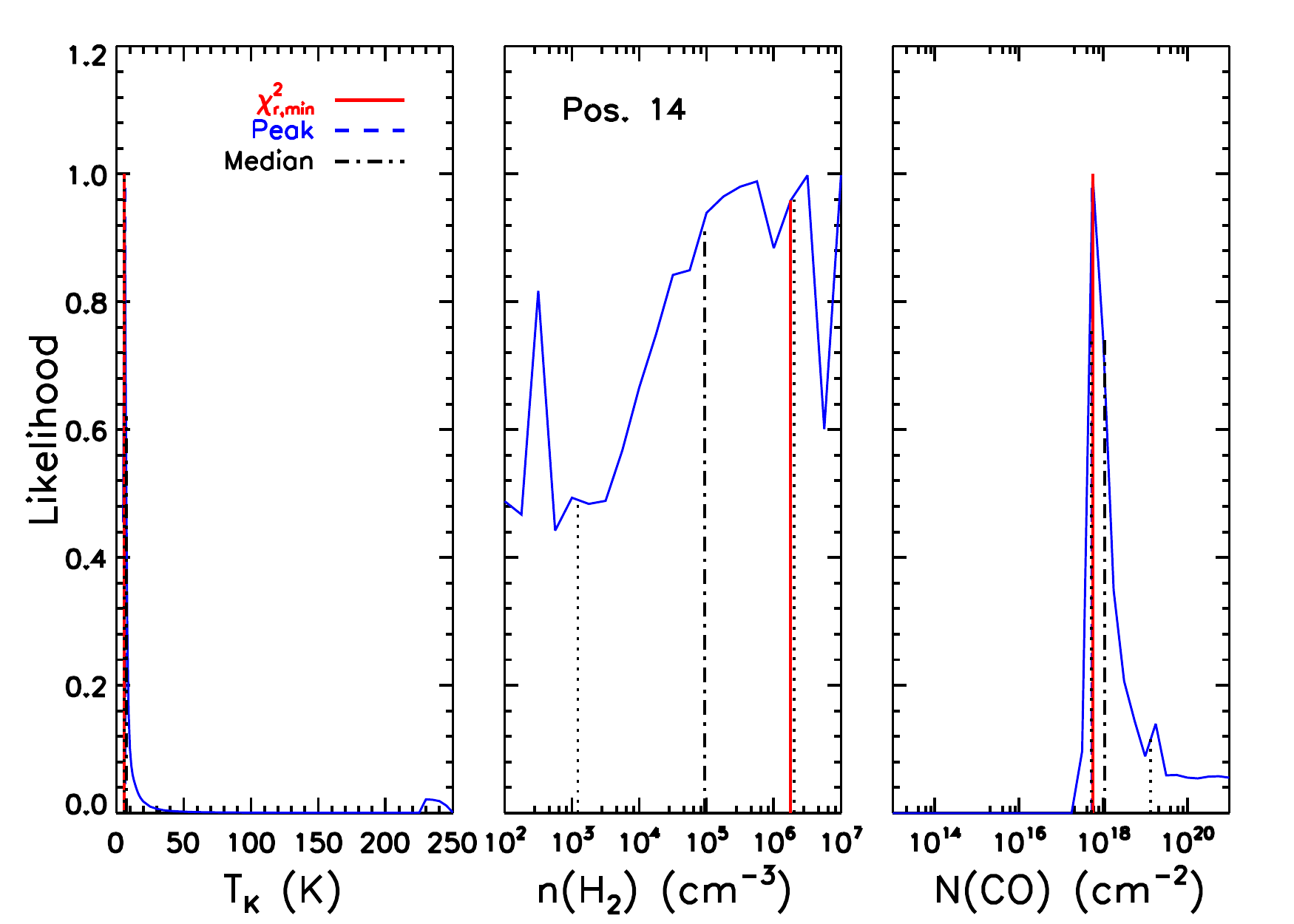}\\
  \includegraphics[width=7.0cm,clip=]{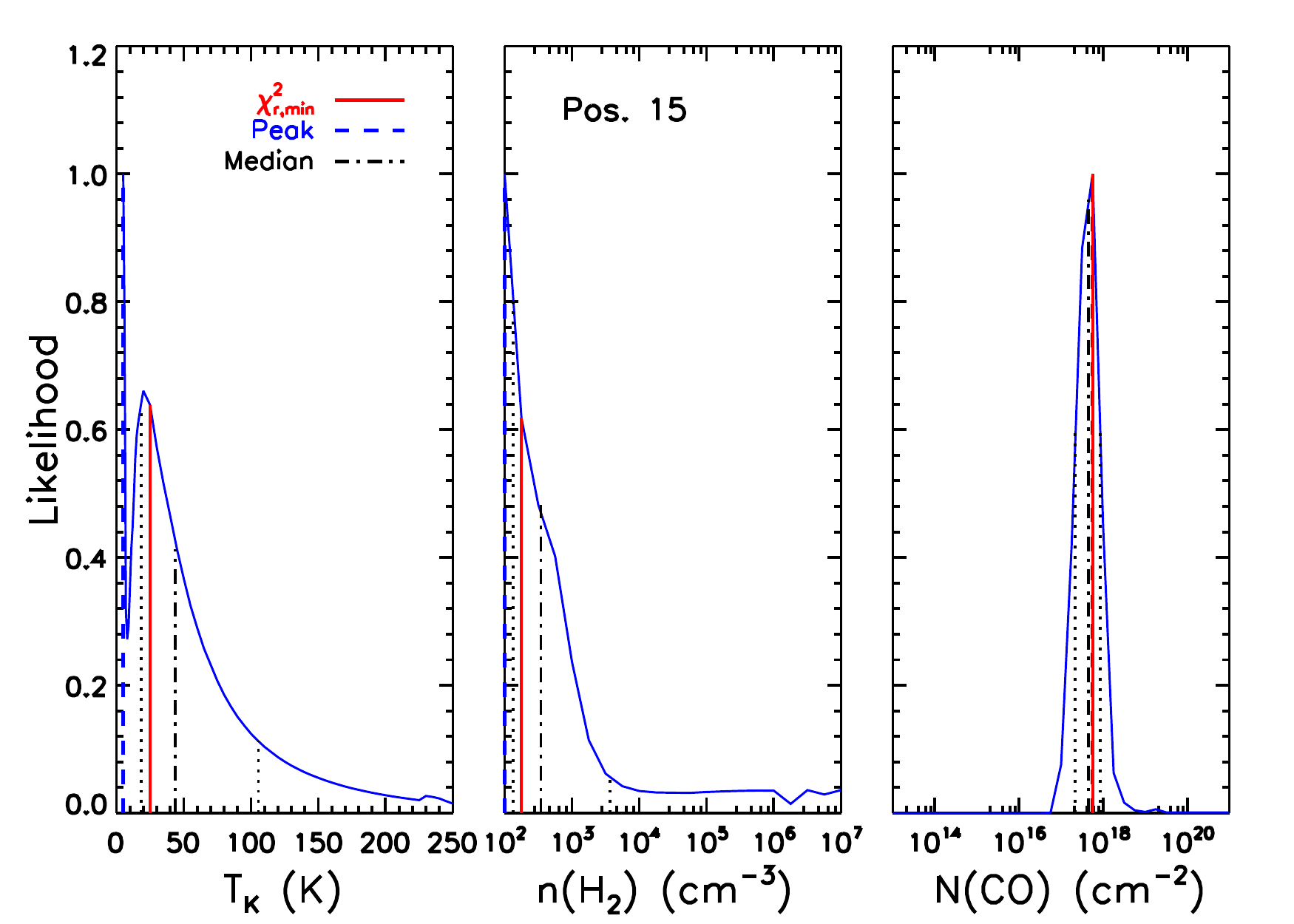}
   \includegraphics[width=7.0cm,clip=]{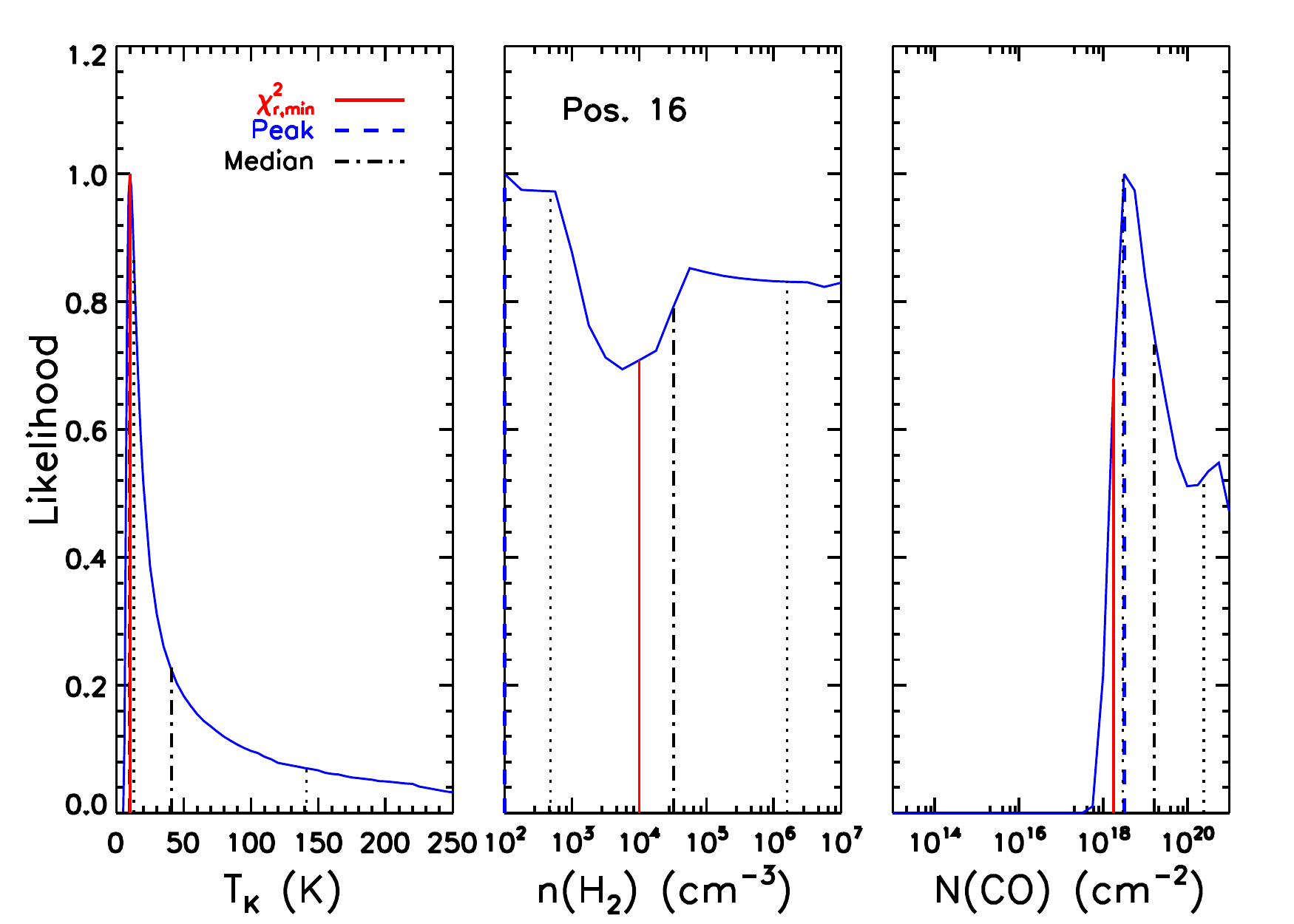}\\
    \includegraphics[width=7.0cm,clip=]{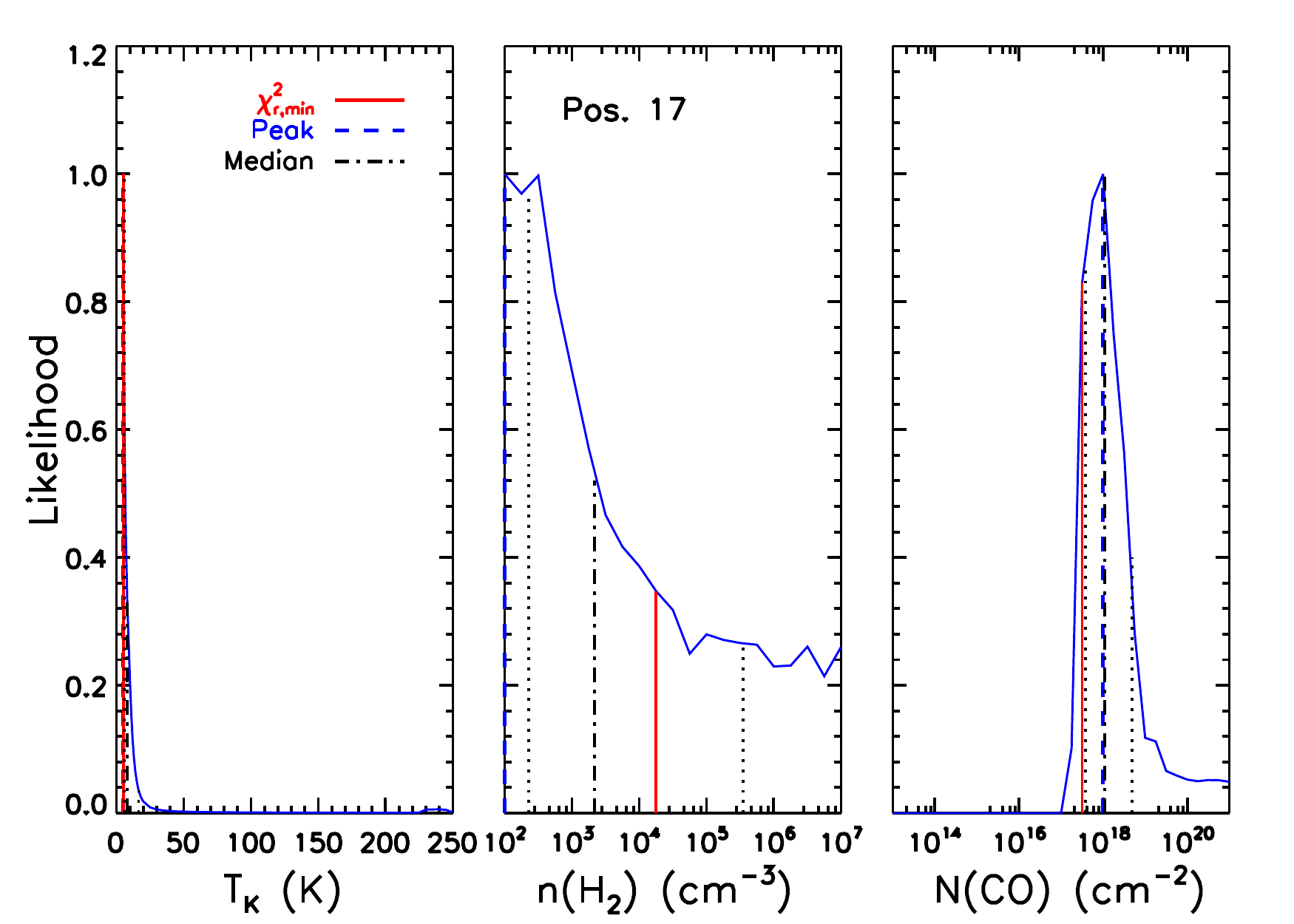} \\
  \caption{{\bf Continued}. Same plots as Fig.\ref{fig:like} but for positions $9$, $14$, $15$, $16$, and $17$.}
  \label{fig:like}
\end{figure*}

\end{document}